\documentclass{emulateapj}
\usepackage{apjfonts}
\usepackage[colorlinks,linkcolor=red,anchorcolor=black,citecolor=blue]{hyperref}
\usepackage{threeparttable}
\pdfoutput=1

\usepackage{color}
 \usepackage{graphicx}
\usepackage{epstopdf}
\usepackage{subfigure}
\usepackage{amsmath}
\usepackage{hyperref}
\usepackage{natbib}
\usepackage{appendix}
\usepackage{verbatim}
\usepackage{hyperref}
\begin{document}

\newcommand{\kpc}{\,{\rm kpc}}
\newcommand{\pc}{\,{\rm pc}}

\bibliographystyle{apj}
\title{No evidence for feedback: Unexceptional Low-ionization winds in Host galaxies of Low Luminosity Active Galactic Nuclei at Redshift $\mathcal{z} \sim 1$}
\author{Hassen M. Yesuf\altaffilmark{1}, David C. Koo\altaffilmark{1}, S. M. Faber\altaffilmark{1}, J. Xavier Prochaska\altaffilmark{1}, Yicheng Guo\altaffilmark{1}, F. S. Liu\altaffilmark{2}, Emily C. Cunningham\altaffilmark{1}, Alison L. Coil\altaffilmark{3}, Puragra Guhathakurta\altaffilmark{1}}
\shortauthors{Yesuf et al.}

\altaffiltext{1}{University of California Observatories and the Department of Astronomy and Astrophysics, University of California, Santa Cruz, CA 95064, USA}
\altaffiltext{2}{College of Physical Science and Technology, Shenyang Normal University, Shenyang 110034, China}
\altaffiltext{3}{Center for Astrophysics and Space Sciences, Department of Physics, University of California, San Diego, 9500 Gilman Drive, La Jolla, CA 92093, USA}

\begin{abstract}

We study winds in 12 X-ray AGN host galaxies at $z \sim 1$. We find, using the low-ionization \ion{Fe}{2} $\lambda$2586 absorption in the stacked spectra, that the probability distribution function (PDF) of the centroid velocity shift in AGN has a median, 16th and 84th percentiles  of  (-87, -251, +86) km s$^{-1}$ respectively. The PDF of the velocity dispersion in AGN has a median, 84th and 16th percentile of (139, 253, 52) km s$^{-1}$ respectively. The centroid velocity and the velocity dispersions are obtained from a two component (ISM+wind) absorption line model. The equivalent width PDF of the outflow in AGN has median, 84th and 16th percentiles of (0.4, 0.8, 0.1)  {\AA}. There is a strong ISM component in \ion{Fe}{2} $\lambda 2586$ absorption with (1.2, 1.5, 0.8) {\AA}, implying presence of substantial amount cold gas in the host galaxies. For comparison, star-forming and X-ray undetected galaxies at a similar redshift, matched roughly in stellar mass and galaxy inclination, have a centroid velocity PDF with percentiles of (-74, -258, +90) km s$^{-1}$, and a velocity dispersion PDF percentiles of (150, 259, 57) km s$^{-1}$. Thus, winds in the AGN are similar to star-formation-driven winds, and are too weak to escape and expel substantial cool gas from galaxies. Our sample doubles the previous sample of AGN studied at $z \sim 0.5$ and extends the analysis to $z \sim 1$. A joint reanalysis of the  $z \sim 0.5$ AGN sample and our sample yields consistent results to the measurements above.

\end{abstract}
\section{Introduction}

Galactic scale winds are one of the most fundamental, yet least understood, facets of galaxy evolution. They are recognized to be fundamental in shaping the baryonic growth, dark matter density profile, star formation and metallicity of galaxies as well as the enrichment of the intergalactic medium \citep[e.g.,][]{Aguirre+01,Veilleux+05,Oppenheimer+10,Dave+11,PontzenGovernato12}. Galaxy formation models that do not include feedback processes form stars too efficiently and fail to reproduce even basic observed galaxy properties. 

High velocity winds are predicted manifestations of the AGN feedback process invoked to reproduce observed properties of massive galaxies \citep{Silk+98,Fabian99,Granato+04,DiMatteo+05,Springel+05,Hopkins+08,Debuhr+12}. In AGN feedback process, a tremendous energy output from accretion onto a black hole, if somehow harnessed, removes or heats gas in the host galaxy  and shuts down subsequent star formation. The consequence of the black hole's action in turn limits gas accretion onto the black hole and stunts its growth. AGN feedback is an essential ingredient in current theoretical models of massive galaxy evolution \citep[for recent reviews, see][]{AlexanderHickox12,Fabian12, Kormendy+13,Heckman+14}. Many semi-analytical models and theoretical simulations require AGN feedback to correctly predict the observed color bi-modality of galaxies and the lack of extremely luminous galaxies \citep[e.g.,][]{Benson+03,Croton+06,Hopkins+06,Cattaneo+07,Somerville+08,Gabor+11}. 

It is predicted that outflows driven by stellar feedback alone are less likely to reach typical velocities higher than 500\,km\,s$^{-1}$ \citep{Thacker+06,Bower+12,Hopkins+13}. In their simulations of stellar feedback (without AGN feedback) in major mergers, \citet{Hopkins+13} found that in all cases the winds have a broad velocity distribution extending up to $\sim 1000$\,km\,s$^{-1}$, but most of the wind mass is near the circular velocity ($\sim 100-200$\,km\,s$^{-1}$), with relatively little ($\ll 1\%$ of the wind mass) at $v \ge 500$\,km\,s$^{-1}$. \citet{Muratov+15} also found winds with similar properties in their analysis of galaxy-scale outflows from the Feedback in Realistic Environments (FIRE) cosmological simulations. In contrast, considerably higher bulk outflow velocities $v \sim 1000-3000$\,km\,s$^{-1}$ are predicted from AGN feedback \citep[e.g.,][]{King+11,Zubovas+12,Choi+12,Debuhr+12,Gabor+14}. In observations, in particular in those that indirectly infer the bulk velocity, the high velocity tail of the star formation-driven wind could be confused for a wind powered by AGN.

However, other models predict that AGN feedback affects a galaxy very little, despite the large outflow velocities \citep{Gabor+14,Roos+15} or it could even enhance star formation \citep{Silk+10, Ishibashi+12,Bourne+15}.  \citet{Bourne+15} claim that the mass resolution of a simulation significantly affects the inferred AGN feedback. At resolutions typical of cosmological simulations, they found that simulated AGN are artificially more efficient in gas removal.  Yet at a higher resolution, the authors found that simulated AGN expel only diffuse gas, and a denser gas falls in towards the black hole and forms stars. Thus, it is not clear whether AGN have negative or positive feedback or both happening simultaneously \citep{Cresci+15} or on different timescales. A consistent and unified theoretical picture on the role of AGN outflows in galaxy evolution is still lacking and observations need to inform and aid the theoretical developments.

Recent observational studies at $z \sim 0.5-2.5$, using the background light of star forming galaxies in self-absorption, have observed ubiquitous velocity offsets from the the systemic zero velocities of the galaxies, indicative of galactic winds \citep{Weiner+09,Erb+12,Law+12,Tang+14,Martin+12,Bordoloi+14,Rubin+14,Zhu+15}. Using the background light of galaxies gives a clearer indication of inflow or outflow unlike using bright background QSOs to probe gas associated with the foreground galaxies \citep[e.g.,][]{Churchill+00}. Since galaxies are much fainter, the analysis is performed on stacks of hundreds of short exposure galaxy spectra or on very deep spectra of a modest sample of individual galaxies ($3-8$ hr integration on Keck for instance).  The outflows studied in both ways show asymmetric absorption profiles with a typical velocity offset of $\sim 200$\,km\,s$^{-1}$ and a high velocity tail which may reach up to $\sim 1000$\,km\,s$^{-1}$ \citep[e.g.,][]{Weiner+09,Martin+12}. Note that this high velocity tail in star-forming non-AGN galaxies is observed more prominently in \ion{Mg}{2} $\lambda2796$ and it is consistent with the theoretical expectation \citep[e.g.,][]{Hopkins+13,Muratov+15}. 

In the latest works that used deep spectroscopic data of individual galaxies, the wind speed is best correlated with SFR surface density but it is not significantly correlated with either galaxy stellar mass or inclination \citep{Martin+12,Kornei+12,Rubin+14}. The wind detection rate on the other hand is highly dependent on inclination \citep{Rubin+14}. While the recent works have made important advances in characterizing such outflows and in establishing their relationships to host galaxy properties, many basic properties of these winds and their driving physics remain uncertain. One such uncertainty is the wind velocity of AGN host galaxies.

The evidence for AGN hosts having ubiquitous high mean galaxy-wide velocity outflows with the potential to impact star formation is sparse (more detailed discussion in \S \ref{sec:disc}). Ionized outflows have been studied in emission using large samples both at low-redshift \citep[e.g.,][]{Zakamska+14} and high-redshift \citep[e.g.,][]{Harrison+16}. Even though, convincing evidence for ubiquitous, ionized outflows exists, details on the interpretation of the observed wind properties are debatable. Most of the emission-line studies have found high-velocity, extended outflows on several kpc scale, resulting in very large inferred mass outflow rates and kinetic power, in support of AGN feedback models \citep[e.g.,][]{Liu+13,Harrison+14,McElroy+15}. However, recent studies have questioned these results and have argued that the apparently very extended emission is a consequence of seeing smearing \citep{Husemann+15,Karouzos+16,Villar-Martin+16}. Accounting for the seeing effect, these later works found much smaller and weaker winds, which may not significantly impact the star formation in their host galaxies.

On the other hand, absorption line wind studies are hard to undertake in distant galaxies but are relatively easy to interpret. Existing absorption line studies of winds in AGN host galaxies have small sample sizes $\sim 10-30$ at $z \sim 0.5-2.5$ \citep{Coil+11, Hainline+11} and the aforementioned deep spectroscopic wind absorption studies did not primarly target AGN. This may be because AGN hosts are rare and are generally fainter than the targeted star-forming galaxies. \citet{Coil+11} studied a sample of 10 low-luminosity, narrow-line AGN ($L_\mathrm{X} \sim 10^{41-42}$\,erg\,s$^{-1}$) at $0.2 <z < 0.6$. Five of the ten X-ray AGN host galaxies show a wind in \ion{Fe}{2} $\lambda 2586$ absorption, with a typical mean outflow velocity signatures of only $\sim 200$ to $300$\,km\,s$^{-1}$. The velocity widths are generally unresolved but are, $\sim 100-300$\,km\,s$^{-1}$. On the other hand, \citet{Hainline+11} qualitatively studied a stacked spectrum of 33 UV-selected narrow-line AGN at $z \sim 2.5$ and reported a detection of a highly blue-shifted ($v \sim -850$\,km\,s$^{-1}$) and weak \ion{Si}{4}$ \lambda1393,1402$ absorption line, which is different from the \ion{Si}{4} absorption in the composite spectrum of non-AGN Lyman break galaxies at a similar redshift.

It should be noted that \citet{Prochaska+11} have found that atomic transitions that are only coupled to the ground state (e.g., \ion{Si}{4}, \ion{Mg}{2}, \ion{Na}{1}) have line emission preferentially at the systemic velocity and their observed absorption profiles are significantly reduced in depth as well as are significantly offset in velocity from the intrinsic profile. On the hand, they also found that resonance transitions that are strongly coupled to fine-structure transitions (e.g., \ion{Fe}{2} and \ion{Si}{2}) dominantly produce florescent emissions at longer wavelength which do not affect the absorption profiles. These resonance absorption lines offer the best characterization of the opacity of the wind as well as the opacity of the gas near the systemic velocity. Therefore, the discrepancy between the two previous works on AGN winds may be due to this effect. The AGN sample in \citet{Hainline+11} show stronger \ion{Si}{4} $ \lambda1393,1402$ emission near the systemic and have much weaker absorption than do their non-AGN star-forming counterparts. In the follow up work, the same authors showed that the stellar population properties of their AGN sample are consistent with those of the mass-matched control sample of star-forming galaxies. They inferred that the presence of an AGN is not connected with the cessation of star formation activity in star-forming galaxies at $z \sim 2-3$ \citep{Hainline+12}. In other words, the observed high winds in these AGN have not yet affected star formation even if the measured mean velocity is reliable.

The work in this paper bridges the gap in redshift between the two major previous studies of AGN winds in absorption and aims to independently confirm the previously reported wind velocities in AGN. We examine winds in a composite spectrum of 12 X-ray selected AGN at  $z=0.9 - 1.5$ or that of a comparison sample of star-forming galaxies using \ion{Fe}{2} $\lambda2586$, a preferred  wind diagnostic. Our AGN sample has a comparable X-ray luminosity to that of \citet{Coil+11}. Our spectral resolution is two times higher than theirs and we have three times more wavelength sampling. Our sample also has more extensive multi-wavelength deep HST photometry and other ancillary data to better characterize the host galaxy properties. This has enabled a first attempt to have a control sample for star-forming galaxies matched both in stellar mass and galaxy inclination. Furthermore, we use similar wind model and methods that have been adopted in the most recent wind studies \citep{Martin+12,Rubin+14}. These methods were not used in the two previous studies of AGN. Thus, for a fair comparison, we present a reanalysis of the \citet{Coil+11} data using the new approach, which also has better quantified model uncertainties.

The rest of the paper is organized as follows: Section \ref{sec:data} presents the data and sample selection. Section \ref{sec:res} presents the analysis and results on winds in AGN at $z \sim 1$, AGN at $z \sim 0.5$ and the comparison sample. Section \ref{sec:disc} extensively discusses previous works to put the result of this work in a larger context. A brief summary and conclusion is given in section \ref{sec:conc} . $(\Omega_m,\Omega_\Lambda,h) = (0.3,0.7,0.7)$ cosmology is assumed and AB magnitude is adopted. A wavelength measured in air is given throughout the paper. We use words ``wind'' and ``outflow'' interchangeably to mean outward movement of gas without making subtle distinctions in some previous works.

\section{Data} \label{sec:data}
\subsection{Observation \& data reduction}

The spectroscopic data are taken from our on going deep (8--24 hr) Keck/DEIMOS \citep{Faber+03} spectroscopic survey in CANDELS fields \citep{Grogin+11,Koekemoer+11,Guo+13} called HALO7D. This multi-semester program will survey faint halo stars with HST-measured proper motions, to measure their line-of-sight velocities and chemical abundances, giving 6D phase-space information and chemical abundances for hundreds of Milky Way halo stars. The deep exposures necessary to reach the faintest stars in the Galaxy halo is an opportunity for a novel synergy of extragalactic and Galactic science. In addition to the primary halo star targets, which only occupy about a quarter of slitlets on a given DEIMOS mask, we are conducting a survey of galactic winds in star-forming galaxies at $z \sim 1$, and stellar populations of quiescent galaxies at redshifts $0.4 < z < 0.8$. A total of about $\sim 1500$ deep spectra of galaxies are expected with survey completion in a year. 

The HALO7D survey uses the 600 line/mm grating on DEIMOS centered around 7200{\AA} with the GG455 order-blocking filter. This setup gives a nominal wavelength coverage of 4600-9500{\AA} at a resolution (FWHM) of $\sim 3.5${\AA} for a 1$^{\prime\prime}$ slit width and 0.65{\AA/}pixel dispersion. The slit position angles are set to within $\pm 30^{\circ}$ of the parallactic angle to minimize light loss in the blue due to atmospheric dispersion. The exposure times for AGN studied in this work range between $5-12$ hr and the observations were taken over the course of two years under variable seeing ($0.8-1.2^{\prime\prime}$) and fair to good transparency conditions. Despite the long integration times, poor seeing has significantly lowered the signal-to-noise for almost half of the AGN sample in the current work. Additional make-up observations of these AGN are scheduled.

The observations were reduced using the automated DEEP2/DEIMOS \emph{spec2d} pipeline developed by DEEP2 team \citep{Newman+13}. Calibrations were done using a quartz lamp for flat fielding and both red NeKrArXe lamps and blue CdHgZn lamps for wavelength calibration. The spectroscopic redshifts were measured from the reduced spectra using the  \emph{spec1d} pipeline. All spectra were visually inspected using the interactive $zspec$ tool to access the quality of the redshift estimated by \emph{spec1d} \citep[for software details, see][]{Newman+13}. Almost all galaxies studied in this work have previous spectroscopic measurements and the new spectroscopic redshifts imply minor changes if any.

Based on the available redshifts, stellar masses and other stellar population properties (age, extinction $A_V$, etc.) were computed with FAST \citep{Kriek+09} using a combination the newly obtained CANDELS HST/WFC3 multi-wavelength photometric data with existing ground-based and space-based multi-wavelength data\footnote{The following filters are used in the SED fitting : \\ In EGS: CFHT (u, g, r, i, z), ACS (F606W, F814W), WFC3 (F125W, F140W, F160W), WIRCAM (J, H, K), NEWFIRM (J1, J2, J3, H1, H2, K), IRAC (CH1, CH2, CH3,CH4).\\ In GOODS-N : KPNO\_U, LBC\_U, ACS (F435W, F606W, F775W, F814W, F850LP), WFC3 (F105W, F125W, F140W, F160W, F275W), MOIRCS\_K, CFHT\_K, IRAC (CH1,CH2,CH3,CH4). \\ In GOODS-S: CTIO\_U,VIMOS\_U, ACS (F435W, F606W, F775W, F814W, F850LP), WFC3 (F098M, F105W, F125W, F160W), ISAAC\_KS, HAWAKI\_KS, IRAC (CH1,CH2,CH3,CH4).} as inputs \citep[e.g.,][]{Guo+13}. The modeling is based on a \citet{BC03} stellar population synthesis model and assumes a \citet{Chabrier03} IMF, exponentially declining star formation histories, solar metallicity, and a \citet{Calzetti+00} dust extinction law. The typical uncertainty in stellar masses is $\sim 0.1$ dex. The star-formation rates (SFR) are the sum of the SFR$_{UV}$, derived from the rest-frame near UV luminosity at 2800{\AA}, and the SFR$_{IR}$, derived from the total infrared luminosity. If a galaxy is not detected in infrared, its dust-corrected UV-based star formation rate is used \citep{Barro+11,Wuyts+11}. The SFR estimates are uncertain by a factor of $\lesssim 2$. The axis-ratio measurements were done on HST/WFC3 F160W (H) band imaging using GALFIT \citep{Peng+02}. 

For comparison, we also used data of 6 previously studied low-luminosity AGN at $0.3 < z < 0.6$ with \ion{Fe}{2} coverage \citep{Coil+11}. These AGN were observed using Keck LRIS \citep{Oke+95} for roughly one half hour to one hour. The average apparent brightness for this sample is $B \sim 20.9$ and our AGN sample has about 9 times fainter average apparent brightness. \citet{Coil+11} have stellar mass and star-formation rate measurements for four of the six AGN. We adopted their measurements. In comparison plots that use the stellar mass and star-formation rate measurements, we only show their four AGN with such measurements, but we reanalyzed the spectra of all six AGN.

\subsection{Sample selection}

For the wind study, we primarily targeted sources that are brighter in the V band ($V < 23.5$) and above $z > 0.7$, such that we would detect near UV continuum levels in the individual spectra in eight hours at a signal-to-noise ratio per Angstrom (SNR/{\AA}) of 5 in good observing conditions. We gave higher priority to sources that are brighter and are above $z > 0.9$, which likely have both \ion{Fe}{2} and  \ion{Mg}{2} coverage. Galaxies with $V > 23.5$ were targeted at lower priority as fillers. AGN are a small fraction ( $< 10\%$) of all galaxies in our survey. In some cases, we targeted bright X-ray sources \citep{Alexander03,Laird+09,Xue+11} and gave them highest weights in the mask design process. In other cases, the AGN were selected by chance (i.e., were selected for other reasons). So far, there are about one hundred X-ray sources in the observed sample. Only about a third of them are above $z > 0.9$ and, therefore, have coverage of \ion{Fe}{2} $\lambda2586$. 

Out of the $\sim 100$ targeted/serendipitous X-ray sources, we selected all 12 AGN candidates at $z> 0.9$ that have X-ray luminosity $L_\mathrm{X} \gtrsim 5 \times 10^{41}$ erg s$^{-1}$, SNR/{\AA} $> 2.7$ around \ion{Fe}{2} $\lambda2586$ and uncontaminated \ion{Fe}{2} $\lambda2586$ by skylines. We also require that they have HST WFC3 imaging for axis-ratio measurements and have highly reliable redshift measurements (show clear \ion{O}{2} emission and/or \ion{Ca}{2} H \& K absorption lines). 

\begin{figure*}
\includegraphics[scale=0.65,angle=270]{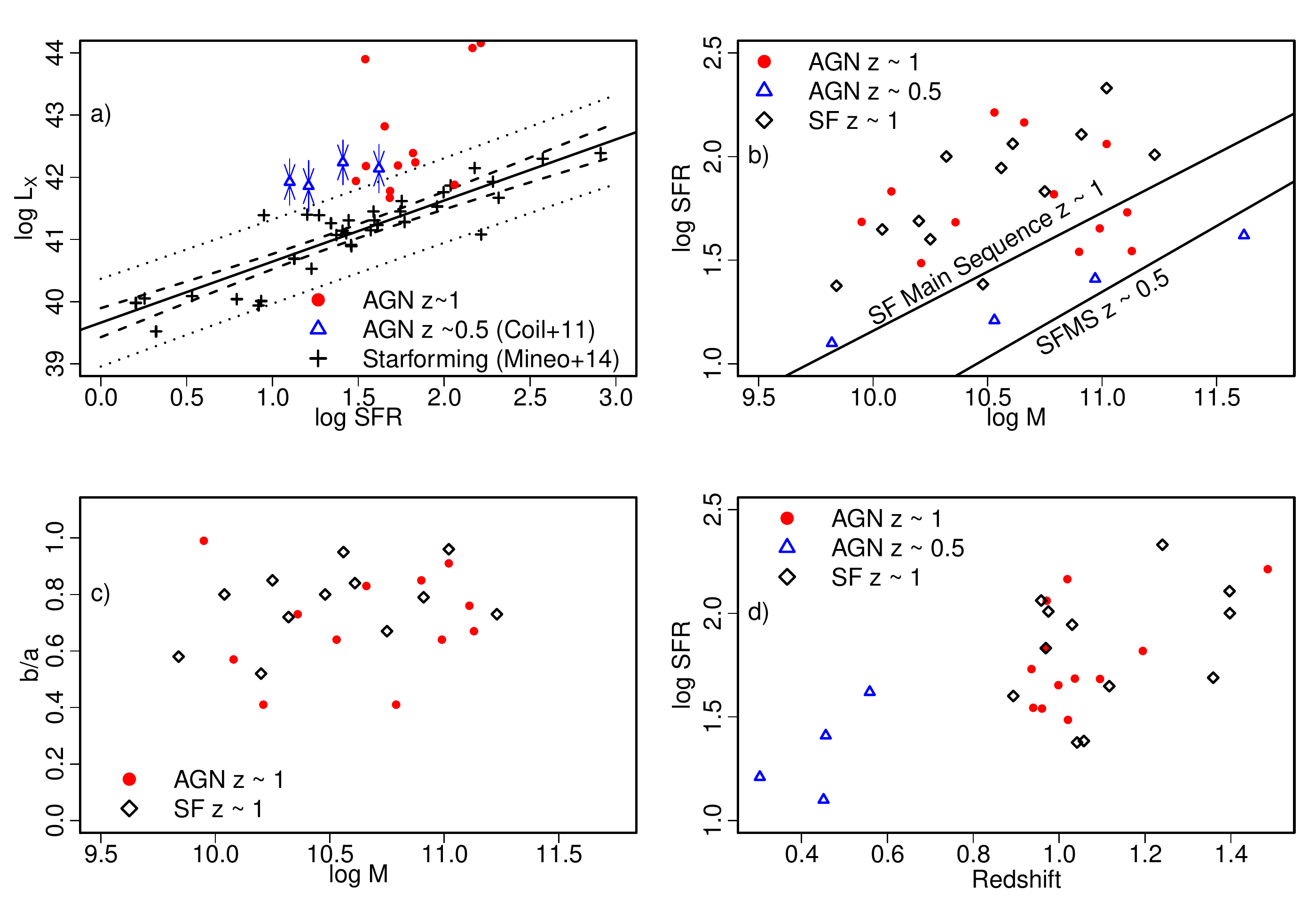}
\caption{Sample properties: in all panels the red dots are the low-luminosity AGN at $z \sim 1$, the diamonds are the X-ray undetected star-forming comparison sample at  $z \sim 1$ and the blue triangles are AGN at $z \sim 0.5 $ from \citet{Coil+11}. Panel (a) shows the star-formation rates (M$_\odot$ yr$^{-1}$) versus the full X-ray Luminosities (erg s$^{-1}$) of AGN and non-AGN. The blue triangles are shown with arrows to indicate upper and lower limits of their X-ray luminosities. The + signs denote X-ray detected star forming galaxies in Chandra Deep Field South (CDFS) from \citet{Mineo+14}. The solid line shows the line of best linear fit for the relationship between star formation rate and X-ray luminosity in the star forming non-AGN galaxies. The dashed lines and the dotted lines indicate the 95\% confidence and predictive bands respectively for these galaxies. Panel (b) plots the stellar mass (M$_\odot$) against the star-formation rate. The black lines indicate the star-formation main sequence of galaxies at $z=1$ and $z=0.5$ \citep{Whitaker+14}. Panel (c) shows the stellar mass versus the H-band axis ratio. Panel (d) depicts the redshifts and the star-formation rates of the samples.} 
\label{fig:sample}
\end{figure*}

Figure~\ref{fig:sample}a plots star formation rate against X-ray luminosity. Nine of the selected AGN candidates are $2\sigma$ outliers from the relationship between X-ray luminosity and star formation rates for normal star-forming galaxies \citep{Mineo+14}. Of these, 3 show broad \ion{Mg}{2} emission. When we have previous spectroscopic data, we targeted objects with broad \ion{Mg}{2} emission at lower priority. Quasars have hard X-ray luminosities $\gtrsim 10^{44}$\,erg\,s$^{-1}$ \citep[e.g.,][]{Ku+80,Piconcelli+05}. Except for the 3 AGN with broad \ion{Mg}{2} emissions, the AGN studied in this work have significantly lower X-ray luminosities compared to quasars. We thus refer to them as low-luminosity AGN to differentiate them from quasars.

Furthermore, 3 of the 12 X-ray sources lie within $2\sigma$ of the mean relation between X-ray luminosity and star formation rate. One of them has strong \ion{Ne}{5} $\lambda3426$ emission, a strong signature of AGN. Therefore, 10 of the AGN sample have robust AGN identifications. Excluding the two X-ray sources which may not be AGN does not change the main results of this work. We have excluded from our sample some AGN candidates which satisfy the SNR cut but their spectra around \ion{Fe}{2} are contaminated by a skyline or possible absorption from a foreground galaxy.

The three broad-line AGN in our sample show no or very weak and narrow \ion{Fe}{2} and/or \ion{Mg}{2} absorption but have higher X-ray luminosity $L_\mathrm{X} > 10^{43}$ erg s$^{-1}$. \citet{Coil+11} did not target objects with broad \ion{Mg}{2} emission in their spectra, as strong broad emission may affect one's ability to detect or interpret blue-shifted absorption features because of emission infill or ionization to a higher state. The current estimates of stellar masses and star-formation rates of broad-line AGN do not properly account for the presence of the luminous AGN and thus they may be biased compared to the measurements for the star-forming comparison sample. However, for low-luminosity narrow-line AGN, the stellar mass and SFR measurements are not significantly affected by presence of  AGN \citep{Ciesla+15}. So, we present analyses both with and without the inclusion of the three broad-line AGN.

To test for the effects of the presence of AGN on winds in their host galaxies, here we define a comparison galaxy sample which lacks AGN but has a similar distribution of stellar masses and axis ratios. For each AGN host, we select the X-ray undetected object which has the most similar stellar mass and axis ratios. The masses and axis-ratios agree within a factor of two for all AGN and their comparison galaxies. All galaxies except one have matches with similar SFRs within a factor of two.

Figure~\ref{fig:sample}b$-$d show the stellar mass, star formation rate and axis-ratio of the AGN and the comparison sample of star-forming X-ray undetected galaxies at $z \sim 1$. Due to the limited sample size of the parent sample from which the comparison sample of X-ray undetected galaxies were drawn ($< 100$ galaxies with $z > 0.9$ and $V < 23.5$), our matching is crude It may be sufficient since winds are expected to depend weakly on galaxy properties \citep{Coil+11,Martin+12,Rubin+14,Balmaverde+16}. We think the comparison galaxies do not host AGN because they are located in regions of the sky with X-ray coverage by Chandra observations yet they are not detected in X-ray. However, it has been argued that a substantial fraction of AGN are missed by the X-ray selection \citep[e.g.,][]{Juneau+13}. Future near-IR spectroscopic data for rest-frame optical line-ratio AGN diagnostic are needed to completely rule out the presence AGN in the comparison sample.

\begin{figure*}
\subfigure[GDS 20168]{%
\includegraphics[width=0.25\linewidth]{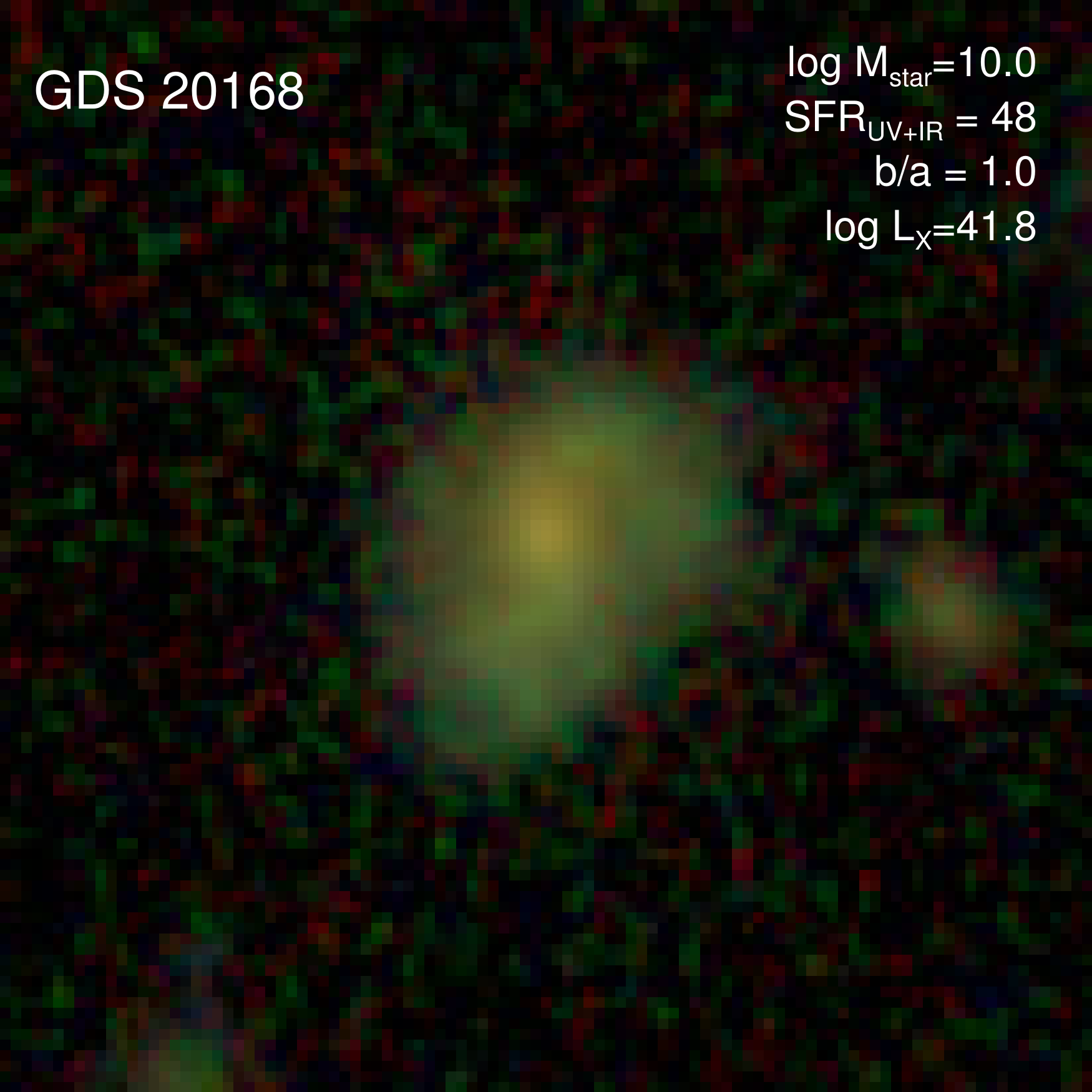}%
}%
\hfill
\subfigure[GDS 23803]{%
\includegraphics[width=0.25\linewidth]{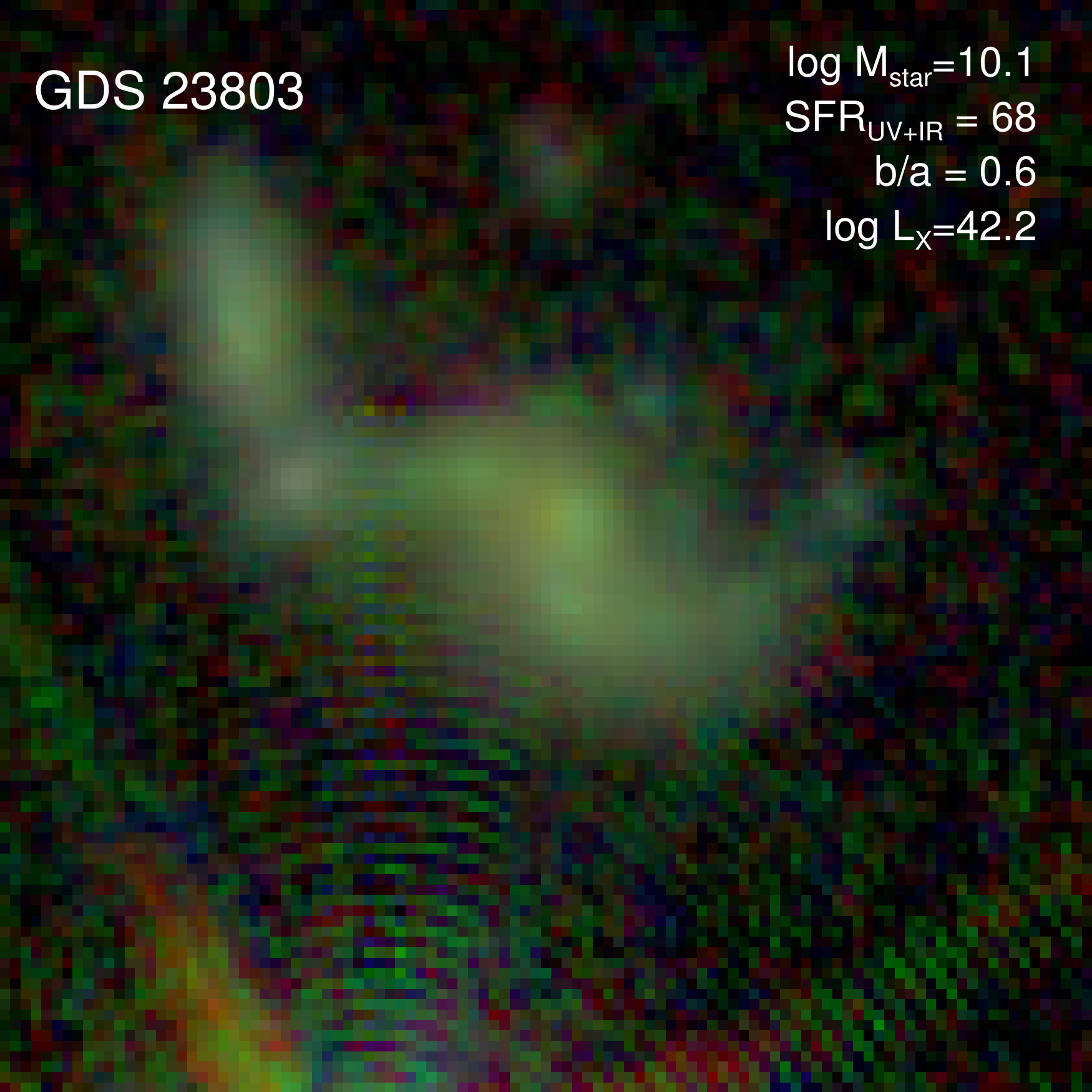}%
}%
\hfill
\subfigure[GDS 15870]{%
\includegraphics[width=0.25\linewidth]{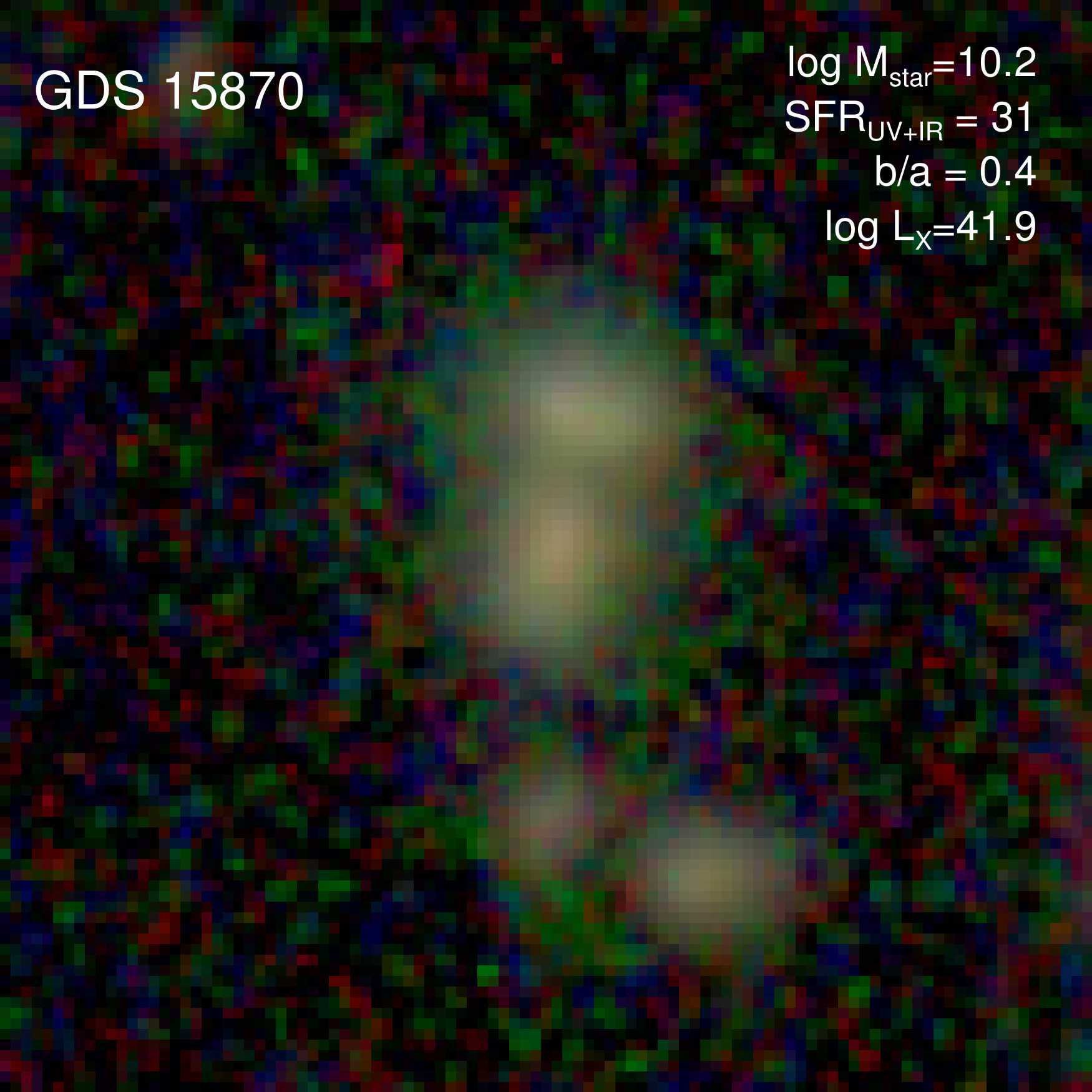}%
}%
\hfill
\subfigure[EGS 10518]{%
\includegraphics[width=0.25\linewidth]{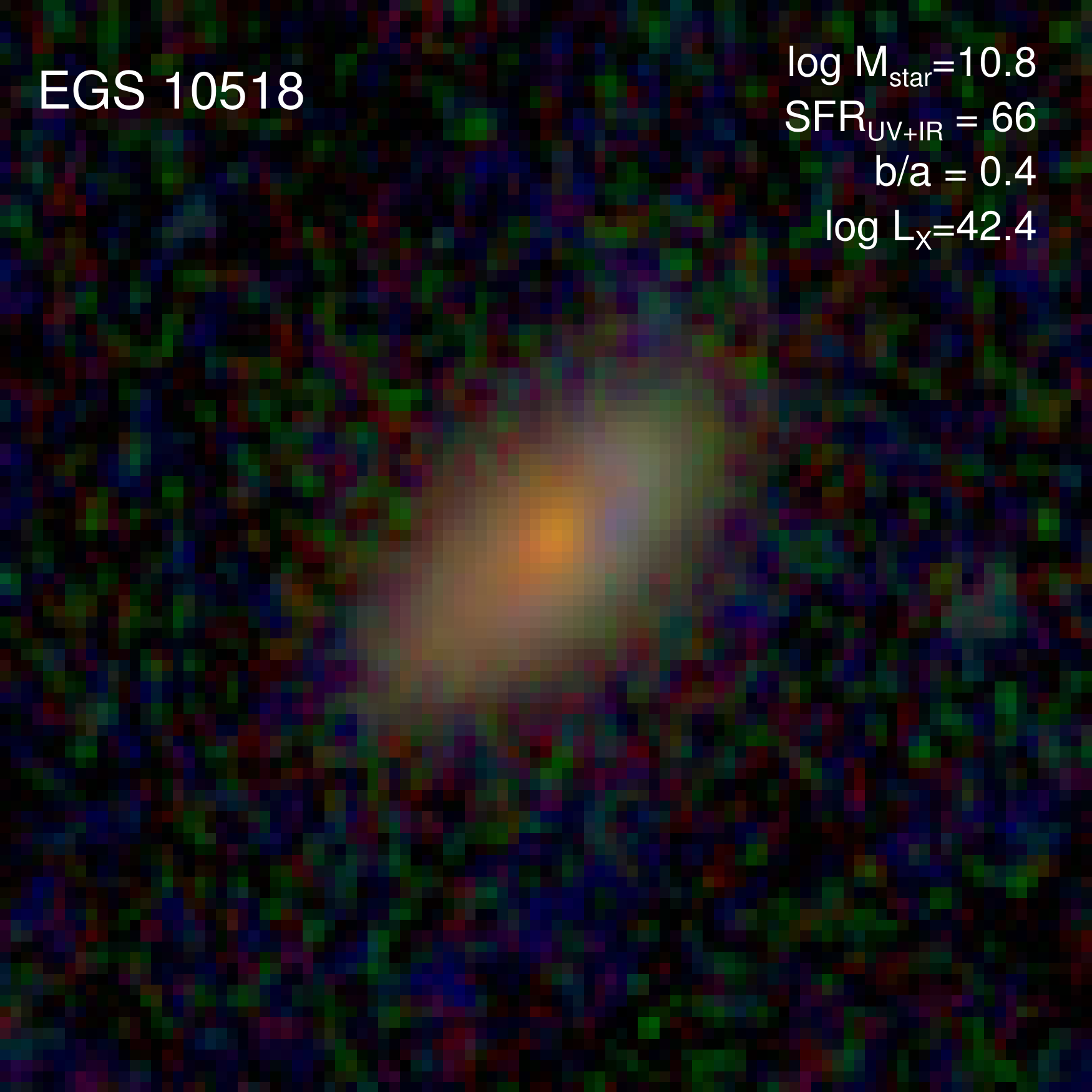}%
}%
\hfill
\subfigure[GDS 7837]{%
\includegraphics[width=0.25\linewidth]{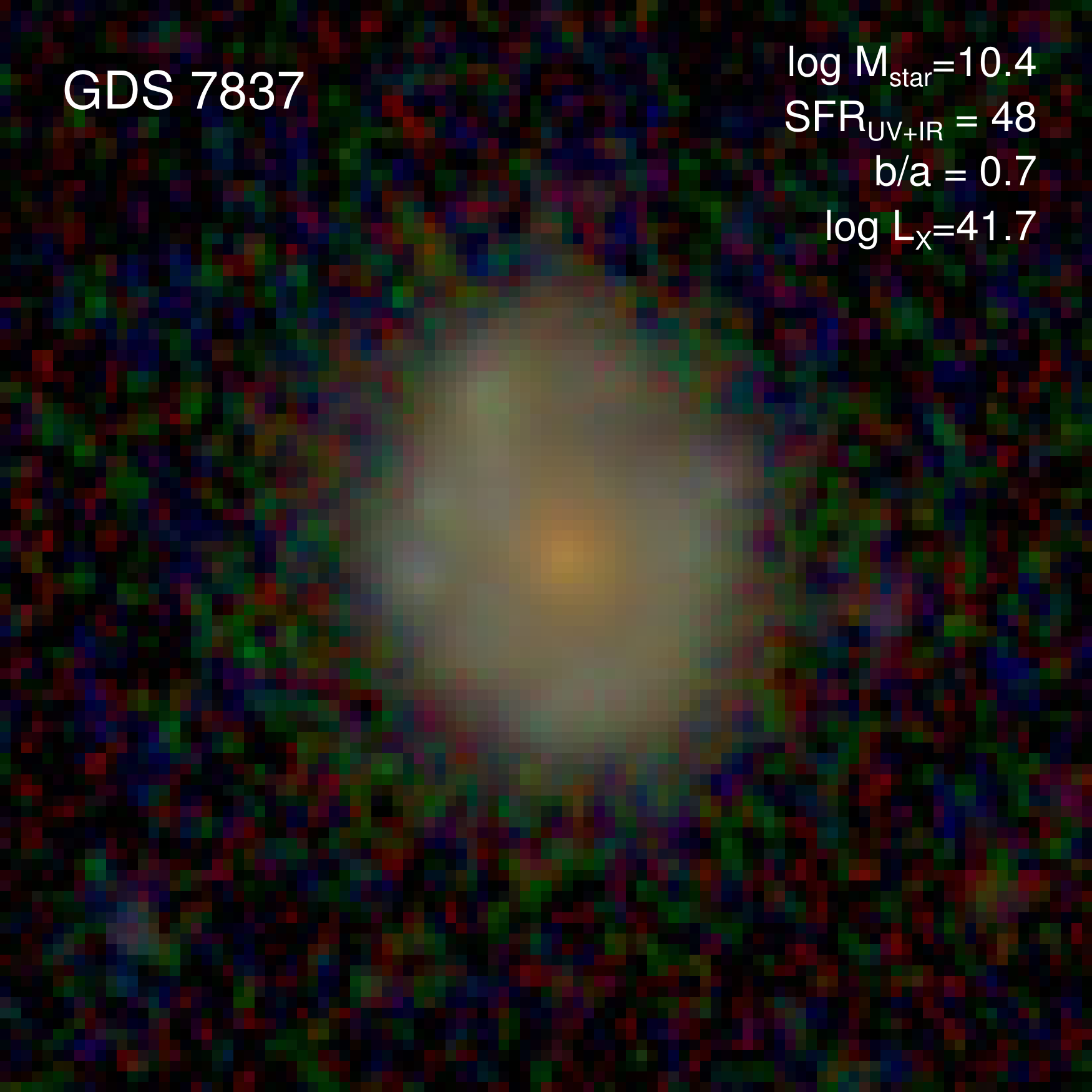}%
}%
\hfill
\subfigure[EGS 30572]{%
\includegraphics[width=0.25\linewidth]{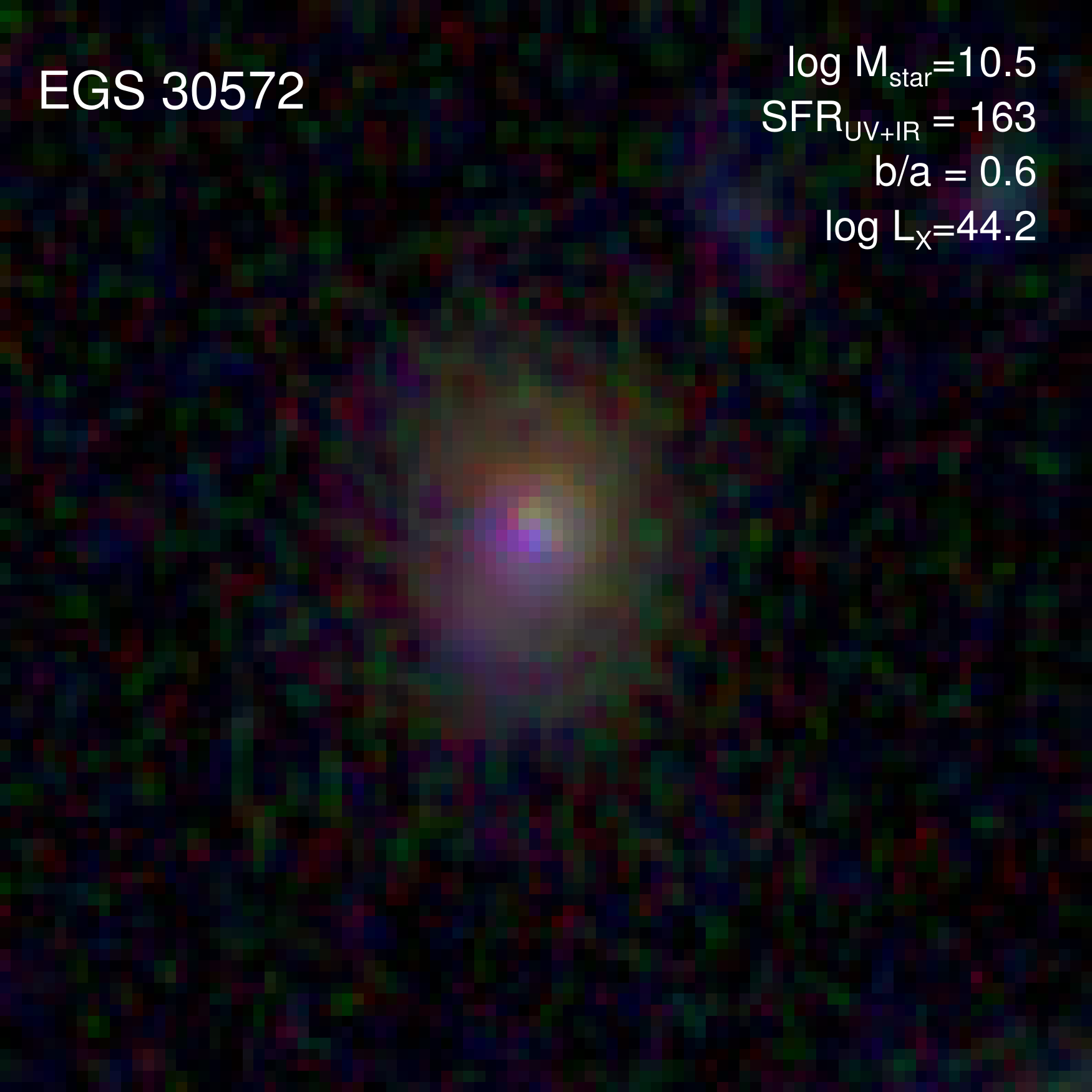}%
}%
\hfill
\subfigure[GDN 6274]{%
\includegraphics[width=0.25\linewidth]{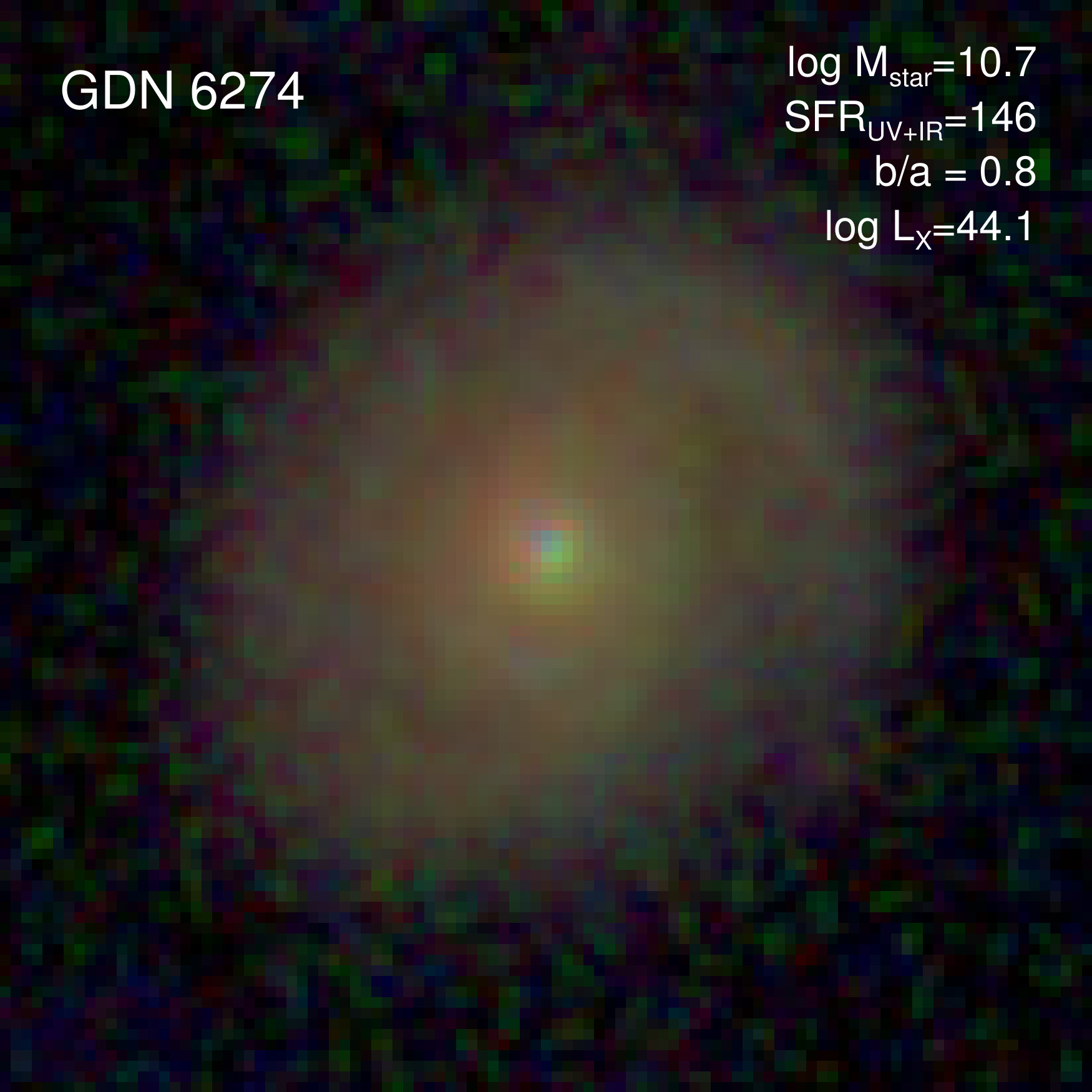}%
}%
\hfill
\subfigure[GDN 12523]{%
\includegraphics[width=0.25\linewidth]{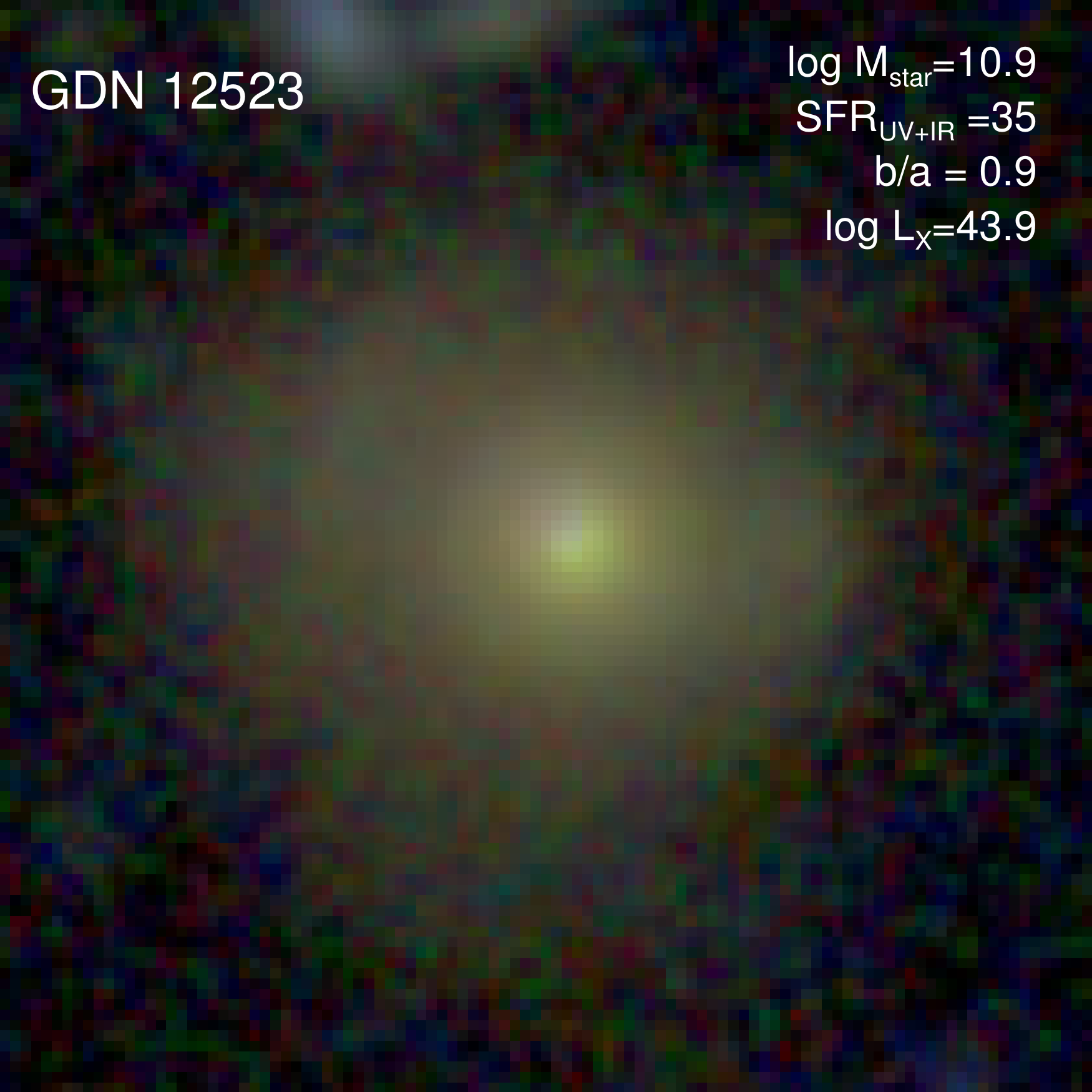}%
}%
\hfill
\subfigure[GDS 21627]{%
\includegraphics[width=0.25\linewidth]{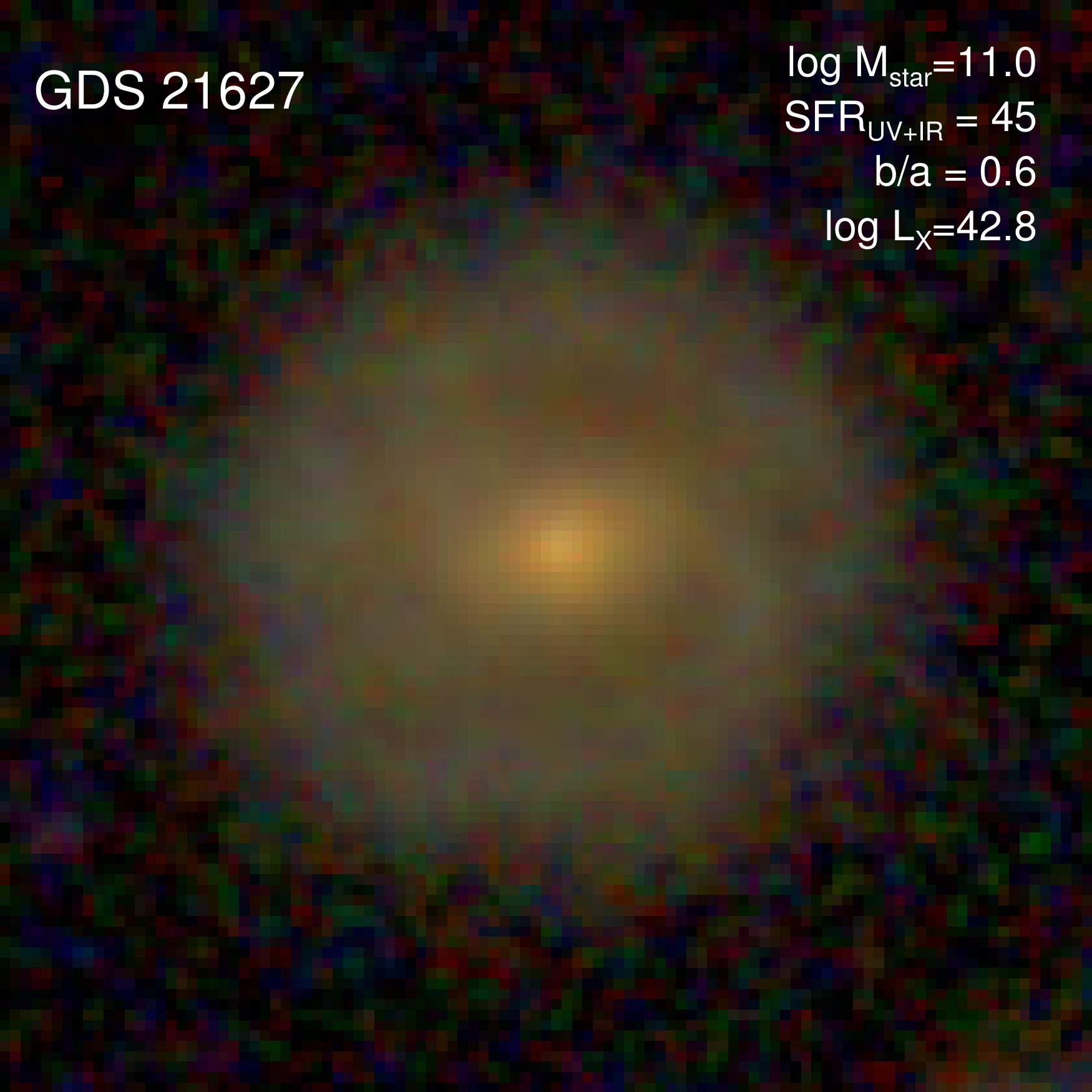}%
}%
\hfill
\subfigure[GDS 21627]{%
\includegraphics[width=0.25\linewidth]{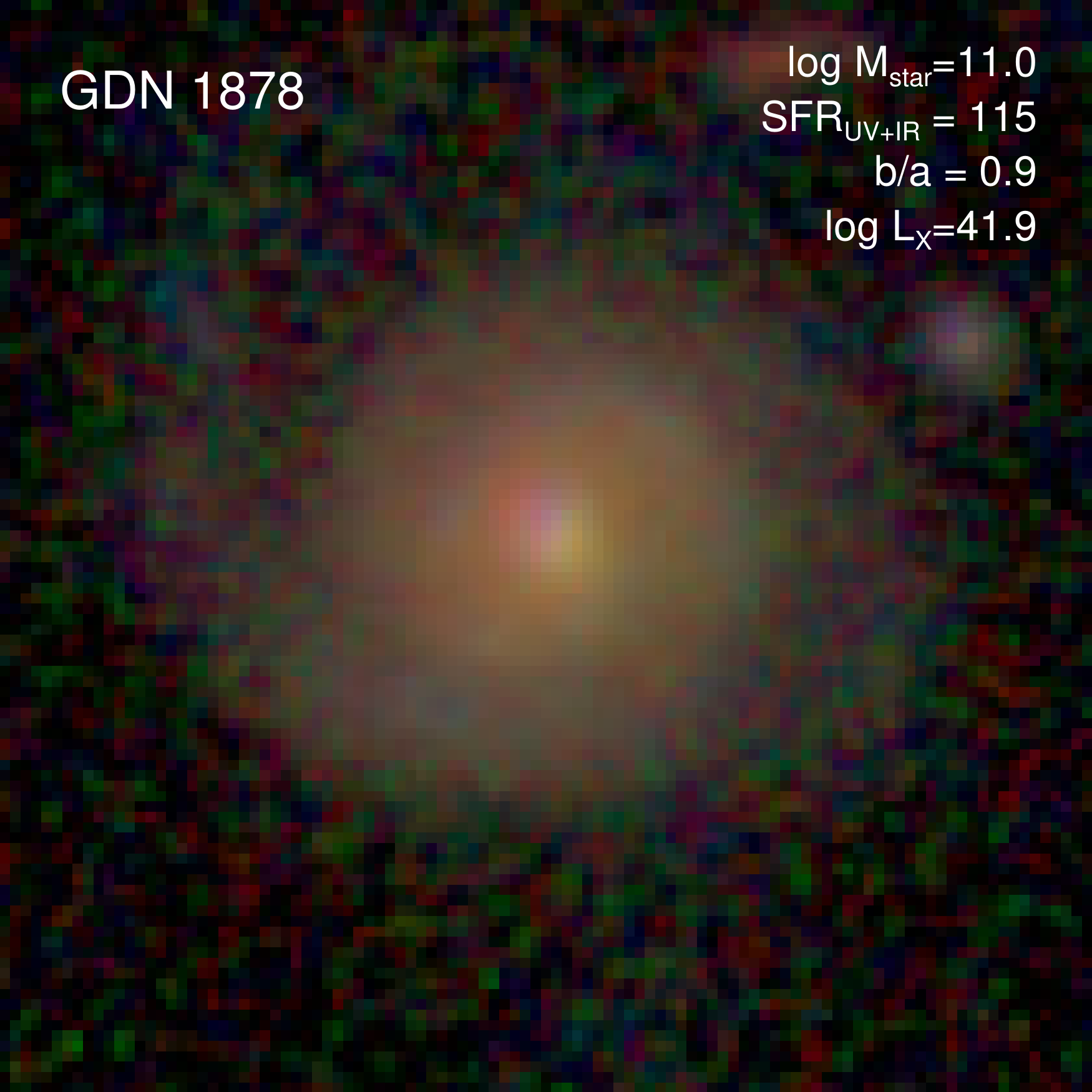}%
}%
\hfill
\subfigure[GDN 17041]{%
\includegraphics[width=0.25\linewidth]{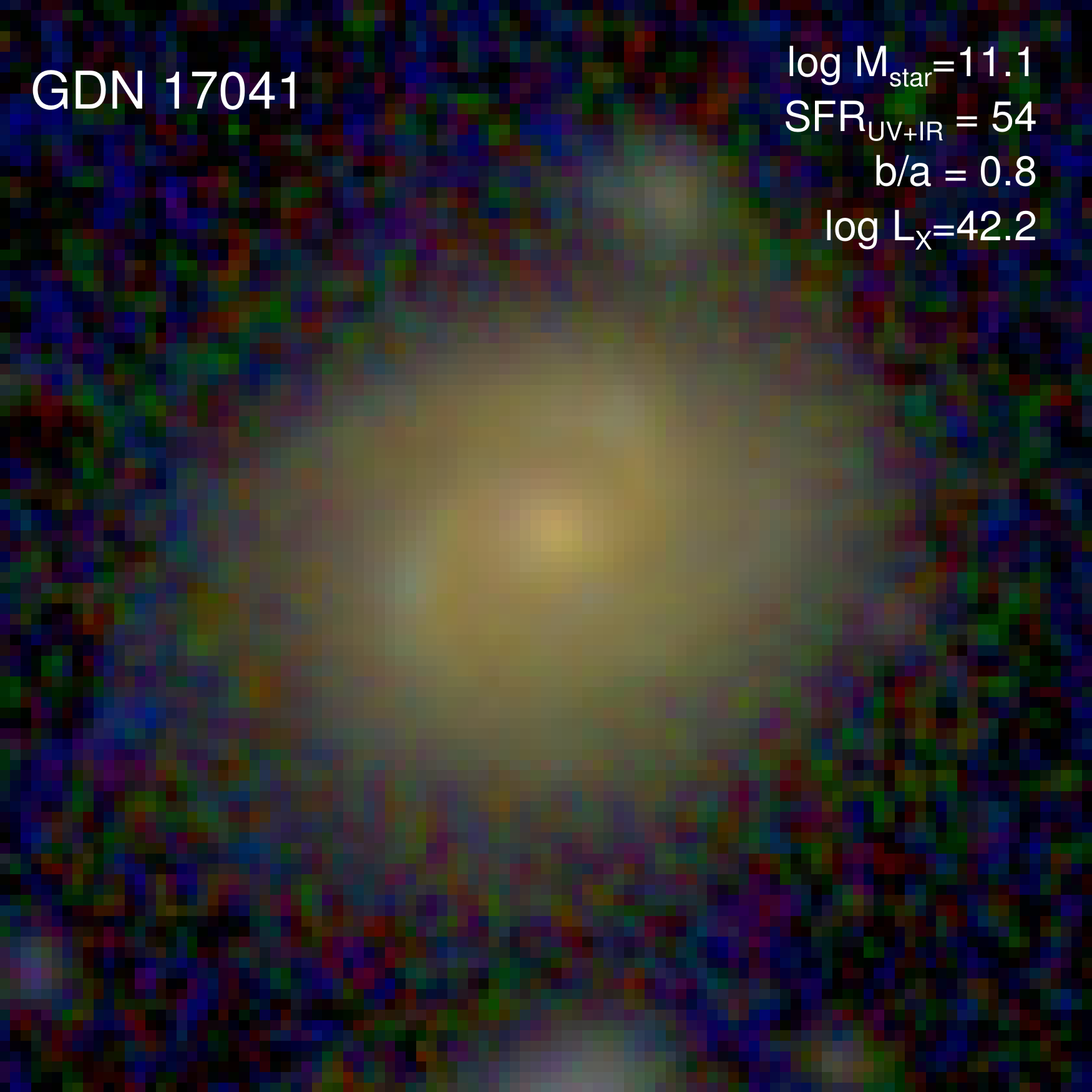}%
}%
\hfill
\subfigure[GDN 17389]{%
\includegraphics[width=0.25\linewidth]{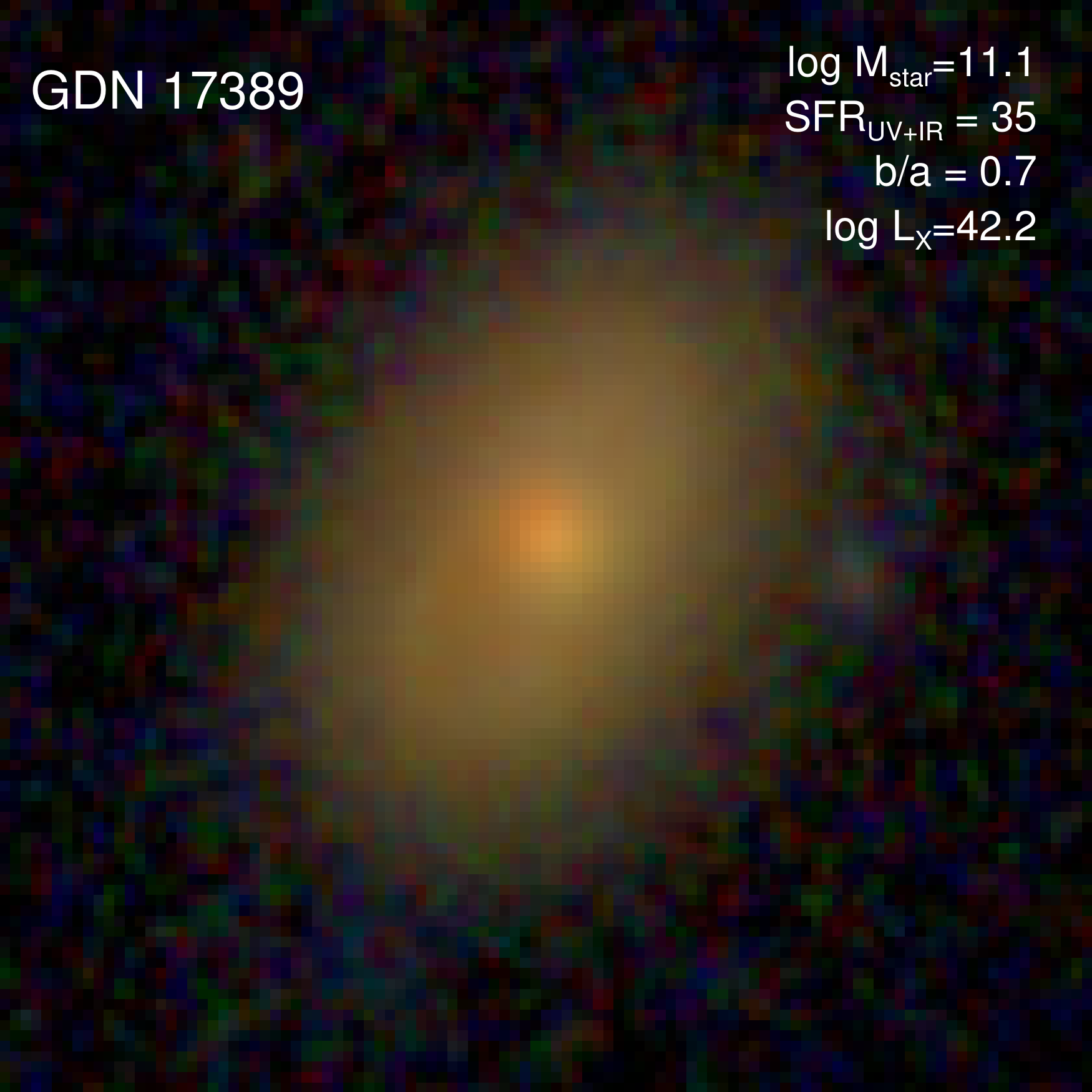}%
}%
\caption{Rest-frame pseudo RGB composite Hubble Space Telescope images of the AGN at $z \sim 1$. The number at the top of the image indicates galaxy id, stellar mass, total star formation rate, the axis-ratio and X-ray luminosity of the host galaxy. Each image is $50 \times 50$\,kpc. North is to the top and East is to the left.}
\label{fig:imageAGN}
\end{figure*}

\begin{figure*}
\subfigure[GDS 26255]{%
\includegraphics[width=0.25\linewidth]{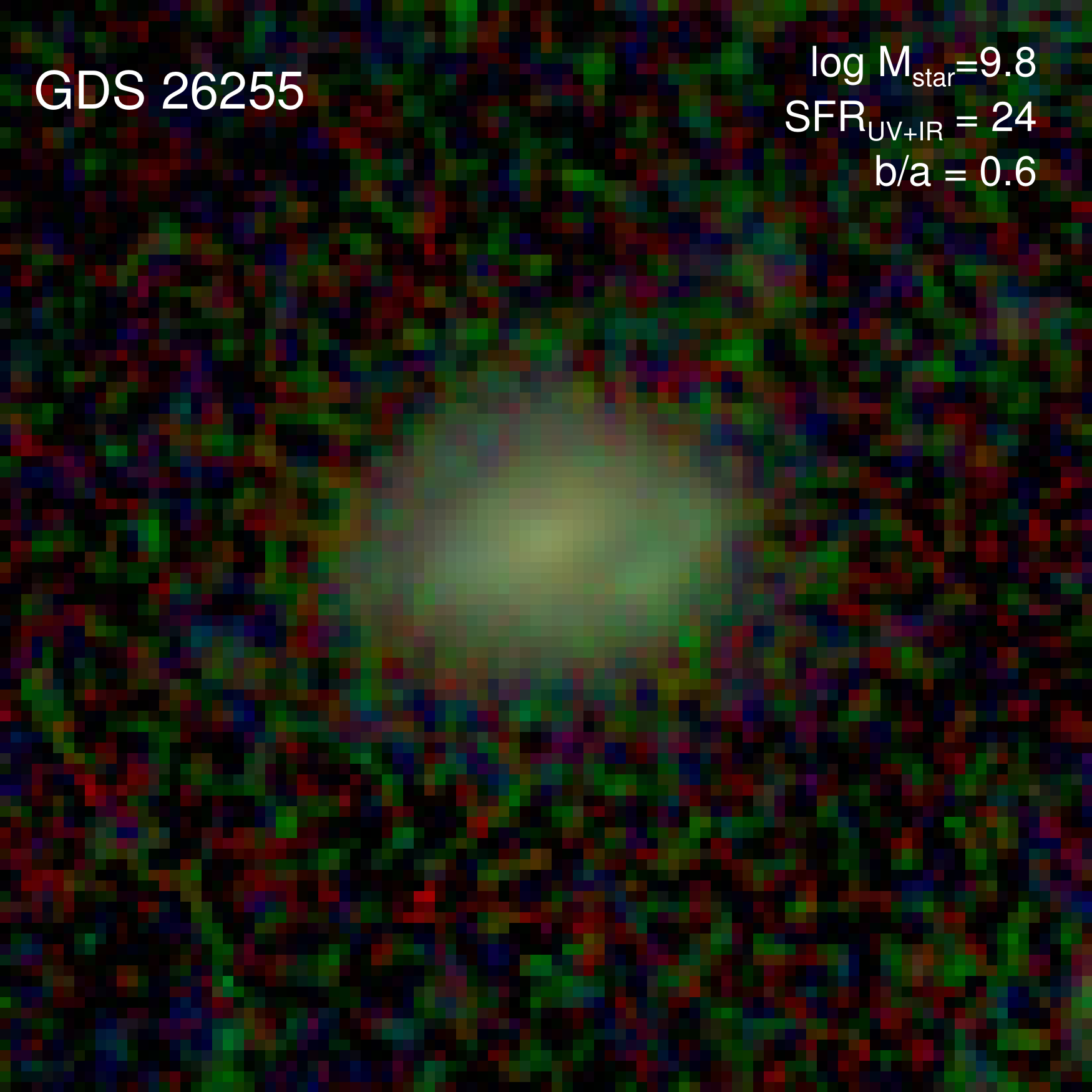}%
}%
\hfill
\subfigure[EGS 15131]{%
\includegraphics[width=0.25\linewidth]{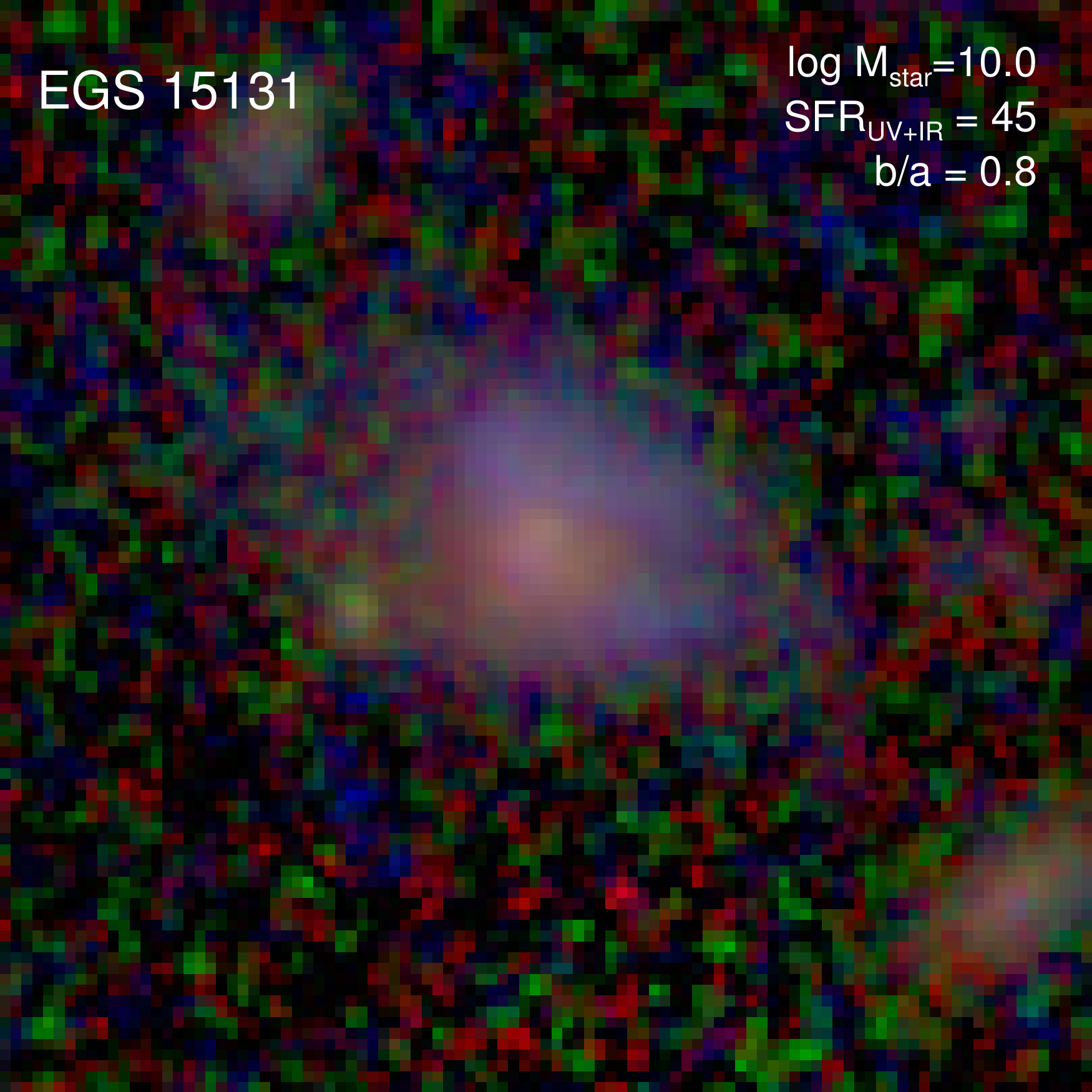}%
}%
\hfill
\subfigure[GDS 11799]{%
\includegraphics[width=0.25\linewidth]{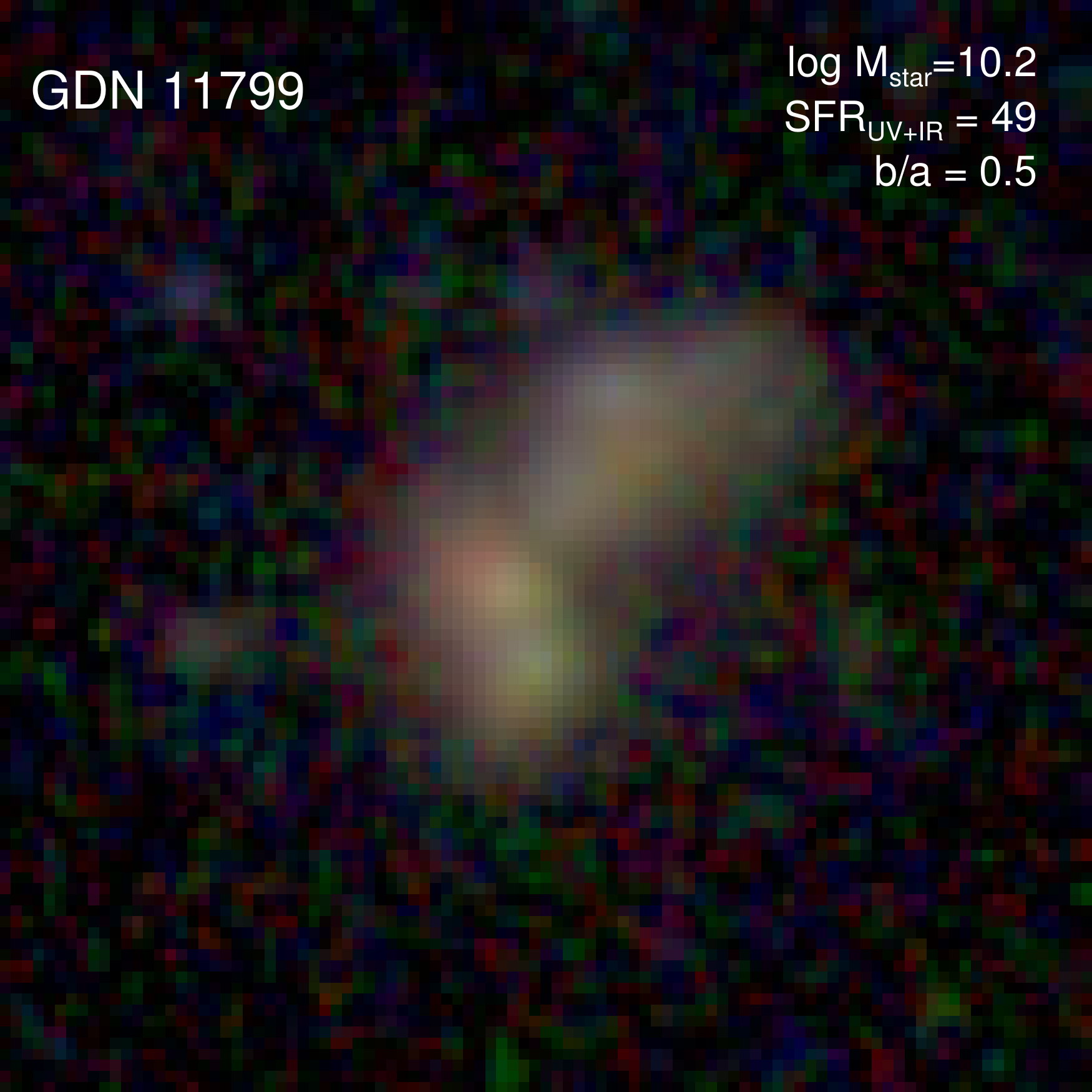}%
}%
\hfill
\subfigure[EGS 27292]{%
\includegraphics[width=0.25\linewidth]{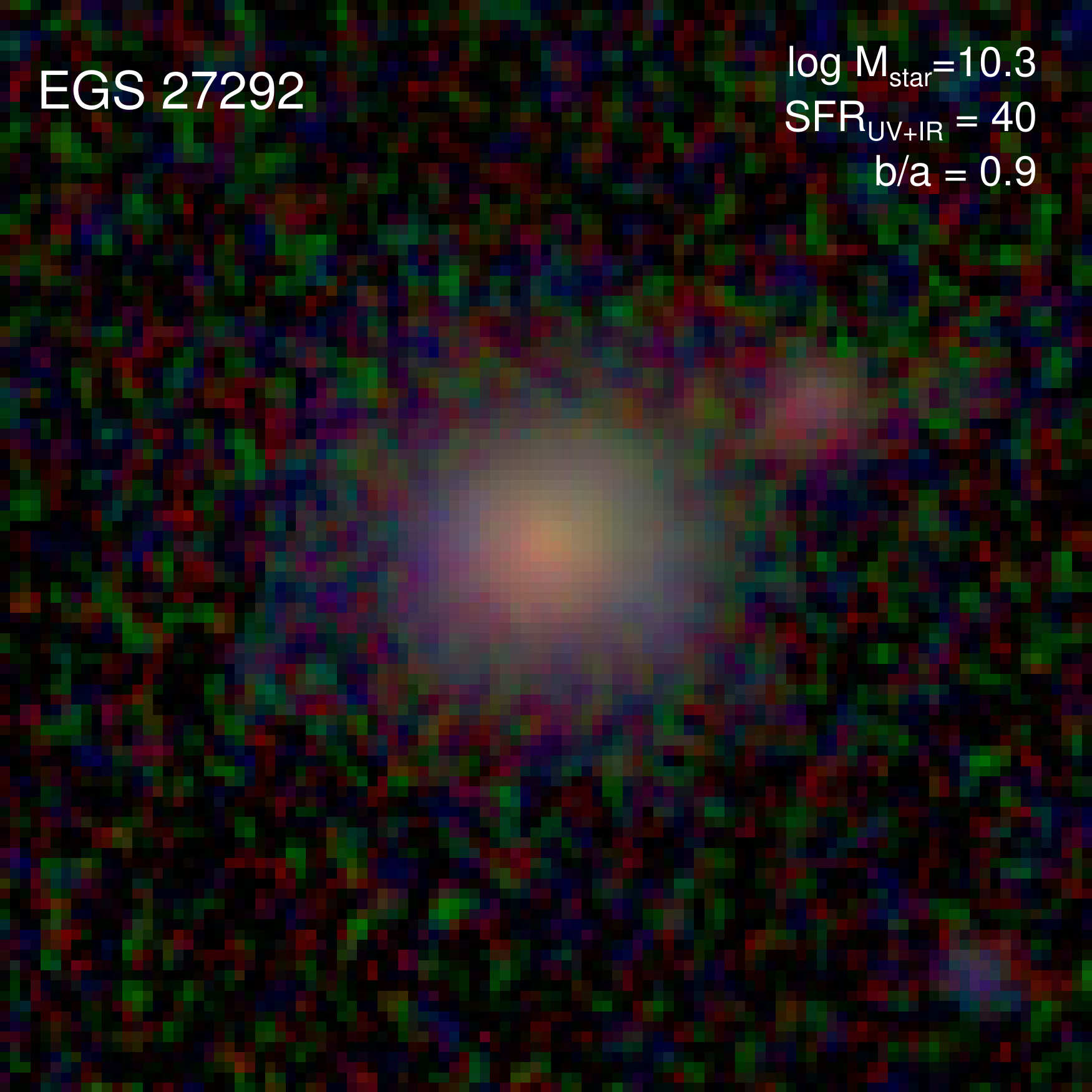}%
}%
\hfill
\subfigure[EGS 13622]{%
\includegraphics[width=0.25\linewidth]{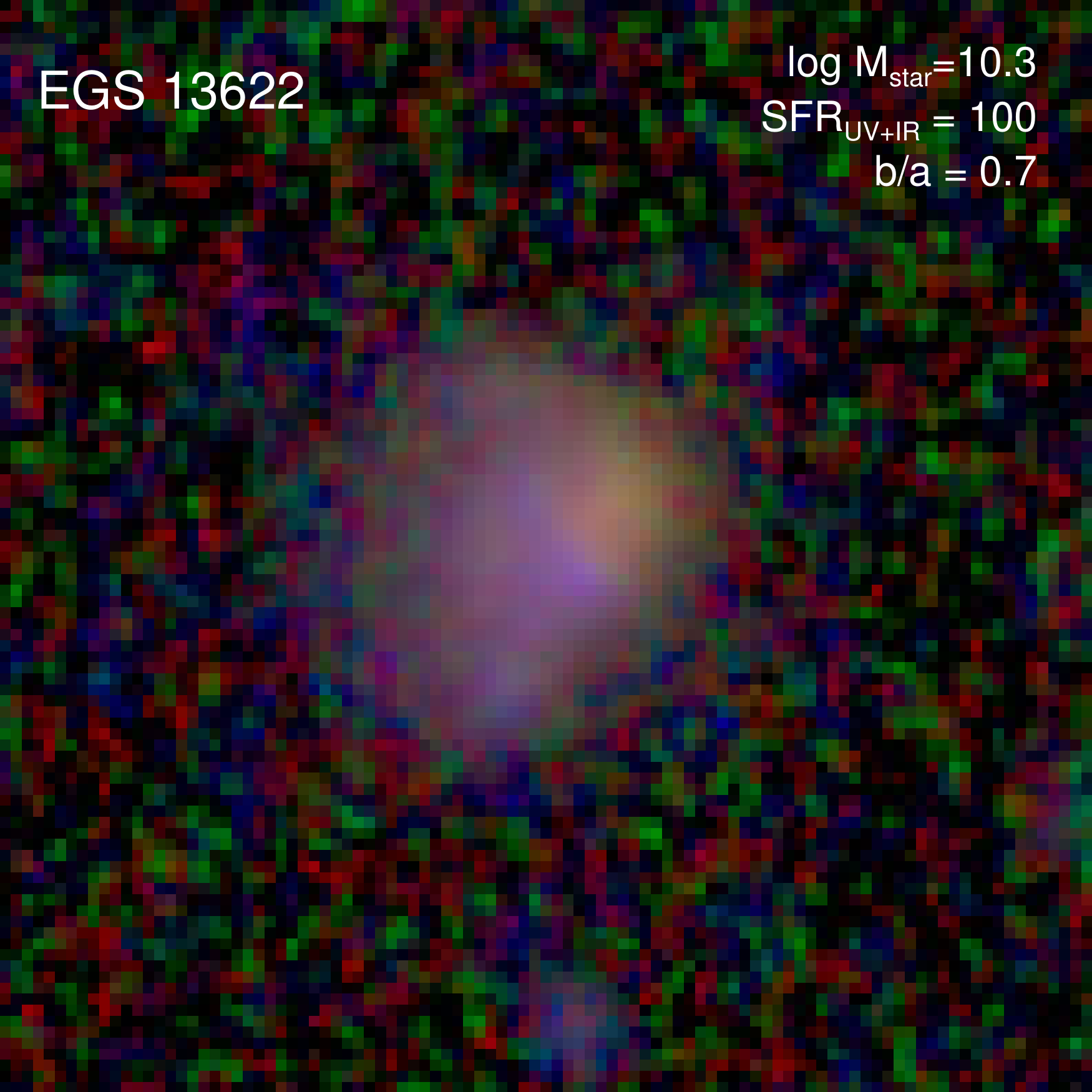}%
}%
\hfill
\subfigure[GDS 25246]{%
\includegraphics[width=0.25\linewidth]{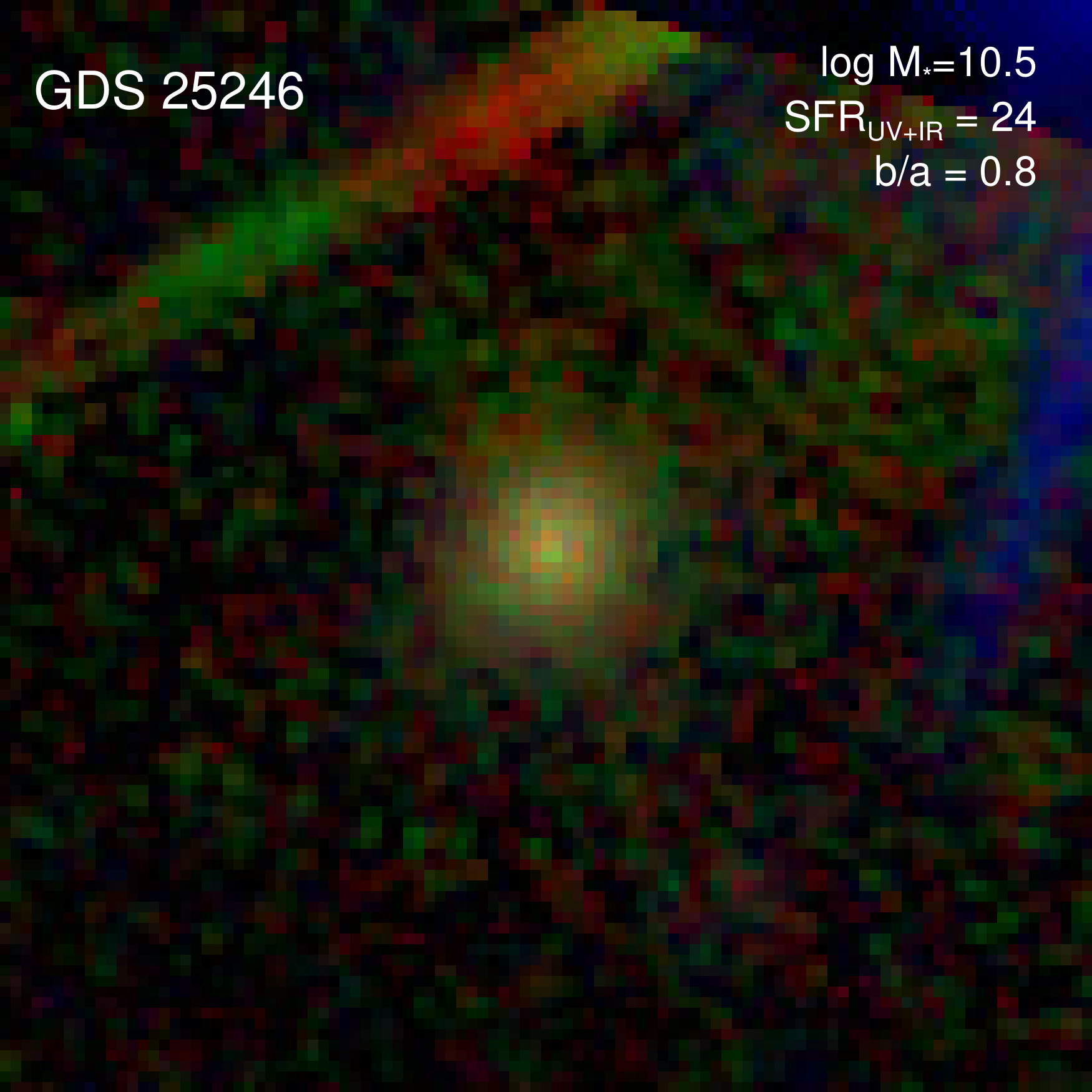}%
}%
\hfill
\subfigure[EGS 9240]{%
\includegraphics[width=0.25\linewidth]{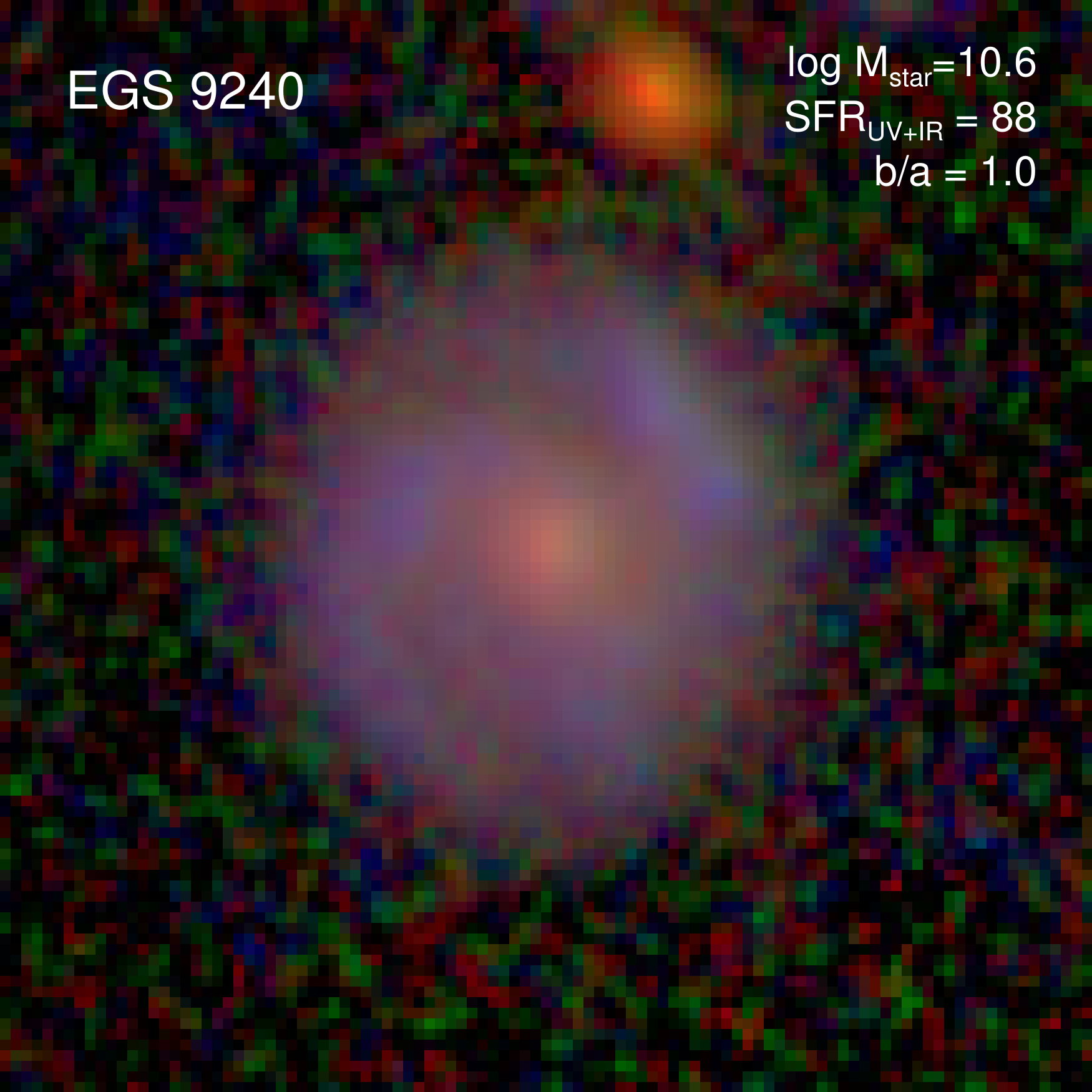}}%
\hfill
\subfigure[EGS 31460]{%
\includegraphics[width=0.25\linewidth]{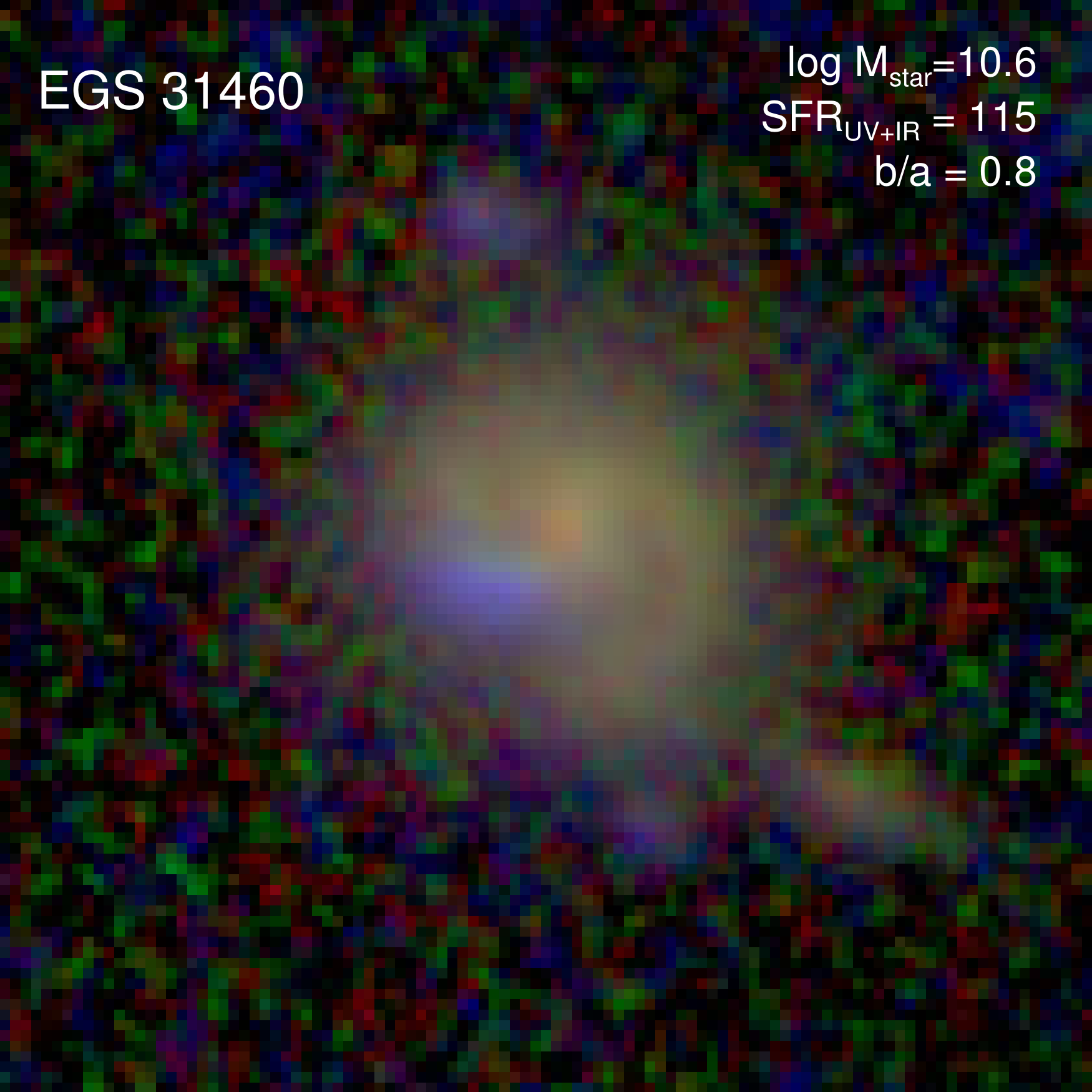}%
}%
\hfill
\subfigure[GDS 19443]{%
\includegraphics[width=0.25\linewidth]{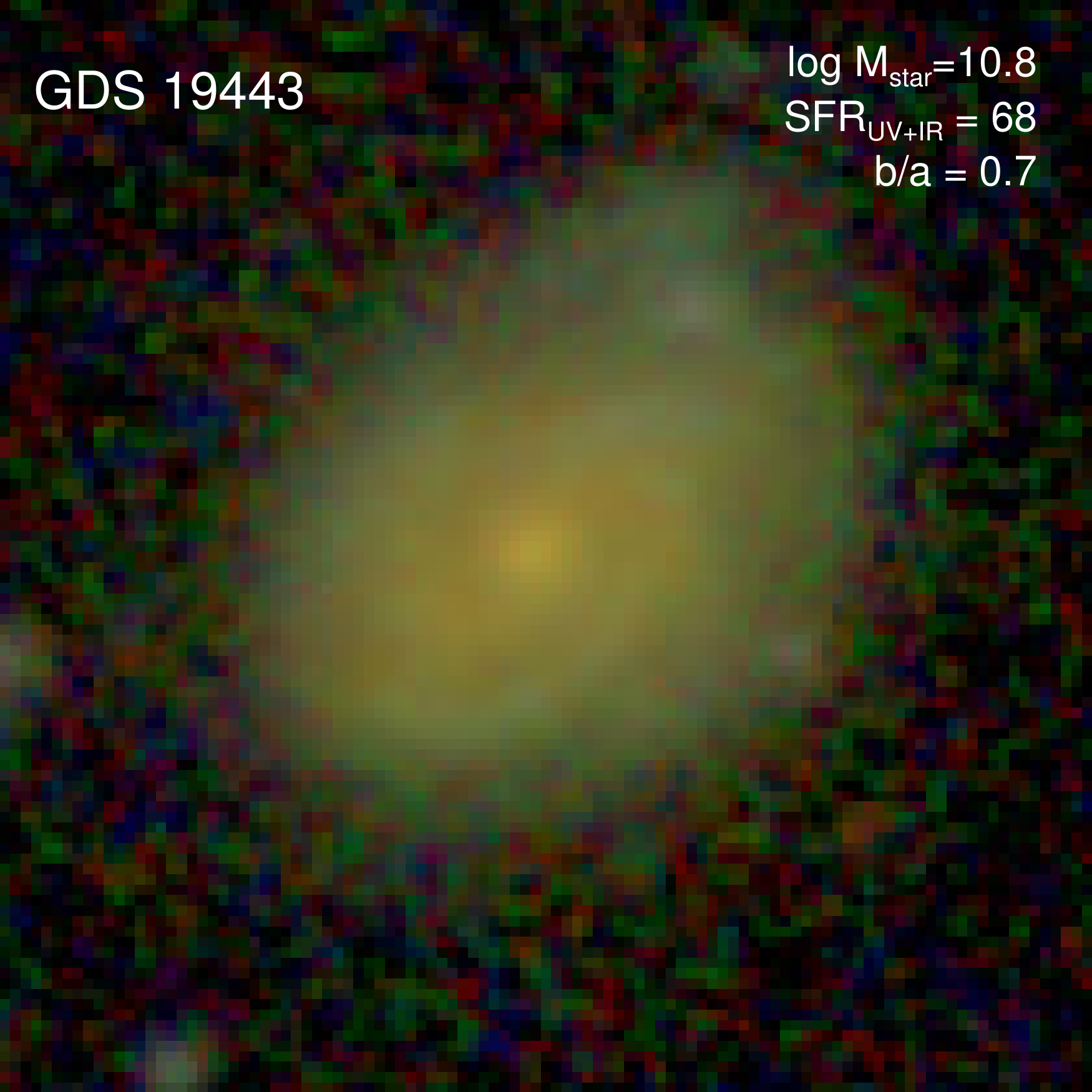}%
}%
\hfill
\subfigure[EGS 25589]{%
\includegraphics[width=0.25\linewidth]{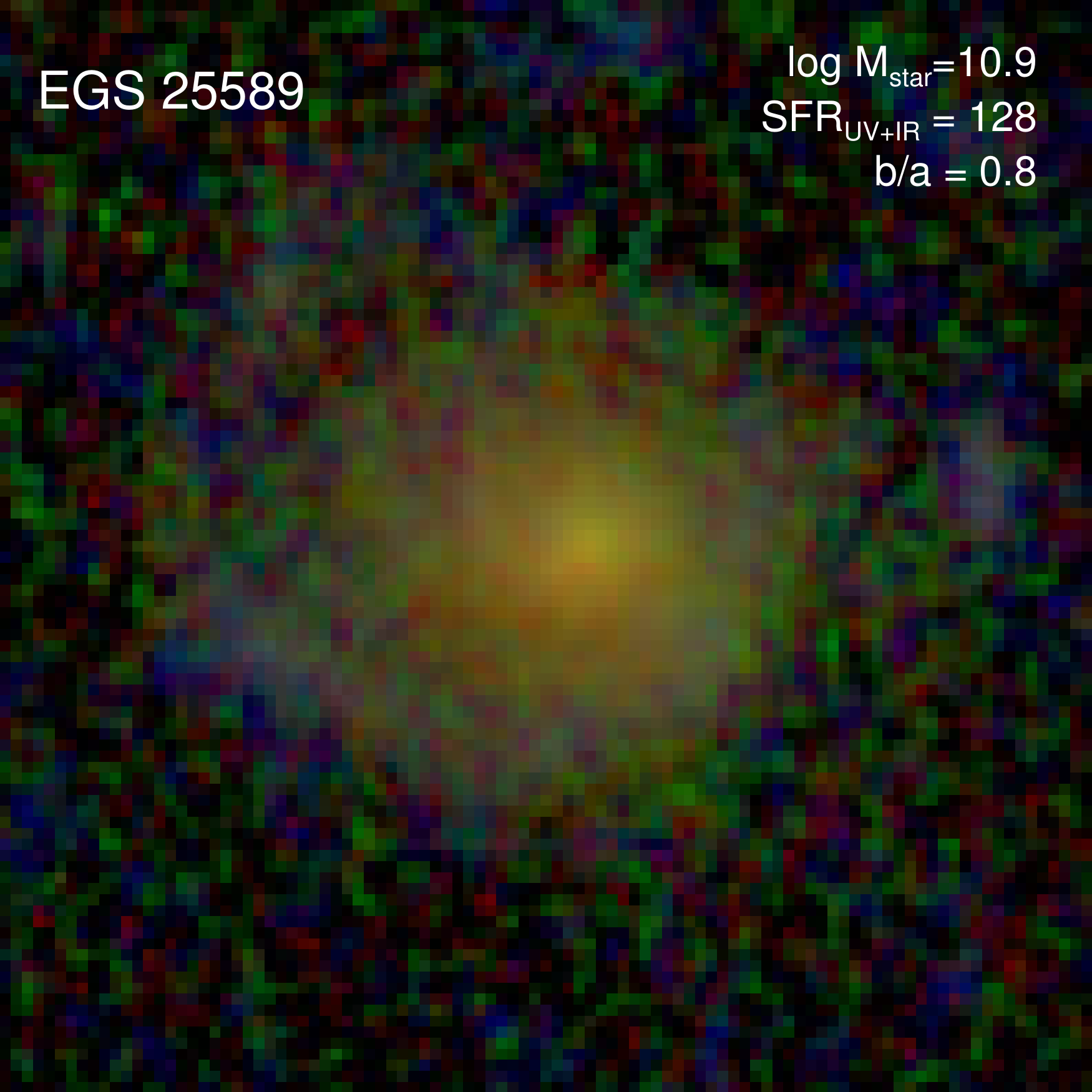}%
}%
\hfill
\subfigure[EGS 12671]{%
\includegraphics[width=0.25\linewidth]{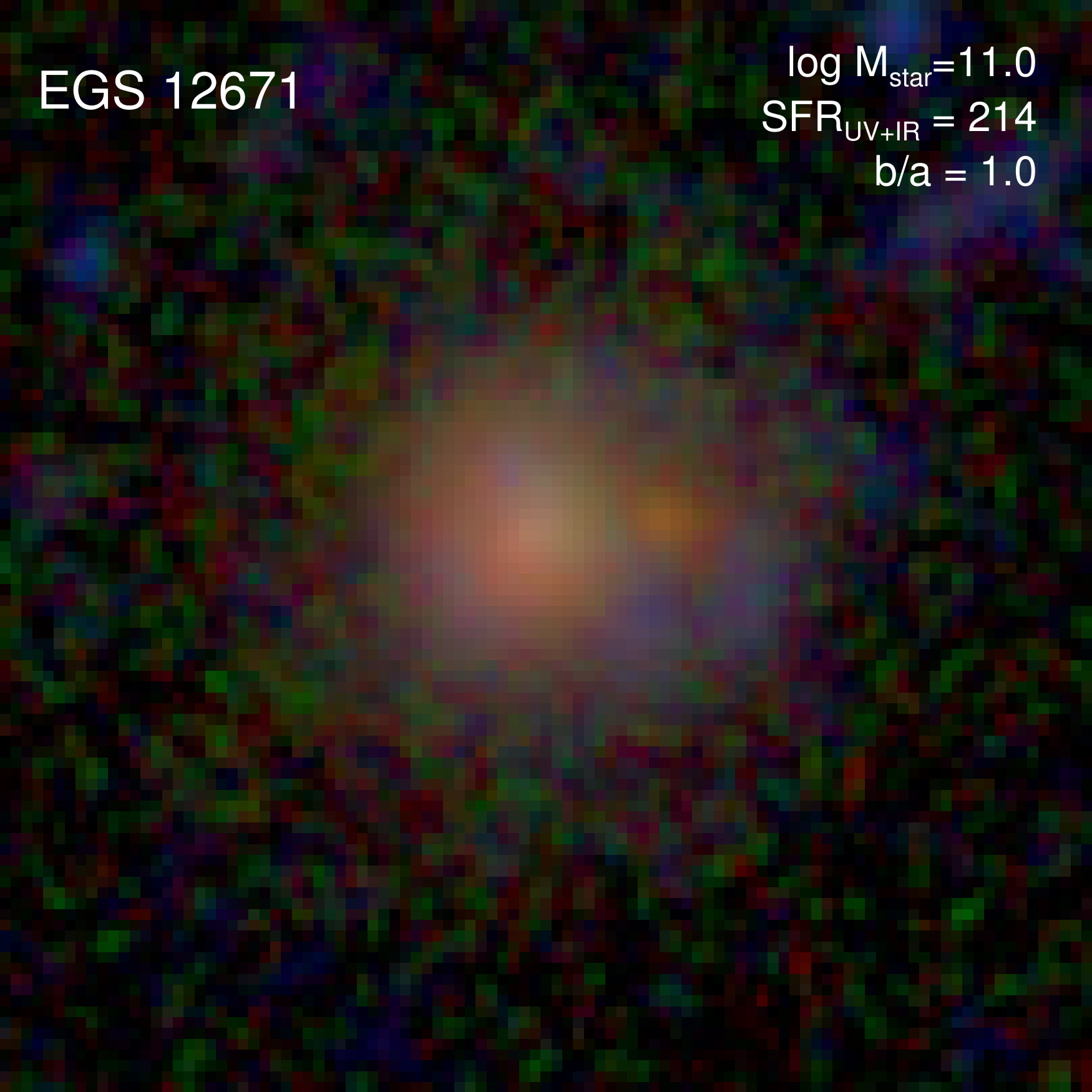}%
}%
\hfill
\subfigure[GDN 17481]{%
\includegraphics[width=0.25\linewidth]{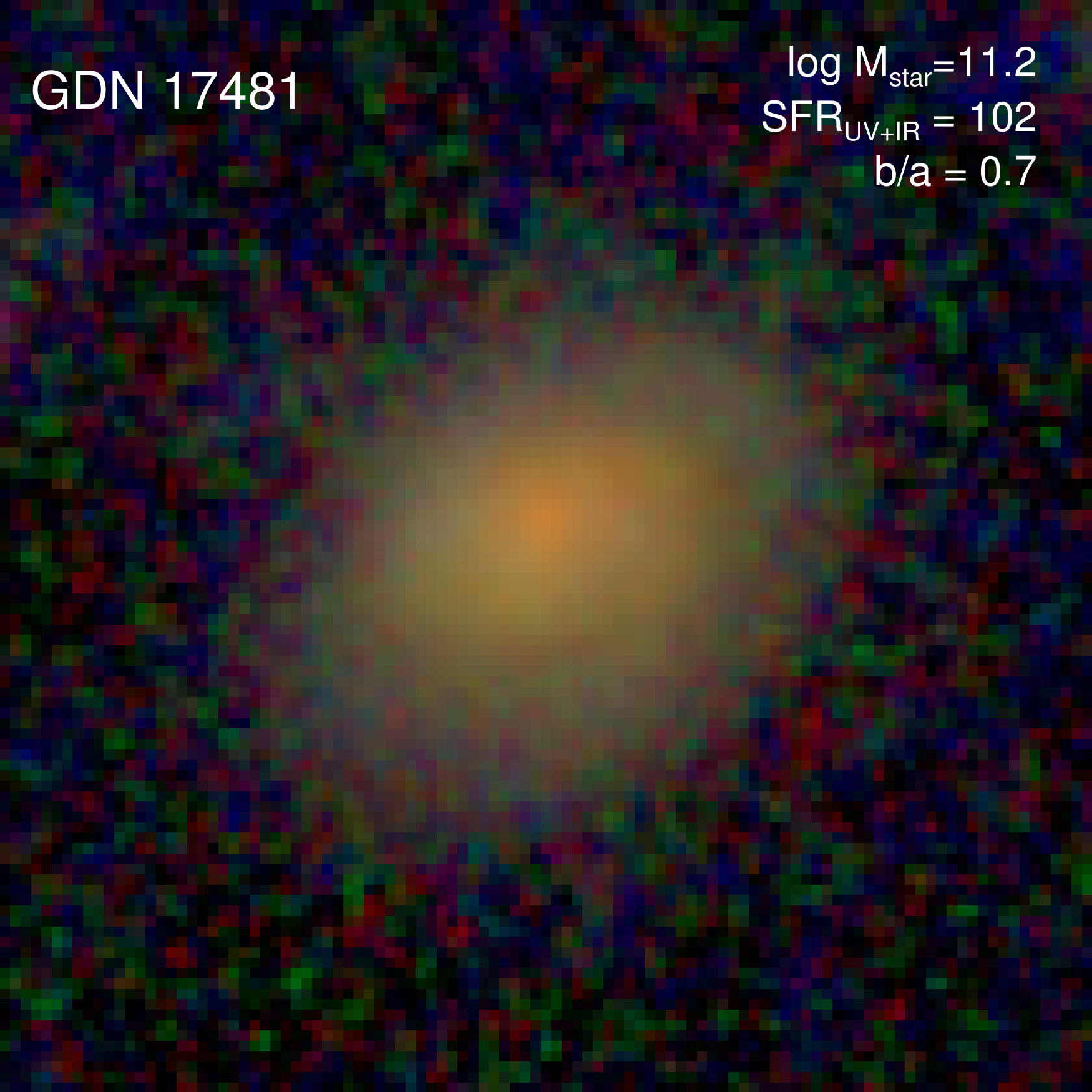}%
}%
\caption{Similar to Figure~\ref{fig:imageAGN} but showing the pseudo RGB composite HST images of the comparison sample of X-ray undetected star-forming galaxies at $z \sim 1$.}
\label{fig:imageSF}
\end{figure*}

The galaxy properties of the AGN and the star-forming X-ray undetected galaxies are summarized in Tables~\ref{tbl:AGN} \&~\ref{tbl:SF}. The rest-fame pseudo RGB images of the two samples are shown in Figure~\ref{fig:imageAGN} and Figure~\ref{fig:imageSF}. Pseudo-color images are created by simply combining three high-resolution HST ACS/WFC3 band cut-out images \citep[e.g.,][]{Koekemoer+11} that have central wavelength closest to R (700nm), G (546.1nm) and B (435.8nm) after correcting for redshift (i.e., $\lambda/(1+z),\,z \sim 1$). The images are normalized and combined with the ratio of 1(R):6(G):3(B).

\begin{deluxetable*}{lrrrcccccccccc} 
\tablewidth{0in}
\tabletypesize{\footnotesize}

\tablecaption{$z \sim 1$ AGN sample \label{tbl:AGN} }
\tablehead{ \colhead{Field} & \colhead{ID} & \colhead{RA} & \colhead{Dec.} & \colhead{$z$} &  \colhead{log L$_\mathrm{X}$}  & \colhead{log M}  & \colhead{SFR} & \colhead{b/a} & \colhead{B} & \colhead{V} & \colhead{SNR} & \colhead{t$_{\rm obs}$} \\
\colhead{} & \colhead{} & \colhead{(degree)} & \colhead{(degree)}&  & \colhead{(erg s$^{-1}$)}  & \colhead{(M$_\odot$)}  & \colhead{(M$_\odot$yr$^{-1}$)} &  \colhead{} & \colhead{mag}  &  \colhead{mag}  & \colhead{(per {\AA})}  & \colhead{(hr)}}\\

\startdata 
EGS   &       10518 &    214.877075   & 52.819477   & 1.19500       &    42.4   & 10.8  & 66  & 0.4 & 24.6 & 24.1 & 3 & 6\\
EGS   &       30572  &   214.671600   &   52.773415  & 1.48599    & 44.2   & 10.5 & 163 &  0.6\tablenotemark{a} &23.9 & 23.1& 8 & 9\\
GDN  &       17041 &   189.282730    &  62.268250   &   0.93573     & 42.2   & 11.1  & 54   & 0.8  & 23.8 & 22.6 & 3  & 9 \\
GDN  &       17389  &  189.282242    & 62.271099    &   0.93971     &  42.2  & 11.1  & 35   & 0.7  & 23.1 & 23.0 & 3  & 9\\
GDN  &        1878\tablenotemark{b}  &  189.194458     & 62.142590    &   0.97115      & 41.9  & 11.0  & 115  & 0.9 &  22.9& 22.1 & 12  & 9\\
GDN  &       12523  &  189.193115    &  62.234634   &   0.96023    & 43.9 & 10.9 & 35 & 0.9  &23.8 & 22.6& 10& 11\\
GDN  &         6274  &  189.077515    & 62.187534    &  1.01933  &  44.1  & 10.7 & 146  & 0.8 & 22.1 & 21.5 & 16 &   0 \\
 GDS   &      21627  &  53.045460    & -27.728624     &   0.99829      &  42.8  & 11.0  & 45   & 0.6\tablenotemark{a} & 23.1 & 22.6 & 3  & 9\\
 GDS   &      7837    & 53.107758     & -27.838812     &   1.09525      & 41.7  &  10.4  & 48   & 0.7 & 23.2 & 22.7 & 9 &  12 \\ 
 GDS   &      15870 &  53.065838     & -27.775131     &   1.02060      & 41.9  & 10.2  & 31   & 0.4 & 23.7 & 23.4 & 10  & 9\\
 GDS   &       23803   & 53.150719   &  -27.716194    &   0.96708      & 42.2  & 10.1  & 68  &  0.6 & 22.3 & 22.1 & 7 & 5\\
 GDS   &       20168 &   53.150126     &  -27.739926      &   1.03670       &  41.8  & 10.0   & 48   & 1.0 & 23.2 & 22.9 & 3 & 5\\
\enddata

\tablenotetext{a}{has a bad or suspicious GALFIT flag}
\tablenotetext{b}{has a strong \ion{Ne}{5} emission}
\end{deluxetable*}

\begin{deluxetable*}{lrrrcccc}
\tablewidth{0in}
\tabletypesize{\footnotesize}
\tablecaption{$z\sim 1$ inactive star-forming comparison sample \label{tbl:SF}}

\tablehead{ \colhead{Field} & \colhead{ID} & \colhead{RA} & \colhead{Dec.} & \colhead{$z$} & \colhead{log M}  & \colhead{SFR} & \colhead{b/a}\\
 \colhead{} & \colhead{} & \colhead{(degree)} & \colhead{(degree)} & \colhead{}  & \colhead{(M$_\odot$)}  & \colhead{(M$_\odot$ yr$^{-1}$)} & \colhead{} }\\

\startdata
 EGS & 9240  &           215.054153 &    52.937103  & 1.03030   & 10.6   &   88   &  1.0 \\
 EGS &13622 &         214.953293   &   52.890411    & 1.39770   & 10.3    &  100   &  0.7 \\
 EGS &31460 &         214.708572   &   52.791351   &  0.95830  & 10.6   &   115  &  0.8 \\
 EGS & 25589 &         214.655823  &   52.738495   & 1.39700   & 10.9     &  128  &  0.8 \\
 EGS & 15131 &        215.095734   &  52.999435    & 1.11690   & 10.0    &  45   &  0.8\\
 EGS  & 27292 &        214.931488  &   52.945190  & 0.89330  & 10.3    &  40   &  0.9 \\
 EGS & 12671  &        215.110123  &  52.994293   & 1.24070   & 11.0     &  214   &  1.0 \\
 GDN & 17481  &        189.379868 &  62.272289 &  0.97500  & 11.2   &   102   &  0.7 \\
GDN & 11799  &           189.219743    &   62.231889     &  1.35896    & 10.2        &  49       & 0.5 \\
GDS & 25246  &            53.046933   &   -27.690841     &  1.05765    & 10.5        &  24       & 0.8 \\
GDS & 26255  &            53.144206   &   -27.700337     &  1.04180    & 9.8         & 24        & 0.6 \\
GDS & 19443  &            53.086326   &  -27.748261    &  0.96900      & 10.8        & 68        & 0.7\\
\enddata

\end{deluxetable*}

\begin{figure*}
\centering
\subfigure[GDN 1878]{%
\includegraphics[width=0.3\linewidth]{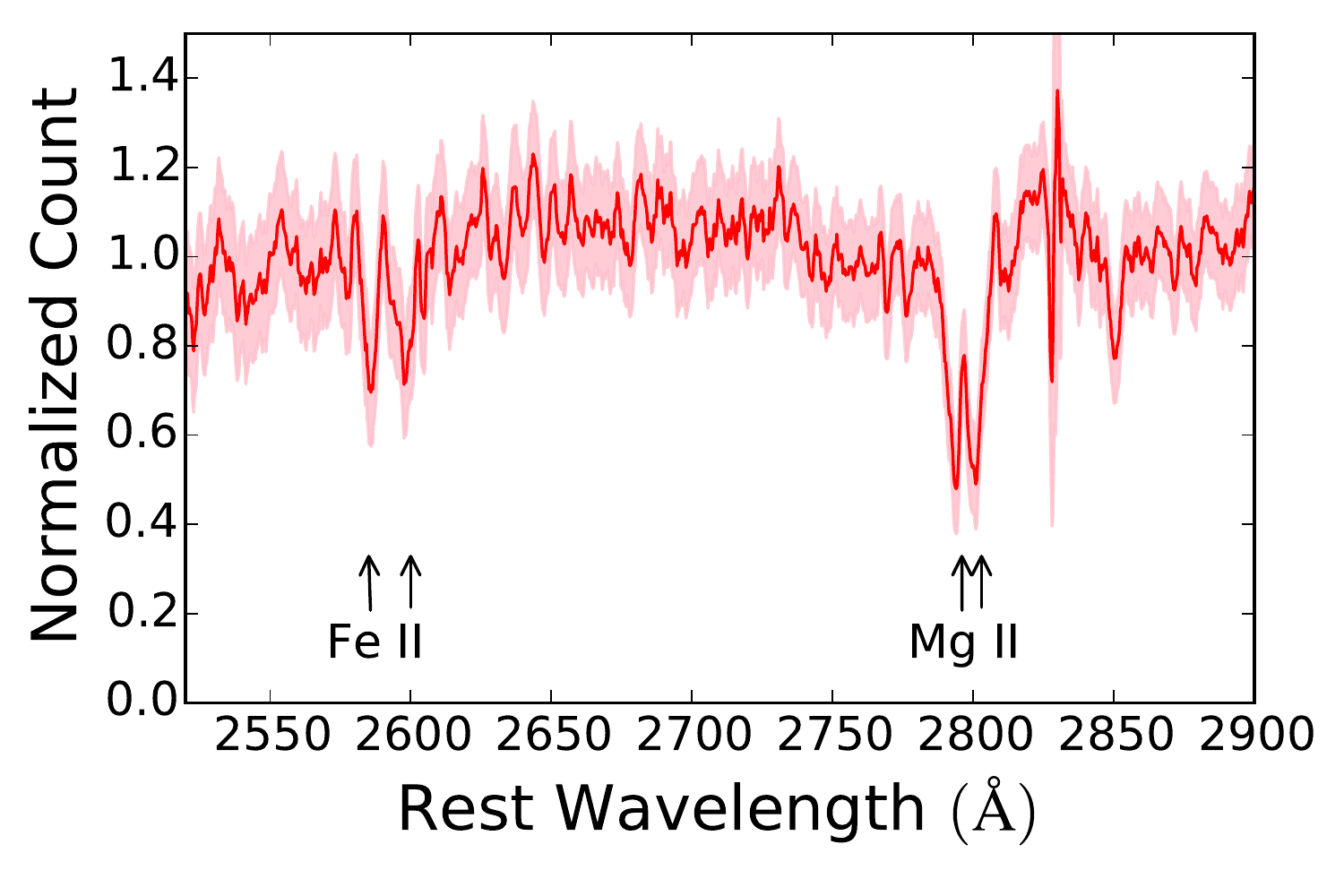}%
}%
\hfill
\subfigure[GDS 23803]{%
\includegraphics[width=0.3\linewidth]{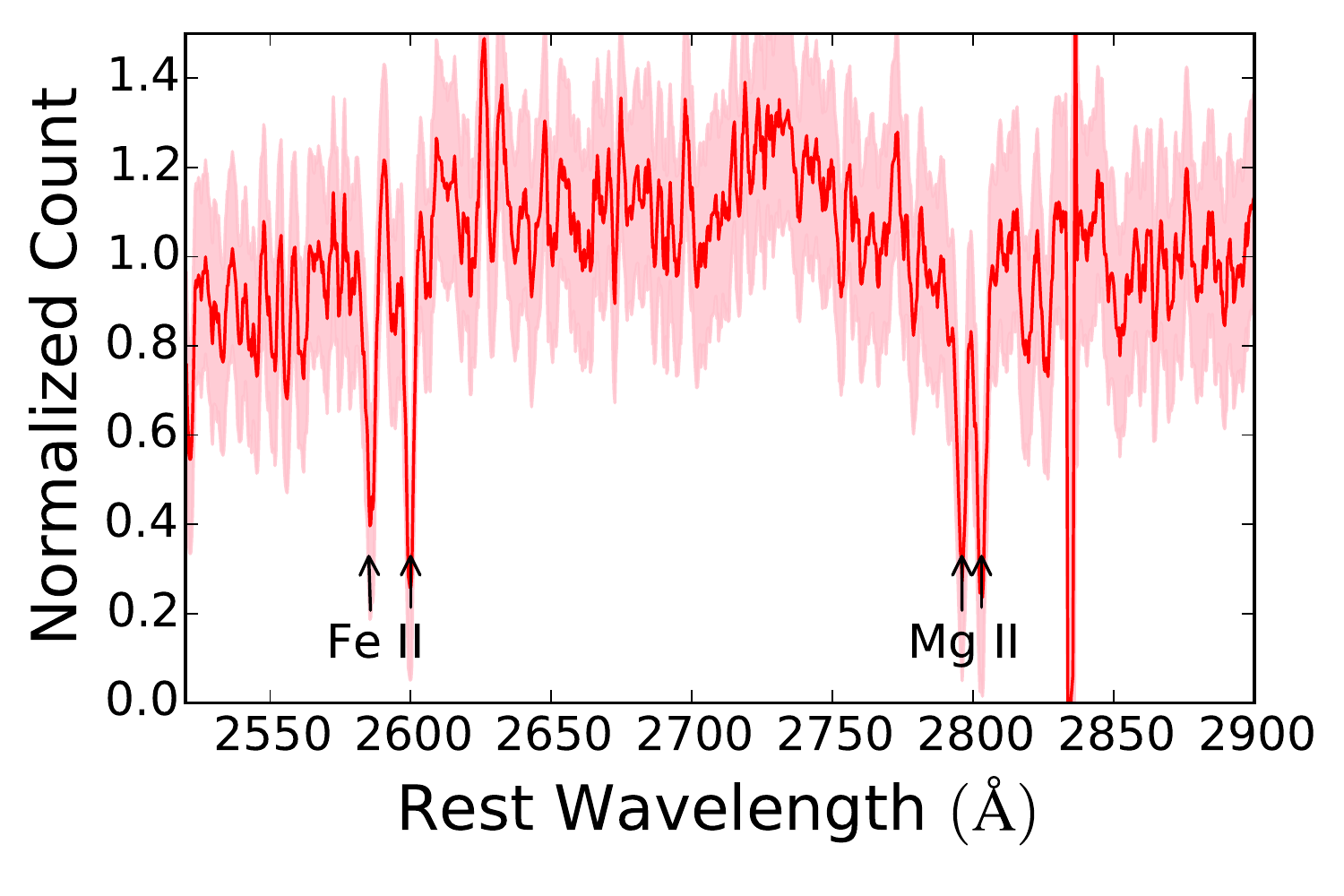}%
}%
\hfill
\subfigure[GDS 15870]{%
\includegraphics[width=0.3\linewidth]{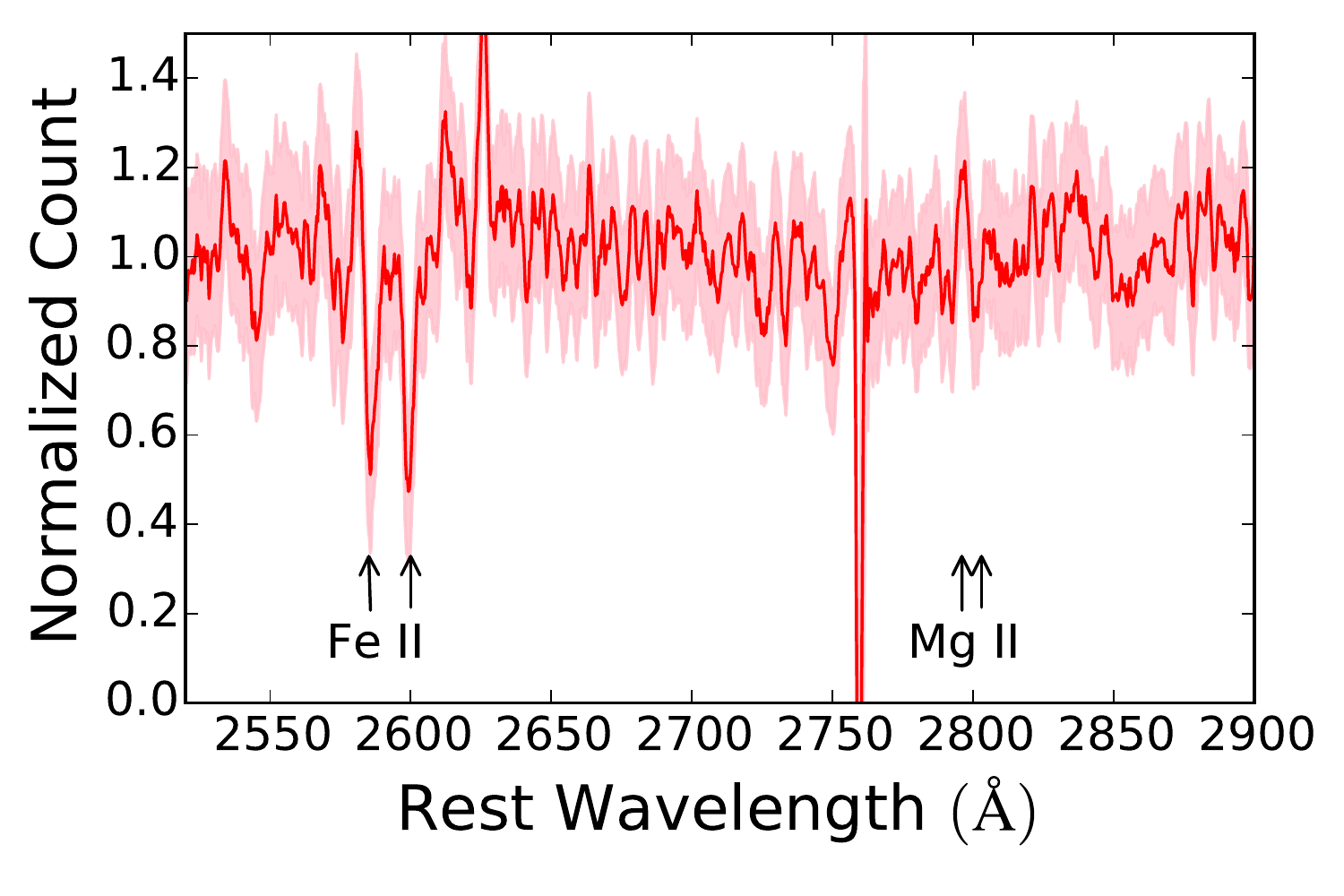}%
}%
\hfill
\subfigure[GDN 6274]{%
\includegraphics[width=0.3\linewidth]{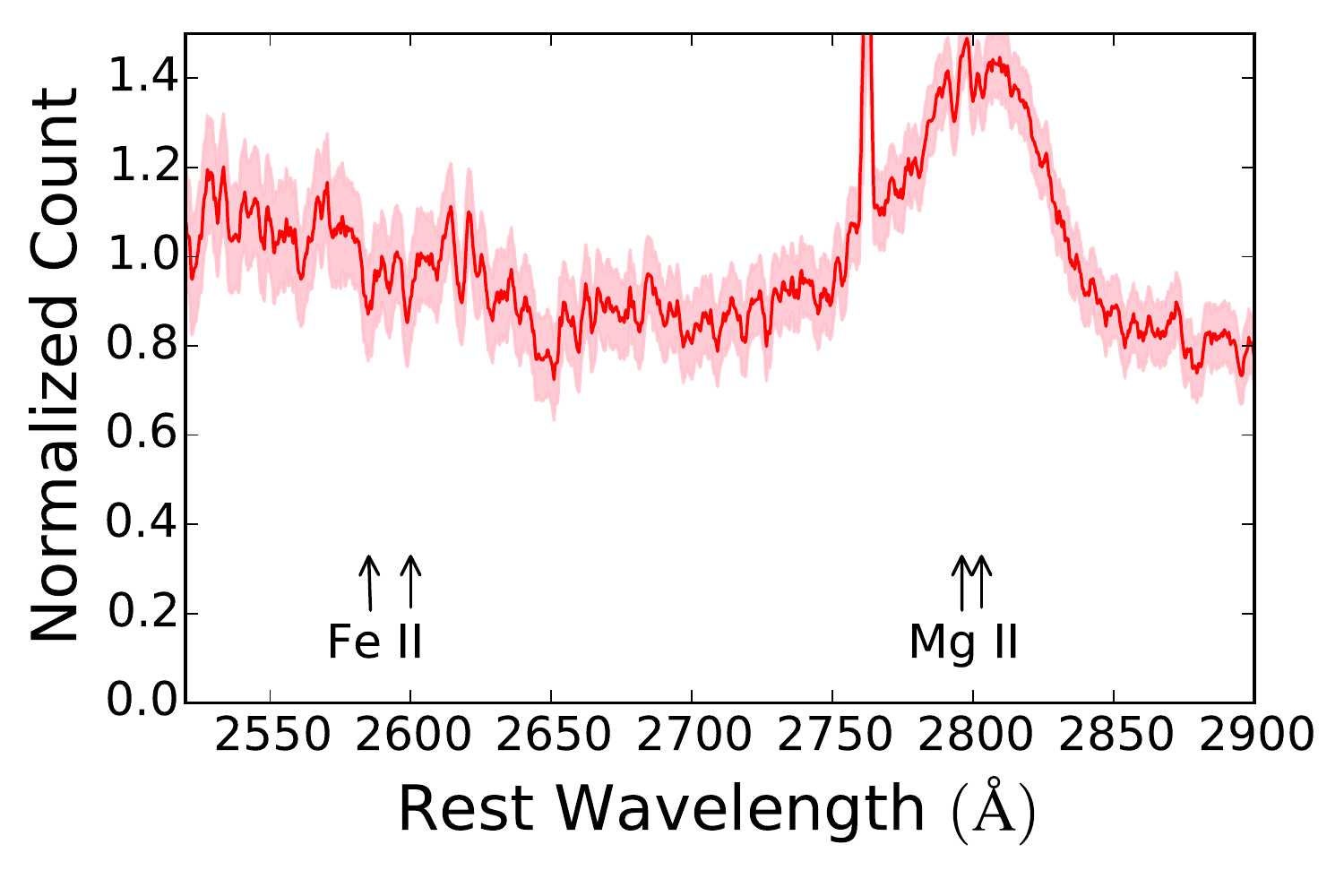}%
}%
\hfill
\subfigure[EGS 30572]{%
\includegraphics[width=0.3\linewidth]{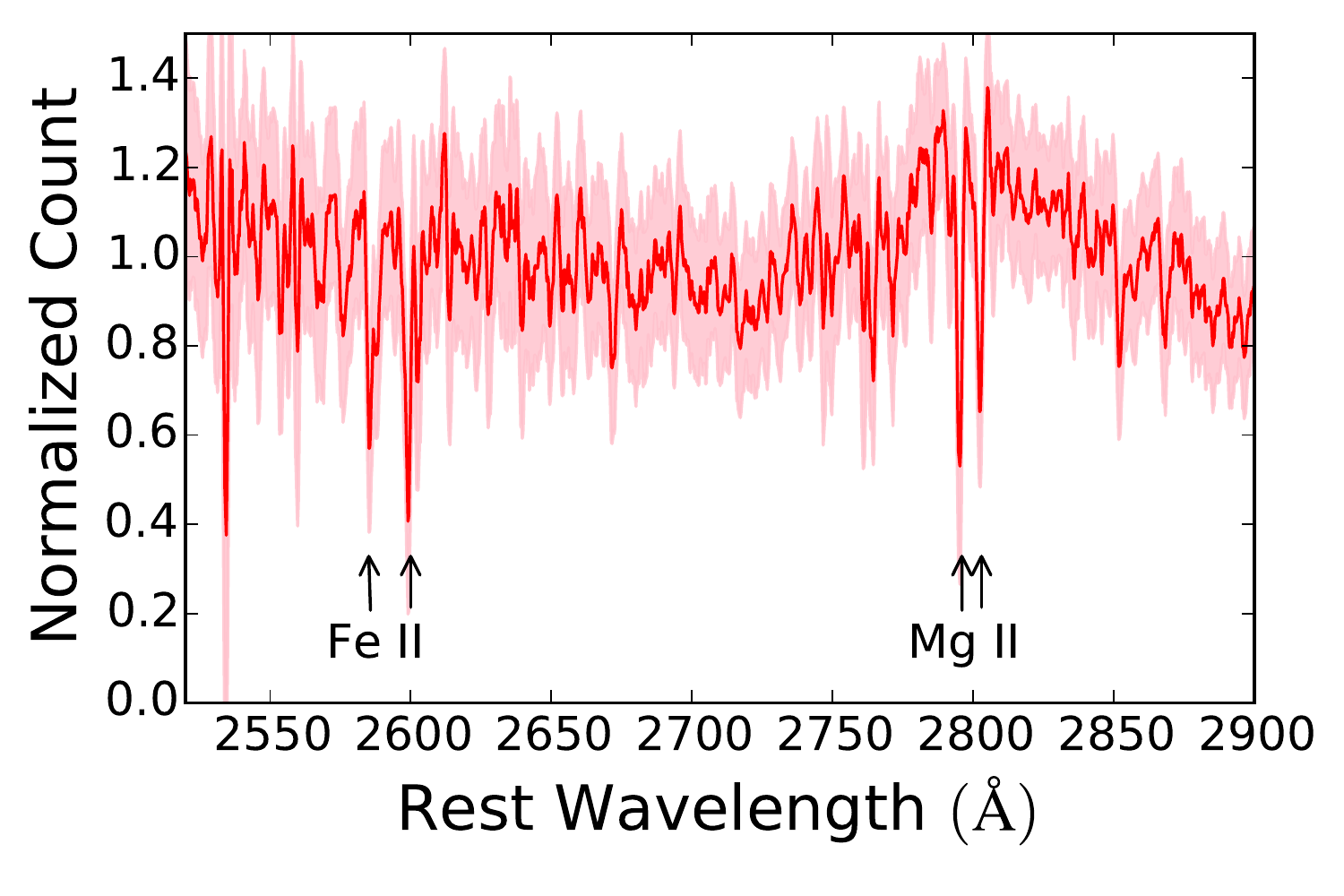}%
}%
\hfill
\subfigure[GDN 12523]{%
\includegraphics[width=0.3\linewidth]{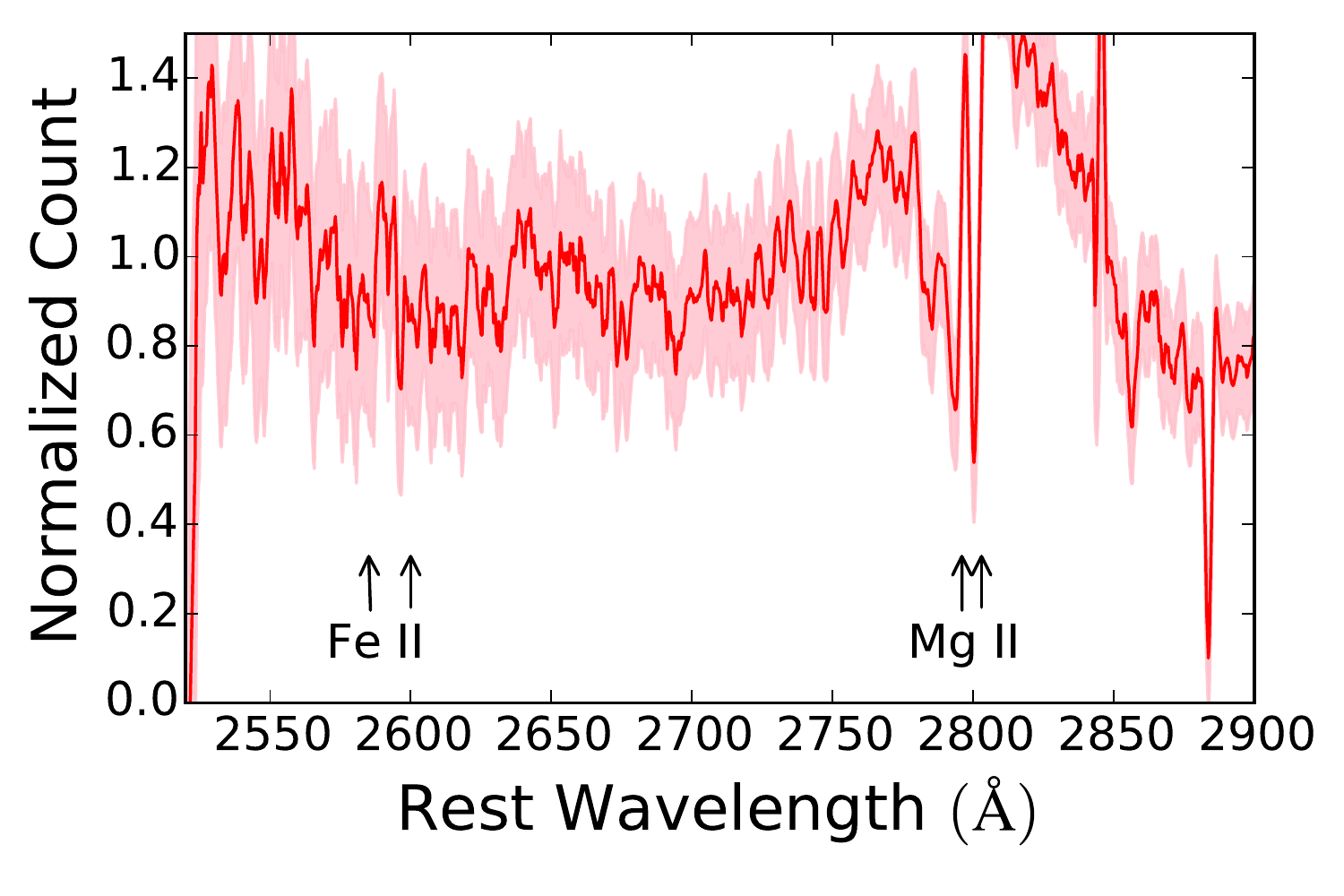}%
}%
\hfill
\subfigure[GDS 7837]{%
\includegraphics[width=0.3\linewidth]{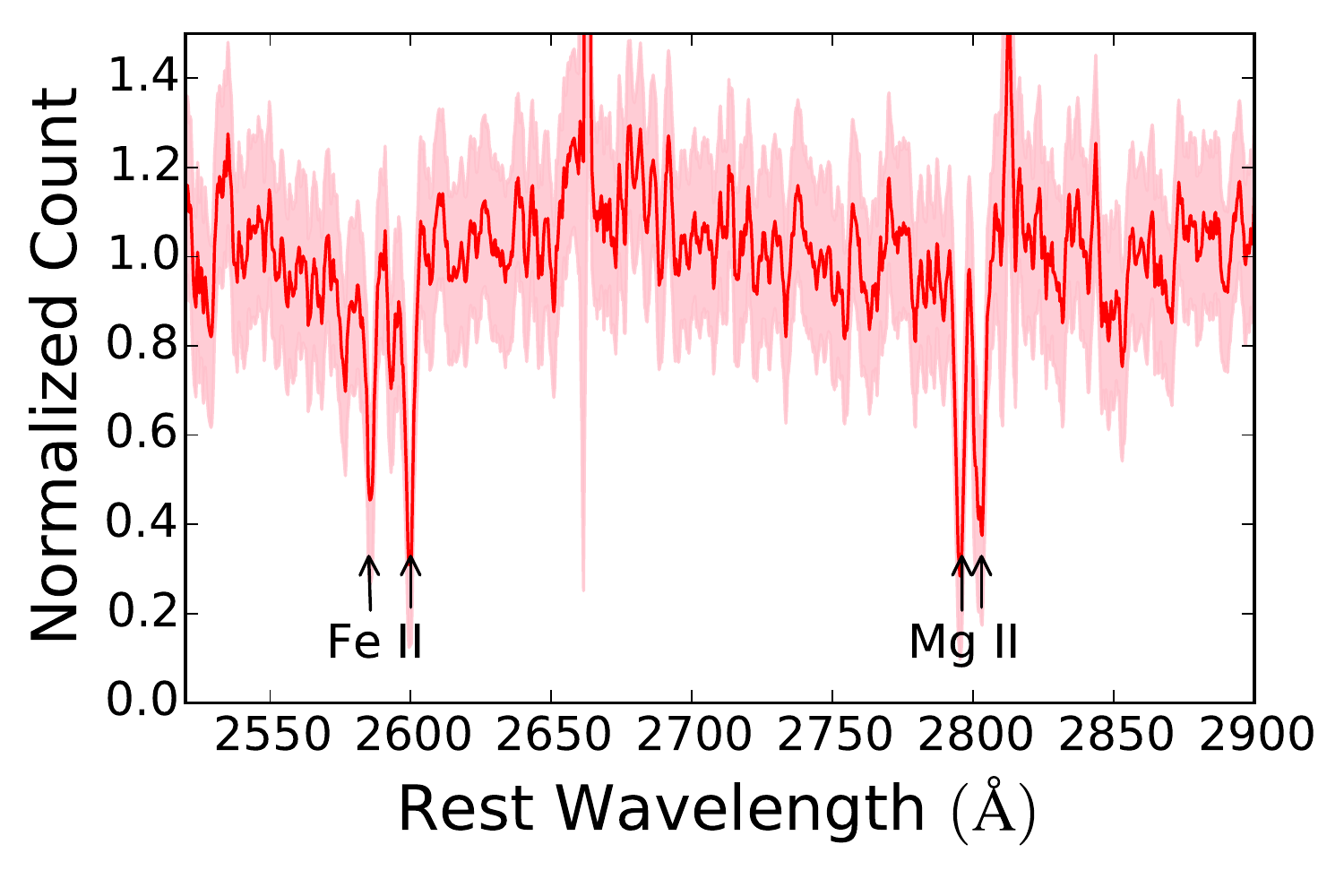}%
}%
\hfill
\subfigure[EGS 10518]{%
\includegraphics[width=0.3\linewidth]{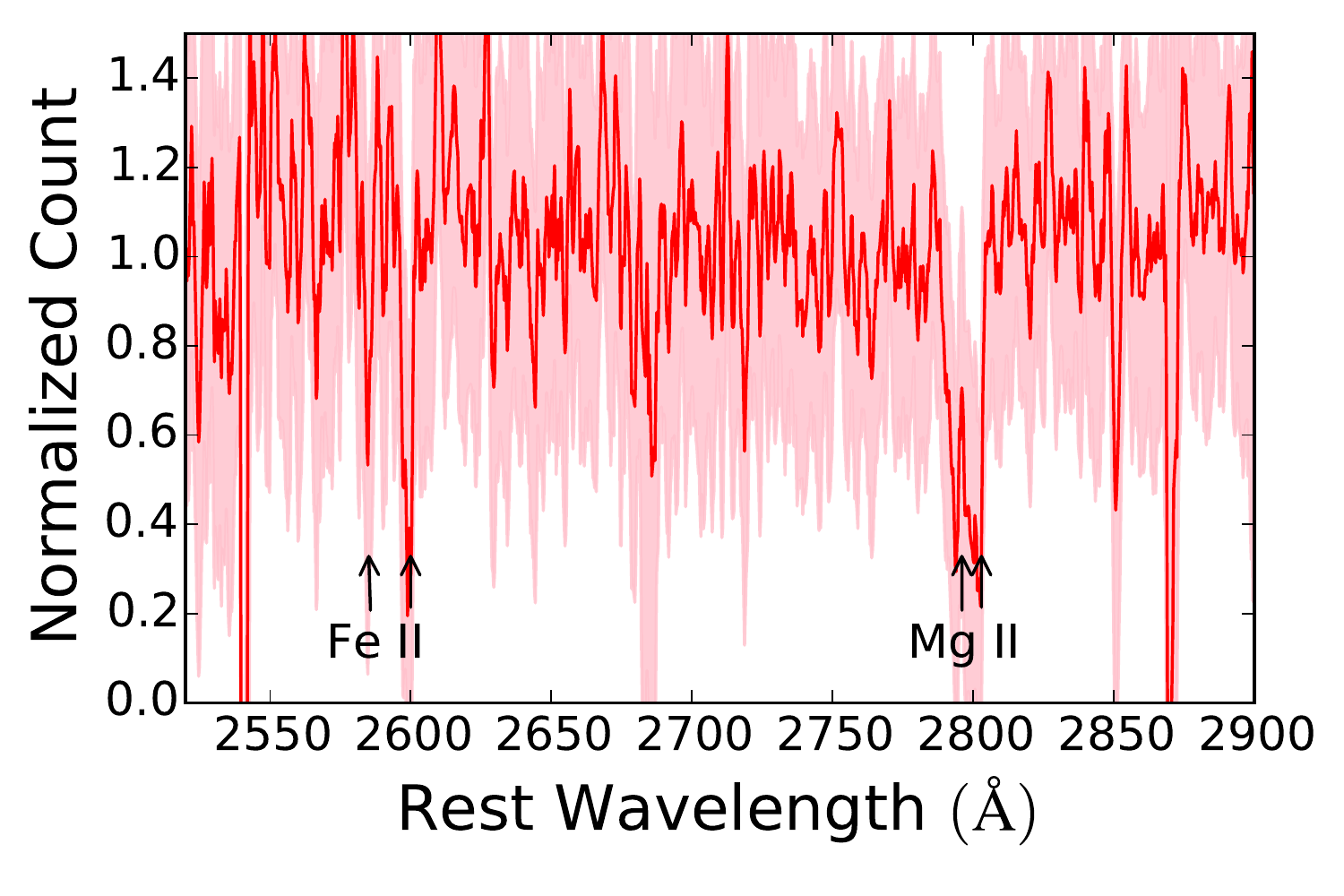}%
}%
\hfill
\subfigure[GDN 17041]{%
\includegraphics[width=0.3\linewidth]{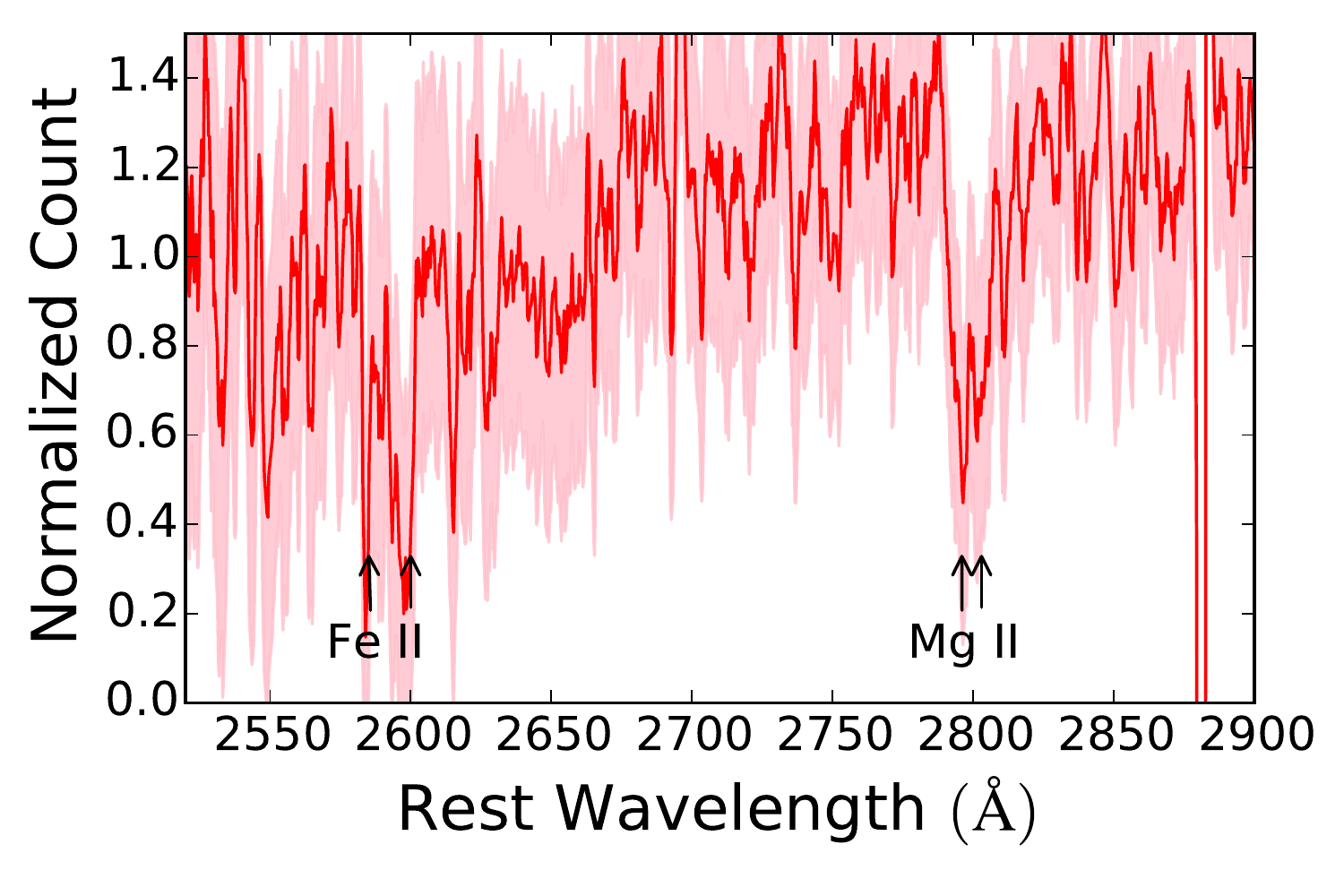}%
}%
\hfill
\subfigure[GDN 17389]{%
\includegraphics[width=0.3\linewidth]{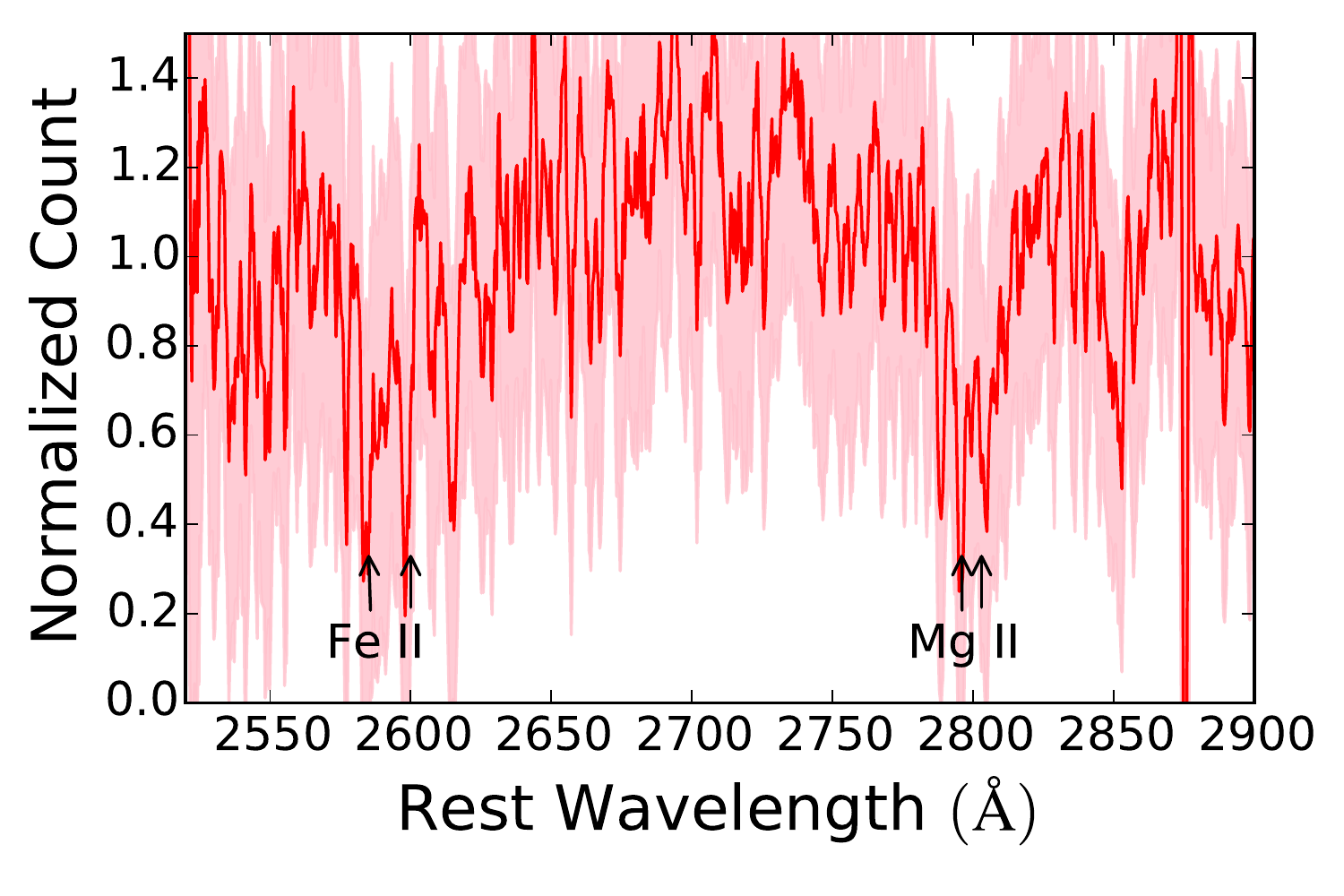}%
}%
\hfill
\subfigure[GDS 21627]{%
\includegraphics[width=0.3\linewidth]{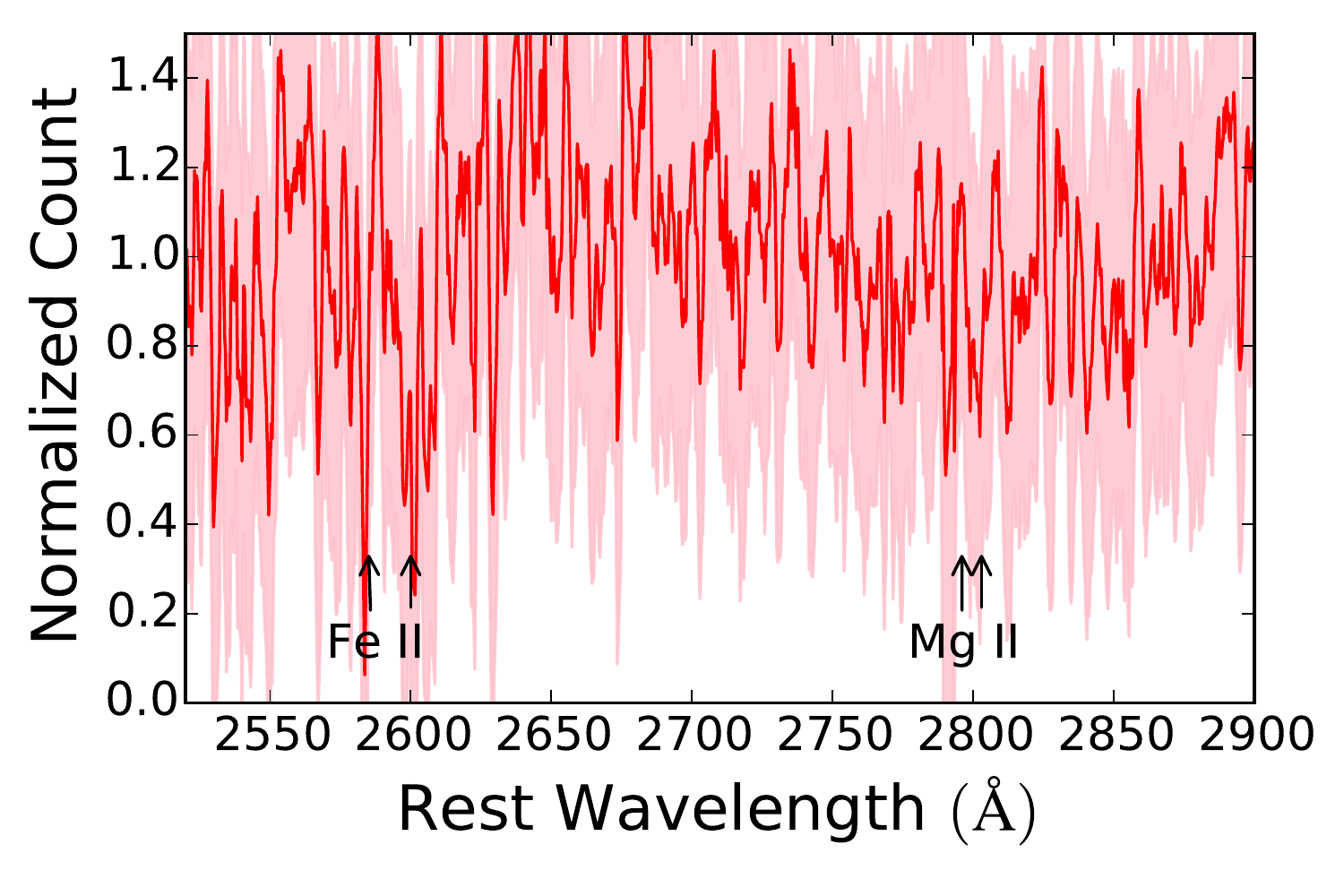}%
}%
\hfill
\subfigure[GDS 20168]{%
\includegraphics[width=0.3\linewidth]{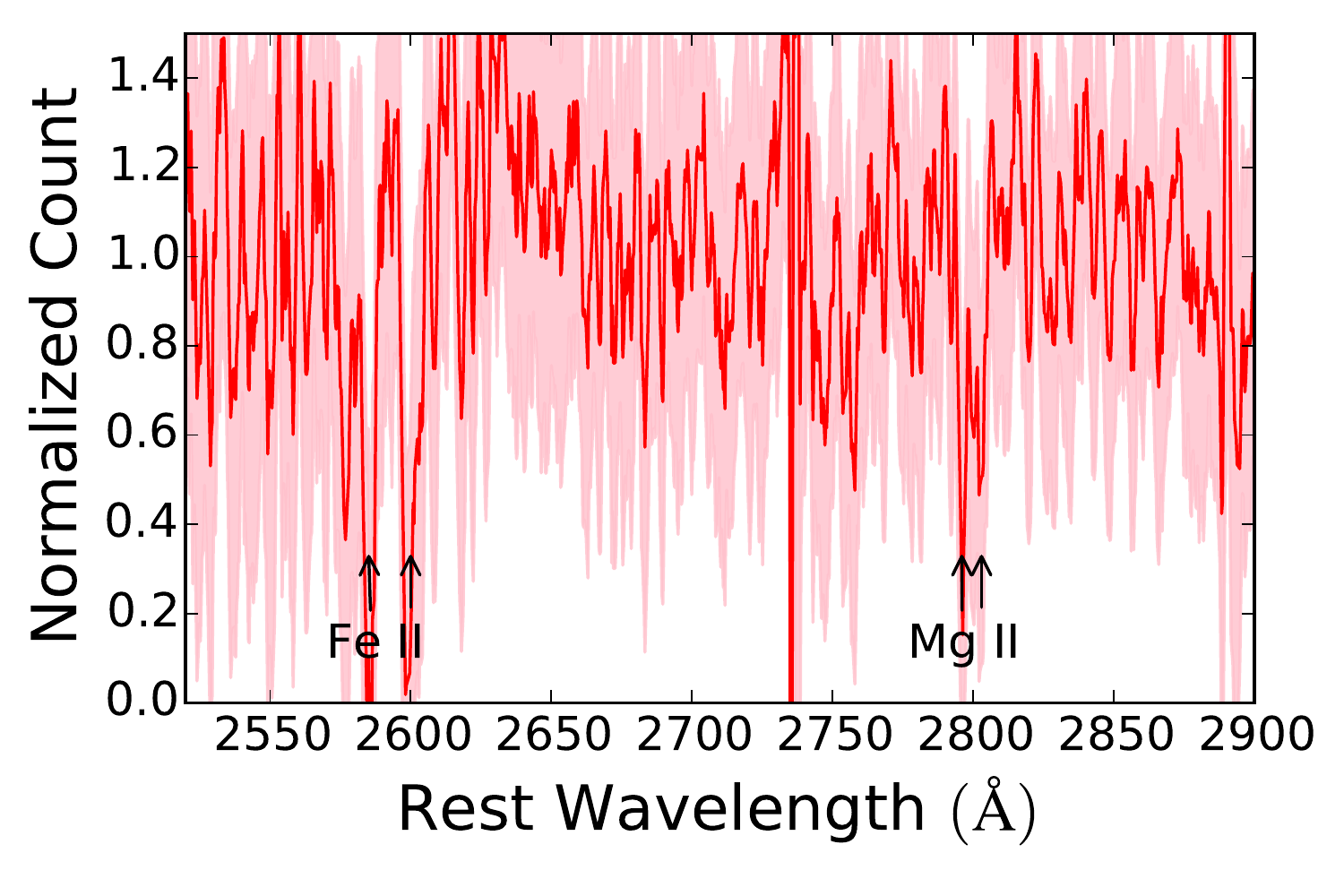}%
}%
\caption{Near UV spectra of the AGN sample at $z \sim 1$. The pink shadings are 1$\sigma$ error bands}. The spectra are boxcar smoothed by 7 pixels (2 {\AA}). Absorption lines of \ion{Fe}{2} and  \ion{Mg}{2} are evident in most AGN. Spectra in the second row are for the three broad-line AGN. The \ion{Fe}{2} absorption are very weak in two of the AGN with broad \ion{Mg}{2} emission. These galaxies have high X-ray luminosities in Figure~\ref{fig:sample}.
\label{fig:samplespec}
\end{figure*}

\section{Analysis \& Results} \label{sec:res}

\subsection{Coadding spectra}

First, multiple exposures of the same object were averaged with inverse-variance weighting per pixel, which minimizes the variance of the weighted average spectrum. Then to coadd the spectra of AGN or their comparison sample, each spectrum was shifted to its rest frame wavelength. The rest-frame spectra were then interpolated on a linear wavelength grid with $\Delta \lambda = 0.3${\AA} bin, which is close to the pixel size of the DEIMOS spectrograph at $z\sim 1$ for our setup. We co-added the observed photon counts at a given wavelength bin for each rest-frame spectrum with inverse-variance weighting.

\begin{figure*}
\centering
\subfigure[All AGN]{%
\includegraphics[width=0.45\linewidth]{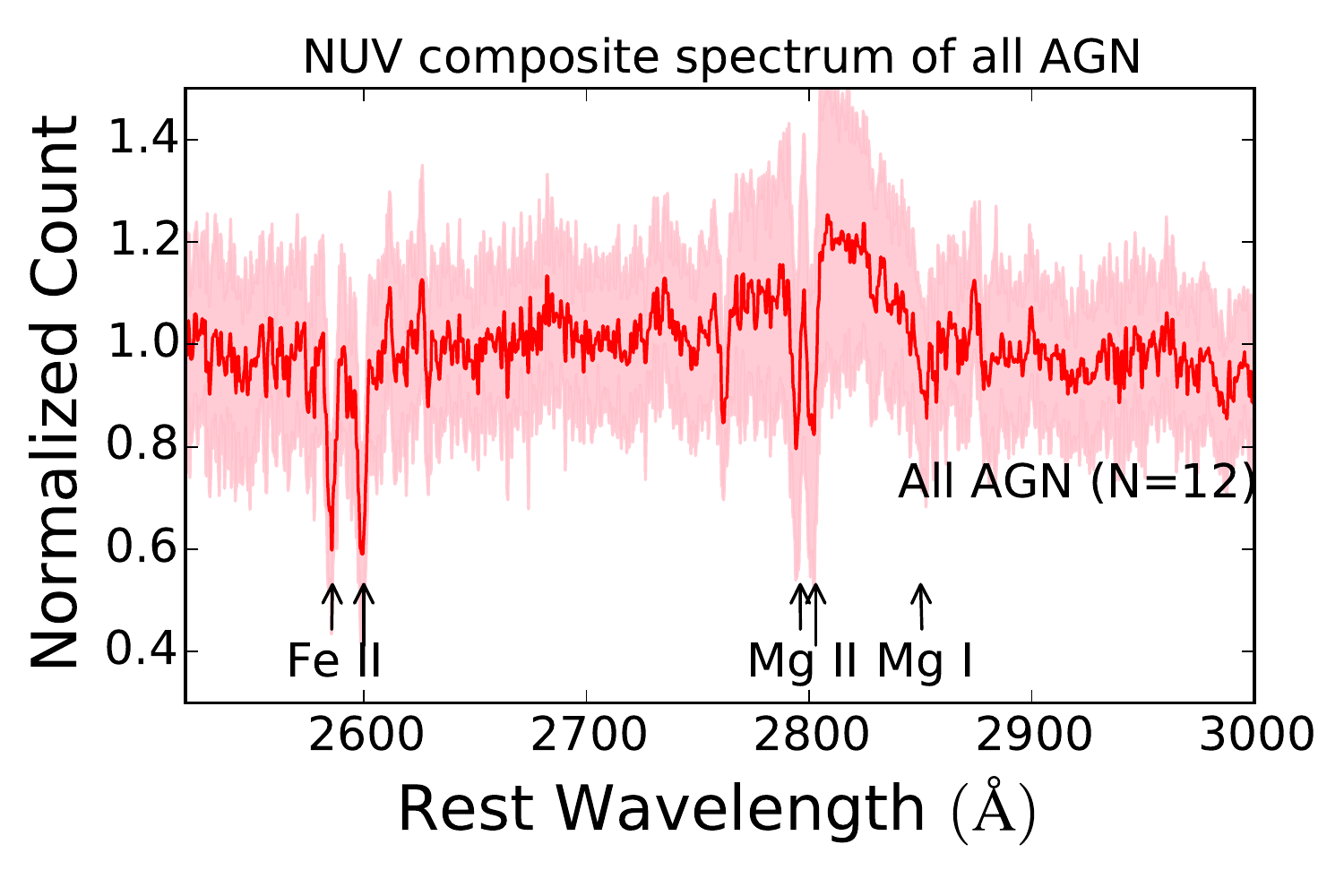}%
}%
\hfill
\subfigure[All narrow-line AGN candidates]{%
\includegraphics[width=0.45\linewidth]{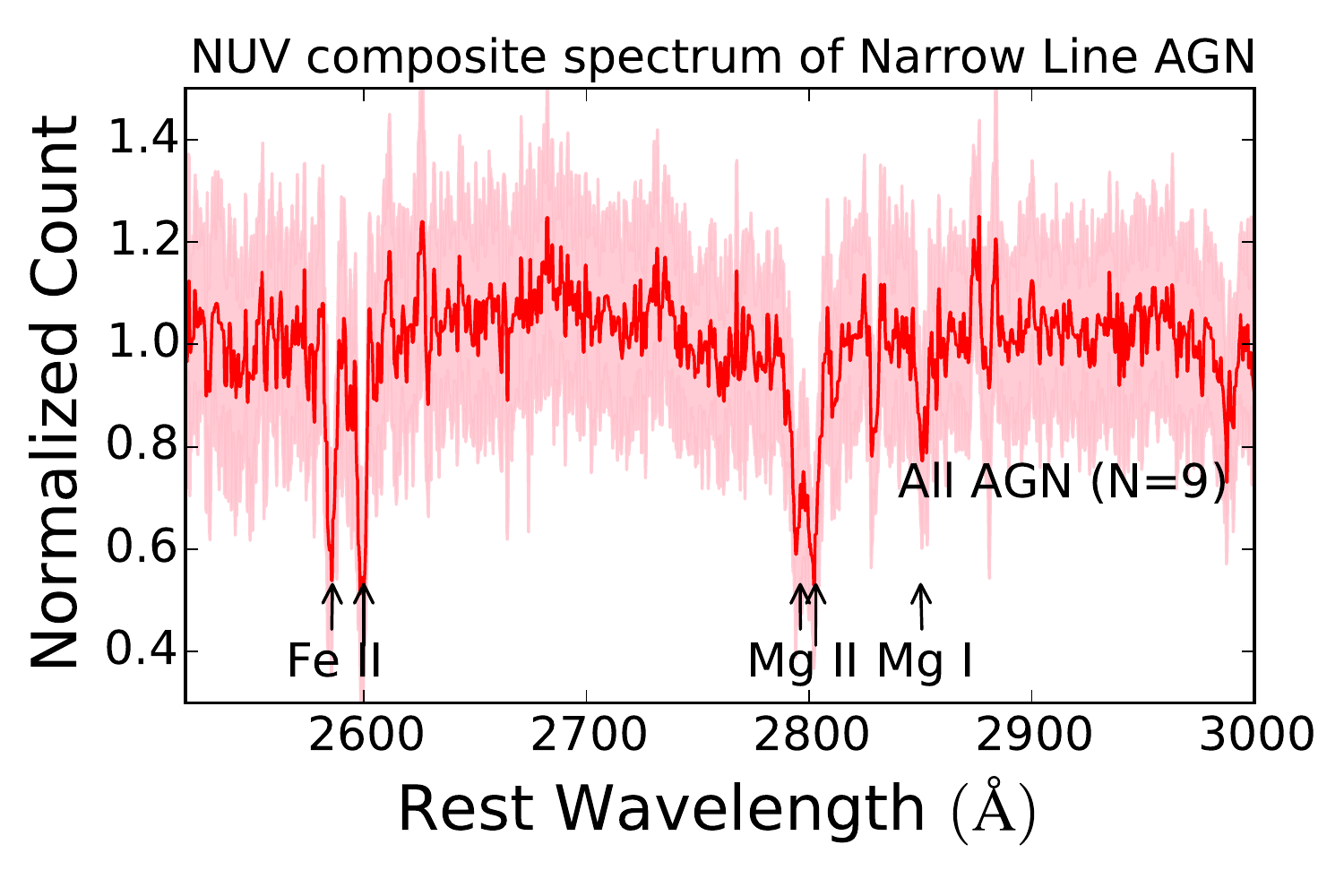}%
}%
\hfill
\subfigure[Narrow-line AGN with robust AGN identification]{%
\includegraphics[width=0.45\linewidth]{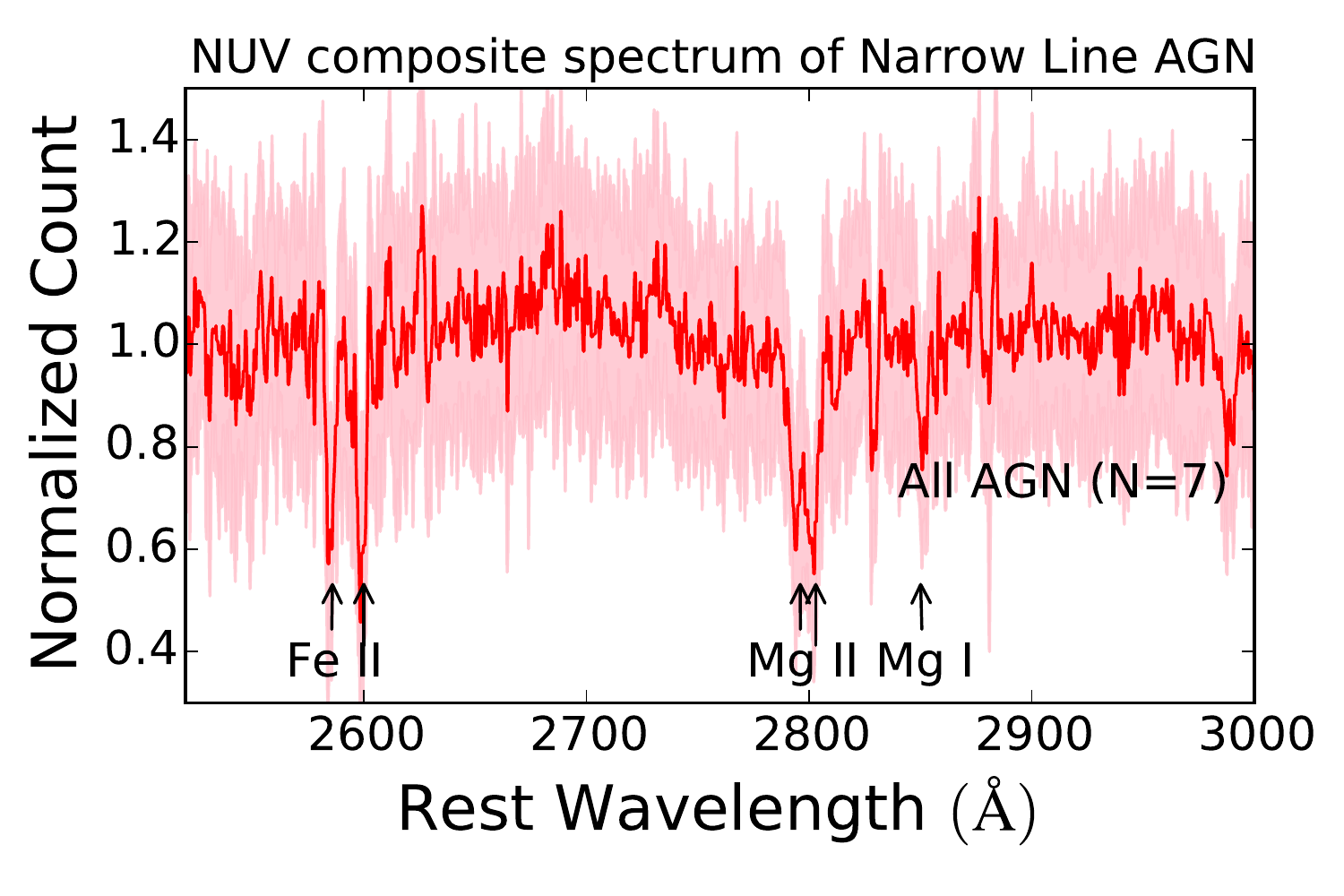}%
}%
\hfill
\subfigure[Broad-line AGN]{%
\includegraphics[width=0.45\linewidth]{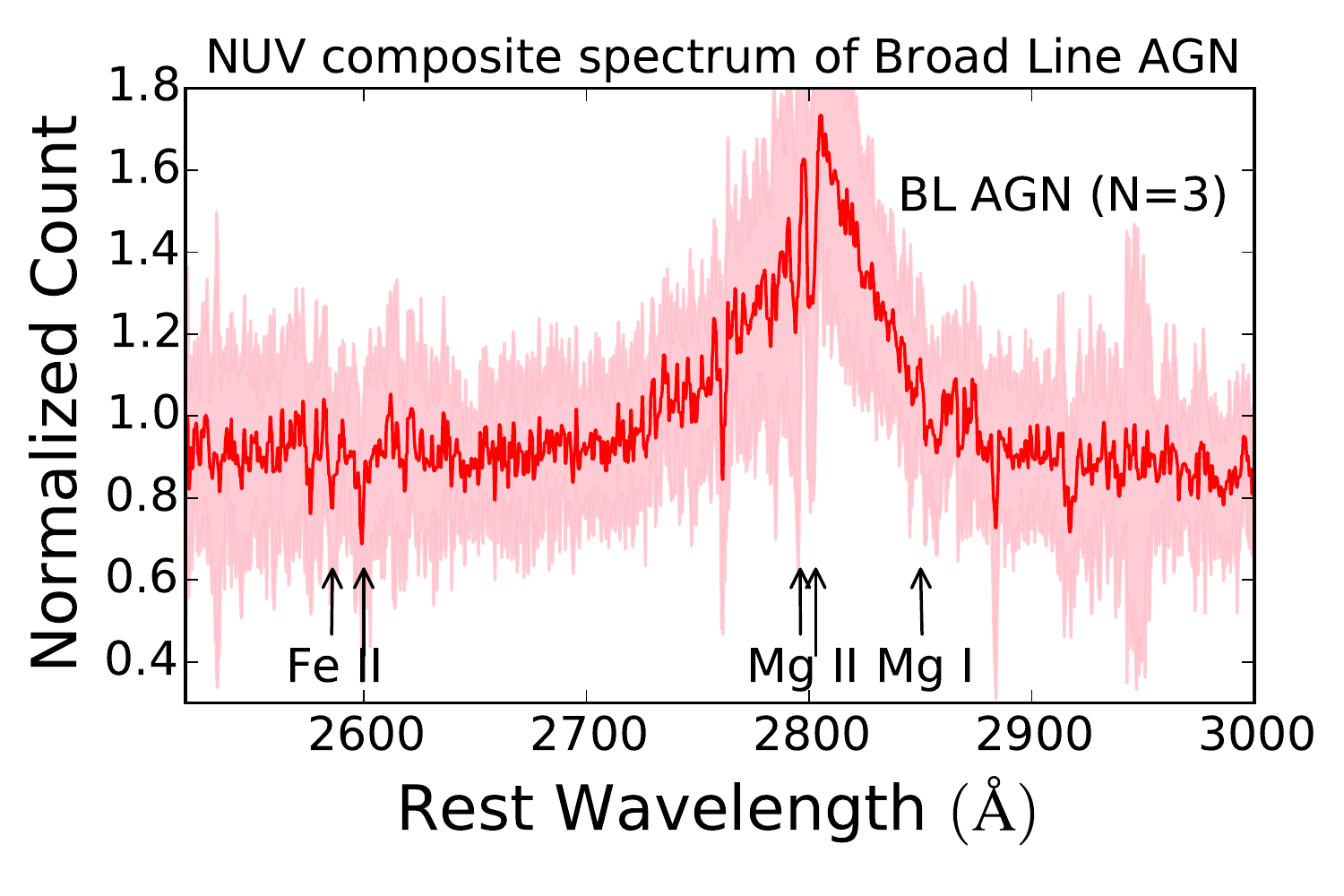}%
}%
\caption{Near UV composite spectra of the AGN sample and subsamples at $z \sim 1$. The pink shadings are 1$\sigma$ bootstrap error bands.} \label{fig:nuvspec}

\end{figure*}

\begin{figure*}
\centering
\subfigure[Narrow-line AGN compared to inactive star-forming galaxies]{%
\includegraphics[width=0.45\linewidth]{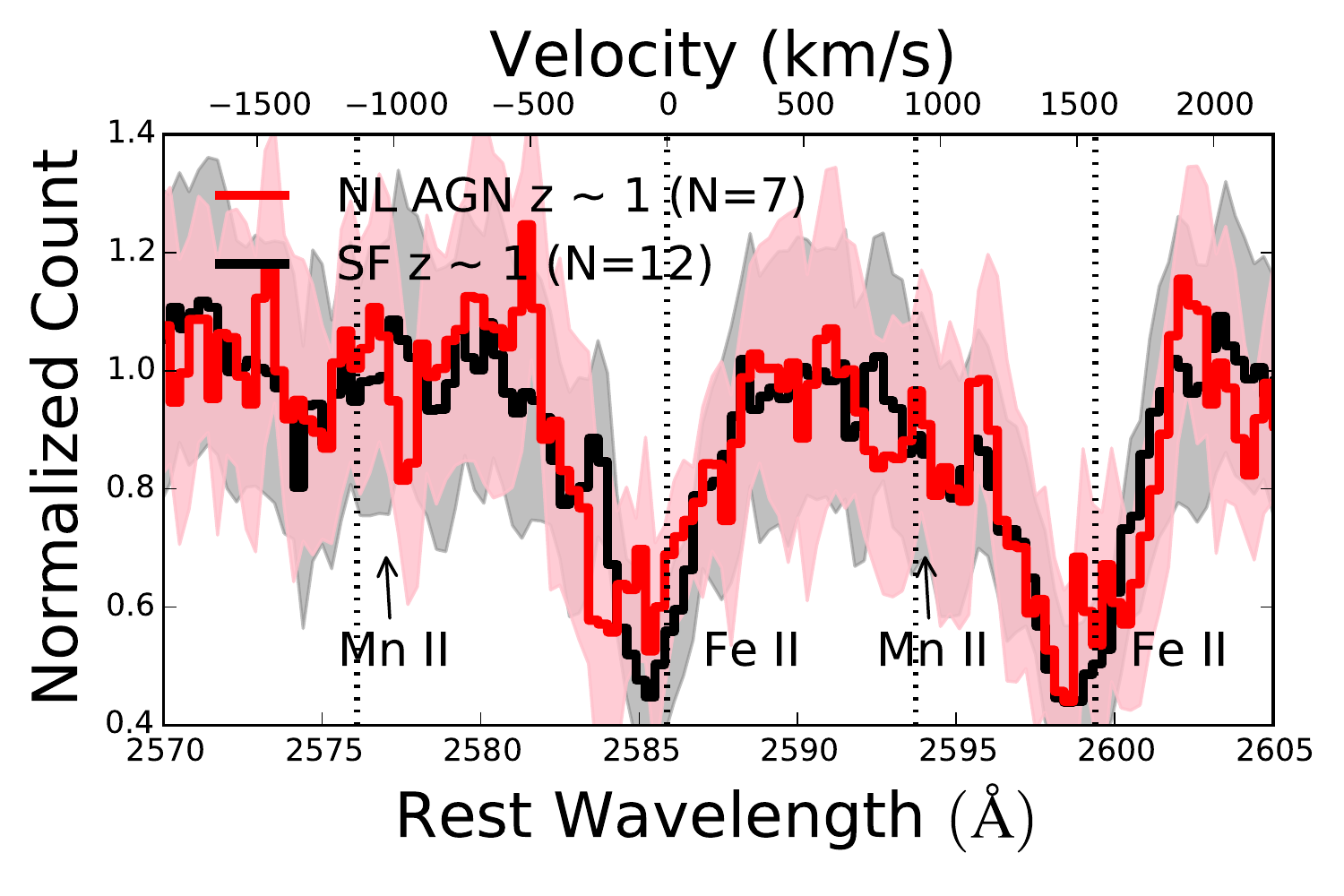}}
\hfill
\subfigure[Narrow-line AGN compared to inactive star-forming galaxies]{%
\includegraphics[width=0.45\linewidth]{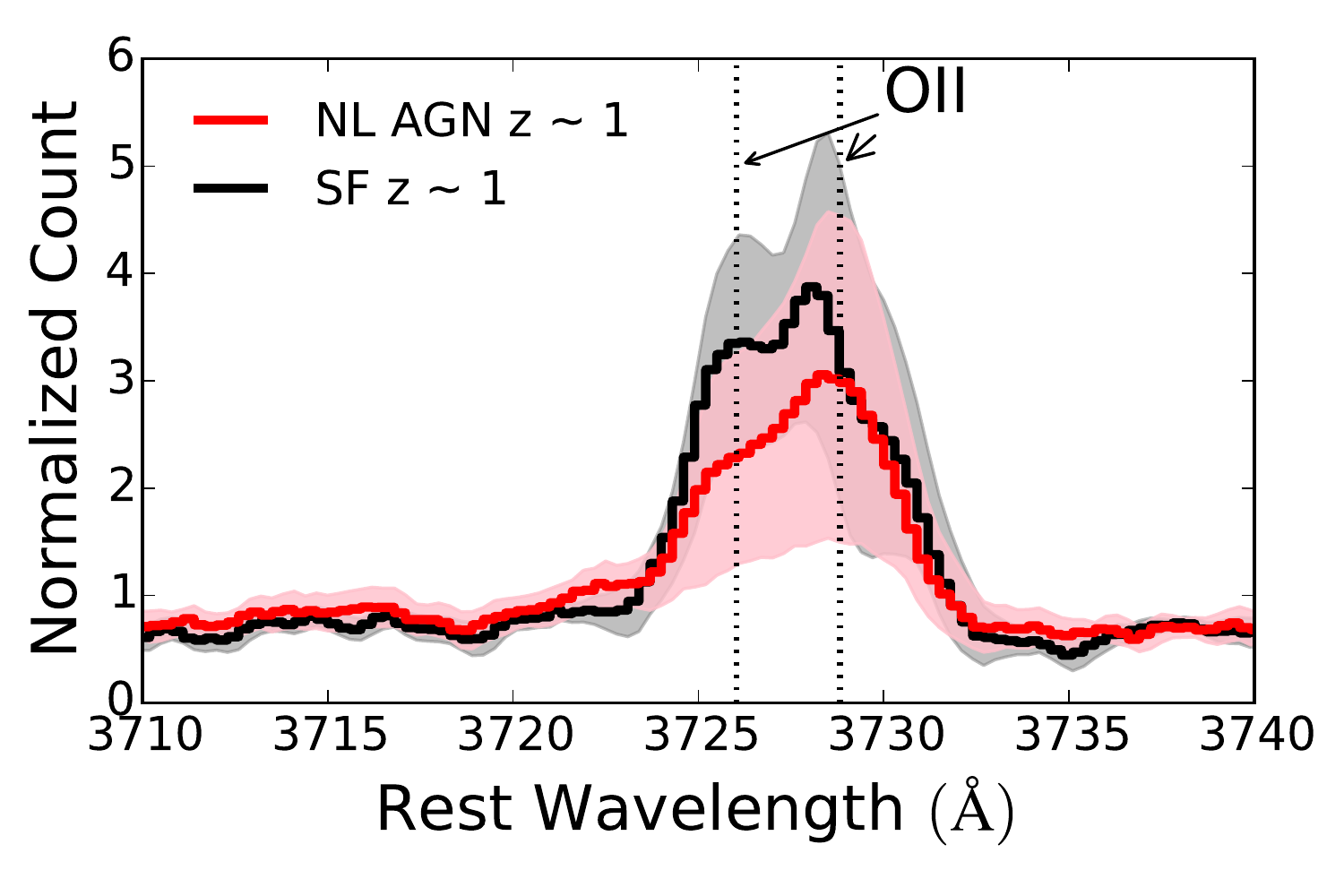}}
\hfill
\subfigure[Narrow-line AGN at $z \sim 1$ and $z \sim 0.5$ compared]{%
\includegraphics[width=0.45\linewidth]{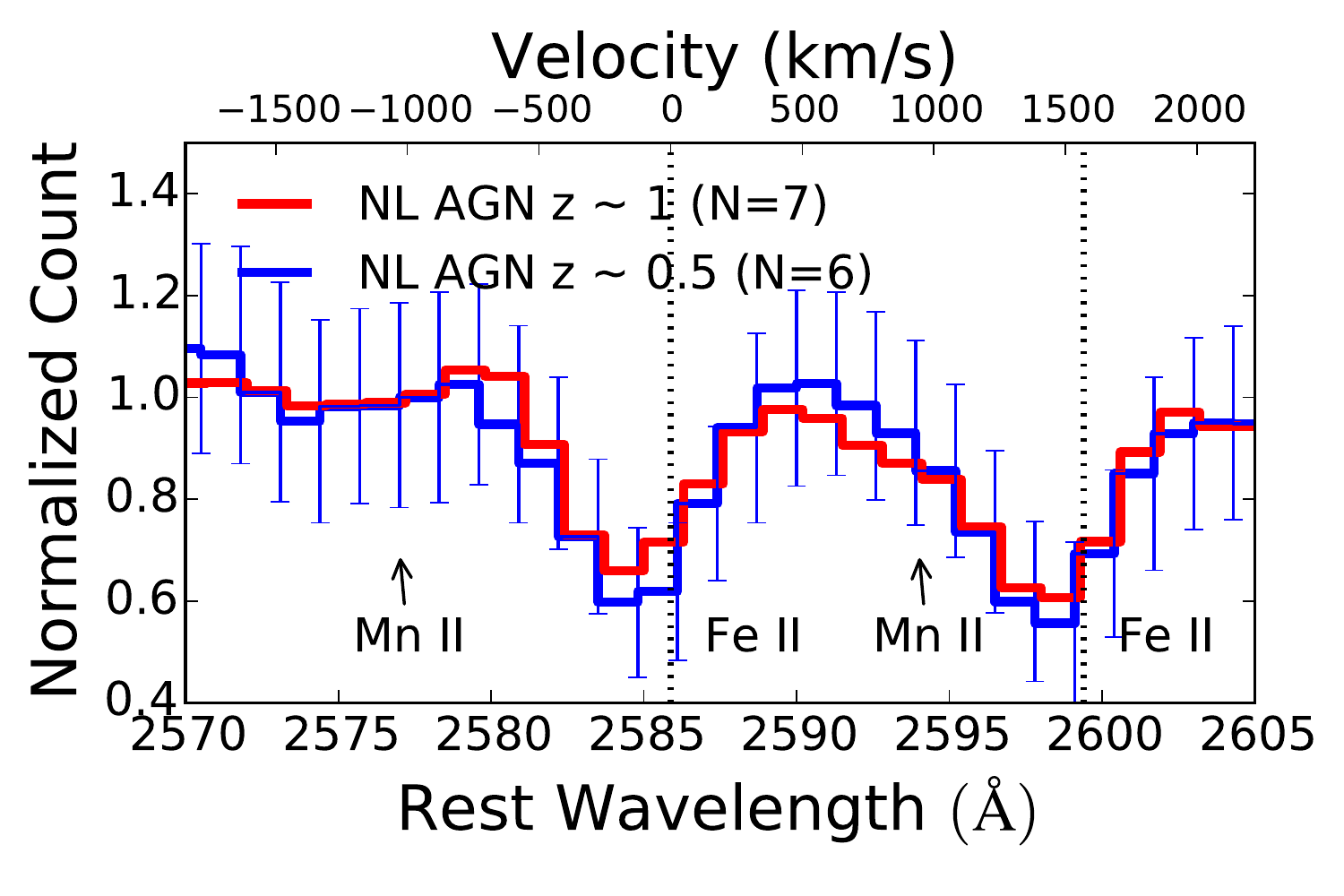}}
\hfill
\subfigure[Narrow-line AGN at $z \sim 1$ and $z \sim 0.5$ compared]{%
\includegraphics[width=0.45\linewidth]{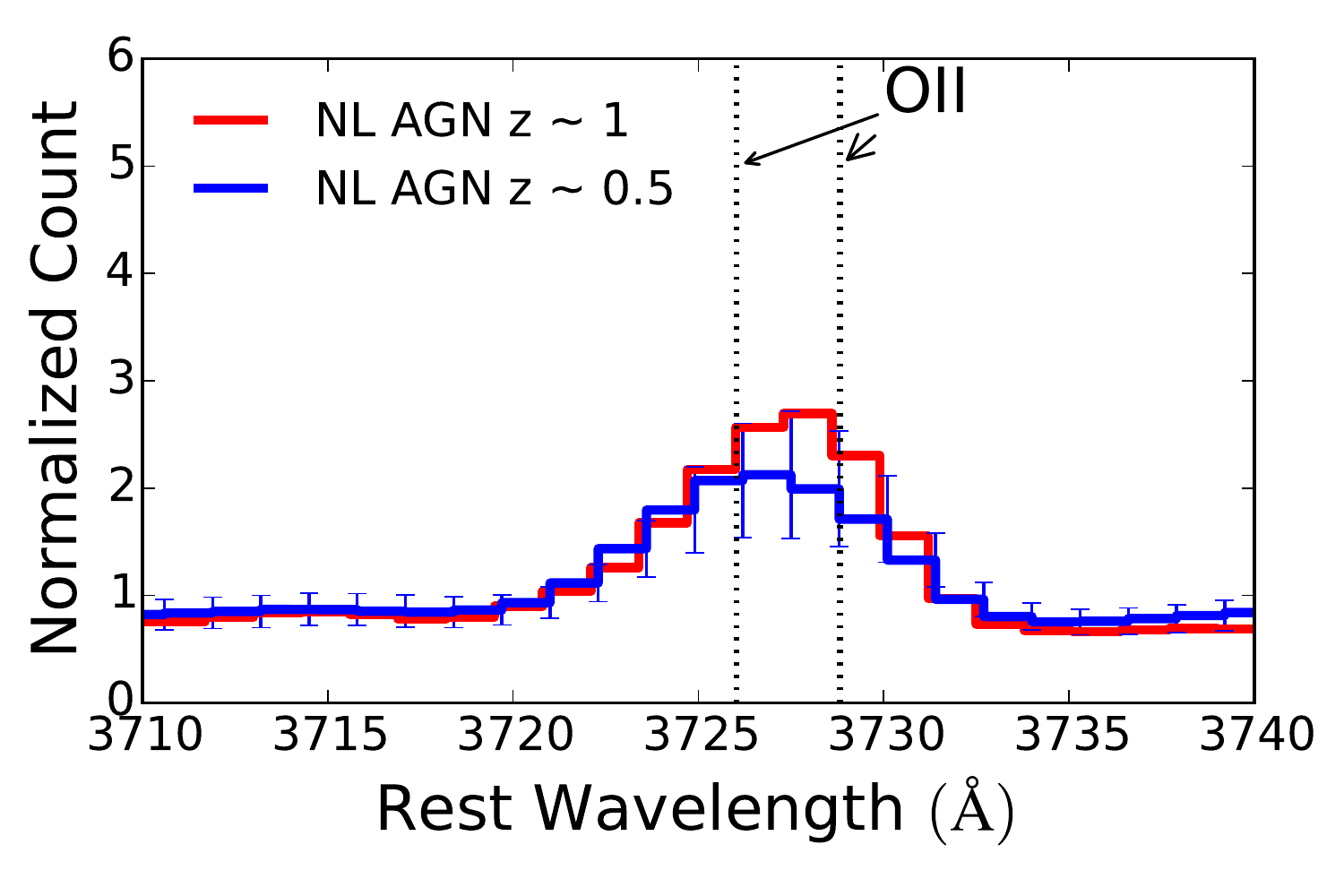}}
\caption{The near UV composite spectrum around \ion{Fe}{2} $\lambda2586$ of the narrow-line AGN at $z \sim 1$ compared to that of X-ray undetected galaxies at a similar redshift and that of $z \sim 0.5$ AGN \citep{Coil+11}. Panel (a) The near UV composite spectrum of the 7 bona fide narrow-line AGN at $z \sim 1$ compared to composite spectrum of X-ray undetected star-forming sample at a similar redshift. The errors in this and the other panels are estimates by bootstrap analysis. A variant of this figure for other subsamples of AGN can be found in the Appendix. Panel (b) A comparison between \ion{O}{2} emission of AGN at $z \sim 1$ with the \ion{O}{2} emission of the inactive star-forming sample. The  \ion{O}{2} line is vital in determining the redshift and the systemic zero velocity of the galaxies. Panel (c) A comparison between the composite spectra around \ion{Fe}{2}$ \lambda2586$ of AGN at $z \sim 1$ and $z \sim 0.5$. The composite spectrum of $z \sim 1$ AGN is rebinned and convolved to match the spectrum of $z \sim 0.5$ AGN. Panel (d) A comparison between \ion{O}{2} emission of AGN at $z \sim 1$ and of AGN at $z \sim 0.5$. The main point that the data shows is that there are robust systematic blue shifts in the \ion{Fe}{2}$ \lambda2586$ absorption profiles of AGN at the two redshifts and of the comparison sample. The blueshift are less than 500 km s$^{-1}$. Interpreting the blueshifts as wind signatures, one can conclude that winds with average velocity above 500 km s$^{-1}$ are not observed in AGN as well as the comparison sample. Therefore, the winds in the low-luminosity AGN are most likely driven by star formation activity.\label{fig:data}}
\end{figure*}

The individual near UV spectra of the redshift $z \sim 1 $ AGN sample are plotted in Figure~\ref{fig:samplespec}. The \ion{Fe}{2} and \ion{Mg}{2} absorption lines are observed in most of the spectra. Figure~\ref{fig:nuvspec} shows the normalized near UV composite spectra of all AGN and subsets of the AGN sample separated into narrow and broad-line AGN. The normalization was determined by a linear fit to the continuum around both sides of the \ion{Fe}{2} absorption doublet. We used the wavelength ranges $2520-2578${\AA}, $2640-2770${\AA}, and $2900-2970${\AA} to fit the continuum level. Figure~\ref{fig:data}a shows the normalized composite spectrum near the \ion{Fe}{2} absorption line of the 7 narrow-line AGN at $z \sim 1$ with robust AGN identifications. The composite spectrum of the comparison sample of X-ray undetected, star-forming galaxies is overplotted on the same figure. It is clear from this figure that both AGN and normal star-forming galaxies have asymmetrically blue-shifted \ion{Fe}{2} absorption lines. The absorption profiles of AGN and the comparison sample have very similar width and depth. Therefore, without detailed modeling, one can infer that they have similar wind velocity and strength. The wind velocities, in both samples, are in the order of $100-200$ km s$^{-1}$ and extend to $\sim 500$ km s$^{-1}$. A variant of this figure for all 12 AGN candidates can be found in the Appendix Figure~\ref{fig:nuvapp}. Figure~\ref{fig:data}b plots the \ion{O}{2} emission lines of AGN and the comparison sample. \ion{O}{2} is vital in determining the redshift and the systemic zero velocity of the galaxies.

Figure~\ref{fig:data}c \& d are similar to Figure~\ref{fig:data}a \& b. In these later figures, the comparison is made between the composite spectrum of our AGN at $z \sim 1$ and that of the 6 AGN studied by \citet{Coil+11} at $z<0.6$. We note that these authors studied winds in their AGN using individual spectra. Since we coadd the spectra of our AGN for improved signal-to-noise average spectrum, we also coadd the AGN at $z<0.6$. We interpolate the individual AGN spectra onto a linear wavelength grid with $\Delta \lambda = 1.3${\AA}, which is about the pixel size of the LRIS spectrograph at $z\sim 0.6$ for their setup. We convolve our composite spectrum to the instrumental resolution of LRIS and rebin it to match the resolution and bin size of the composite spectrum of their data. There is a very good agreement between the \ion{Fe}{2} absorption profiles of AGN at $z \sim 1$ and that of the AGN at $z\sim 0.5$. The strength of the \ion{O}{2} emission lines of the two AGN samples also agree with each other.

We reanalyze the \citet{Coil+11} data both separately and jointly with our data. For the joint analysis, we linearly interpolate the composite spectrum of their data to a wavelength grid with $\Delta \lambda = 1.2${\AA} bin, convolve our AGN composite spectrum to the LRIS resolution but here our spectrum is not rebinned (has $\Delta \lambda = 0.3${\AA}). The two component model decomposition may be sensitive to binning and therefore we do not rebin our composite spectrum. To combine the two datasets, we average the normalized counts in the bins where the two datasets coincide (i.e., every fourth bin in our spectrum). The averaging uses inverse-variance weighting per pixel. In the bins where the two datasets do not overlap, we use values of our composite spectrum. The joint analysis is done using the composite spectrum of our 7 narrow-line AGN with robust AGN identification, which are more similar to the \citet{Coil+11} sample, as well as with the composite spectrum of all AGN in our sample.

To give a quantitative gauge of both the dispersion intrinsic to a sample (to account for the effects caused by the outlier galaxies within a sample) and the measurement errors, we use the bootstrapping scheme to estimate the standard errors of  the average composite spectrum at different wavelengths. In the bootstrapping scheme, we randomly resample, with replacement, the ID of galaxies of a sample size equal to the size of the original sample size at each iteration. This resampling was done 1000 times. At each iteration, we averaged the spectra of the selected galaxies using inverse-variance weighting at each wavelength pixel. We used the standard deviation of the average counts in a given wavelength bin of all 1000 composite spectra as the error of the original averaged spectrum at that given bin. We add the bootstrap standard deviations of the composite spectra in quadrature when we combine the two AGN sample at  $z \sim 1$ and $z \sim 0.5$. An alternative analysis which may better minimize the effects of poor measurements, provided that the dispersion intrinsic to sample is small, is to simply use the errors of inverse-variance weighted average. The results from this alternative analysis is presented in the Appendix section~\ref{sec:invmodel}. Because the errors from this method are significantly smaller than the errors from the bootstrapping scheme, the wind parameters are better constrained in the results presented in the Appendix. In the next section, we quantitatively show that the winds in the two samples are similar using a standard wind model.

\subsection{The simple wind model}

We adopt the partial covering wind model of \citet{Rupke+05a} to model the observed \ion{Fe}{2} $\lambda 2585.876$ \footnote{More information and references for this and other atomic transition used in the paper can be found in NIST Atomic Spectra Database at \url{https://www.nist.gov/pml/atomic-spectra-database} or in \citet{Morton03}.} absorption profile, similar to recent works that studied galactic winds in star-forming galaxies \citep{Kornei+12,Martin+12,Rubin+14,Tang+14}. Due to signal-to-noise and spectral resolution limitations of observational data to fully constrain the wind model, the following simplifying assumptions were customarily made in the previous works and are also adopted in this work: 1) the covering factor of the wind is independent of velocity. Some studies suggest that this may not be a good assumption \citep{Martin+09}; 2) the stellar continuum emission is fully covered by a uniform screen of ISM absorption. If this assumption is not valid, the inferred ISM column density is lower than the actual column density since the covering fraction anti-correlates with column density. In this work, we do not aim to constrain these two values independently. We show in the Appendix section~\ref{sec:cf} that the wind velocities are not significantly affected by this assumption when using a covering fraction of 50\% for the ISM;  3) The line profile shape is due entirely to the absorption of the stellar continuum. However, scattered emission infill may also affect the absorption profile. This effect is expected to be less significant for \ion{Fe}{2} $\lambda 2586$ compared to resonant lines without florescent emission lines such as \ion{Mg}{2} \citep[see][]{Prochaska+11}. \citet{Zhu+15} estimated that the equivalent width of the \ion{Fe}{2} $\lambda2586$ absorption profile is affected by only 10\% due to emission infill; 4) Two absorption components, an ISM component centered at zero systemic velocity and a wind component, are sufficient to characterize the observed absorption line profiles. This assumption may result in inaccurate column densities and line widths, if the profiles are composed of multi-components from multiple clouds. Higher-resolution galaxy spectra are required to test the effects of this assumption. Studies using high resolution spectra of background quasars find the absorption lines are composed of multiple clouds with more complex kinematics \citep[e.g.,][]{Churchill+00}; 5) the velocity distribution of absorbing atoms within a component is Maxwellian such that each absorption optical depth is modeled as a Gaussian, $\tau(\lambda) = \tau_c \exp\,-(\lambda-\lambda_c)^2/( \lambda_c\,b/c)^2$ where  $\tau_c$  is a central optical depth at the line center, $\lambda_c$, $c$ is the speed of light, and $b=\sqrt2\,\sigma$ is the Doppler parameter. This assumption is likely to be an over-simplification but it is reasonable given that the observed shape of the absorption trough is strongly influenced by the instrumental resolution.

According to this simple model, the normalized continuum can be described as a product of the line intensity of the galaxy component and of the wind component. Each component has the form
 $1-C (1-\exp\,-\tau(\lambda))$. C is the covering fraction and it is set to 1 for the galaxy ISM component. We can express $\lambda_c$ in terms of centroid velocity $v=c(\lambda_c-\lambda_0)/\lambda_0$, where $\lambda_0$ is the rest wavelength of the transition. For the galaxy component, $ \lambda_c=\lambda_0$ (i.e., no velocity shift). The central optical depth is expressed in term of the column density, $N$, oscillator strength, $f_0$, and $b$ using the relation $\tau_c = 1.45 \times 10^{-15} \lambda_0\,f_0N/b$ for $N$ in unit of cm$^{-2}$, $\lambda$ in {\AA} and $b$ in km\,s$^{-1}$.

To summarize, the 6 free parameters of this two component model are: the covering fraction of the wind, $C_{w}$, the velocity centroid shifts of the wind, $v_{w}$, the Doppler broadening parameter of the wind, $b_{w}$, the Doppler parameter of the ISM in the galaxy, $b_g$, column density of the wind, $N_{w}$, and the column density of the gas in the galaxy, $N_{g}$. The model is convolved with the instrumental resolution and rebinned to match the observed data before comparing the two. 

The model is fit to the data using a Bayesian method with custom Python code. Only the \ion{Fe}{2} $\lambda 2585.876$ absorption line is fitted. The entire wavelength ranges $\lambda=(2572, 2578)$ and $\lambda=(2591,2632)$ are masked out prior to fitting because they are contaminated by \ion{Mn}{2} $\lambda 2576.877, 2594.499, 2606.462$ absorptions or \ion{Fe}{2} $\lambda 2599.395$ absorption or the \ion{Fe}{2} $\lambda 2611.873, 2625.489, 2631.047$ emissions. The posterior probability densities (PDFs) of the model parameters were computed using the affine-invariant ensemble Metropolis-Hastings sampling algorithm \citep{Foreman-Mackey+13} assuming uniform priors: $v_{w} =(-450, 450)$, $b_{w}=(20, 450)$,  $b_{g}=(20,450)$, $\log N_{w}=(14, 17.5)$ and $\log N_{g}=(14, 17.5)$. A centroid velocity shift greater than 500\,km\,s$^{-1}$ is ruled out by the data without the model, so we have restricted the velocity prior to estimate the velocity shift which is supported by the data. To compute the likelihood of the data given the model parameters, we assumed that each data point is drawn from independent Gaussians centered around the model profile with a dispersion given by the measurement errors. This is equivalent to assuming a $\chi^2$ distribution for the sum of squares of flux differences between the model and the data, with degrees of freedom equal to the number of observed data points.

\subsection{The wind velocities of  X-ray AGN at $z \sim 1$ are similar to those of star-forming X-ray undetected galaxies at similar redshifts.}

Table~\ref{tbl:fit_param} summarizes the marginalized PDFs of the six model parameters fitted to \ion{Fe}{2} $\lambda2586$ absorption lines for both the AGN and the comparison samples . We used the median, 16th and 84th percentiles as summary statistics for the marginalized PDFs. For convenience, we express the percentiles as $\pm$ deviations from the median throughout the text. For instance, $X^{+Y}_{-Z}$ denotes that X is the median, $X+Y$ is the 84th percentile and $X-Z$ is the 16th percentile. For a Gaussian PDF, Y and Z equal to its standard deviation but note that the PDFs of the wind centroid velocity shift and of the Doppler dispersion parameter are non-Gaussian in almost all cases.

\begin{figure*}
\centering
\subfigure[\ion{Fe}{2} profile of all AGN $z \sim 1$ (N $=12$)]{%
\includegraphics[width=0.3\linewidth,height=0.25\linewidth]{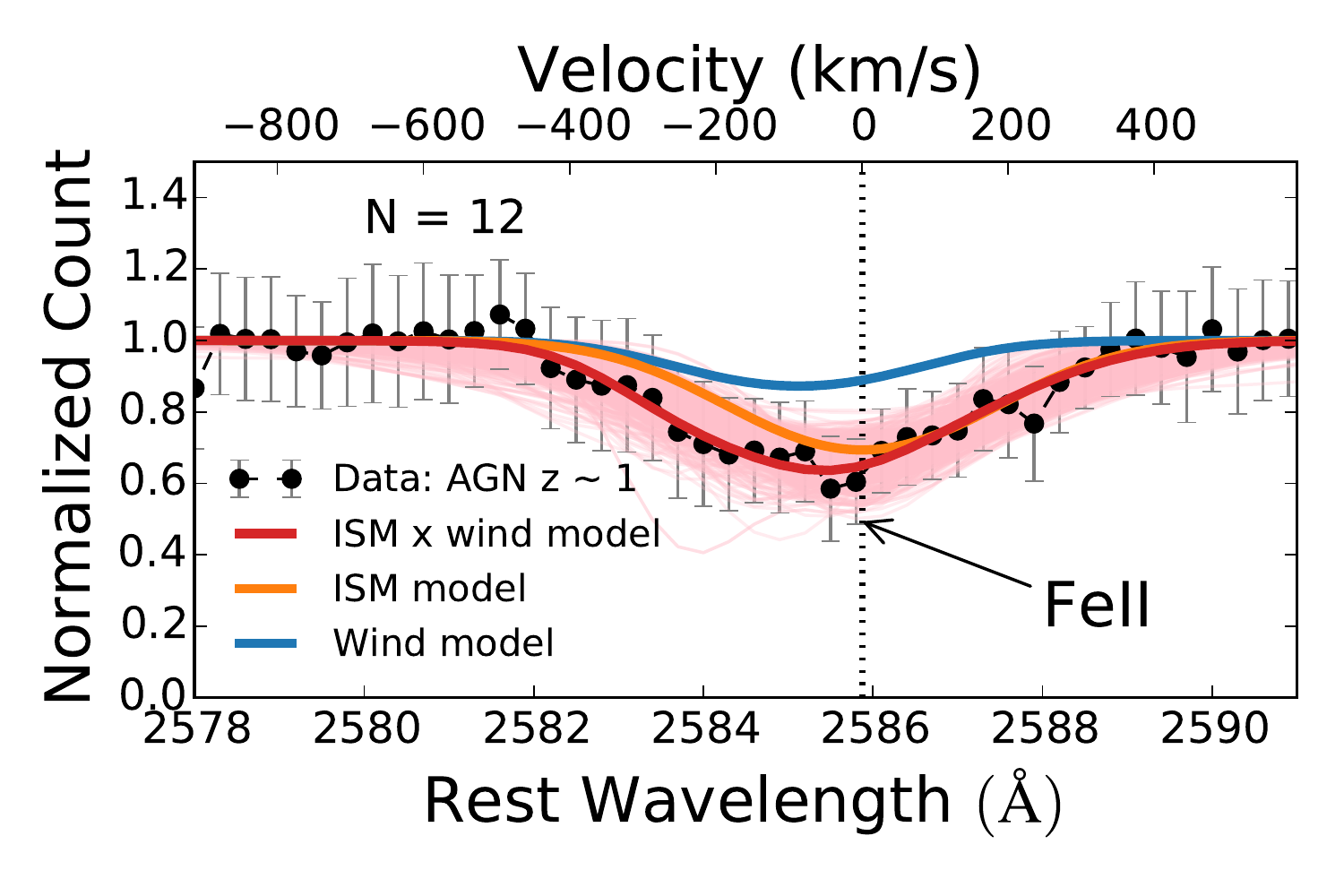}}
\hfill
\subfigure[Narrow-line AGN $z \sim 1$ (N $=9$)]{%
\includegraphics[width=0.3\linewidth,height=0.25\linewidth]{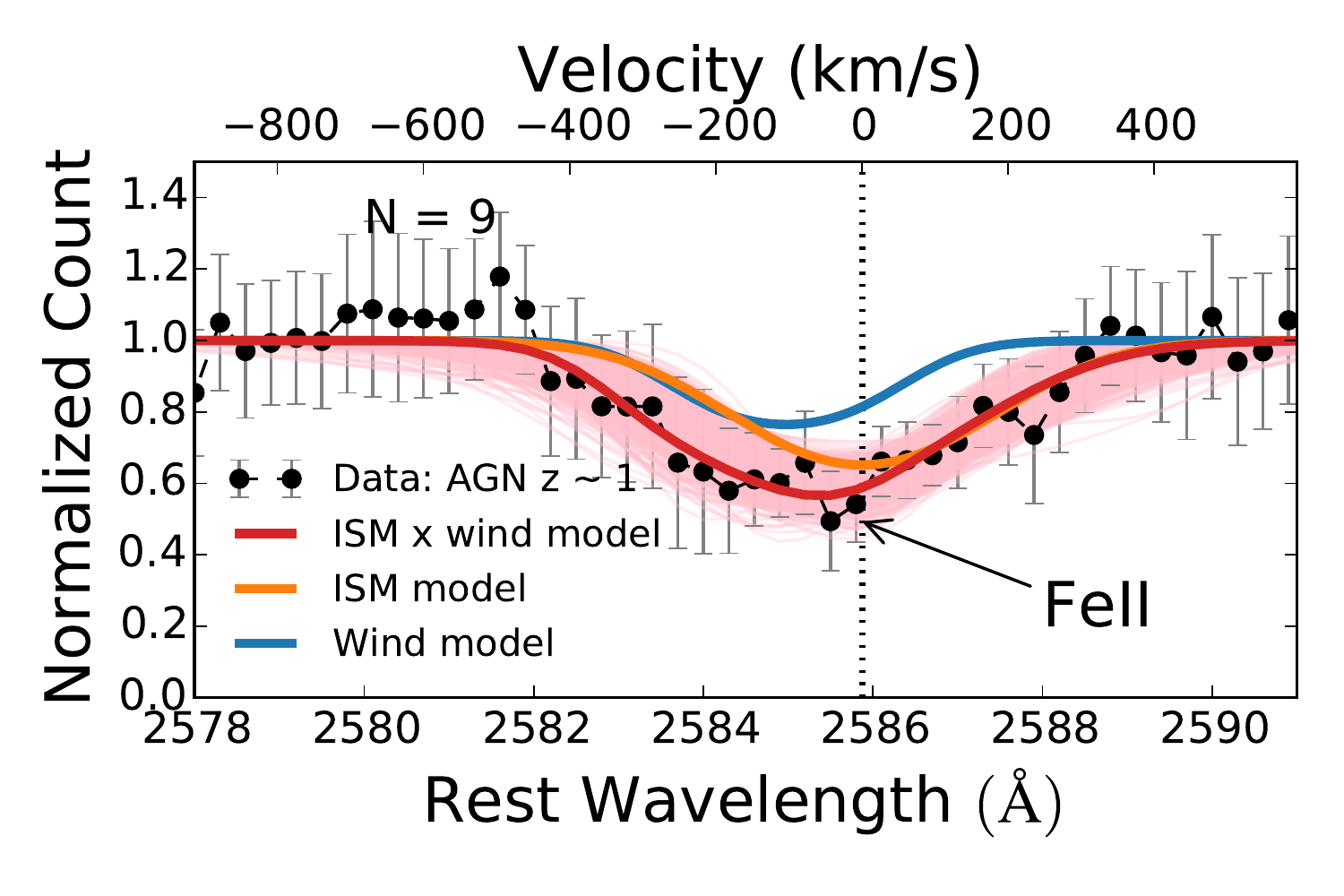}}
\hfill
\subfigure[Very reliable narrow-line AGN $z \sim 1$ (N $=7$)]{%
\includegraphics[width=0.3\linewidth,height=0.25\linewidth]{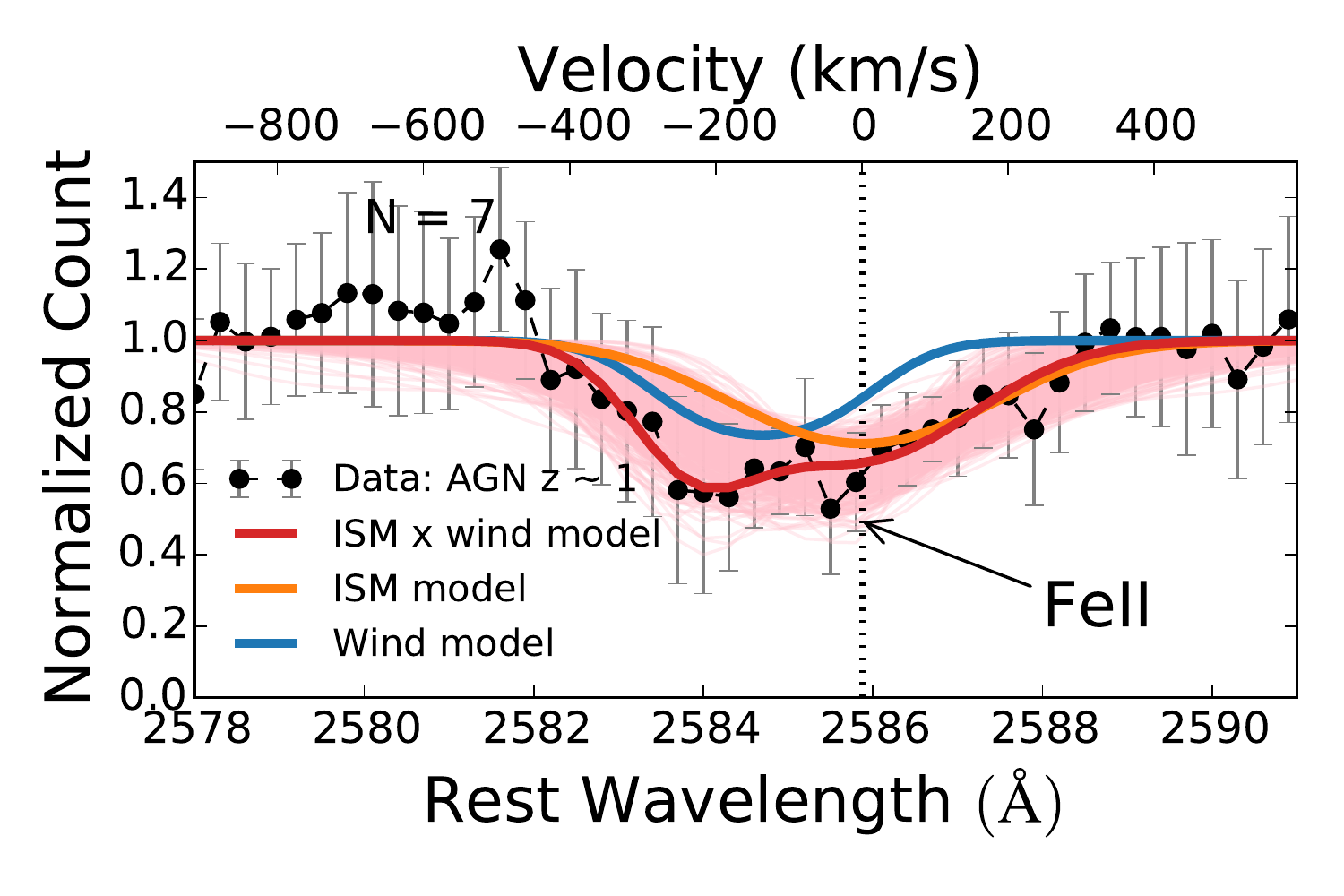}}
\hfill
\subfigure[Wind velocity of all AGN $z \sim 1$ (N $=12$)]{%
\includegraphics[width=0.3\linewidth,height=0.25\linewidth]{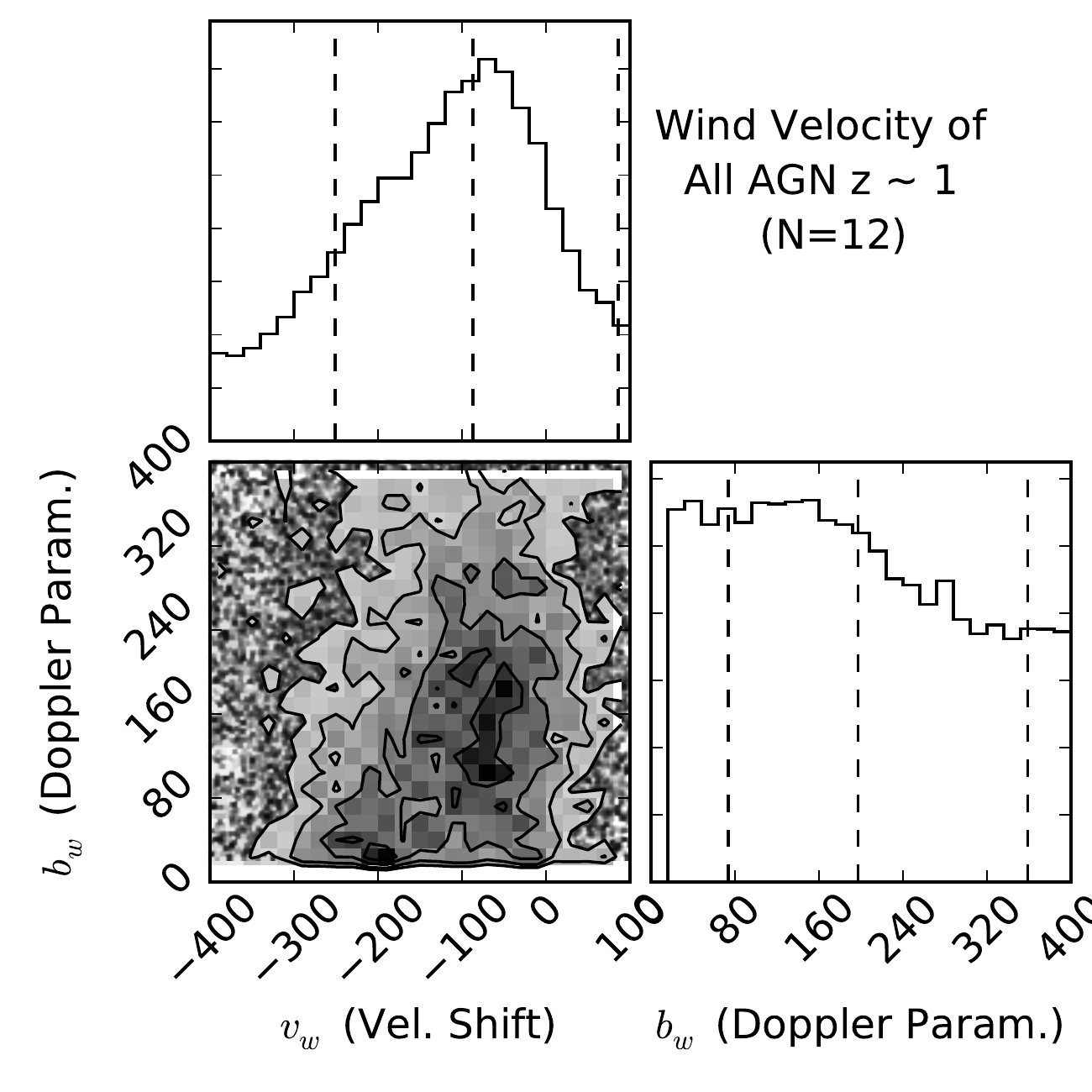}}
\hfill
\subfigure[Narrow-line AGN $z \sim 1$ (N $=9)$]{%
\includegraphics[width=0.3\linewidth,height=0.25\linewidth]{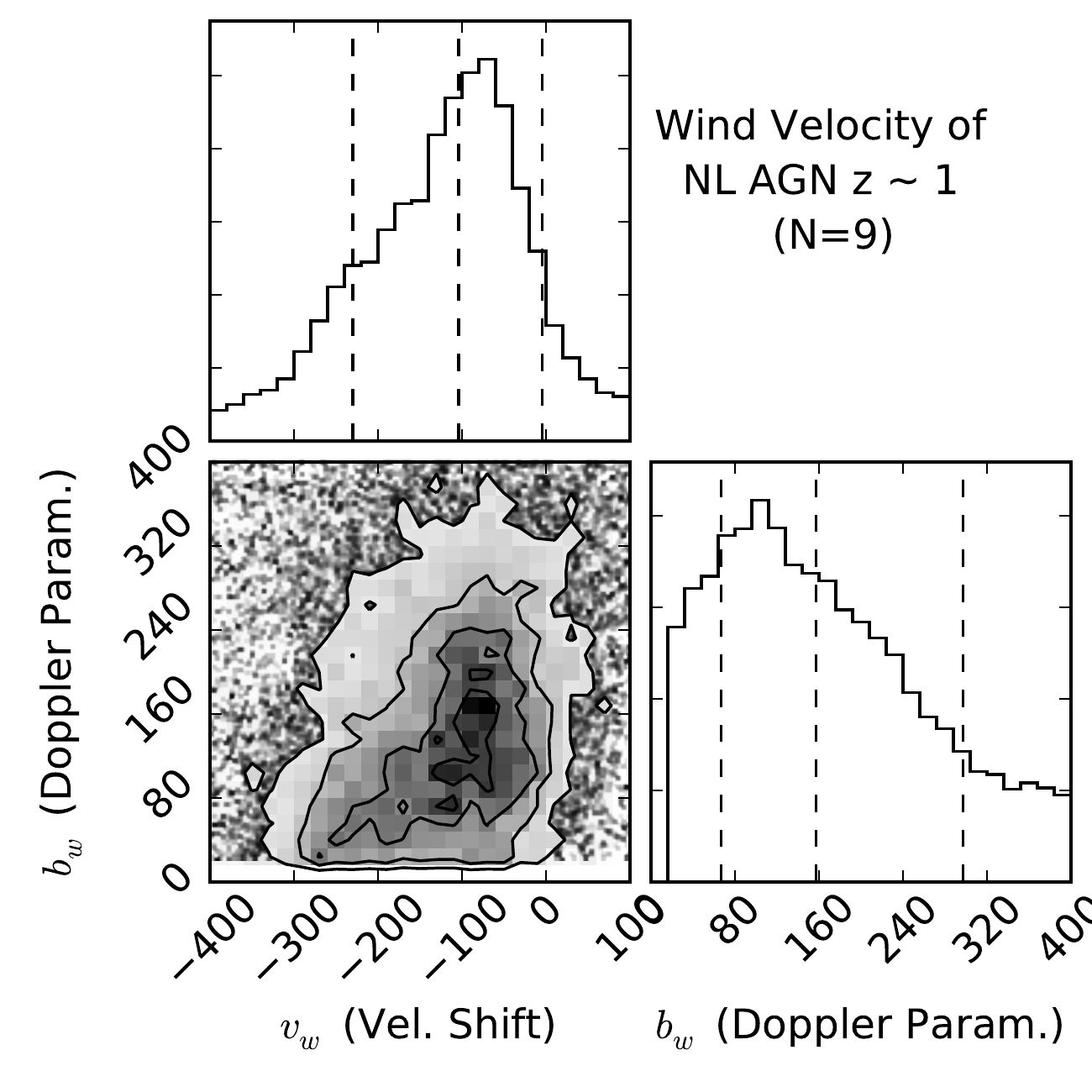}}
\hfill
\subfigure[Very reliable narrow-line AGN $z \sim 1$ (N $=7)$]{%
\includegraphics[width=0.3\linewidth,height=0.25\linewidth]{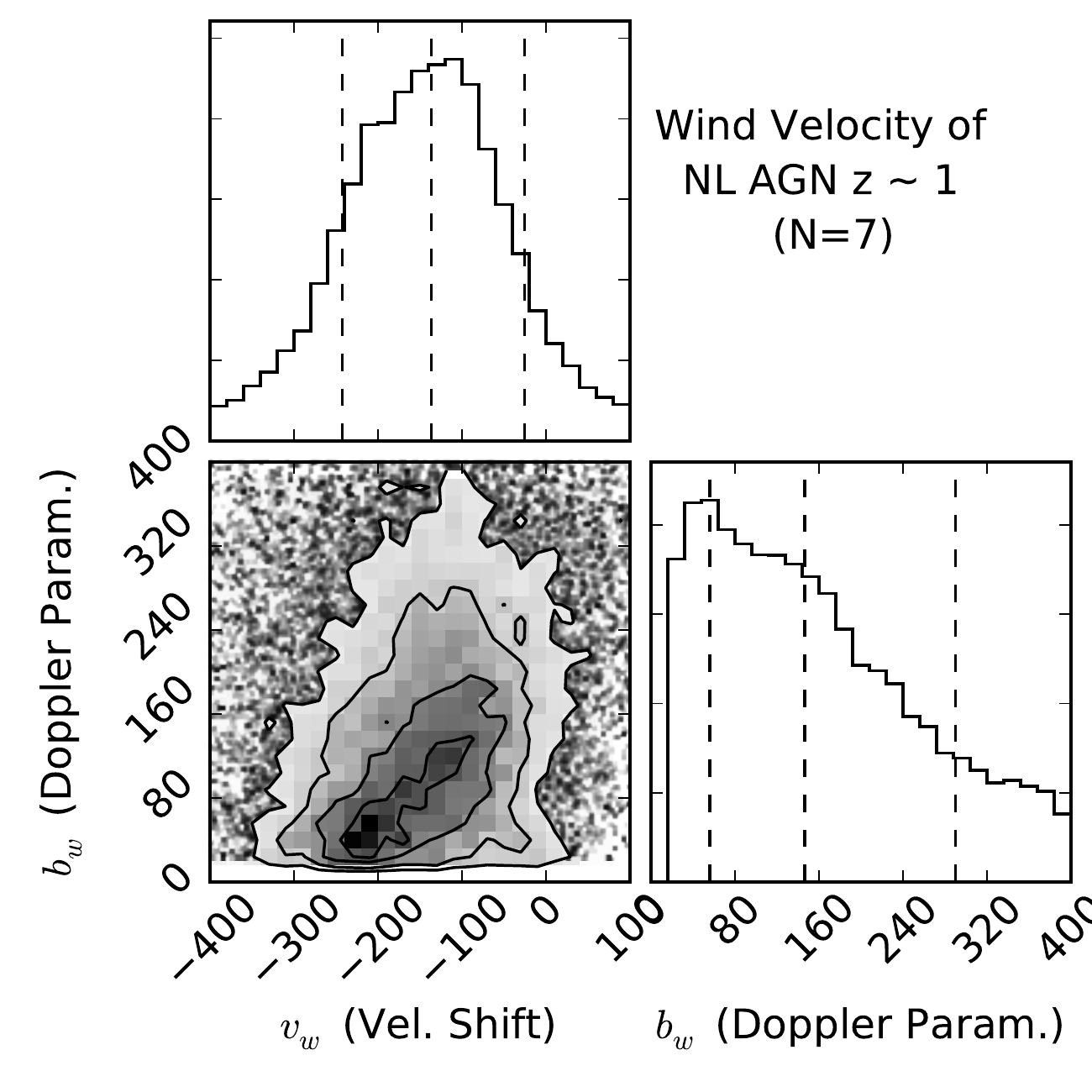}}
\hfill
\subfigure[\ion{Fe}{2} profile of AGN $z \sim 0.5$ \citep{Coil+11}]{%
\includegraphics[width=0.3\linewidth,height=0.25\linewidth]{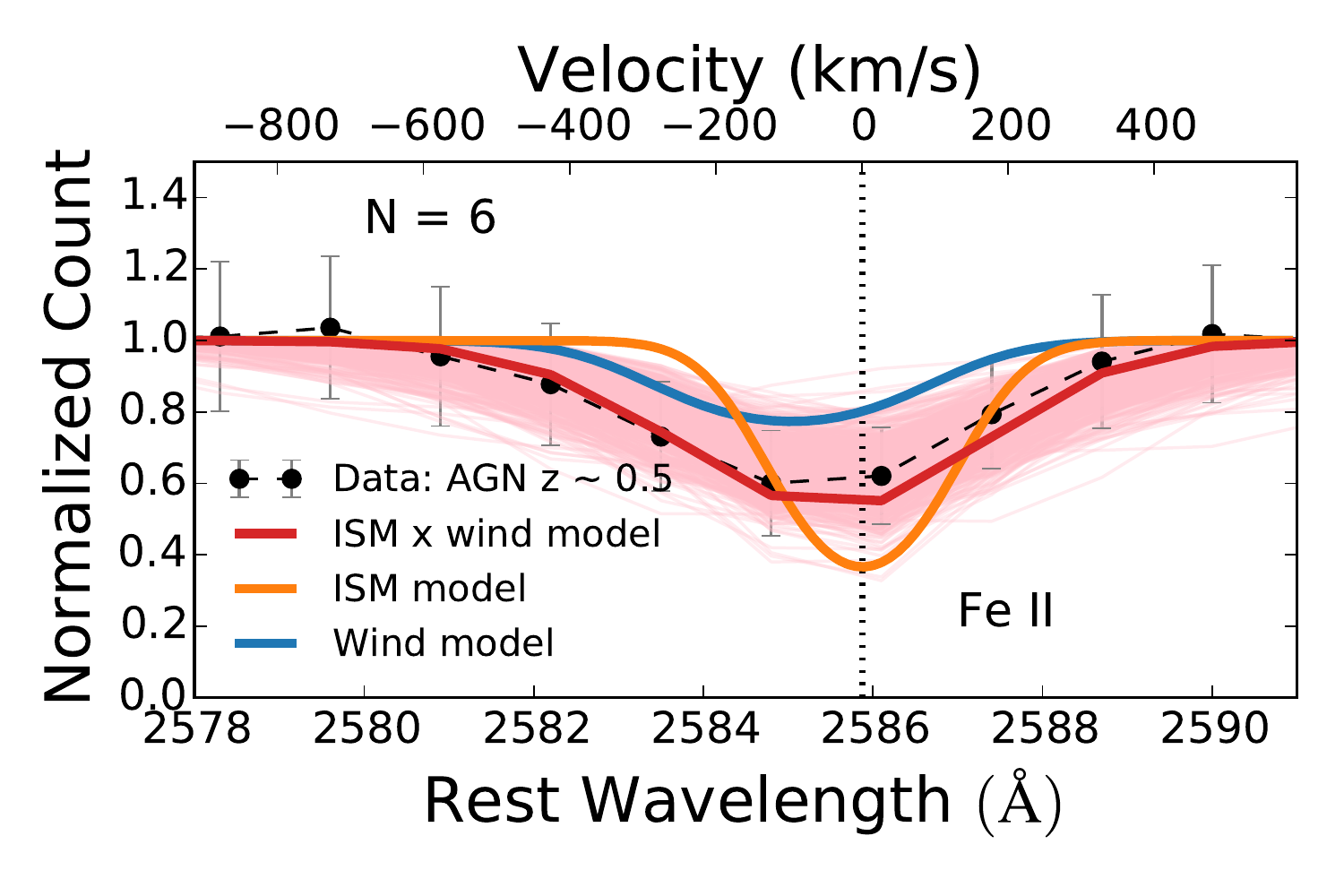}}
\hfill
\subfigure[Combined sample of AGN $z \sim 0.5$ \& $z \sim 1$]{%
\includegraphics[width=0.3\linewidth,height=0.25\linewidth]{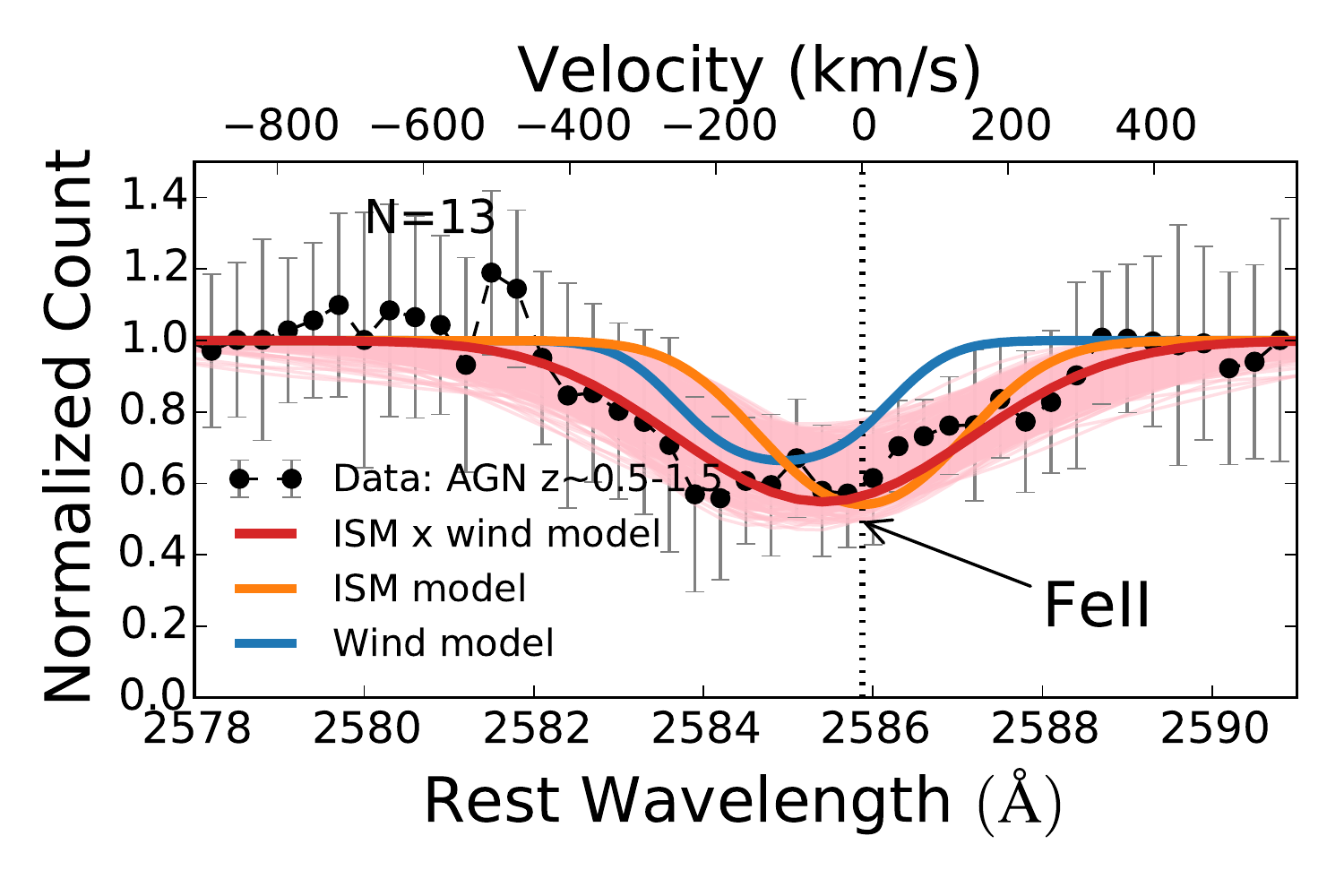}}
\hfill
\subfigure[Star-forming comparison sample $z \sim 1$]{%
\includegraphics[width=0.3\linewidth,height=0.25\linewidth]{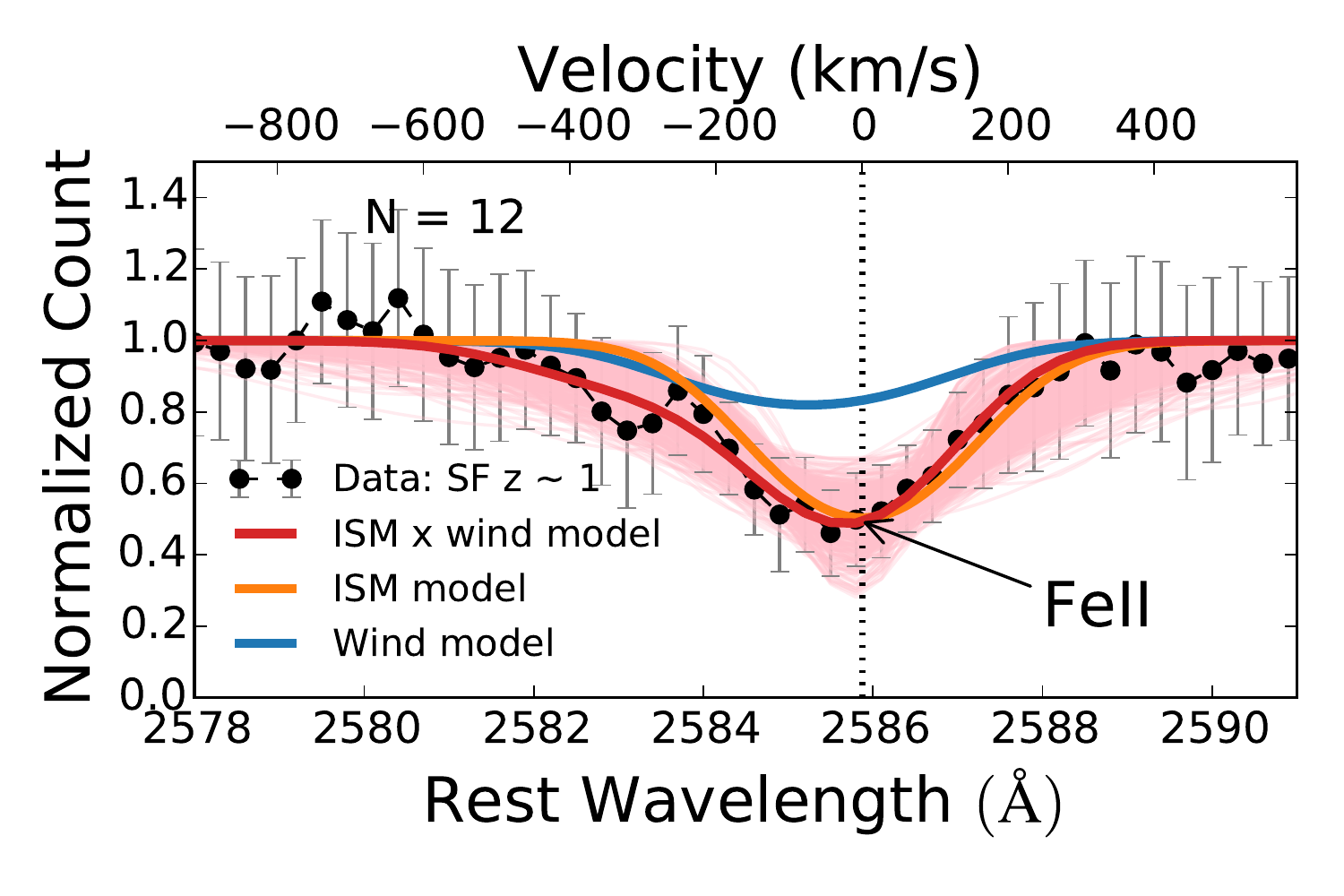}}
\hfill
\subfigure[Velocity of AGN $z \sim 0.5$ \citep{Coil+11}]{%
\includegraphics[width=0.3\linewidth,height=0.25\linewidth]{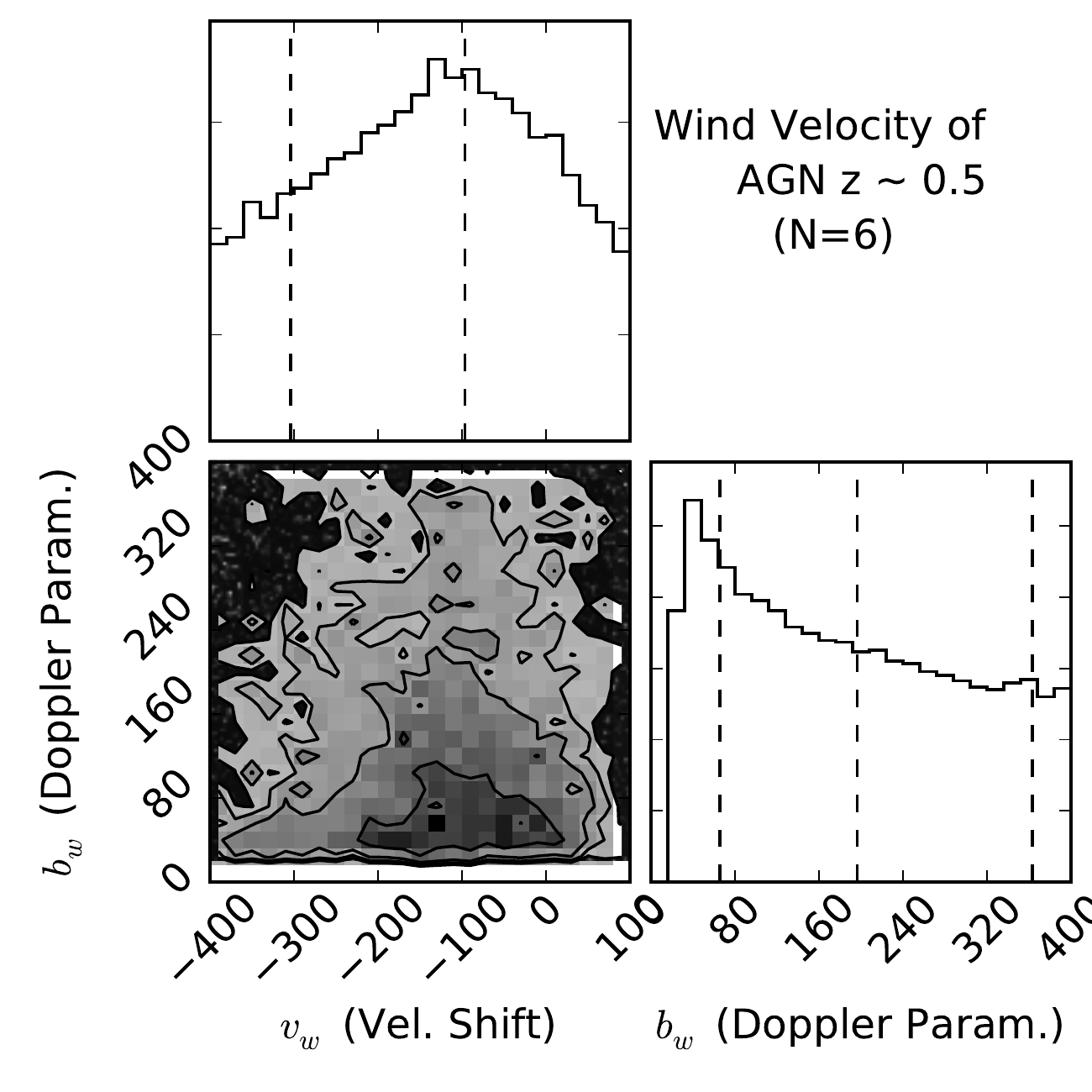}}
\hfill
\subfigure[Combined sample of AGN $z \sim 0.5$ \& $z \sim 1$]{%
\includegraphics[width=0.3\linewidth,height=0.25\linewidth]{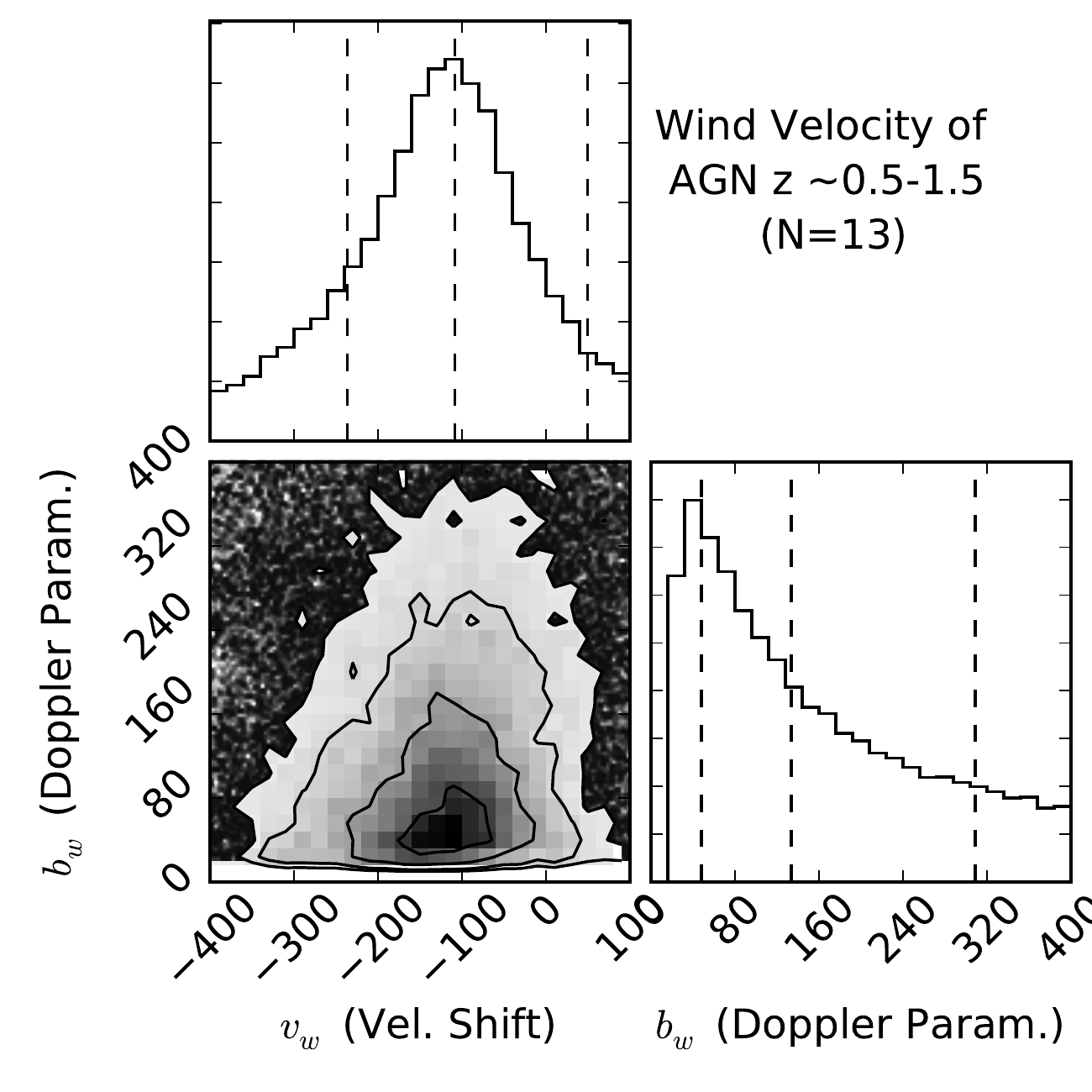}}
\hfill
\subfigure[Star-forming comparison sample $z \sim 1$]{%
\includegraphics[width=0.3\linewidth,height=0.25\linewidth]{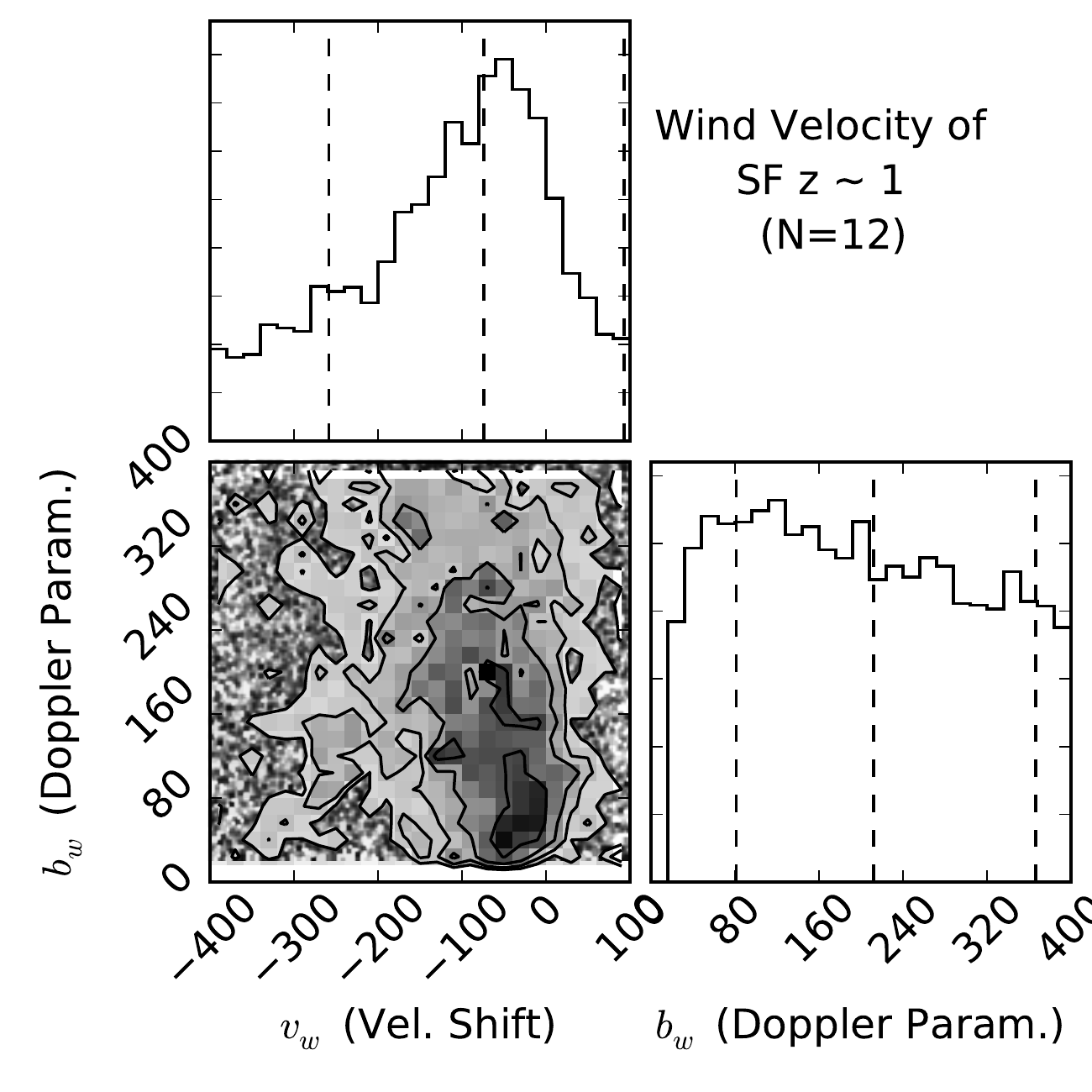}}

\caption{Fitting the wind model to \ion{Fe}{2} $\lambda2586$ absorption lines of AGN and a comparison sample of X-ray undetected galaxies. The black points are data with bootstrap errors. The red curves are convolved wind models constructed from the medians of marginalized posterior PDFs of the model parameters. The pink shadings are regions spanned by 500 randomly-drawn wind models from the posterior PDFs of the model parameters. The orange curves are the unconvolved ISM components of the models while blue curves are the unconvolved wind components. The histograms show the posterior PDFs of wind velocities and wind Doppler dispersion parameters. The dashed lines dividing the histograms indicate 16th, 50th (median) and 84th percentile values. AGN have a wind centroid velocity of $\sim -150$\,km\,s$^{-1}$ which extends only to $\sim 300$\,km\,s$^{-1}$. Similar wind velocities are observed in star-forming X-ray undetected galaxies.\label{fig:model1}} 
\end{figure*}

Figure~\ref{fig:model1} shows, for both AGN and the comparison sample at $z \sim 1$, the observed \ion{Fe}{2} $\lambda2586$ absorption profiles and the fitted model profiles. In the top row, we show the fit to the composite spectrum of all 12 AGN candidates (Figure~\ref{fig:model1}a) or of only the 9 AGN without broad-line AGN (Figure~\ref{fig:model1}b) or only the 7 narrow-line AGN with robust AGN identifications (Figure~\ref{fig:model1}c). In the second row, Figure~\ref{fig:model1}d-f show the corresponding velocity centroid and Doppler parameter PDFs for the three AGN subsamples. Figure~\ref{fig:model1}g \& h show similar figure for \citet{Coil+11} $z \sim 0.5$ AGN sample analyzed separately or jointly with our data. Figure~\ref{fig:model1}i shows observed  \ion{Fe}{2} $\lambda2586$ absorption profiles and the model profiles for the comparison sample of X-ray undetected star-forming galaxies at $z \sim 1$. In last row, Figure~\ref{fig:model1}j-l show the velocity centroid and Doppler parameter PDFs of two reanalyses of the \citet{Coil+11} data and of the comparison sample. In each plot of the \ion{Fe}{2} profile, the black points with error bars are the observed data points.  The blue curve is the wind component while the orange curve is the galaxy absorption component. Both the blue and orange curves are shown before the convolution with the instrumental line-spread-function and thus represent the true components. The red curve is the product of the two components after convolution. All three curves are constructed from medians of the six marginalized model parameters. The marginalized median model fits the data well. The randomly drawn 500 model profiles from the PDFs of the model parameters are shown in pink. They also characterize the flux uncertainties very well. Cutting the histograms depicting PDFs of the centroid velocities and Doppler parameters, the dashed vertical lines mark the 16th, 50th (median), and 84th percentiles of the PDFs.

To summarize results presented in Table \ref{tbl:fit_param} and Figure~\ref{fig:model1}, the wind centroid-velocity for all 12 AGN is \ion{Fe}{2} $\lambda 2586$, is $v_w = -87^{+173}_{-164}$\,km\,s$^{-1}$ and its Doppler dispersion parameter is $b_w = 197^{+161}_{-124}$\,km\,s$^{-1}$. For the 9 narrow-line AGN the centroid-velocity is,  $v_w = -109^{+111}_{-129}$\,km\,s$^{-1}$ and its Doppler parameter is $b_w = 160^{+142}_{-98}$\,km\,s$^{-1}$. These two parameters anti-correlate and their joint PDF is asymmetric. $v_w$ may also be degenerate with the covering factor of the wind (see Figure~\ref{fig:fit_param6}). The value of $v_w$ quoted above is after integrating over $b_w$. The maximum, centroid, velocity-shift ($v_w-2\sigma_{v_w}$, where $\sigma_{v_w}$ is the dispersion of $v_w$ estimated from its PDF) in the narrow-line AGN is likely less than $\sim 370$\,km\,s$^{-1}$. Similarly, the star-forming comparison sample has a wind centroid velocity of $-74^{+167}_{-184}$\,km\,s$^{-1}$ and a Doppler parameter of $212^{+154}_{-131}$\,km\,s$^{-1}$. Therefore, the velocity profiles of winds in AGN at $z \sim 1$ overlap those of the comparison sample at similar redshifts.

Even though the comparison sample has higher SFRs than typical galaxies at a similar redshift, its wind properties are similar to what was found in typical galaxies at $z \sim 1$.  For instance, \citet{Martin+12} found an \ion{Fe}{2} centroid velocity of $-119 \pm 6$\,km\,s$^{-1}$ in several tens of star-forming galaxies at at $z \sim 1$. 

The escape velocity from a galaxy is approximately $5-6 \times$ its \ion{O}{2} emission line velocity dispersion \citep{Weiner+09}, which is $122 \pm 16$ km s$^{-1}$ for the AGN sample and and is $131 \pm 18$ km s$^{-1}$ for the comparison sample (see Appendix section \ref{sec:o2fit} and Figure~\ref{fig:o2fit} therein for detailed information on the measurement of the velocity dispersion). Therefore, most of the outflowing gas does not escape from the host galaxies. Incidentally, the galaxy (ISM) velocity dispersion inferred from fitting the \ion{Fe}{2} $\lambda 2586$ absorption profile is consistent with the \ion{O}{2} emission-line, velocity-dispersion in both AGN and the comparison samples. 

The total equivalent width (the combined contribution of the ISM and the wind components) is $1.6^{+0.2}_{-0.3}${\AA} for all 12 AGN, $1.8^{+0.2}_{-0.3}${\AA} for the 9 narrow-line AGN and $1.9^{+0.4}_{-0.3}${\AA} for the star-forming sample. About a third of the total equivalent width is due to the wind component in both samples (i.e., $0.4^{+0.4}_{-0.3}${\AA} for all AGN and $0.5^{+0.7}_{-0.3}${\AA} for star-forming sample). The presence of strong ISM component ($1.2^{+0.3}_{-0.4}${\AA}) in the AGN implies that substantial amount cold gas is present in the host galaxies and it has not been affected by AGN feedback. The maximum wind velocity ($v_w-2\,b_w/\sqrt 2$) is $-234^{+153}_{-194}$\,km\,s$^{-1}$ for the 12 AGN while it is $-228^{+177}_{-227}$ \,km\,s$^{-1}$ for the comparison sample. The equivalent width and maximum velocity measurements for the other samples of AGN are found in Table~\ref{tbl:eqw}.

We also find weak winds in AGN at $z \sim 0.5$ upon reanalysis of \citet{Coil+11} data. The wind centroid velocity for this sample  is $-93^{+248}_{-209}$\,km\,s$^{-1}$ and its Doppler dispersion parameter is $190^{+166}_{-130}$\,km\,s$^{-1}$. \citet{Coil+11} have measured the \ion{Fe}{2} $\lambda 2586$ centroid velocities and velocity widths for four of their AGN. Their measurements for both these two quantities range roughly between $130$ to $330$\,km\,s$^{-1}$. Averaging their four measurements with inverse-variance weighting gives, a centroid velocity shift of $-180 \pm 9$\,km\,s$^{-1}$ and a Doppler parameter of $274 \pm 13$\,km\,s$^{-1}$. The total equivalent width of \ion{Fe}{2} in this sample is $2.0^{+0.5}_{-0.5}${\AA} and the equivalent width due to the wind component is $0.6^{+0.7}_{-0.4}${\AA} . The average wind equivalent width measured by \citet{Coil+11} is 1.2{\AA}.

\begin{deluxetable*}{lccccccc}
\tablewidth{0in}
\tabletypesize{\footnotesize}
\tablecaption{Wind model parameter fits to \ion{Fe}{2} $\lambda2586$ profiles. \label{tbl:fit_param}}

\tablehead{ \colhead{Model} & \colhead{AGN $z \sim 1$} & \colhead{NLAGN $z \sim 1$} & \colhead{NLAGN $z \sim 1$} & \colhead{SF $z \sim 1$} & \colhead{AGN $z \sim 0.5$\tablenotemark{a}} & \colhead{AGN $z \sim 0.5-1.5$\tablenotemark{b}} & \colhead{AGN $z \sim 0.5-1.5$\tablenotemark{c}} \\
\colhead{Param.} & \colhead{(N = 12)} & \colhead{(N = 9)} & \colhead{(N = 7)} & \colhead{(N = 12)} & \colhead{(N = 6)} & \colhead{(N = 13)} & \colhead{(N=18)\tablenotemark{d}}}\\
\startdata

$v_w$ (km s$^{-1}$) &$ -87^{+173}_{-164} $& $-109^{+111}_{-129}$  & $-143^{+109}_{-98}$ & $-74^{+167}_{-184}$ & $-93^{+248}_{-209}$ &$-119^{+144}_{-135}$ &$-98^{+153}_{-180}$ \\
$b_w$ (km s$^{-1}$) & $197^{+161}_{-124} $&$160^{+142}_{-98}$ & $134^{+145}_{-84}$ & $212^{+154}_{-131}$ & $190^{+166}_{-130}$ & $131^{+175}_{-84}$ &$171^{+178}_{-116}$ \\
$C_w $& $0.2^{+0.4}_{-0.2}$ & $0.3^{+0.4}_{-0.2}$ &$ 0.3^{+0.4}_{-0.2} $& $0.3^{+0.4}_{-0.2}$ &$ 0.4^{+0.2}_{-0.1}$ & $0.4^{+0.4}_{-0.3}$ &$ 0.3^{+0.4}_{-0.2}$\\
$\log N_w$ (cm$^{-2}$)&$ 14.7^{+1.3}_{-0.5}$ &$ 14.9^{+1.4}_{-0.6}$ &$14.9^{+1.5}_{-0.6}$ &$14.9^{+1.3}_{-0.6}$&$ 15.2^{+1.3}_{-0.6}$ &$15.0^{+1.4}_{-0.7}$  &$14.9^{+1.4}_{-0.6}$\\
$b_g$ (km s$^{-1}) $&$ 240^{+78}_{-66} $&$ 228^{+74}_{-61}$ & $238^{+98}_{-83}$&$187^{+81}_{-89}$ & $137^{+189}_{-97}$ & $162^{+135}_{-115}$ &$158^{+113}_{-112}$\\
$\log N_g$ (cm$^{-2}$) &$ 14.5 ^{+0.1}_{-0.2} $&$14.6^{+0.1}_{-0.2}$ & $14.5^{+0.1}_{-0.2}$  &$ 14.7^{+0.1}_{-0.2}$ &$ 14.7^{+0.9}_{-0.3}$ &$14.6^{+0.2}_{-0.3}$ &$14.6^{+0.2}_{-0.2}$\\
\enddata

\tablenotetext{a}{Renalysis of \citet{Coil+11} data}
\tablenotetext{b}{A joint renalysis of \citet{Coil+11} data with our 7 narrow-line AGN with highly reliable AGN identification.}
\tablenotetext{c}{A joint renalysis of \citet{Coil+11} data with all AGN including 3 broad-line AGN and 2 with less reliable AGN identification.}
\tablenotetext{d}{N denotes the number of galaxies in the stacked spectra.}
\tablecomments{The parameters of the two component wind model are:  the velocity centroid shift of the wind, $v_{w}$, the Doppler broadening parameter of the wind, $b_{w}$, the covering fraction of the wind, $C_{w}$, column density of the wind, $N_{w}$, the Doppler parameter of the ISM in the galaxy, $b_g$, and the column density of the gas in the galaxy, $N_{g}$. The median, the 84th and 16th percentile deviations of the PDFs of the model parameters are given in the table. In our notation, $X^{+Y}_{-Z}$ denotes that X is the median, $X+Y$ is the 84th percentile and $X-Z$ is the 16th percentile.}
\end{deluxetable*}

\begin{deluxetable*}{lccccccc}
\tablewidth{0in}
\tabletypesize{\footnotesize}
\tablecaption{Wind equivalent width (EW) and maximum wind velocity derived from \ion{Fe}{2} $\lambda2586$ model profiles. \label{tbl:eqw}}

\tablehead{ \colhead{Derived} & \colhead{AGN $z \sim 1$} & \colhead{NLAGN $z \sim 1$} & \colhead{NLAGN $z \sim 1$} & \colhead{SF $z \sim 1$} & \colhead{AGN $z \sim 0.5$} & \colhead{AGN $z \sim 0.5-1.5$} & \colhead{AGN $z \sim 0.5-1.5$} \\
\colhead{Quantity} & \colhead{(N = 12)} & \colhead{(N = 9)} & \colhead{(N = 7)} & \colhead{(N = 12)} & \colhead{(N = 6)} & \colhead{(N = 13)} & \colhead{(N=18)}}\\
\startdata
Wind EW ({\AA}) & $0.4^{+0.4}_{-0.3} $& $0.5^{+0.6}_{-0.3}$  & $0.5^{+0.5}_{-0.3}$ & $0.5^{+0.7}_{-0.3}$ & $0.6^{+0.7}_{-0.4}$ &$0.6^{+0.5}_{-0.4}$ &$0.5^{+0.5}_{-0.3}$ \\
ISM EW ({\AA})  & $1.2^{+0.3}_{-0.4} $&$1.3^{+0.4}_{-0.5}$ & $1.3^{+0.4}_{-0.5}$ & $1.4^{+0.5}_{-0.5}$ & $1.4^{+0.6}_{-0.5}$ & $1.1^{+0.3}_{-0.5}$ &$1.2^{+0.3}_{-0.4}$ \\
Total EW ({\AA}) & $1.6^{+0.2}_{-0.3}$ & $1.8^{+0.2}_{-0.3}$ &$ 1.8^{+0.2}_{-0.3} $& $1.9^{+0.4}_{-0.3}$ &$2.0^{+0.5}_{-0.5}$ & $1.7^{+0.3}_{-0.3}$ &$ 1.7^{+0.2}_{-0.3}$\\
Max. Velocity (km s$^{-1}$)  &$ -234^{+153}_{-194}$ &$ -236^{+114}_{-107}$ &$-242^{+108}_{-124}$ &$-228^{+177}_{-227}$&$ -204^{+214}_{-266}$ &$-202^{+112}_{-166}$  &$-203^{+140}_{-243}$\\
\enddata
\tablecomments{The median, the 84th and 16th percentile deviations of the PDFs of the derived quantities from the wind model are given in the table. In our notation, $X^{+Y}_{-Z}$ denotes that X is the median, $X+Y$ is the 84th percentile and $X-Z$ is the 16th percentile.}
\end{deluxetable*}

\section{Discussion} \label{sec:disc}

A key physical manifestation of active galactic nuclei (AGN) feedback is predicted to be powerful galactic winds. However, the relative roles between AGN activity and star formation in driving such winds remain largely unexplored at redshifts $z \sim 1$, near the peak of cosmic activity for both. We study winds in 12 X-ray AGN host galaxies at $z \sim 1$ in the CANDELS fields using deep Keck rest-frame UV spectroscopy. We find that winds in the AGN are similar to those found in star-formation-driven winds, and are too weak to escape and expel substantial cool gas from their host galaxies.

Despite theoretical appeal, confirming evidence of star-formation quenching by powerful winds in AGN remains elusive. Here, we discuss some of the evidence reported in the literature, with emphasis on those with larger samples when multiple, similar studies exist. Our aim is to show that most wind studies using cold, warm or molecular gas are either in agreement or consistent with our finding that low luminosity AGN ($L_\mathrm{X} \sim 10^{42}$ erg s$^{-1}$) have similar winds as those from star-forming galaxies of similar galaxy properties, especially if the comparison is made at similar AGN luminosities as those of our sample.

\subsection{Most previous cold gas absorption studies also found low wind velocities}

Neutral gas outflows in low redshift $z \sim 0.1$ AGN have been extensively studied using \ion{Na}{1} \citep[e.g.,][]{Rupke+05,Krug+10}. The clearest result from these studies is that the wind velocities of narrow-line (type 2) AGN are similar to the wind velocities of starburst and star-forming galaxies. \citet{Krug+10} studied outflows in 35 infrared-faint (i.e., low star-forming) Seyferts in an effort to disentangle the starburst effects on the winds from the AGN effects. The authors compared the outflow properties of these Seyferts with that of infrared-bright composite Seyferts in which both starbursts and AGN co-exist. The wind detection rates for the infrared-faint Seyfert 1s (6\%) and Seyfert 2s (18\%) are lower than previously reported for infrared-luminous Seyfert 1s (50\%) and Seyfert 2s (45\%). In addition, the outflow velocities of both high and low SFR Seyfert 2s are similar to those of starburst galaxies, while the outflow velocity in only one out of eighteen Seyfert1s  is significantly higher. The measured average wind velocity for infrared-faint Seyferts 2 galaxies ($v =-137\pm  8$ km s$^{-1}$, $b = 250 \pm 214$\,km\,s$^{-1}$) and the authors' conclusion that AGN do not play a significant role in driving the outflows in most local infrared-faint and infrared-bright Seyferts 2s is consistent with our result. The particular Seyfert1 with the strong wind has an average wind velocity -600\,km\,s$^{-1}$ and very small velocity dispersion ($b = 21 \pm 6$\,km\,s$^{-1}$). It is likely that this object's high velocity measurement is affected by emission infill at systemic velocity. 

Likewise, \citet{Rupke+05} studied a sample of 26 Seyfert ULIRGs using \ion{Na}{1}. They found no significant differences between the velocities of Seyfert 2s which are ultra-Infrared galaxies (ULIRGs) ($v =-456^{+330}_{-191}$\,km\,s$^{-1}$, $b = 232^{+244}_{-119}$\,km\,s$^{-1}$) and starbursts of comparable infrared luminosity ($v =-408^{+224}_{-191}$\,km\,s$^{-1}$, $b = 232^{+244}_{-119}$\,km\,s$^{-1}$).  They also found very high velocities ($\sim 5000$\,km\,s$^{-1}$) in two Seyfert 1 AGN  and argued that they are likely small-scale ($\sim 10$\,pc) disk winds. They stressed that large-scale, lower velocity outflows certainly exist in Seyfert 1 ULIRGs, since such winds are common in general infrared bright galaxies, but the wind signatures are likely rendered unobservable by the intense nuclear radiation in Seyfert 1s due to infilling of the absorption profile by scattered emission or due to ionization to higher states of the absorbing atoms.

Recently, \citet{Sarzi+16} studied a sample of 456 nearby galaxies of which 103 exhibit compact radio emission indicating radio AGN activity. They found that only 23 objects (5\%) out of their entire sample exhibited outflow signatures in \ion{Na}{1}. Not even a single object showed evidence of AGN activity in radio and of cold-gas outflow simultaneously. Radio-AGN activity was found predominantly in early-type galaxies, while cold-gas outflows were mainly observed in late-type galaxies with central star formation or with composite galaxies of star formation and AGN activities. The authors emphasized that their work supports a picture in which the onset of AGN activity appears to lag behind the peak of starburst activity \citep[e.g.,][]{Wild+10,Yesuf+14}, and in which the gas reservoir has been significantly depleted by star-formation or stellar feedback before the AGN had a chance to couple to it. 

Similarly, \citet{Sato+09} found \ion{Na}{1} outflow velocities of $\sim 100$\,km\,s$^{-1}$ in fading post-starburst galaxies with low-level nuclear activity at $0.1 < z < 0.5$. Within a similar redshift range,  \citet{Coil+11} also found low velocity winds ($\sim 200$\,km\,s$^{-1}$) in 13 post-starbursts galaxies by using \ion{Mg}{2} and \ion{Fe}{2} absorption lines. This result is in addition to the low-velocity outflows they found in low-luminosity AGN. 

In contrast, \citet{Tremonti+07} observed high-velocity winds (with median $v \sim 1100$\,km\,s$^{-1}$) in massive transitional post-starburst galaxies and concluded that AGN likely played a major role in the abrupt truncation of star-formation in these systems. But, in subsequent works, they argued that these fast outflows are most likely driven by feedback from extremely-compact, obscured, star formation rather than AGN \citep{Diamond-stanic+12,Geach+14,Sell+14}. But, it remains possible that the outflows were driven by AGN activity that has been recently switched off or are driven by extremely obscured AGN. \citet{Sell+14} found low-luminosity AGN in half of their post-starburst sample. Nevertheless, it should be noted that the authors concluded that the fast outflows are most likely driven by feedback from star formation rather than AGN. Generally, AGN are known to be common among post-starburst galaxies but are not directly linked to quenching starbursts \citep{Yesuf+14}.

To summarize the discussion so far, to our knowledge, all AGN-host galactic-wind studies using (near-)UV or optical absorption lines, with the exception of \citet{Hainline+11}, mentioned in the introduction, found moderate wind velocities in AGN which are similar to those from star-forming galaxies. Attributing emission infill for the difference with \citet{Hainline+11}'s absorption profiles, \citet{Coil+11} concluded that their finding, namely, AGN host galaxies at $z \sim 0.2-0.5$ do not have significantly faster winds than star-forming galaxies at similar redshifts, was not strongly at odds with results from lower and higher redshifts. Our work and those discussed above affirm this conclusion.

\subsection{Some previous ionized gas emission-line studies found high wind velocities and some did not}

Next, we discuss AGN winds detected in emission lines of ionized gas. Emission lines as wind diagnostics are much more difficult to interpret compared to absorption lines. For instance, to infer the mean velocity of the wind from emission lines, detailed understanding of the geometry of the wind, velocity distribution of the gas, and dust extinction in the host galaxy is needed. In a spherically-symmetric optically-thin outflow, the emission-line profile is symmetric and peaks at the systemic velocity of the galaxy (i.e, zero velocity). In contrast, in absorption lines which are minimally affected by resonant emission, the wind velocity profile is significantly offset from the systemic velocity. Regardless of how the emission line observations are interpreted, the result that low luminosity AGN at redshifts $z \sim 0.2-1$ do not have significantly faster winds (in absorption) than star-forming galaxies is consistent with the wind speeds inferred from emission lines in AGN of comparable luminosities.

\citet{Rupke+13} have explored the multiphase structure of galactic winds in six local ULIRGs using deep integral-field spectroscopy. Three of the ULIRGs host obscured quasars. Despite its small sample size, this work is unique by studying winds in the same objects in both emission and absorption, and it serves as a benchmark for interpreting myriads of emission-line only wind studies. Both the neutral and ionized gas of the six ULIRGs were studied by fitting the \ion{Na}{1} absorption line and multiple Gaussian components to strong, nebular, emission lines ([\ion{O}{1}], H$\alpha$ and [\ion{N}{2}]). In all systems, high-velocity, collimated, multiphase kiloparsec-scale outflows were reported and the neutral phase dominates the mass outflow rate. The spatially-averaged, mean, wind velocities were found to be similar ($v \sim 200-400$ km s$^{-1}$) in AGN and non-AGN, both for cold, neutral and warm, ionized gas. While the maximum wind velocities reach $\sim 1000$ km s$^{-1}$ in neutral gas for both AGN and non-AGN, the highest gas velocities ($2000-3000$  km s$^{-1}$) were only observed in ionized gas in the obscured quasars. 

Several spatially-resolved, spectroscopic studies at both low and high redshifts have shown that broad, ionized emissions are common in luminous AGN 
\citep{Liu+13, ForsterSchreiber+14, Genzel+14, Liu+14, McElroy+15, Harrison+16, Zakamska+14}. For example, \citet{Harrison+16} studied ionized gas kinematics in a representative sample of 89 X-ray AGN using [\ion{O}{3}] at $z =1.1-1.7$ or H$\alpha$ emission at  $z=0.6-1.1$. The authors found high-velocity emission-line features in about half of the targets studied using [\ion{O}{3}] . The velocity-width containing 80 \% ($W_{80}$) of the [\ion{O}{3}] line flux was found to mildly correlate with X-ray luminosity. For a Gaussian velocity distribution, $W_{80}=2.56\, \sigma$, where $\sigma$ is the velocity dispersion. \citet{Liu+13} found $W_{80} \sim 1.3-1.6 \,v_0$ for their wind models, where $v_0$ is the initial velocity of the wind. \citet{Liu+13} modelled the observed extended emission lines of their quasars as ensembles of narrow-line-emitting clouds embedded in the wind. Their model relates the observed projected surface brightness of the [\ion{O}{3}] emission line to the unobserved three dimensional outflow velocity profile assuming power-law luminosity density and velocity distributions, which depend on the three-dimensional radius vector of the outflow. They consider both a spherically symmetric outflow and a biconical outflow with or without the effect of dust extinction from the host galaxy. \citet{Harrison+16} found that 70\% of higher-luminosity AGN ($L_\mathrm{X} > 6 \times 10^{43}$ erg s$^{-1}$) have line widths of $W_{80} > 600$ km s$^{-1}$, while only 30\% of the lower-luminosity AGN have velocity widths above 600 km s$^{-1}$. If we use the trend in \citet{Harrison+16}, the 9 low-luminosity AGN studied in this work would have $W_{80} \sim 300$ km s$^{-1}$ based on their X-ray luminosity. This value is roughly consistent with the velocity estimated using the \ion{Fe}{2} absorption line. 

AGN are known to exhibit jet-ISM interactions that accelerate gas to high velocities \citep{Morganti+05,Nesvadba+06,Nesvadba+08,Dasyra+15}. Based on their current sample, \citet{Harrison+16} could not conclusively determine whether the radio luminosity or the X-ray luminosity is more fundamental in driving the highest-velocity outflows. They found marginal evidence that a higher fraction of the radio-luminous AGN have $W_{80} > 600$ km s$^{-1}$ compared to the non-radio-luminous AGN. Thus, high-velocity outflows may be due to small-scale, compact radio jets instead of radiation from the quasar \citep{VillarMartin+14,Mullaney+13}. Spatially-resolved studies have observed very-broadened [\ion{O}{3}] emission-line that are co-spatial with kiloparsec scale jets \citep{Holt+08,Muller-Sanchez+11, Husemann+13,Shih+13}. At $z \sim 0.4$, using a large sample from SDSS, \citet{Mullaney+13} have shown that the highest velocity outflows are better linked to the mechanical radio luminosity of the AGN rather than to the radiative luminosity of the AGN. Alternatively, \citet{Zakamska+14} have proposed that radio emission in radio-quiet quasars could be due to relativistic particles accelerated in the shocks initiated by the quasar-driven outflow. The authors also found that the velocity width of [\ion{O}{3}] is positively correlated with mid-infrared luminosity, suggesting that outflows are linked to the radiative output of the quasar (i.e., are ultimately radiation-driven.)

Furthermore, \citet{Husemann+13} performed a detailed analysis of the extended ionized gas around 31 low-redshift QSOs and found only 3 QSOs have outflows with velocities greater than 400 km s$^{-1}$. In all three cases, they found a radio jet that is most likely driving the outflows, and they argued that jet-cloud interactions are the most likely cause of disturbances in the kinematics of the quasars. \citet{Husemann+15} have argued that disagreements between their work and the aforementioned previous works, that claimed high velocity outflows in luminous AGN, are likely due to the effects of beam smearing of unresolved emission lines caused by seeing \citep[see also][]{Karouzos+16,Villar-Martin+16}. They reanalyzed the unobscured QSO sample of \citet{Liu+14} and found that the widths of [\ion{O}{3}] lines on kiloparsec scales are significantly narrower after PSF deblending. The estimated kinetic power of the outflow is reduced by two orders of magnitude ($< 0.1$\% of the quasar bolometric luminosity) after the correction. Thus, the feedback efficiency is smaller than required by some numerical simulations of AGN feedback. As the authors pointed out, the majority of previous works have not carefully taken into account the effects of beam smearing. The incidence and energetics of large-scale AGN-driven outflows still remain an unsolved issue, especially in spatially unresolved observations of ionized gas outflows beyond the local universe.

\subsection{Some previous molecular gas studies found high wind velocities.}

Molecular outflows have been reported in several AGN both in absorption and emission \citep{Fischer+10, Feruglio+10,Sturm+11,Spoon+13,Veilleux+13,Cicone+14,Garcia-Burillo+14,Sun+14}. \citet{Veilleux+13} studied molecular OH 119$\mu$m outflows in a sample of 43, $z < 0.3$, galaxy mergers, which are mostly ULIRGs and QSOs.  The OH 119$\mu$m feature is observed in emission, absorption, or both depending on the AGN strength. The OH emission is stronger relative to OH absorption in quasar-dominated systems and the feature is seen in pure emission in the most luminous quasars. The authors found that the median outflow velocities are typically $\sim 200 $ km s$^{-1}$ but the maximum velocities may reach $\sim 1000$ km s$^{-1}$ in some objects. For even the most AGN-dominated systems with pure OH emissions, the emission line widths and shifts are $\sim 200$ km s$^{-1}$. The authors also reported that the absorption line centroids are distinctly more blue-shifted among systems with large AGN fractions and luminosities. It not clear how much this trend is due to emission infill of the absorption profile.  

A recent X-ray observation of a mildly relativistic accretion disk wind in a local Seyfet 1 ULIRG, which also shows high velocity molecular OH 119$\mu$m  outflow have been hailed as providing direct connections between large-scale molecular outflows and the small-scale, AGN accretion wind in ULIRGs \citep{Tombesi+15}. A review of the powerful and highly ionized accretion winds observed in X-ray spectra of luminous AGN can be found in \citet{KingPounds15}.

\citet{Cicone+14} have studied CO emission in 19 local ULIRGs and quasars hosts. They found that starburst-dominated galaxies can have an outflow rate which are $\sim 2-4$ times their star formation rates and the presence of AGN may enhance the outflow rates by a large factor depending on its luminosity.  The maximum velocities reach up to $\sim 750$ km s$^{-1}$. The authors estimated that the outflow kinetic power for galaxies with the most powerful AGN is about 5\% of the AGN luminosity,  as expected from some numerical models of AGN feedback. 

In contrast, recent, local studies of molecular gas in recently quenched or quenching post-starbursts surprisingly found that these galaxies have large molecular gas reservoirs comparable to star-forming galaxies \citep{French+15, Rowlands+15}. Therefore, they did not find evidence that the global gas reservoir is expelled by stellar winds or active galactic nuclei feedback. Similarly, at $z \sim 2$, \citet{Prochaska+14} observed that quasar halos are have abundant, cool gas which is sufficient to fuel the observed SFR for at least 1 Gyr. These authors note that the current AGN feedback models remove too much gas from galactic halos and, therefore, under-predict the gas observed within quasar halos at $z \sim 2$.

To summarize, most absorption-line studies found that the wind velocities in AGN are moderate ($\sim 200-400$) km s$^{-1}$ and are similar to velocities in star-formation-driven winds.
Most high-velocity AGN winds reported to date are controversial and can be attributed to observational complications such as emission infill and PSF smearing or they may not be due to radiation from luminous AGN, or they may just be spatially-unresolved, small-scale wind confined to the vicinity of black holes.

\subsection{The feedback efficiency in the low-luminosity AGN at $z \sim 1$ and in other AGN samples.}

\begin{figure*}
\centering
\includegraphics[scale=0.8]{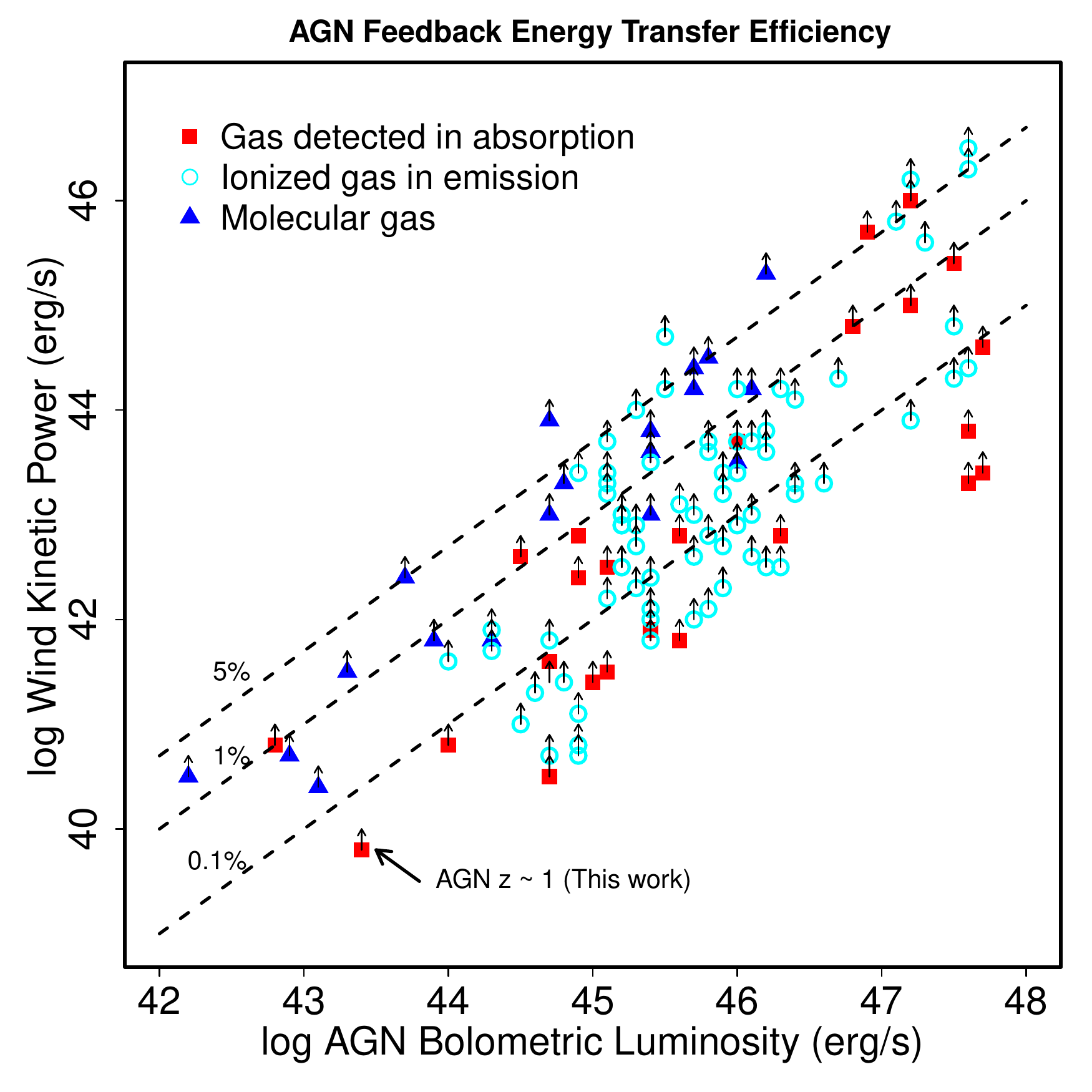}
\caption[AGN bolometric luminosity against the wind kinetic power for our sample and other AGN wind samples in the literature]{AGN bolometric luminosity against the wind kinetic power for our sample and other AGN wind samples in the literature, covering a wind range in redshift $z \sim 0.1-2$. Our compilation include AGN which are known to show wind signatures in cool gas absorption \citep{Edmonds+11,Rupke+05,Krug+10,Borguet+12,Arav+13,Borguet+13,Rupke+13,Chamberlain+15}, in ionized gas emission \citep{Liu+13,Rupke+13,Harrison+14,Harrison+16,McElroy+15,Carniani+15,Zakamska+16} and in molecular gas \citep{Combes+13,Cicone+14,Sun+14,Garcia-Burillo+15,Tombesi+15}. The dashed lines are lines of constant feedback efficiencies. }5\% feedback efficiency is used in popular feedback models \citep[e.g.,][]{Scannapieco+04,DiMatteo+05,Choi+12, Zubovas+12}. We caution that the systematic uncertainties in wind power are very high. The trends in this figure should be interpreted qualitatively.
 \label{fig:feedeff}
\end{figure*}

The wind kinetic power can be parametrized in terms of the radiative luminosity of an AGN  as $\dot{E} = \epsilon_f L_\mathrm{AGN}$, where  $\epsilon_f $ is the fraction of the radiative luminosity transferred to the wind. Popular AGN feedback models invoke feedback efficiency, $\epsilon_f \sim 5\%$ to reproduce the properties of massive galaxies \citep[e.g.,][]{Scannapieco+04,DiMatteo+05,Choi+12, Zubovas+12}. Following the thin, partially-filled shell wind model of \citet{Rupke+05b}, we estimate $\dot{E}$ from their equation given below, which includes a contribution from both bulk flow and turbulent energy. 

\begin{equation*}
\begin{aligned}
dE/dt  & = 1.4 \times 10^{41} {\rm erg\,s^{-1}} \left(\frac{C_\Omega}{0.4}C_f \right)\left( \frac{r}{10 \,{\rm kpc}}  \right) \\
          & \times \left( \frac{N(H)}{10^{21}\,{\rm cm^{-2}}}  \right) \left( \frac{v_w}{200\,{\rm km\, s^{-1}}}  \right) \\
          & \times \left[ \left( \frac{v_w}{200\,{\rm km\, s^{-1}}}  \right)^2+1.5\left( \frac{b}{200\,{\rm km \, s^{-1}}}  \right)^2\right]
\end{aligned}
\end{equation*}

In the above estimate, we use the wind measurements in Table~\ref{tbl:fit_param} and assume a wind radius of $r=3$ kpc and global covering factor $C_\Omega=0.3$ (which is reasonable for the average opening angle of winds in star-forming galaxies $z \sim 1$ as estimated by \citet{Rubin+14}). To convert the Iron column density to Hydrogen column density (N(H), which is a lower limit), we adopt solar abundance ratio and a dust depletion factor of 0.1 and no ionization correction \citep{Rubin+14}. The sample bolometric luminosity of our AGN is estimated by multiplying the mean X-ray luminosity of our sample with bolometric correction of 20 \citep{Vasudevan+07}. With these assumptions, we find feedback efficiency lower limit estimate of $\epsilon_f \gtrsim 0.02\%$ for our nine narrow-line AGN sample.

Note that that the metal abundance ratio, dust depletion factor, and ionization correction are completely unconstrained by the current data. 
From the mass-metallicity relation at $z \sim 1$ \citep{Zahid+11}, the mean oxygen gas abundance, 12 + log(O/H), of our sample ranges between 8.9 -- 9.1 and the typical 1$\sigma$ scatter in of the relation is $\sim 0.15$ dex. Therefore, the gas phase metallicity of our sample is not very inconsistent with the assumed solar value of 8.7. In the local ISM of the Milky Way, the dust depletion factor for Iron is 0.005 -- 0.1 \citep{Jenkins09}. The ionization correction for \ion{Fe}{2} is uncertain and there are not many measured values of it. In the Orion Nebula,  the ionization correction is $\sim 0.08$, and the column densities of N(\ion{Fe}{4}) $\sim 0.87 \times$ N(\ion{Fe}{3}) and N(\ion{Fe}{2}) $\sim 0.16 \times$ N(\ion{Fe}{3}) \citep{Rodriguez+05}. The N(\ion{Fe}{2})/N(\ion{Fe}{3}) ratio ranges between 0.04 -- 0.35 in 8 Galactic H II regions, including the Orion Nebula \citep{Garcia+07}. Since N(\ion{Fe}{4}) is not measured in most of these H II regions, 0.04 -- 0.35 is a crude range for the ionization correction for \ion{Fe}{2}. Since we do not do the ionization correction in our lower-limit estimate of the column density, the true column density may be a factor of 3 -- 10 higher than our limit, other things being equal.

To put the present measurements in larger context, Figure~\ref{fig:feedeff} shows AGN bolometric luminosity against the wind kinetic power for our sample and other AGN wind samples in the literature. A very heterogeneous data, covering a wide range in redshift $z \sim 0.1-2$, are used in the figure. Our compilation include AGN which are known to show wind signatures in absorption \citep{Edmonds+11,Rupke+05,Krug+10,Borguet+12,Arav+13,Borguet+13,Rupke+13,Chamberlain+15}, in ionized gas emission \citep{Liu+13,Rupke+13,Harrison+14,Harrison+16,McElroy+15,Carniani+15,Zakamska+16} and in molecular gas \citep{Combes+13,Cicone+14,Sun+14,Garcia-Burillo+15,Tombesi+15}

All the absorption line and molecular gas based kinetic wind power measurements are taken directly from values published in the literature, while all the emission-line based the wind power measurements except those taken from \citet{Rupke+13} are based on our calculations. Where the wind power measurements for the ionized gas are not available, we estimate them following the standard methods and assumptions \citep[e.g.,][]{Nesvadba+06,Harrison+14, Zakamska+16}. Using the equation below, the wind power for all ionized gas measurements are estimated from nebular emission lines assuming a wind radius of 3 kpc, electron density of 100 cm$^{-3}$ and the velocity-width is 1.3 times the initial wind velocity, $W_{80}=1.3 v_0$ (for a spherically symmetric and constant velocity wind). 

\begin{equation*}
\begin{aligned}
dE/dt & \sim 1/2\,M_{\rm gas} v_0^2/\tau \sim 1/2 M_{\rm gas} v_0^3/r \\
& \sim  6 \times 10^{44}\, {\rm erg\,s^{-1}} \left(\frac{M_{\rm gas}}{2.8 \times 10^9 \,{\rm M_\odot}}\right) \\
& \times \left(\frac{W_{80}}{1300\, {\rm km s^{-1}}}\right)^3\left(\frac{3{\rm\,kpc}}{r}\right)\\
\end{aligned}
\end{equation*}
\begin{equation*}
$$$ {\rm Where } \, \frac{M_{\rm gas}}{2.8 \times 10^9 {\rm M_\odot}}= \frac{L_{\rm H\beta}}{10^{43}\,{\rm erg s^{-1}}} \frac{n_{\rm e}}{100 \,{\rm cm^{-3}}} \\ $$$
$$$ {\rm and } \, L_{\rm H\beta} =  0.1 \times L_{\rm O III}\frac{10}{{\rm OIII/H\beta}}  {\rm or}  L_{\rm H\beta}  = 0.35\times L_{\rm H\alpha}\frac{2.86}{{\rm H\alpha/H\beta}} \\$$$
\end{equation*}

As discussed in \citet{Zakamska+16}, a standard method of estimating the wind power for the ionized gas is to use Hydrogen recombination lines to estimate the mass of the emitting Hydrogen, but [\ion{O}{3}] emission-line may be a better probe of the extended emission. When they are available, \ion{O}{3} velocity width and luminosity are used to estimate wind kinetic power, assuming [\ion{O}{3}]/H$\beta$ ratio of 10. Otherwise, Hydrogen lines are used. Due to unaccounted for dust extinction of emission lines and turbulent kinetic energy, the estimated wind kinetic energy for the ionized gas is a lower limit.

The outflow rate in molecular gas is estimated assuming continuously filled spherical wind ($\dot{M}= 3vM/R$, where M is the mass, v is velocity and R is the radius of the wind). This estimates is three times lower if shell-like geometry is assumed instead \citep{Cicone+14}. The molecular wind kinetic energy is estimated simply as  $\frac{1}{2}\dot{M}v^2$. For 90\% of the galaxies, the molecular wind mass is estimated from CO luminosity, assuming a conversion factors from CO to molecular Hydrogen, $X_{\rm CO}$, which is about one-fifth of the Milky Way value \citep{Bolatto+13}. If the true $X_{\rm CO}$ is higher than assumed, the current wind kinetic energy values underestimate the true values.

For the consistency, we adjust the literature wind power measurements for the cool gas to our assumed wind radius of 3 kpc when the wind radius were not previously measured. When the bolometric luminosities are not provided in the previous works, they are estimated from literature X-ray luminosities using bolometric correction of 20.

In summary, the existing data hint that the wind kinetic energy is correlated with the bolometric luminosity AGN. However, better future data, which can characterize well the geometry of the wind are need to constrain AGN feedback models. We caution that the systematic uncertainties in $\epsilon_f$ are substantial and the values $\epsilon_f$ inferred from Figure~\ref{fig:feedeff} should be considered as lower limits. Because of the uncertain assumptions made and the heterogenous data used to estimate wind kinetic power, our aim is rather to qualitatively show that wind kinetic power increases with AGN bolometric luminosity. This trend is mainly driven by the correlation between the wind velocity and the AGN luminosity. We hope Figure~\ref{fig:feedeff} shows how our study broadly fits into wind characteristics reported in previous studies of AGN, and perhaps explains our finding of low velocity winds in low luminosity AGN.

\section{Summary and Conclusion} \label{sec:conc}

We study winds using the \ion{Fe}{2} $\lambda 2586$ absorption line in 12 AGN host galaxies at $z \sim 1$.  Nine of these galaxies significantly deviate from the relationship between star-formation and X-ray luminosity and one of them has strong \ion{Ne}{5} emission. We find that the probability distribution function (PDF) of the centroid velocity shift in AGN has a median, 16th and 84th percentiles  of -87 km s$^{-1}$, -251 km s$^{-1}$ and +86 km s$^{-1}$ respectively. The PDF of the velocity dispersion in AGN has a median, 84th and 16th percentile  of 139 km s$^{-1}$, 253 km s$^{-1}$ and 52 km s$^{-1}$ respectively. The centroid velocity and the velocity dispersions are obtained from a two component (ISM+wind) absorption line model. The wind velocities in these AGN are significantly lower than their escape velocities. Thus, the bulk of their gas likely remains bound. The equivalent width PDF of the outflow in AGN has a median, 84th and 16th percentiles of 0.4{\AA}, 0.8{\AA}, and 0.1{\AA} respectively. There is a strong ISM component in \ion{Fe}{2} $\lambda 2586$ absorption (has a PDF with a median, 84th and 16th percentiles of 1.2{\AA}, 1.5{\AA}, and 0.8{\AA}), implying substantial amount cold gas is present in the AGN host galaxies. For comparison, star-forming and X-ray undetected galaxies at a similar redshift, matched roughly in stellar mass and galaxy inclination, have a centroid velocity PDF with a median, 84th and 16 percentiles of -74 km s$^{-1}$, -258 and +90 km s$^{-1}$, and a velocity dispersion PDF with a median, 84th and 16th percentiles of 150 km s$^{-1}$, 259 km s$^{-1}$ and 57 km s$^{-1}$ respectively. The equivalent width PDF of the outflow in the comparison sample has  a median, 84th and 16th percentiles of 0.5{\AA}, 1.2{\AA}, and 0.2{\AA}. We have reanalyzed the sample of 6 low-luminosity AGN at $z \sim 0.5$ from \citet{Coil+11}. Our result is consistent with the wind velocities previously reported in these and other lower-redshift low-luminosity AGN. We conclude that the wind-mode AGN feedback is insignificant in low-luminosity AGN hosts. Future, large-sample-size and high signal-to-noise studies of winds in AGN and in a well-matched control sample of non-AGN are needed to significantly advance our knowledge from existing small sample absorption line studies, and to enable detailed modeling of winds which will potentially uncover subtle differences between the winds in AGN and in their control sample.\\
\\
We are very thankful to the anonymous referee for his/her valuable comments and suggestions that have significantly improved the clarity and content of the paper. We also are grateful to Connie Rockosi and Evan Kirby for doing some of the observations to gather the data used in this work. We thank Dale Kocevski for providing us his X-ray catalogs.This work has made use of the Rainbow Cosmological Surveys Database, which is operated by the Universidad Complutense de Madrid (UCM), partnered with the University of California Observatories at Santa Cruz (UCO/Lick,UCSC). We thank Guillermo Barro for his invaluable help with the Rainbow database. H. Yesuf would like to acknowledge support from NSF grants AST-0808133 and AST-1615730 and STScI grant HST-AR 12822.03-A. \\

The data presented herein were obtained at the W.M. Keck Observatory, which is operated as a scientific partnership among the California Institute of Technology, the University of California and the National Aeronautics and Space Administration. The Observatory was made possible by the generous financial support of the W.M. Keck Foundation. The authors wish to recognize and acknowledge the very significant cultural role and reverence that the summit of Mauna Kea has always had within the indigenous Hawaiian community. We are most fortunate to have the opportunity to conduct observations from this mountain.

\section{Appendix}
In this section we give ancillary information to support the results presented in the main sections of the paper.

\subsection{NUV composite spectra of AGN and the comparison sample}

\begin{figure*}
\subfigure[\ion{Fe}{2} profiles for all AGN and star-forming galaxies at $z \sim 1$]{%
\includegraphics[width=0.45\linewidth]{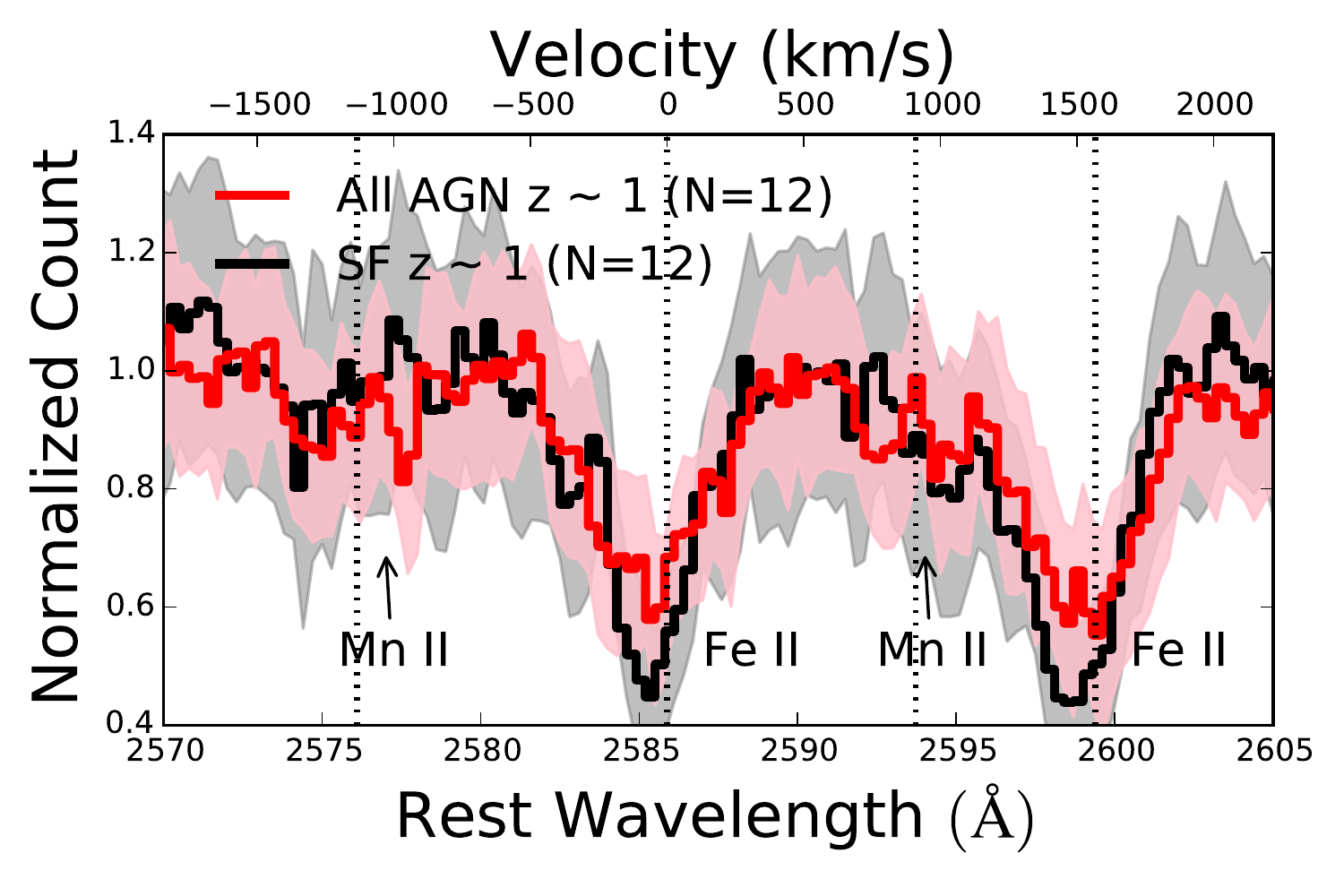}}
\hfill
\subfigure[\ion{Fe}{2} profile for narrow-line AGN (N=9) at $z \sim 1$]{%
\includegraphics[width=0.45\linewidth]{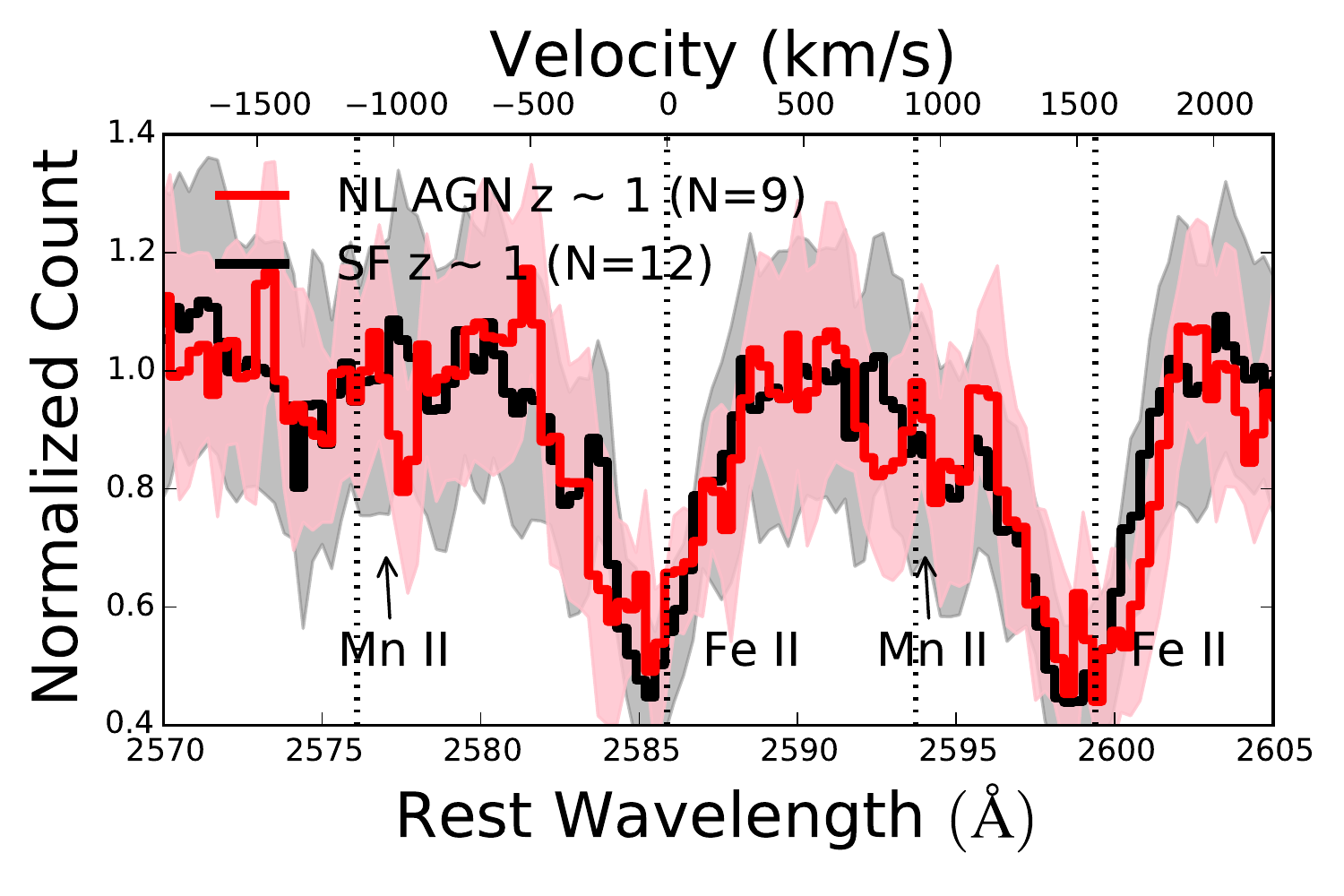}}
\caption[The near UV composite spectrum around \ion{Fe}{2}$ \lambda 2586$ or \ion{Mg}{2}  of AGN at $z \sim 1$.]{The near UV composite spectrum around \ion{Fe}{2}$ \lambda 2586$ of AGN at $z \sim 1$ compared to the composite spectrum of the comparison sample at $z \sim 1$ or to AGN at $z \sim 0.5$ .\label{fig:nuvapp}}
\end{figure*}

Similar to Figure~\ref{fig:data}, Figure~\ref{fig:nuvapp} shows the near UV composite spectra around \ion{Fe}{2} $\lambda 2586$ for AGN subsamples ($N=12$ or $N=9$) and the comparison sample of X-ray undetected galaxies at $z \sim 1$.

\subsection{Re-analysis with simple errors of inverse-variance-weighting (no bootstrapping) \label{sec:invmodel}}
\begin{figure*}
\centering
\subfigure[\ion{Fe}{2} profile of all AGN $z \sim 1$ (N $=12$)]{%
\includegraphics[width=0.3\linewidth,height=0.25\linewidth]{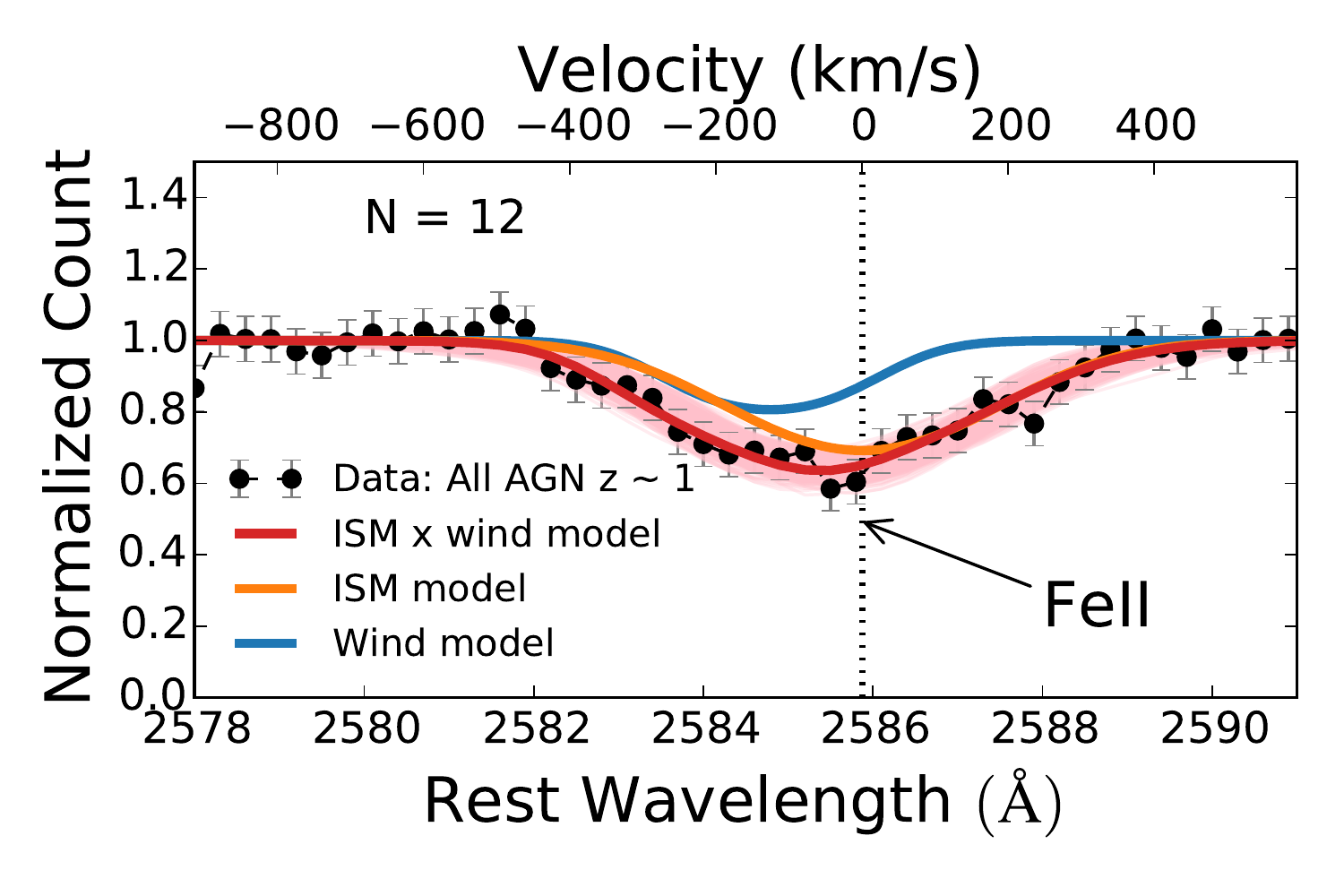}}
\hfill
\subfigure[Narrow-line AGN $z \sim 1$ (N $=9$)]{%
\includegraphics[width=0.3\linewidth,height=0.25\linewidth]{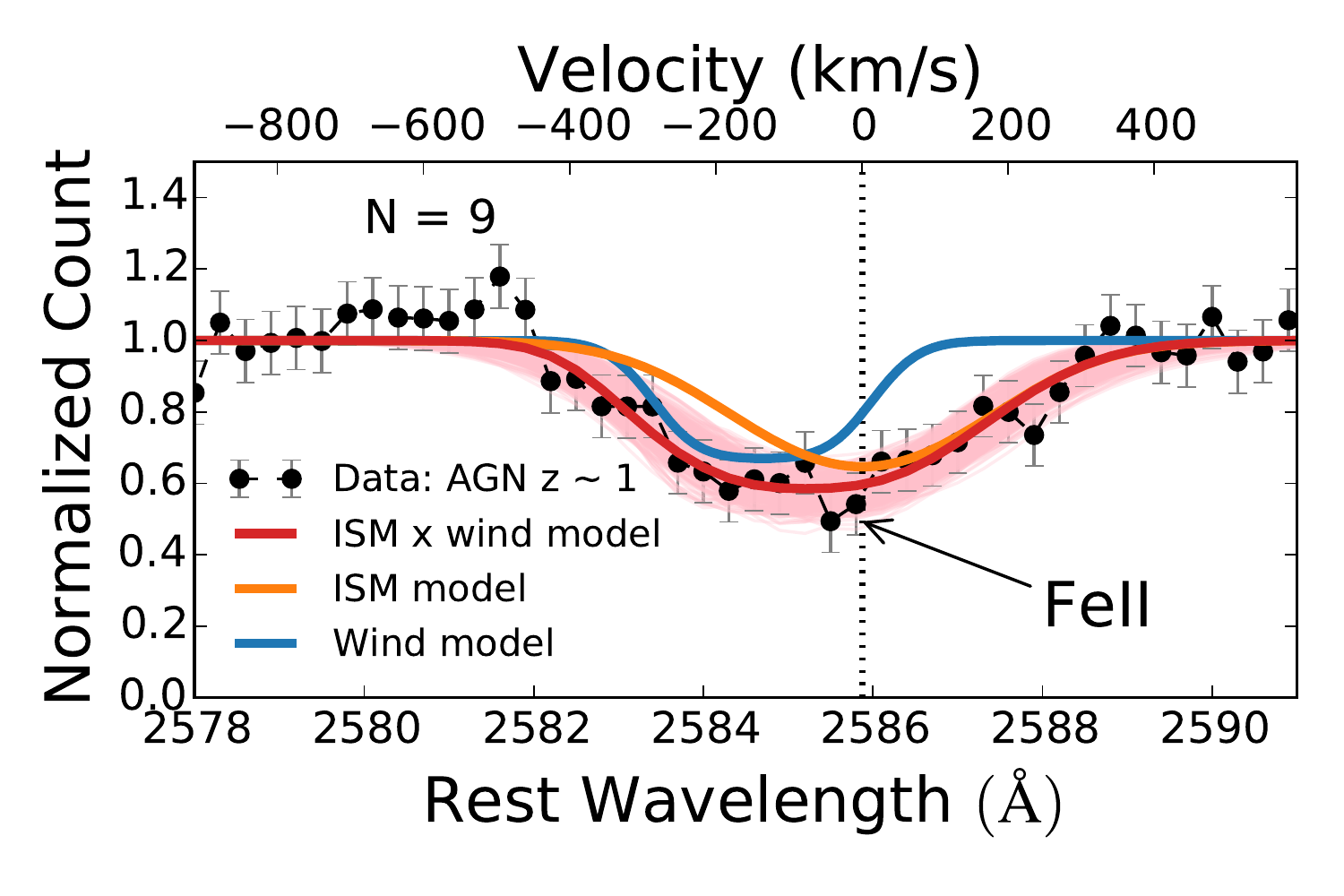}}
\hfill
\subfigure[Very reliable narrow-line AGN $z \sim 1$ (N $=7$)]{%
\includegraphics[width=0.3\linewidth,height=0.25\linewidth]{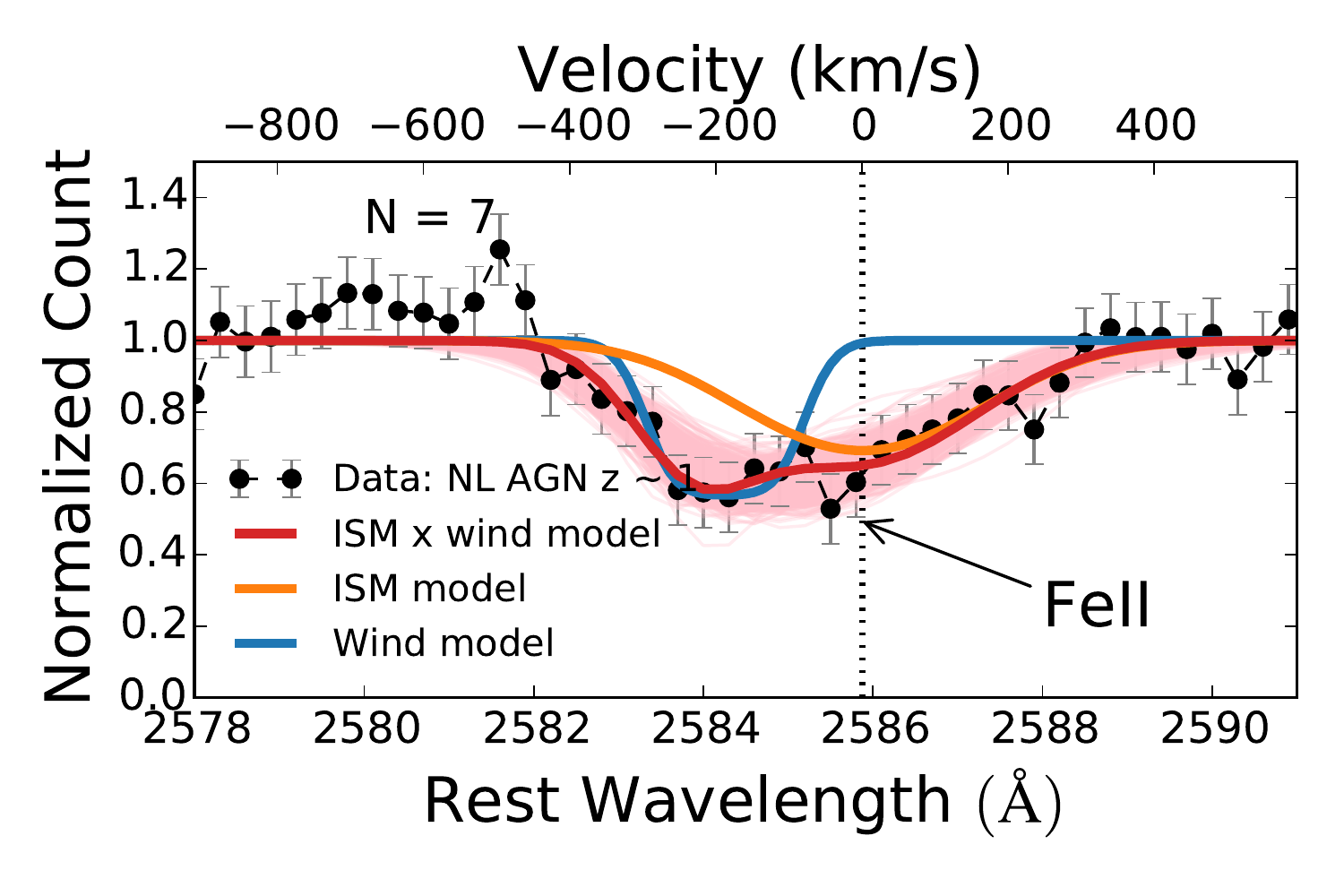}}
\hfill
\subfigure[Wind velocity of all AGN $z \sim 1$ (N $=12$)]{%
\includegraphics[width=0.3\linewidth,height=0.25\linewidth]{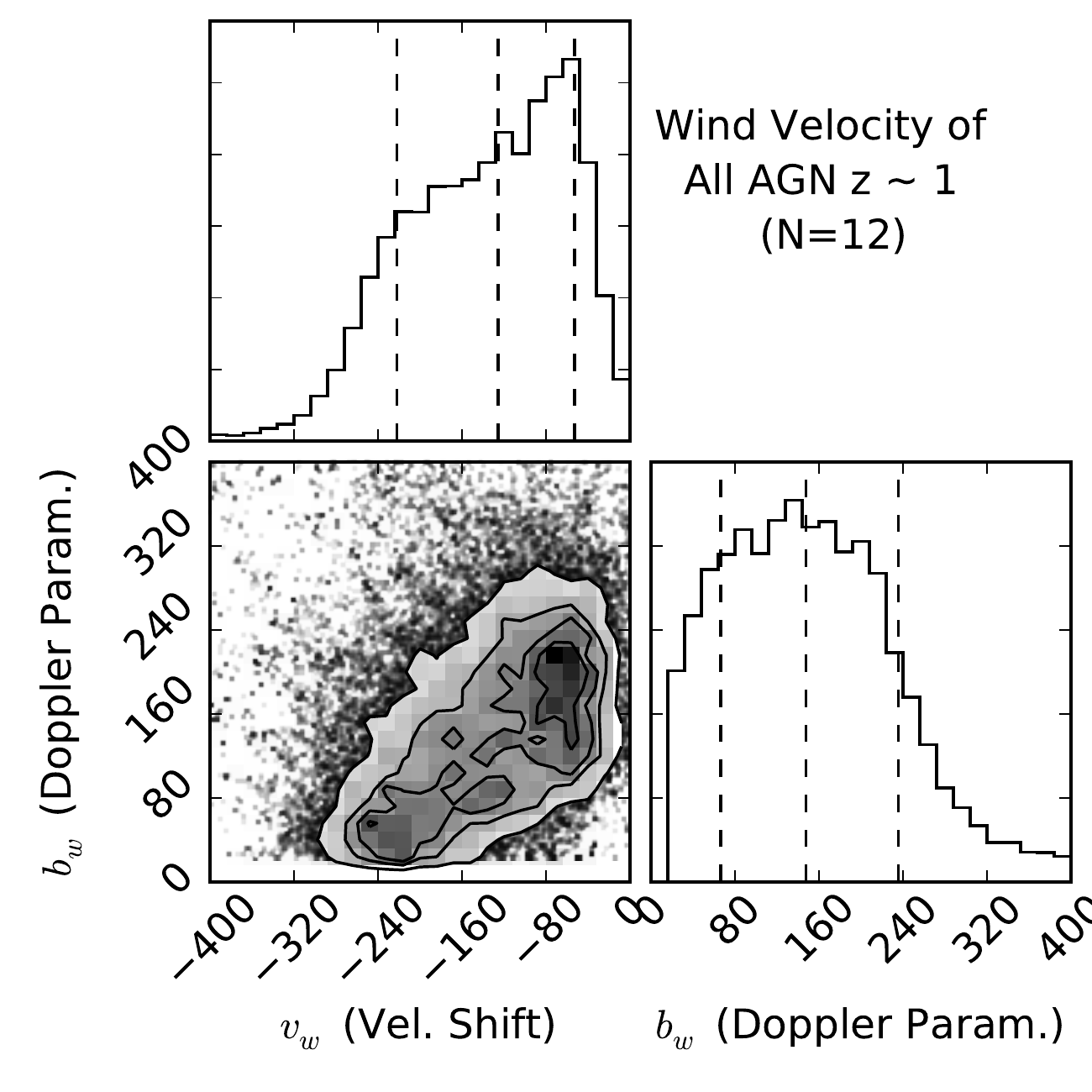}}
\hfill
\subfigure[Narrow-line AGN $z \sim 1$ (N $=9)$]{%
\includegraphics[width=0.3\linewidth,height=0.25\linewidth]{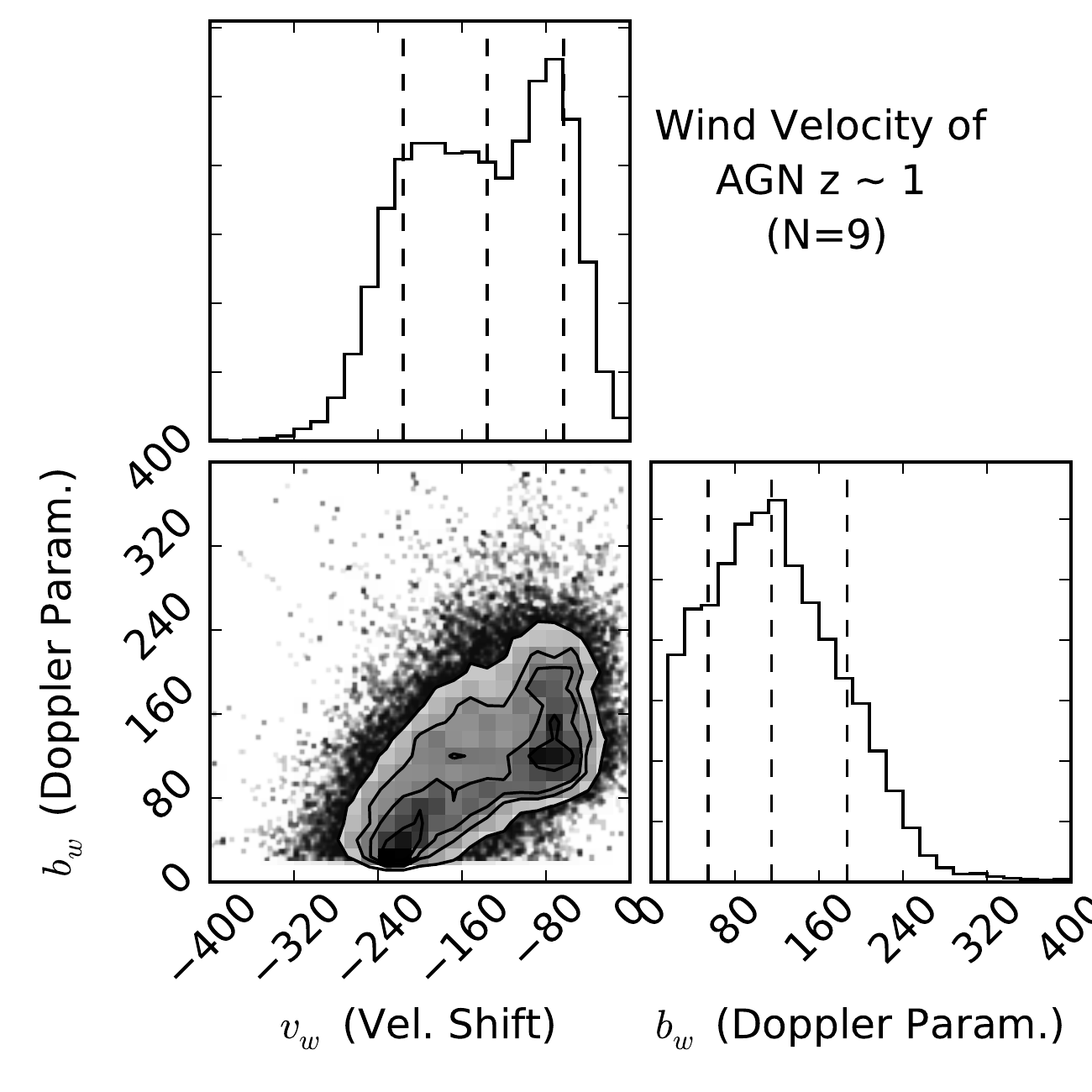}}
\hfill
\subfigure[Very reliable narrow-line AGN $z \sim 1$ (N $=7)$]{%
\includegraphics[width=0.3\linewidth,height=0.25\linewidth]{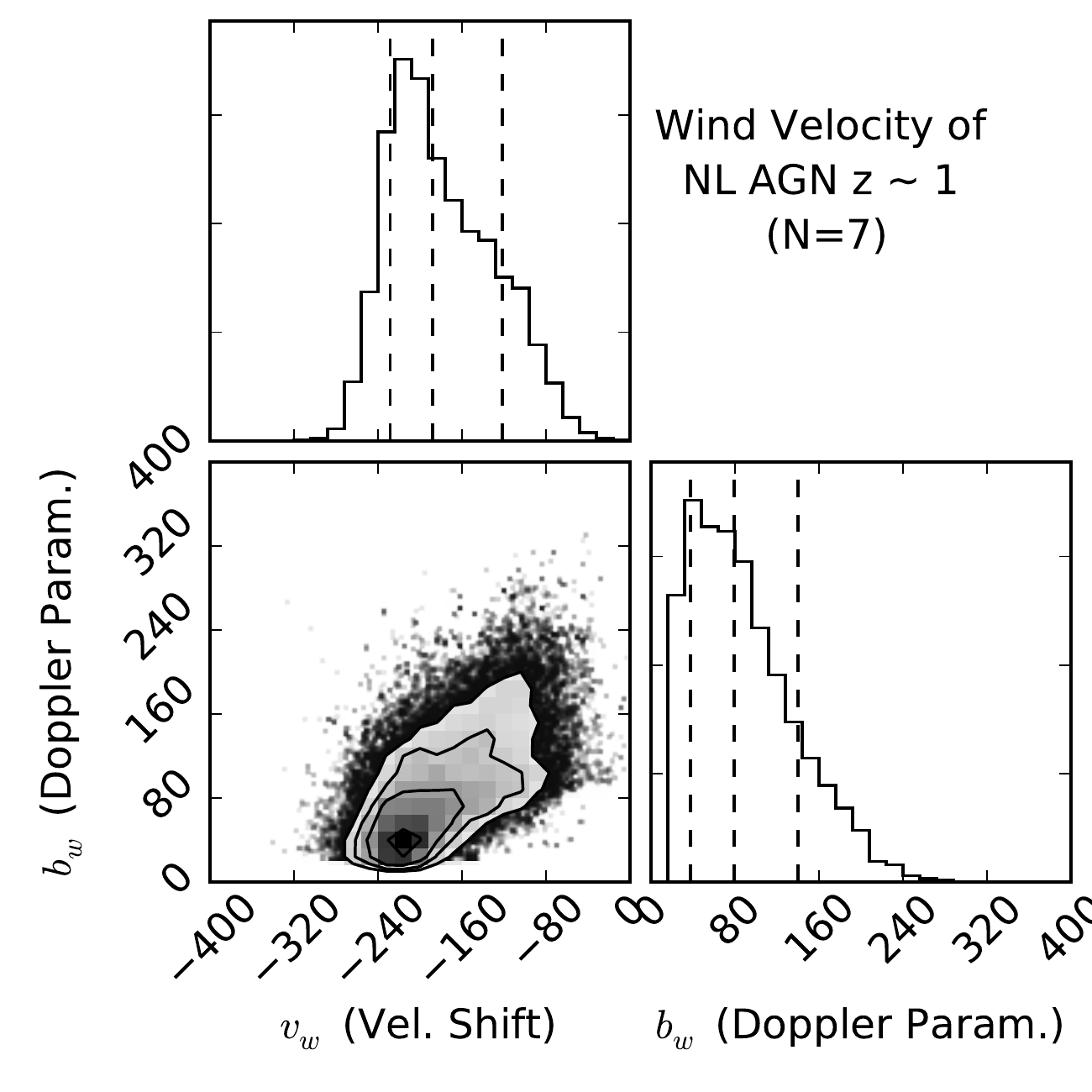}}
\hfill
\subfigure[\ion{Fe}{2} profile of AGN $z \sim 0.5$ \citep{Coil+11}]{%
\includegraphics[width=0.3\linewidth,height=0.25\linewidth]{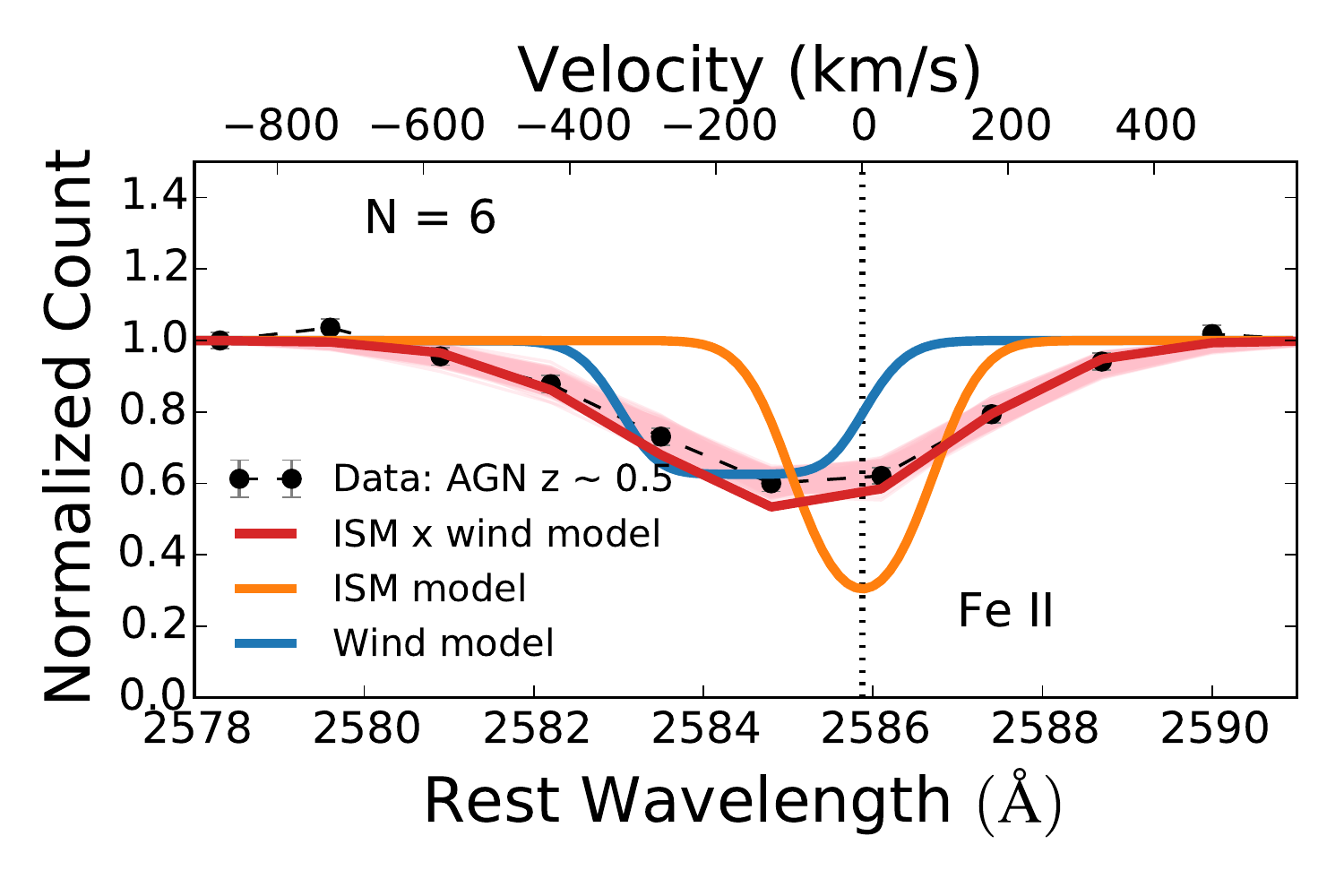}}
\hfill
\subfigure[Combined sample of AGN $z \sim 0.5$ \& $z \sim 1$]{%
\includegraphics[width=0.3\linewidth,height=0.25\linewidth]{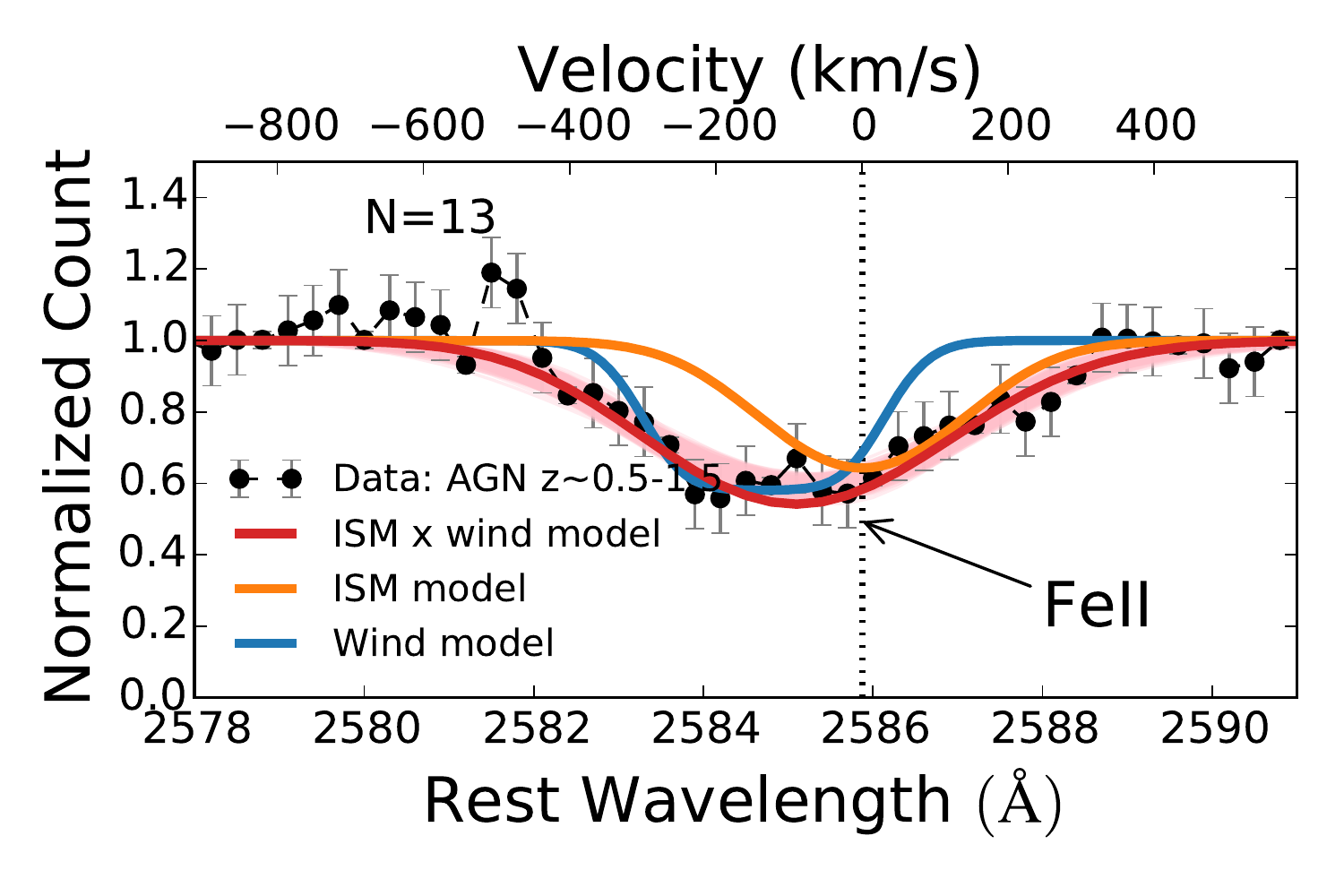}}
\hfill
\subfigure[Star-forming comparison sample $z \sim 1$]{%
\includegraphics[width=0.3\linewidth,height=0.25\linewidth]{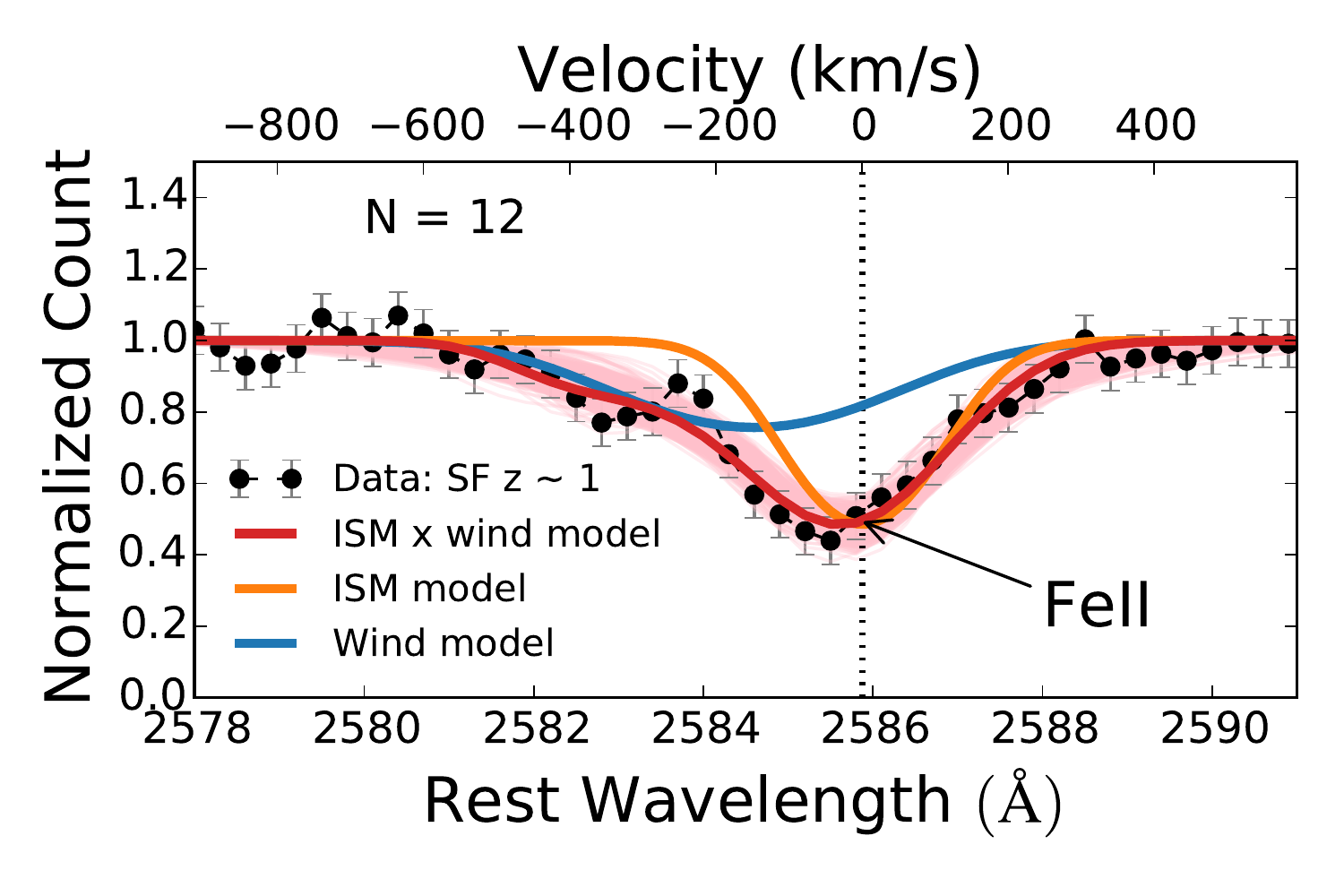}}
\hfill
\subfigure[Velocity of AGN $z \sim 0.5$ \citep{Coil+11}]{%
\includegraphics[width=0.3\linewidth,height=0.25\linewidth]{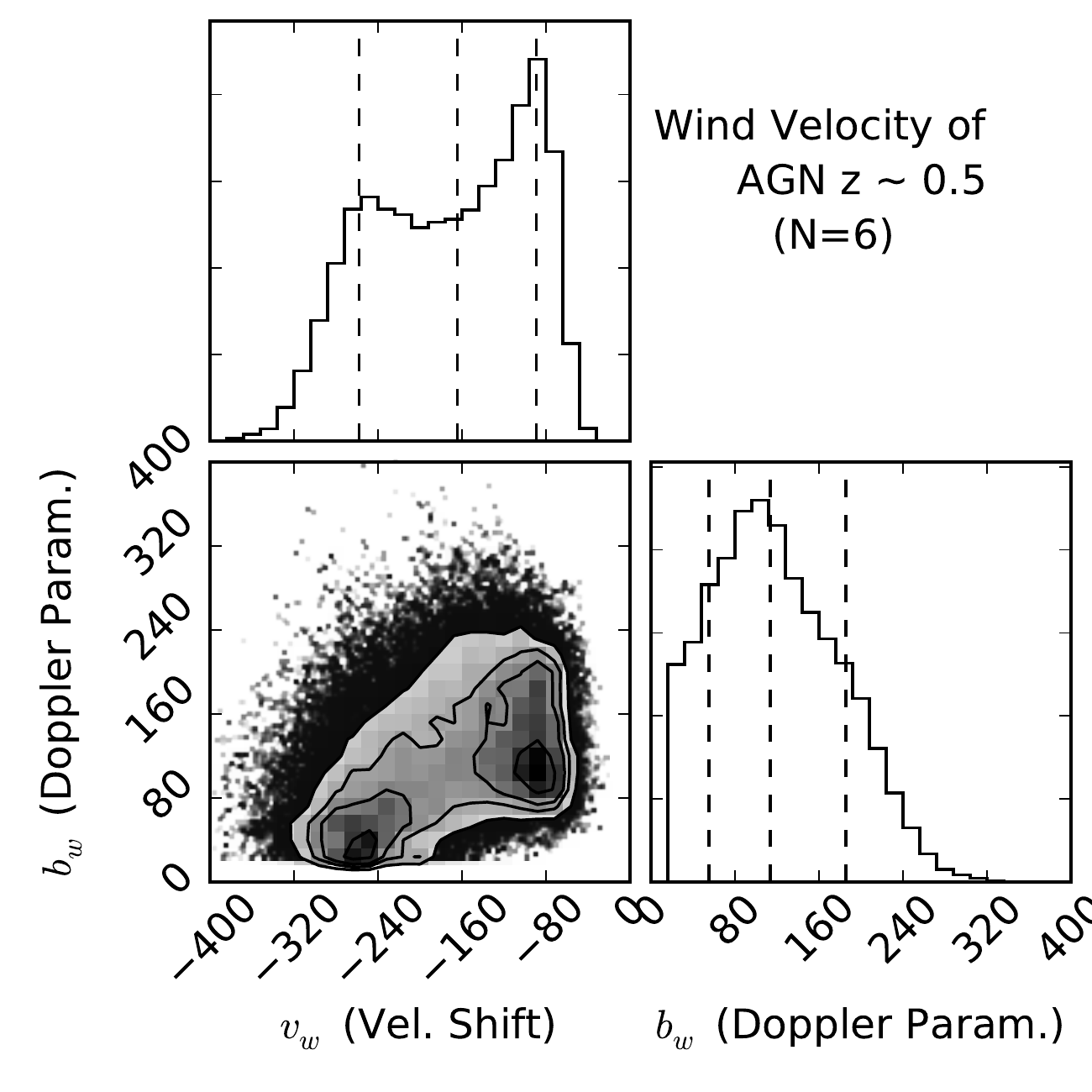}}
\hfill
\subfigure[Combined sample of AGN $z \sim 0.5$ \& $z \sim 1$]{%
\includegraphics[width=0.3\linewidth,height=0.25\linewidth]{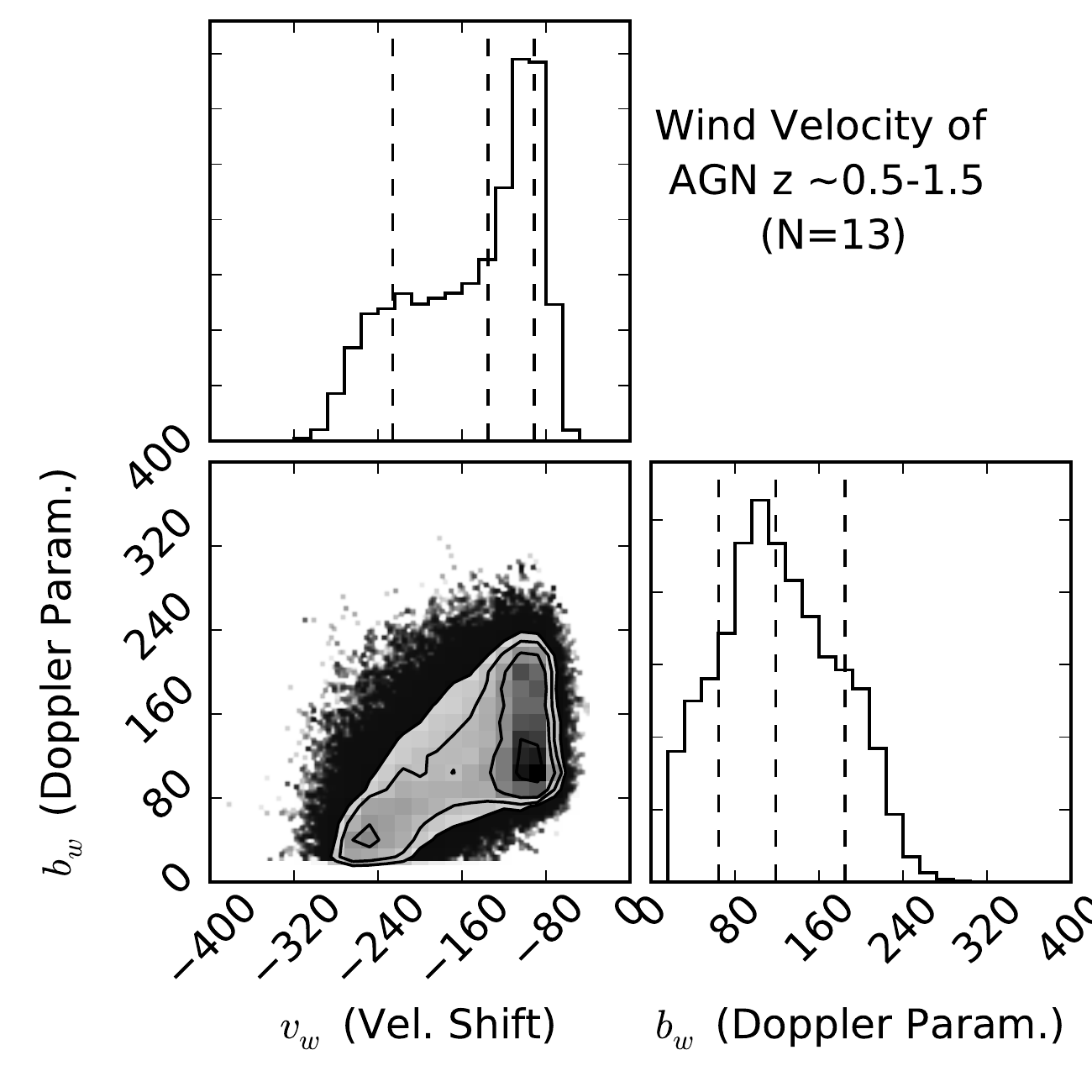}}
\hfill
\subfigure[Star-forming comparison sample $z \sim 1$]{%
\includegraphics[width=0.3\linewidth,height=0.25\linewidth]{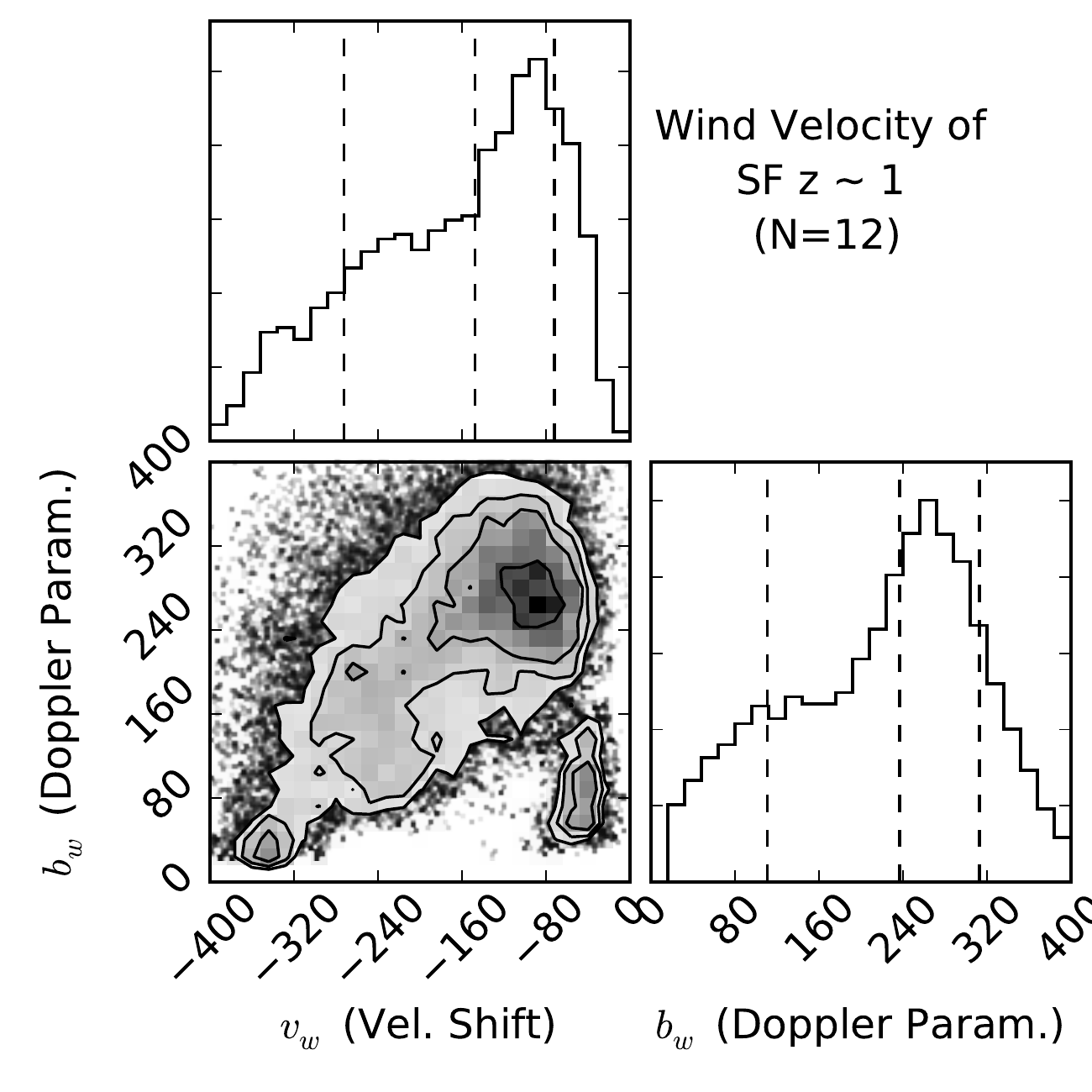}}
\caption{Similar to Figure~\ref{fig:model1} but the errors of the composite spectra are estimated using errors of inverse-variance-weighting as described in \S~\ref{sec:invmodel}. In panel (g) the error bars are invisible because they were too small.\label{fig:invmodel}} 
\end{figure*}

\begin{figure*}
\subfigure[All AGN $z \sim 1$ (N=12)]{%
\includegraphics[width=0.48\linewidth]{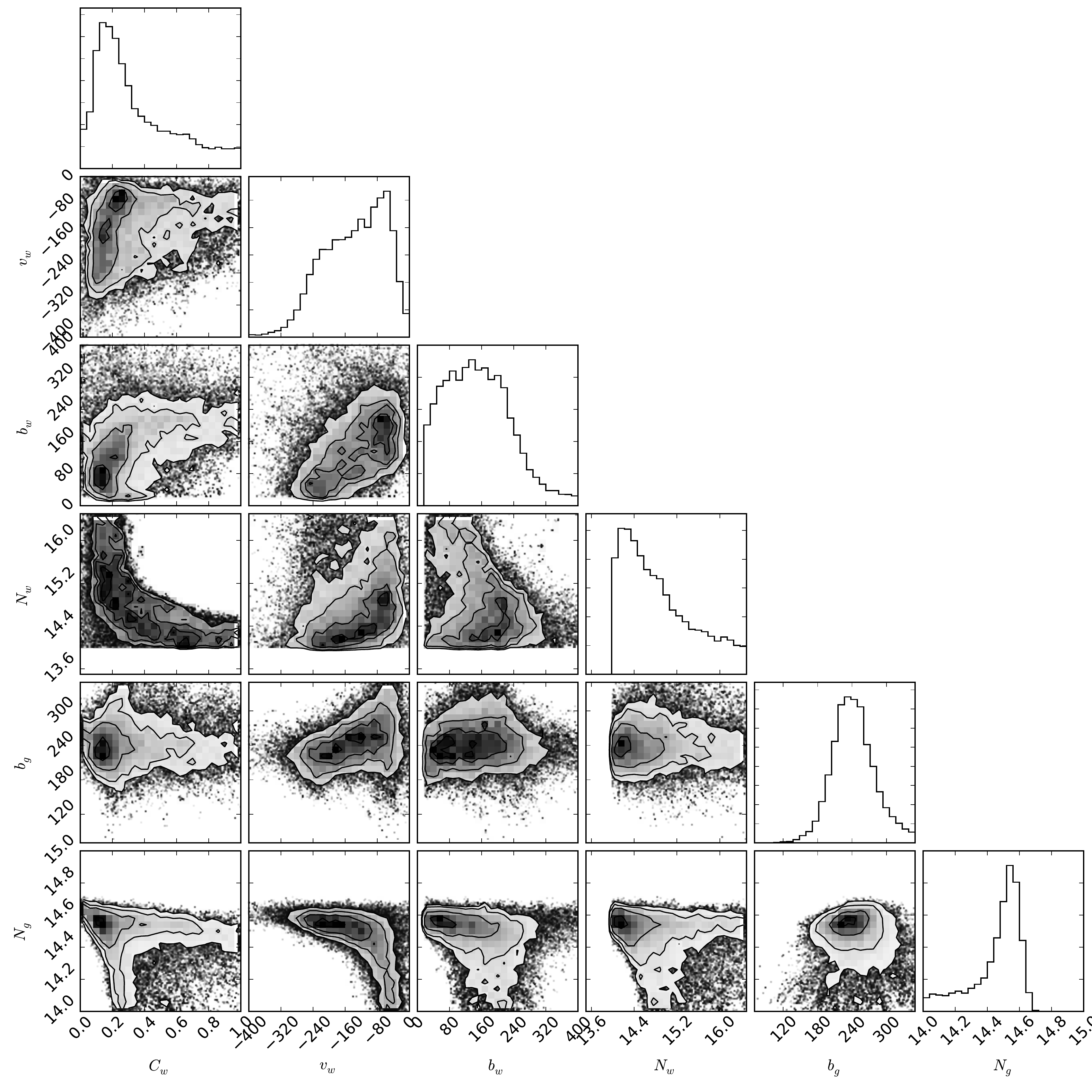}}
\subfigure[Narrow-line AGN $z \sim 1$ (N=7)]{%
\includegraphics[width=0.48\linewidth]{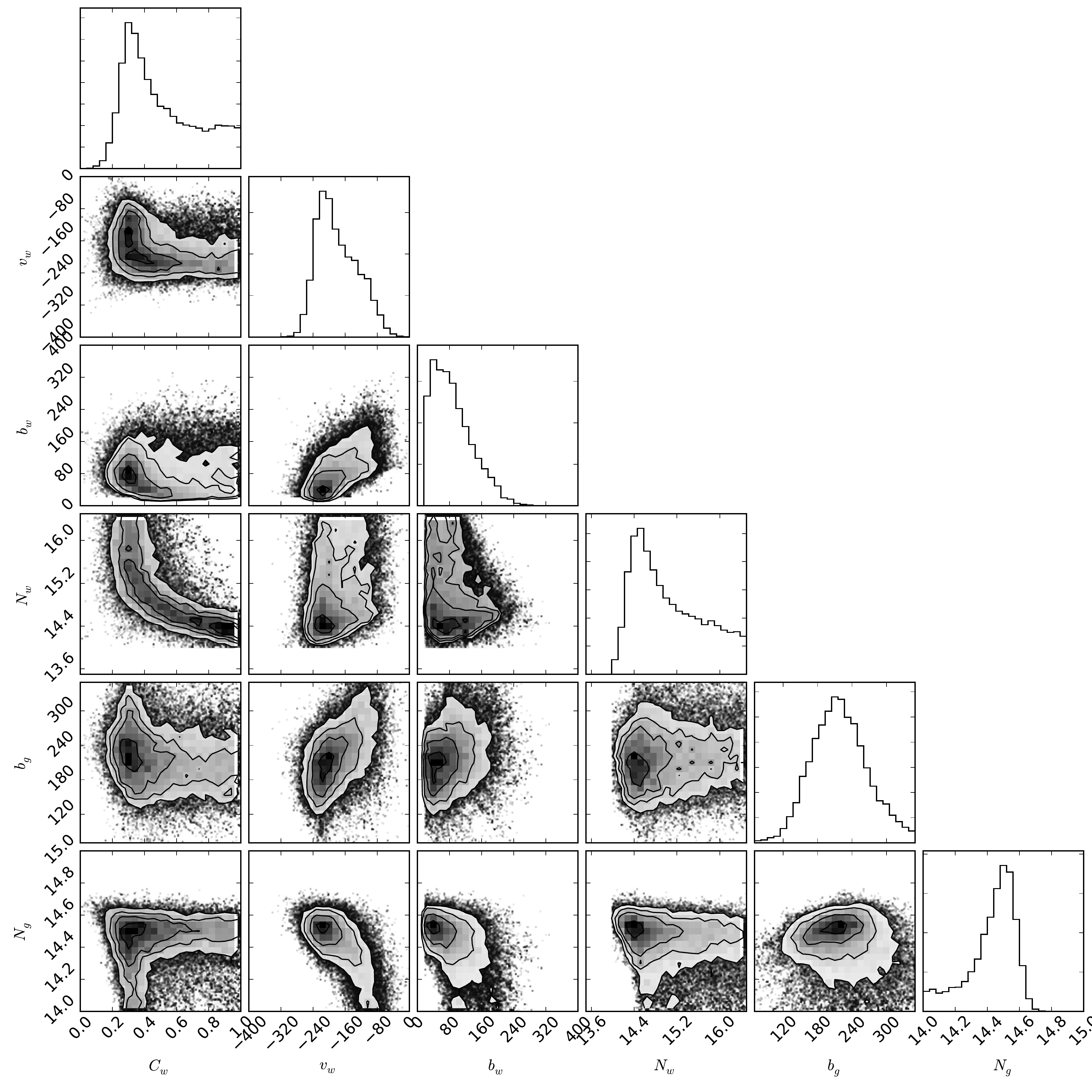}}
\subfigure[AGN $z \sim 0.5$ (N=6)]{%
\includegraphics[width=0.48\linewidth]{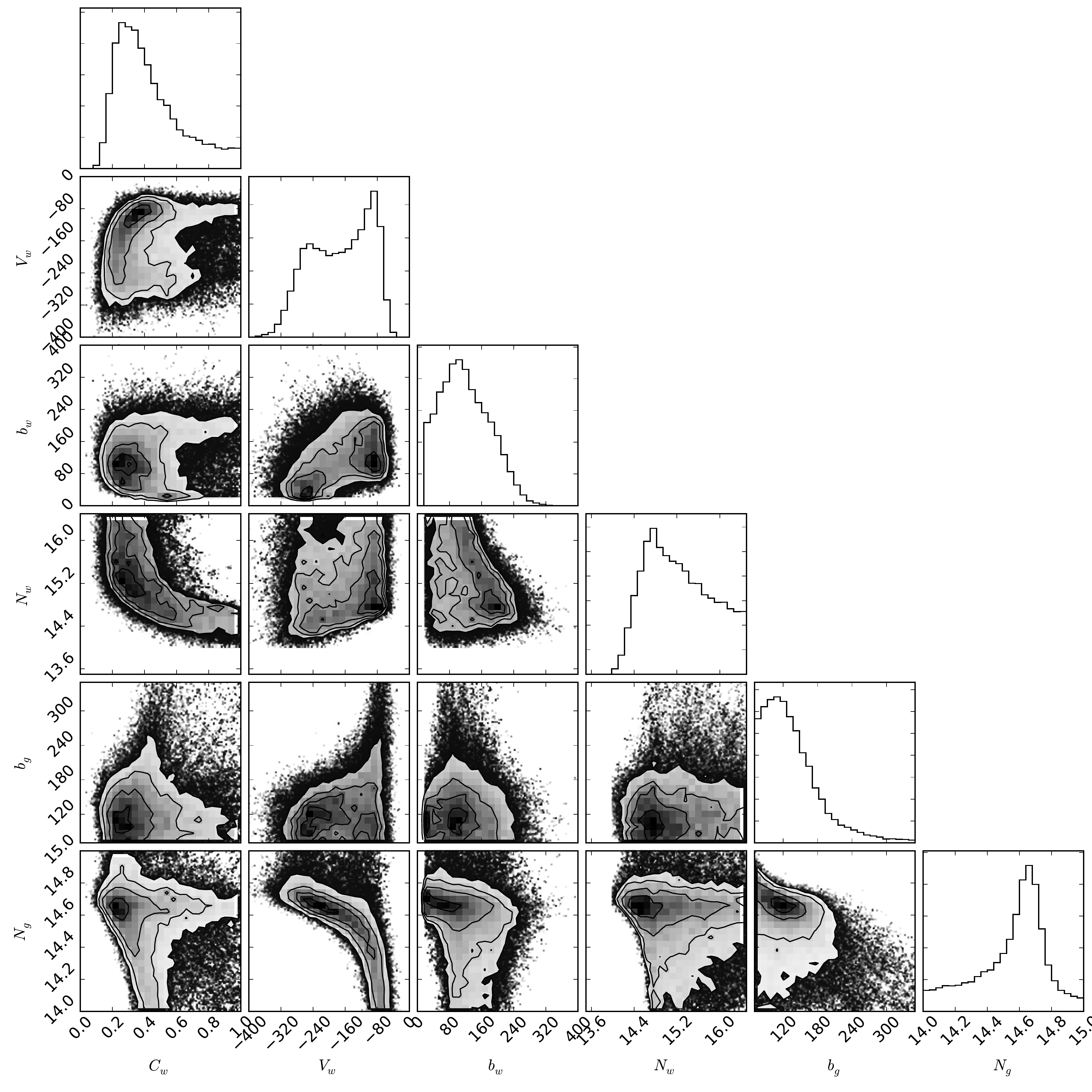}}
\subfigure[SF $z \sim 1$ (N=12)]{
\includegraphics[width=0.48\linewidth]{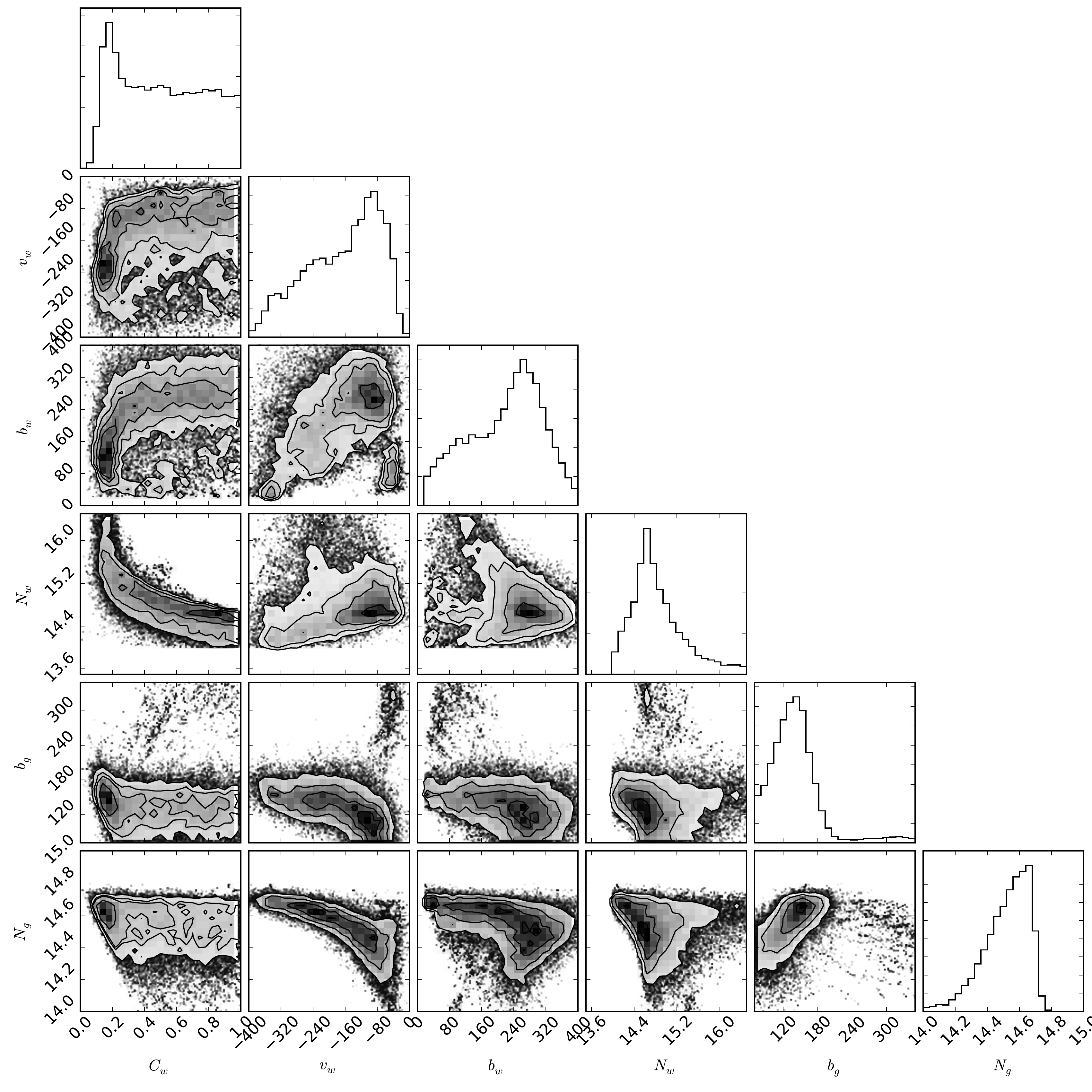}}
\caption{The joint posterior probability distributions of the wind model parameters. They are from top-to-bottom (or left-to-right): the covering fraction of the wind, $C_{w}$, the velocity centroid shifts of the wind, $v_{w}$, the Doppler broadening parameters of the wind, $b_{w}$, column density of the wind, $N_{w}$,  the Doppler parameter of the gas in the galaxy, $b_g$, and the column density of the gas in the galaxy, $N_{g}$. Contours show the joint posterior probability distribution of a given two parameters. The histograms show the marginalized posterior probability distributions the parameters. Notice the correlation between some of the parameters. The results of this figure are summarized in Table~\ref{tbl:invmodel}. The errors of the observed composite spectra  analyzed here are errors of inverse-variance-weighting (no bootstrapping)\label{fig:fit_param6}} 
\end{figure*}

The errors of each individual spectrum are outputs by DEEP2 spec1d pipeline. If $\sigma_{ij}$ is the error of photon count, $C_i(\lambda_j)$, at wavelength pixel $\lambda_j$ of a galaxy $i$, averaging over all $N$ galaxies, the inverse-variance-weighted mean count at pixel $\lambda_j$ is $\hat{C}(\lambda_j) = \frac{\sum_{i=1}^{N}C_i(\lambda_j)/\sigma_{ij}^{2}}{\sum_{i=1}^{N}1/\sigma_{ij}^{2}}$. The standard error of the mean count at pixel $\lambda_j$  is , $\hat{\sigma_{j}} = \sqrt{\frac{1}{\sum_{i=1}^{N} 1/\sigma_{ij}^{2}}}$. The inverse-variance weighting analysis gives lower weights to galaxies having low signal-to-noise spectra in the sample and may not capture well the dispersion intrinsic to the sample. On the other hand, it has the advantage of down-weighting poorly measured data. In the case where the intrinsic sample dispersion is negligible, it may be preferred than bootstrapping scheme. In the latter case, poor data may be resampled instead of good data thereby increase the inferred dispersion. This section presents results of the reanalysis using $\hat{\sigma_j}$ errors of the inverse-variance-weighting as shown in Table~\ref{tbl:invmodel}, Table~\ref{tbl:inveqw}, Figure~\ref{fig:invmodel} \&~\ref{fig:fit_param6}. In the latter figure, we plot the 
posterior PDFs of all the six wind model parameters for the AGN and the comparison samples. Note that the some of the parameters strongly correlate with each others and their PDFs are very asymmetric (non-Gaussian). For example, the covering fraction and the column density of the wind are degenerate with each other. Thus, the column density estimates are uncertain by more than factor of ten.

\begin{deluxetable*}{lccccccc}
\tablewidth{0in}
\tabletypesize{\footnotesize}
\tablecaption{Model parameter fits to \ion{Fe}{2} $\lambda2586$ profiles using errors of  inverse-variance-weighting. \label{tbl:invmodel}}

\tablehead{ \colhead{Model} & \colhead{AGN $z \sim 1$} & \colhead{NLAGN $z \sim 1$} & \colhead{NLAGN $z \sim 1$} & \colhead{SF $z \sim 1$} & \colhead{AGN $z \sim 0.5$\tablenotemark{a}} & \colhead{AGN $z \sim 0.5-1.5$\tablenotemark{b}} & \colhead{AGN $z \sim 0.5-1.5$\tablenotemark{c}} \\
\colhead{Param.} & \colhead{(N = 12)} & \colhead{(N = 9)} & \colhead{(N = 7)} & \colhead{(N = 12)} & \colhead{(N = 6)} & \colhead{(N = 13)} & \colhead{(N=18)\tablenotemark{d}}}\\
\startdata

$v_w$ (km s$^{-1}$) &$ -137^{+76}_{-87} $& $-152^{+76}_{-76}$  & $-194^{+68}_{-41}$ & $-135^{+75}_{-133}$ & $-169^{+78}_{-92}$ &$-139^{+48}_{-87}$ &$-132^{+51}_{-96}$ \\
$b_w$ (km s$^{-1}$) & $150^{+91}_{-79} $&$115^{+71}_{-60}$ & $78^{+60}_{-40}$ & $198^{+90}_{-102}$ & $112^{+67}_{-54}$ & $116^{+67}_{-53}$ &$135^{+72}_{-61}$ \\
$C_w $& $0.3^{+0.4}_{-0.2}$ & $0.4^{+0.3}_{-0.1}$ &$ 0.4^{+0.3}_{-0.1} $& $0.4^{+0.4}_{-0.2}$ &$ 0.4^{+0.2}_{-0.1}$ & $0.4^{+0.3}_{-0.1}$ &$ 0.4^{+0.3}_{-0.1}$\\
$\log N_w$ (cm$^{-2}$)&$ 14.8^{+1.3}_{-0.5}$ &$ 15.1^{+1.3}_{-0.7}$ &$15.2^{+1.2}_{-0.7}$ &$14.9^{+1.1}_{-0.4}$&$ 15.2^{+1.3}_{-0.6}$ &$15.4^{+1.2}_{-0.7}$  &$15.2^{+1.3}_{-0.6}$\\
$b_g$ (km s$^{-1}$) &$ 243^{+45}_{-34} $&$ 224^{+48}_{-38}$ & $214^{+56}_{-48}$&$154^{+52}_{-48}$ & $98^{+64}_{-55}$ & $181^{+67}_{-46}$ &$179^{+59}_{-45}$\\
$\log N_g$ (cm$^{-2}$)&$ 14.5 ^{+0.1}_{-0.2} $&$14.5^{+0.1}_{-0.3}$ & $14.5^{+0.1}_{-0.2}$  &$ 14.6^{+0.1}_{-0.2}$ &$ 14.6^{+0.4}_{-0.3}$ &$14.5^{+0.1}_{-0.3}$ &$14.5^{+0.1}_{-0.3}$\\
\enddata

\tablenotetext{a}{Renalysis of \citet{Coil+11} data}
\tablenotetext{b}{A joint renalysis of \citet{Coil+11} data with our 7 narrow-line AGN with highly reliable AGN identification.}
\tablenotetext{c}{A joint renalysis of \citet{Coil+11} data with all AGN including 3 broad-line AGN and 2 with less reliable AGN identification.}
\tablenotetext{d}{N denotes the number of galaxies in the stacked spectra.}
\tablecomments{The median, the 84th and 16th percentile deviations of the PDFs of parameters are given in the table. In our notation, $X^{+Y}_{-Z}$: X is the median, $X+Y$ is the 84th percentile and $X-Z$ is the 16th percentile.}
\end{deluxetable*}

\begin{deluxetable*}{lccccccc}
\tablewidth{0in}
\tabletypesize{\footnotesize}
\tablecaption{Wind equivalent width (EW) and maximum wind velocity derived from \ion{Fe}{2} $\lambda2586$ model profiles. \label{tbl:inveqw}}

\tablehead{ \colhead{Derived} & \colhead{AGN $z \sim 1$} & \colhead{NLAGN $z \sim 1$} & \colhead{NLAGN $z \sim 1$} & \colhead{SF $z \sim 1$} & \colhead{AGN $z \sim 0.5$} & \colhead{AGN $z \sim 0.5-1.5$} & \colhead{AGN $z \sim 0.5-1.5$} \\
\colhead{Quantity} & \colhead{(N = 12)} & \colhead{(N = 9)} & \colhead{(N = 7)} & \colhead{(N = 12)} & \colhead{(N = 6)} & \colhead{(N = 13)} & \colhead{(N=18)}}\\
\startdata
Wind EW ({\AA)} & $0.5^{+0.5}_{-0.2} $& $0.6^{+0.7}_{-0.1}$  & $0.7^{+0.4}_{-0.2}$ & $0.8^{+0.6}_{-0.3}$ & $0.8^{+0.5}_{-0.3}$ &$1.1^{+0.5}_{-0.4}$ &$0.9^{+0.5}_{-0.3}$ \\
ISM EW  ({\AA)} & $1.2^{+0.2}_{-0.4} $&$1.3^{+0.3}_{-0.6}$ & $1.0^{+0.2}_{-0.4}$ & $1.3^{+0.3}_{-0.5}$ & $1.2^{+0.2}_{-0.4}$ & $1.1^{+0.3}_{-0.5}$ &$1.1^{+0.3}_{-0.5}$ \\
Total EW  ({\AA)} & $1.6^{+0.1}_{-0.1}$ & $1.8^{+0.1}_{-0.1}$ &$ 1.7^{+0.1}_{-0.1} $& $1.9^{+0.1}_{-0.1}$ &$ 1.9^{+0.1}_{-0.1}$ & $1.9^{+0.1}_{-0.1}$ &$ 1.9^{+0.1}_{-0.1}$\\
Max. Velocity (km s$^{-1}$) &$ -341^{+87}_{-104}$ &$ -318^{+66}_{-77}$ &$-303^{+48}_{-57}$ &$-432^{+122}_{-116}$&$ -263^{+75}_{-68}$ &$-237^{+54}_{-51}$  &$-250^{+70}_{-64}$\\
\enddata
\tablecomments{The median, the 84th and 16th percentile deviations of the PDFs of the derived quantities from the wind model are given in the table.}
\end{deluxetable*}

\subsection{Fitting \ion{O}{2} profile to estimate the escape velocity}\label{sec:o2fit}

\begin{figure*}
\centering
\subfigure[\ion{O}{2} emission line for all AGN at $z \sim 1$]{%
\includegraphics[width=0.45\linewidth,height=0.3\linewidth]{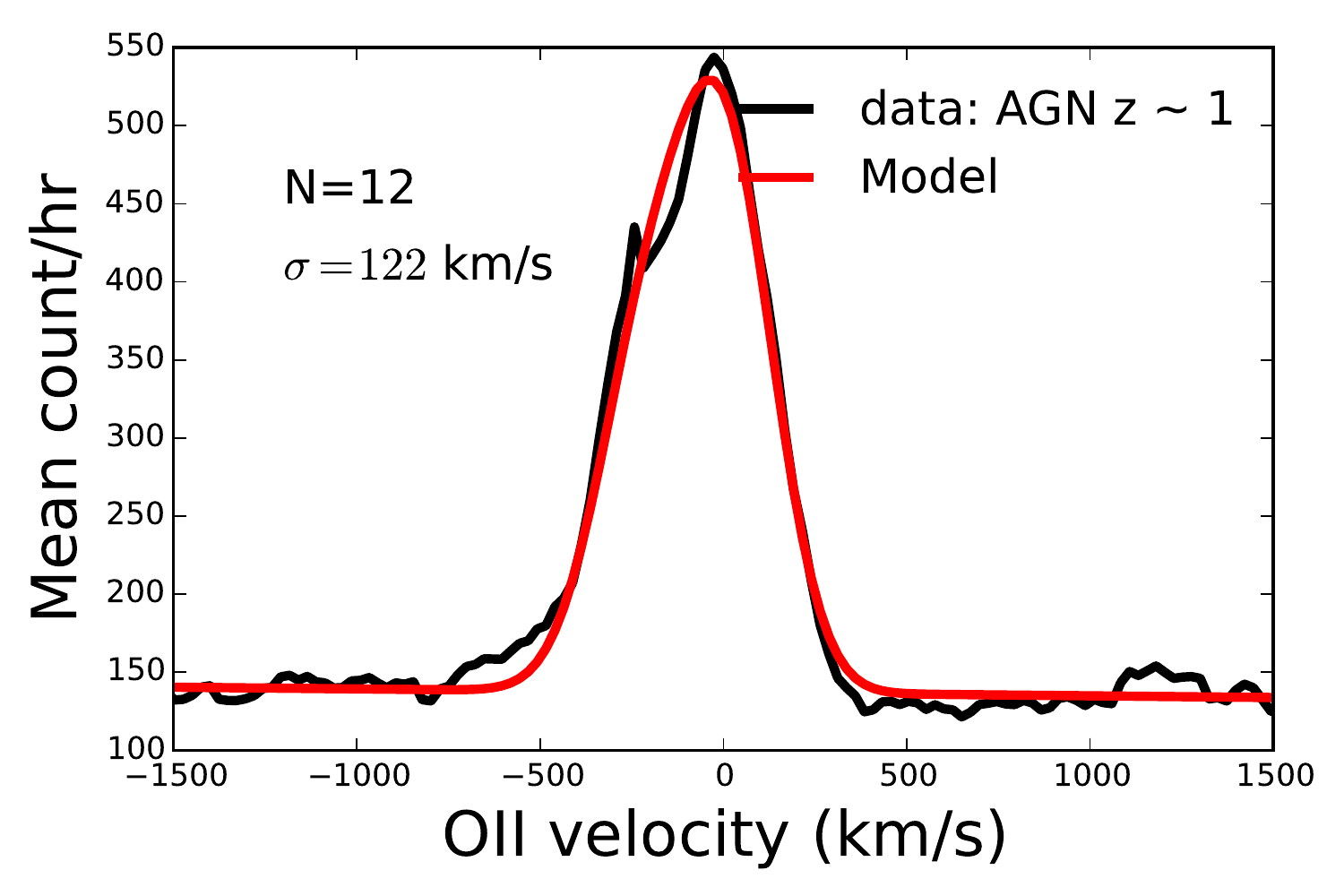}}
\hfill
\subfigure[\ion{O}{2} emission line for star-forming comparison sample at $z \sim 1$]{%
\includegraphics[width=0.45\linewidth,height=0.3\linewidth]{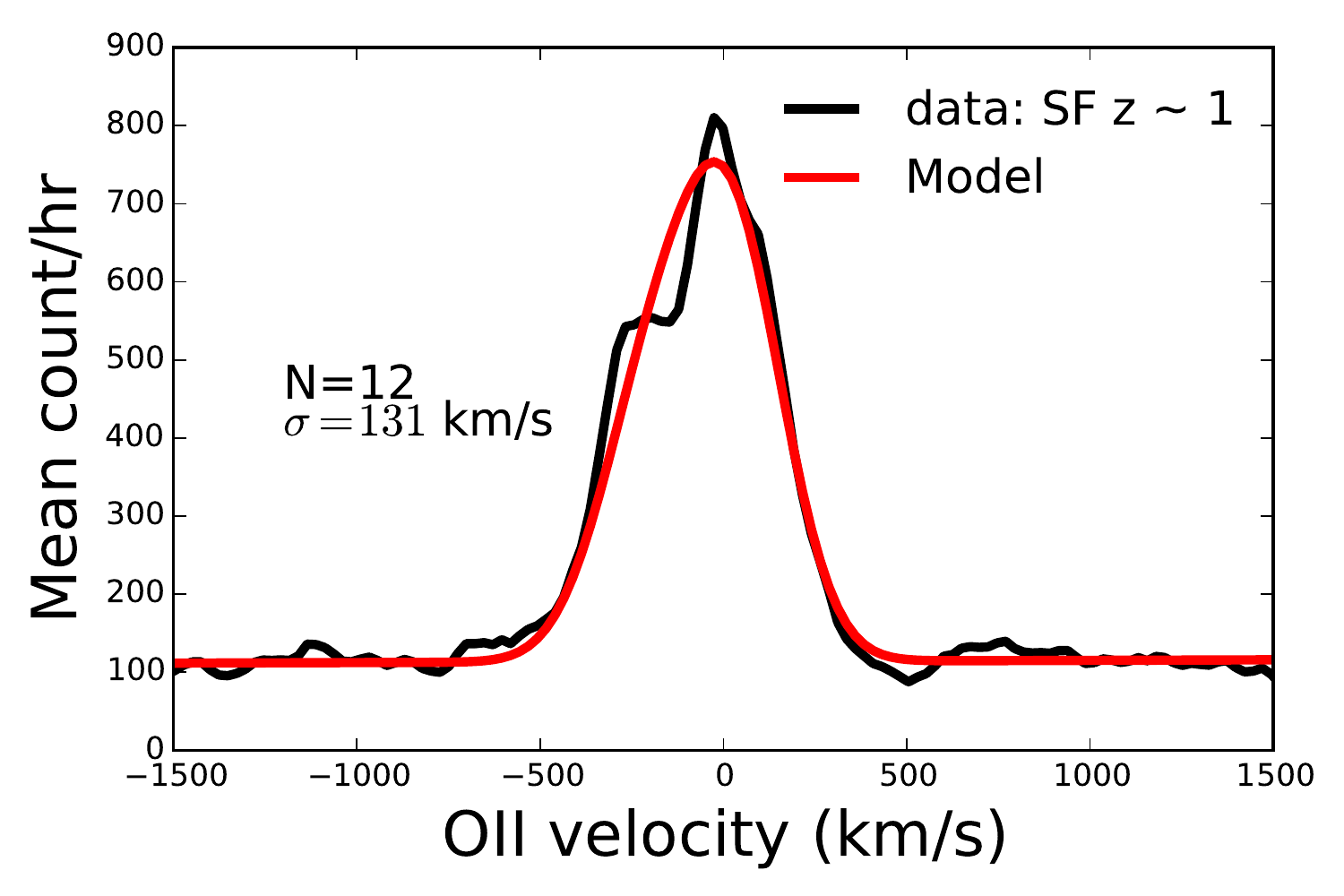}}
\caption{The result of fitting two Gaussian with equal width to the \ion{O}{2} doublet. The velocity dispersions of the fits are shown on the figures. Comparing the observed wind velocity to the \ion{O}{2} width, we conclude that the winds, both in AGN and the comparison sample, are too weak to escape the gravitational potentials of their galaxies.\label{fig:o2fit}}
\end{figure*}

Figure~\ref{fig:o2fit}a shows the result of fitting two Gaussians + a linear continuum model to the \ion{O}{2} $\lambda 3726.03, \lambda3728.82$ doublet to the mean AGN spectra using Levenberg-Marquardt least square minimization. In the model,  the two Gaussians have the same width, and the centroid shifts and the amplitudes of the doublet are also free parameters of the fit. The two Gaussian model is convolved to match the DEIMOS instrumental resolution and rebinned to match the observed data. Figure~\ref{fig:o2fit}b shows the corresponding fit for the star-forming comparison sample. The velocity dispersions of the fits are $131 \pm 18$ km s$^{-1}$ for SF galaxies and $122 \pm 4$ km s$^{-1}$ for AGN. The errors of the velocity dispersions are estimated by repeating the least square fitting procedure for all 1000 bootstrap spectra around \ion{O}{2} (similar to what is shown in Figure~\ref{fig:data}) and then taking the standard deviation of the 1000 velocity dispersions.The centroid shifts of the doublet are consistent with no shift from the the rest wavelengths of the doublet.

\subsection{The effect of ISM covering fraction \label{sec:cf}}

In fiducial wind model in the main text of the paper, we assumed that the ISM fully covers the stellar continuum emission. Table~\ref{tbl:cf_boot} \& ~\ref{tbl:cf} show that the results of the analyses using a covering fraction of 50\%. Our main conclusion does not depend on the assumption of the covering fraction.

\begin{deluxetable*}{lccccc}
\tabletypesize{\footnotesize}
\tablewidth{0in}
\tablecaption{Model parameter fits to \ion{Fe}{2} $\lambda2586$ profiles after adopting a covering fraction of 50\% for the ISM component of the wind model. The error of the composite spectra are estimated by bootstap scheme. \label{tbl:cf_boot}}
    \tablehead{\colhead{Wind Model} & \colhead{All AGN at $z \sim 1$} & \colhead{NL AGN at $z \sim 1$} & \colhead{NL AGN at $z \sim 1$} & \colhead{SF at $z \sim 1$} & \colhead{AGN at $z \sim 0.5$}\\
\colhead{Parameters} & \colhead{(N = 12)} & \colhead{(N = 9)} & \colhead{(N = 7)} & \colhead{(N = 12)} & \colhead{(N = 6)}}\\
\startdata
$v_w $ (km s$^{-1}$) &$ -77^{+139}_{-168} $& $-99^{+116}_{-142} $&$ -127^{+114}_{-112} $& $ -123^{+230}_{-203} $&$-93^{+240}_{-204}$ \\
$b_w $ (km s$^{-1}$) &$ 206^{+144}_{-120} $& $174^{+142}_{-96} $& $152^{+143}_{-90} $& $227^{+148}_{-137} $&$ 190^{+167}_{-128}$ \\
$C_w$ &$ 0.2^{+0.4}_{-0.2} $&$ 0.3^{+0.3}_{0.2}  $&$ 0.4^{+0.4}_{-0.2} $& $0.2^{+0.4}_{-0.2} $&$ 0.3^{+0.4}_{-0.2}$ \\
$\log N_w$ (cm $^{-2}$)&$ 14.9^{+1.6}_{-0.6} $&$14.8^{+1.5}_{-0.6} $& $15.0^{+1.4}_{-0.6} $ &$14.9^{+1.6}_{-0.6} $&$ 15.0^{+1.5}_{-0.7}$ \\
$b_g$ (km s$^{-1}$) &$ 218^{+88}_{-76} $&$ 200^{+81}_{-69} $ & $210^{+106}_{-106}$ & $101^{+73}_{-36}  $&$ 128^{+157}_{-63}$ \\
$\log N_g$ (cm $^{-2}$) &$ 14.9^{+0.2}_{-0.3} $&$ 15.0^{+0.2}_{-0.4}  $& $15.0^{+0.2}_{-0.4} $&$ 15.5^{+1.1}_{-0.4} $&$ 15.3^{+1.2}_{-0.6}$  \\
\enddata
 \tablecomments{The median, the 84th and 16th percentile deviations of the PDFs of parameters from the median are given in the table.}
\end{deluxetable*}

\begin{deluxetable*}{lccccc}
\tabletypesize{\footnotesize}
\tablewidth{0in}
\tablecaption{Model parameter fits to \ion{Fe}{2} $\lambda2586$ profiles after adopting a covering fraction of 50\% for the ISM component of the wind model (no bootstapping). \label{tbl:cf}}
    \tablehead{\colhead{Wind Model} & \colhead{All AGN at $z \sim 1$} & \colhead{NL AGN at $z \sim 1$} & \colhead{NL AGN at $z \sim 1$} & \colhead{SF at $z \sim 1$} & \colhead{AGN at $z \sim 0.5$}\\
\colhead{Parameters} & \colhead{(N = 12)} & \colhead{(N = 9)} & \colhead{(N = 7)} & \colhead{(N = 12)} & \colhead{(N = 6)}}\\
\startdata
$v_w$ (km s$^{-1}$)&$ -121^{+74}_{-102} $& $-148^{+86}_{-86} $&$ -190^{+80}_{-41} $& $ -110^{+93}_{-185} $&$-169^{+78}_{-74}$ \\
$b_w$ (km s$^{-1}$)&$ 153^{+84}_{-84} $& $114^{+74}_{-63} $& $79^{+62}_{-44} $& $197^{+157}_{-131} $&$ 120^{+72}_{-56}$ \\
$C_w$ &$ 0.3^{+0.4}_{-0.2} $&$ 0.3^{+0.3}_{0.1}  $&$ 0.4^{+0.3}_{-0.1} $& $0.3^{+0.4}_{-0.2} $&$ 0.4^{+0.3}_{-0.2}$ \\
$\log N_w$ (cm$^{-2}$) &$ 14.8^{+1.2}_{-0.5} $&$15.2^{+1.3}_{-0.8} $& $15.2^{+1.4}_{-0.8} $ &$15.0^{+1.4}_{-0.6} $&$ 15.2^{+1.4}_{-0.7}$ \\
$b_g$ (km s$^{-1}$) &$ 231^{+49}_{-66} $&$ 204^{+51}_{-42} $ & $204^{+60}_{-47}$ & $165^{+113}_{-71}  $&$ 78^{+48}_{-23}$ \\
$\log N_g$ (cm$^{-2}$) &$ 14.9^{+0.1}_{-0.3} $&$ 15.0^{+0.1}_{-0.4}  $& $14.9^{+0.1}_{-0.5} $&$ 14.6^{+0.1}_{-0.2} $&$ 15.7^{+1.1}_{-0.8}$  \\
\enddata
 \tablecomments{The median, the 84th and 16th percentile deviations of the PDFs of parameters from the median are given in the table.}
\end{deluxetable*}

\clearpage
\bibliography{reference.bib}

\begin{thebibliography}{}
\expandafter\ifx\csname natexlab\endcsname\relax\def\natexlab#1{#1}\fi

\bibitem[{{Aguirre} {et~al.}(2001){Aguirre}, {Hernquist}, {Schaye}, {Weinberg},
  {Katz}, \& {Gardner}}]{Aguirre+01}
{Aguirre}, A., {Hernquist}, L., {Schaye}, J., {et~al.} 2001, \apj, 560, 599

\bibitem[{{Alexander} \& {Hickox}(2012)}]{AlexanderHickox12}
{Alexander}, D.~M., \& {Hickox}, R.~C. 2012, \nar, 56, 93

\bibitem[{{Alexander} {et~al.}(2003){Alexander}, {Bauer}, {Brandt},
  {Schneider}, {Hornschemeier}, {Vignali}, {Barger}, {Broos}, {Cowie},
  {Garmire}, {Townsley}, {Bautz}, {Chartas}, \& {Sargent}}]{Alexander03}
{Alexander}, D.~M., {Bauer}, F.~E., {Brandt}, W.~N., {et~al.} 2003, \aj, 126,
  539

\bibitem[{{Arav} {et~al.}(2013){Arav}, {Borguet}, {Chamberlain}, {Edmonds}, \&
  {Danforth}}]{Arav+13}
{Arav}, N., {Borguet}, B., {Chamberlain}, C., {Edmonds}, D., \& {Danforth}, C.
  2013, \mnras, 436, 3286

\bibitem[{{Balmaverde} {et~al.}(2016){Balmaverde}, {Marconi}, {Brusa},
  {Carniani}, {Cresci}, {Lusso}, {Maiolino}, {Mannucci}, \&
  {Nagao}}]{Balmaverde+16}
{Balmaverde}, B., {Marconi}, A., {Brusa}, M., {et~al.} 2016, \aap, 585, A148

\bibitem[{{Barro} {et~al.}(2011){Barro}, {P{\'e}rez-Gonz{\'a}lez}, {Gallego},
  {Ashby}, {Kajisawa}, {Miyazaki}, {Villar}, {Yamada}, \&
  {Zamorano}}]{Barro+11}
{Barro}, G., {P{\'e}rez-Gonz{\'a}lez}, P.~G., {Gallego}, J., {et~al.} 2011,
  \apjs, 193, 30

\bibitem[{{Benson} {et~al.}(2003){Benson}, {Bower}, {Frenk}, {Lacey}, {Baugh},
  \& {Cole}}]{Benson+03}
{Benson}, A.~J., {Bower}, R.~G., {Frenk}, C.~S., {et~al.} 2003, \apj, 599, 38

\bibitem[{{Bolatto} {et~al.}(2013){Bolatto}, {Wolfire}, \&
  {Leroy}}]{Bolatto+13}
{Bolatto}, A.~D., {Wolfire}, M., \& {Leroy}, A.~K. 2013, \araa, 51, 207

\bibitem[{{Bordoloi} {et~al.}(2014){Bordoloi}, {Lilly}, {Hardmeier}, {Contini},
  {Kneib}, {Le Fevre}, {Mainieri}, {Renzini}, {Scodeggio}, {Zamorani},
  {Bardelli}, {Bolzonella}, {Bongiorno}, {Caputi}, {Carollo}, {Cucciati}, {de
  la Torre}, {de Ravel}, {Garilli}, {Iovino}, {Kampczyk}, {Kova{\v c}},
  {Knobel}, {Lamareille}, {Le Borgne}, {Le Brun}, {Maier}, {Mignoli}, {Oesch},
  {Pello}, {Peng}, {Perez Montero}, {Presotto}, {Silverman}, {Tanaka}, {Tasca},
  {Tresse}, {Vergani}, {Zucca}, {Cappi}, {Cimatti}, {Coppa}, {Franzetti},
  {Koekemoer}, {Moresco}, {Nair}, \& {Pozzetti}}]{Bordoloi+14}
{Bordoloi}, R., {Lilly}, S.~J., {Hardmeier}, E., {et~al.} 2014, \apj, 794, 130

\bibitem[{{Borguet} {et~al.}(2013){Borguet}, {Arav}, {Edmonds}, {Chamberlain},
  \& {Benn}}]{Borguet+13}
{Borguet}, B.~C.~J., {Arav}, N., {Edmonds}, D., {Chamberlain}, C., \& {Benn},
  C. 2013, \apj, 762, 49

\bibitem[{{Borguet} {et~al.}(2012){Borguet}, {Edmonds}, {Arav}, {Dunn}, \&
  {Kriss}}]{Borguet+12}
{Borguet}, B.~C.~J., {Edmonds}, D., {Arav}, N., {Dunn}, J., \& {Kriss}, G.~A.
  2012, \apj, 751, 107

\bibitem[{{Bourne} {et~al.}(2015){Bourne}, {Zubovas}, \&
  {Nayakshin}}]{Bourne+15}
{Bourne}, M.~A., {Zubovas}, K., \& {Nayakshin}, S. 2015, \mnras, 453, 1829

\bibitem[{{Bower} {et~al.}(2012){Bower}, {Benson}, \& {Crain}}]{Bower+12}
{Bower}, R.~G., {Benson}, A.~J., \& {Crain}, R.~A. 2012, \mnras, 422, 2816

\bibitem[{{Bruzual} \& {Charlot}(2003)}]{BC03}
{Bruzual}, G., \& {Charlot}, S. 2003, \mnras, 344, 1000

\bibitem[{{Calzetti} {et~al.}(2000){Calzetti}, {Armus}, {Bohlin}, {Kinney},
  {Koornneef}, \& {Storchi-Bergmann}}]{Calzetti+00}
{Calzetti}, D., {Armus}, L., {Bohlin}, R.~C., {et~al.} 2000, \apj, 533, 682

\bibitem[{{Carniani} {et~al.}(2015){Carniani}, {Marconi}, {Maiolino},
  {Balmaverde}, {Brusa}, {Cano-D{\'{\i}}az}, {Cicone}, {Comastri}, {Cresci},
  {Fiore}, {Feruglio}, {La Franca}, {Mainieri}, {Mannucci}, {Nagao}, {Netzer},
  {Piconcelli}, {Risaliti}, {Schneider}, \& {Shemmer}}]{Carniani+15}
{Carniani}, S., {Marconi}, A., {Maiolino}, R., {et~al.} 2015, \aap, 580, A102

\bibitem[{{Cattaneo} {et~al.}(2007){Cattaneo}, {Blaizot}, {Weinberg}, {Kere{\v
  s}}, {Colombi}, {Dav{\'e}}, {Devriendt}, {Guiderdoni}, \&
  {Katz}}]{Cattaneo+07}
{Cattaneo}, A., {Blaizot}, J., {Weinberg}, D.~H., {et~al.} 2007, \mnras, 377,
  63

\bibitem[{{Chabrier}(2003)}]{Chabrier03}
{Chabrier}, G. 2003, \pasp, 115, 763

\bibitem[{{Chamberlain} \& {Arav}(2015)}]{Chamberlain+15}
{Chamberlain}, C., \& {Arav}, N. 2015, \mnras, 454, 675

\bibitem[{{Choi} {et~al.}(2012){Choi}, {Ostriker}, {Naab}, \&
  {Johansson}}]{Choi+12}
{Choi}, E., {Ostriker}, J.~P., {Naab}, T., \& {Johansson}, P.~H. 2012, \apj,
  754, 125

\bibitem[{{Churchill} {et~al.}(2000){Churchill}, {Mellon}, {Charlton},
  {Jannuzi}, {Kirhakos}, {Steidel}, \& {Schneider}}]{Churchill+00}
{Churchill}, C.~W., {Mellon}, R.~R., {Charlton}, J.~C., {et~al.} 2000, \apj,
  543, 577

\bibitem[{{Cicone} {et~al.}(2014){Cicone}, {Maiolino}, {Sturm},
  {Graci{\'a}-Carpio}, {Feruglio}, {Neri}, {Aalto}, {Davies}, {Fiore},
  {Fischer}, {Garc{\'{\i}}a-Burillo}, {Gonz{\'a}lez-Alfonso},
  {Hailey-Dunsheath}, {Piconcelli}, \& {Veilleux}}]{Cicone+14}
{Cicone}, C., {Maiolino}, R., {Sturm}, E., {et~al.} 2014, \aap, 562, A21

\bibitem[{{Ciesla} {et~al.}(2015){Ciesla}, {Charmandaris}, {Georgakakis},
  {Bernhard}, {Mitchell}, {Buat}, {Elbaz}, {LeFloc'h}, {Lacey}, {Magdis}, \&
  {Xilouris}}]{Ciesla+15}
{Ciesla}, L., {Charmandaris}, V., {Georgakakis}, A., {et~al.} 2015, \aap, 576,
  A10

\bibitem[{{Coil} {et~al.}(2011){Coil}, {Weiner}, {Holz}, {Cooper}, {Yan}, \&
  {Aird}}]{Coil+11}
{Coil}, A.~L., {Weiner}, B.~J., {Holz}, D.~E., {et~al.} 2011, \apj, 743, 46

\bibitem[{{Combes} {et~al.}(2013){Combes}, {Garc{\'{\i}}a-Burillo}, {Casasola},
  {Hunt}, {Krips}, {Baker}, {Boone}, {Eckart}, {Marquez}, {Neri}, {Schinnerer},
  \& {Tacconi}}]{Combes+13}
{Combes}, F., {Garc{\'{\i}}a-Burillo}, S., {Casasola}, V., {et~al.} 2013, \aap,
  558, A124

\bibitem[{{Cresci} {et~al.}(2015){Cresci}, {Mainieri}, {Brusa}, {Marconi},
  {Perna}, {Mannucci}, {Piconcelli}, {Maiolino}, {Feruglio}, {Fiore},
  {Bongiorno}, {Lanzuisi}, {Merloni}, {Schramm}, {Silverman}, \&
  {Civano}}]{Cresci+15}
{Cresci}, G., {Mainieri}, V., {Brusa}, M., {et~al.} 2015, \apj, 799, 82

\bibitem[{{Croton} {et~al.}(2006){Croton}, {Springel}, {White}, {De Lucia},
  {Frenk}, {Gao}, {Jenkins}, {Kauffmann}, {Navarro}, \& {Yoshida}}]{Croton+06}
{Croton}, D.~J., {Springel}, V., {White}, S.~D.~M., {et~al.} 2006, \mnras, 365,
  11

\bibitem[{{Dasyra} {et~al.}(2015){Dasyra}, {Bostrom}, {Combes}, \&
  {Vlahakis}}]{Dasyra+15}
{Dasyra}, K.~M., {Bostrom}, A.~C., {Combes}, F., \& {Vlahakis}, N. 2015, \apj,
  815, 34

\bibitem[{{Dav{\'e}} {et~al.}(2011){Dav{\'e}}, {Finlator}, \&
  {Oppenheimer}}]{Dave+11}
{Dav{\'e}}, R., {Finlator}, K., \& {Oppenheimer}, B.~D. 2011, \mnras, 416, 1354

\bibitem[{{Debuhr} {et~al.}(2012){Debuhr}, {Quataert}, \& {Ma}}]{Debuhr+12}
{Debuhr}, J., {Quataert}, E., \& {Ma}, C.-P. 2012, \mnras, 420, 2221

\bibitem[{{Di Matteo} {et~al.}(2005){Di Matteo}, {Springel}, \&
  {Hernquist}}]{DiMatteo+05}
{Di Matteo}, T., {Springel}, V., \& {Hernquist}, L. 2005, \nat, 433, 604

\bibitem[{{Diamond-Stanic} {et~al.}(2012){Diamond-Stanic}, {Moustakas},
  {Tremonti}, {Coil}, {Hickox}, {Robaina}, {Rudnick}, \&
  {Sell}}]{Diamond-stanic+12}
{Diamond-Stanic}, A.~M., {Moustakas}, J., {Tremonti}, C.~A., {et~al.} 2012,
  \apjl, 755, L26

\bibitem[{{Edmonds} {et~al.}(2011){Edmonds}, {Borguet}, {Arav}, {Dunn},
  {Penton}, {Kriss}, {Korista}, {Costantini}, {Steenbrugge},
  {Gonzalez-Serrano}, {Aoki}, {Bautista}, {Behar}, {Benn}, {Crenshaw},
  {Everett}, {Gabel}, {Kaastra}, {Moe}, \& {Scott}}]{Edmonds+11}
{Edmonds}, D., {Borguet}, B., {Arav}, N., {et~al.} 2011, \apj, 739, 7

\bibitem[{{Erb} {et~al.}(2012){Erb}, {Quider}, {Henry}, \& {Martin}}]{Erb+12}
{Erb}, D.~K., {Quider}, A.~M., {Henry}, A.~L., \& {Martin}, C.~L. 2012, \apj,
  759, 26

\bibitem[{{Faber} {et~al.}(2003){Faber}, {Phillips}, {Kibrick}, {Alcott},
  {Allen}, {Burrous}, {Cantrall}, {Clarke}, {Coil}, {Cowley}, {Davis}, {Deich},
  {Dietsch}, {Gilmore}, {Harper}, {Hilyard}, {Lewis}, {McVeigh}, {Newman},
  {Osborne}, {Schiavon}, {Stover}, {Tucker}, {Wallace}, {Wei}, {Wirth}, \&
  {Wright}}]{Faber+03}
{Faber}, S.~M., {Phillips}, A.~C., {Kibrick}, R.~I., {et~al.} 2003, in
  \procspie, Vol. 4841, Instrument Design and Performance for Optical/Infrared
  Ground-based Telescopes, ed. M.~{Iye} \& A.~F.~M. {Moorwood}, 1657--1669

\bibitem[{{Fabian}(1999)}]{Fabian99}
{Fabian}, A.~C. 1999, \mnras, 308, L39

\bibitem[{{Fabian}(2012)}]{Fabian12}
---. 2012, \araa, 50, 455

\bibitem[{{Feruglio} {et~al.}(2010){Feruglio}, {Maiolino}, {Piconcelli},
  {Menci}, {Aussel}, {Lamastra}, \& {Fiore}}]{Feruglio+10}
{Feruglio}, C., {Maiolino}, R., {Piconcelli}, E., {et~al.} 2010, \aap, 518,
  L155

\bibitem[{{Fischer} {et~al.}(2010){Fischer}, {Sturm}, {Gonz{\'a}lez-Alfonso},
  {Graci{\'a}-Carpio}, {Hailey-Dunsheath}, {Poglitsch}, {Contursi}, {Lutz},
  {Genzel}, {Sternberg}, {Verma}, \& {Tacconi}}]{Fischer+10}
{Fischer}, J., {Sturm}, E., {Gonz{\'a}lez-Alfonso}, E., {et~al.} 2010, \aap,
  518, L41

\bibitem[{{Foreman-Mackey} {et~al.}(2013){Foreman-Mackey}, {Hogg}, {Lang}, \&
  {Goodman}}]{Foreman-Mackey+13}
{Foreman-Mackey}, D., {Hogg}, D.~W., {Lang}, D., \& {Goodman}, J. 2013, \pasp,
  125, 306

\bibitem[{{F{\"o}rster Schreiber} {et~al.}(2014){F{\"o}rster Schreiber},
  {Genzel}, {Newman}, {Kurk}, {Lutz}, {Tacconi}, {Wuyts}, {Bandara}, {Burkert},
  {Buschkamp}, {Carollo}, {Cresci}, {Daddi}, {Davies}, {Eisenhauer}, {Hicks},
  {Lang}, {Lilly}, {Mainieri}, {Mancini}, {Naab}, {Peng}, {Renzini}, {Rosario},
  {Shapiro Griffin}, {Shapley}, {Sternberg}, {Tacchella}, {Vergani},
  {Wisnioski}, {Wuyts}, \& {Zamorani}}]{ForsterSchreiber+14}
{F{\"o}rster Schreiber}, N.~M., {Genzel}, R., {Newman}, S.~F., {et~al.} 2014,
  \apj, 787, 38

\bibitem[{{French} {et~al.}(2015){French}, {Yang}, {Zabludoff}, {Narayanan},
  {Shirley}, {Walter}, {Smith}, \& {Tremonti}}]{French+15}
{French}, K.~D., {Yang}, Y., {Zabludoff}, A., {et~al.} 2015, \apj, 801, 1

\bibitem[{{Gabor} \& {Bournaud}(2014)}]{Gabor+14}
{Gabor}, J.~M., \& {Bournaud}, F. 2014, \mnras, 441, 1615

\bibitem[{{Gabor} {et~al.}(2011){Gabor}, {Dav{\'e}}, {Oppenheimer}, \&
  {Finlator}}]{Gabor+11}
{Gabor}, J.~M., {Dav{\'e}}, R., {Oppenheimer}, B.~D., \& {Finlator}, K. 2011,
  \mnras, 417, 2676

\bibitem[{{Garc{\'{\i}}a-Burillo} {et~al.}(2014){Garc{\'{\i}}a-Burillo},
  {Combes}, {Usero}, {Aalto}, {Krips}, {Viti}, {Alonso-Herrero}, {Hunt},
  {Schinnerer}, {Baker}, {Boone}, {Casasola}, {Colina}, {Costagliola},
  {Eckart}, {Fuente}, {Henkel}, {Labiano}, {Mart{\'{\i}}n}, {M{\'a}rquez},
  {Muller}, {Planesas}, {Ramos Almeida}, {Spaans}, {Tacconi}, \& {van der
  Werf}}]{Garcia-Burillo+14}
{Garc{\'{\i}}a-Burillo}, S., {Combes}, F., {Usero}, A., {et~al.} 2014, \aap,
  567, A125

\bibitem[{{Garc{\'{\i}}a-Burillo} {et~al.}(2015){Garc{\'{\i}}a-Burillo},
  {Combes}, {Usero}, {Aalto}, {Colina}, {Alonso-Herrero}, {Hunt}, {Arribas},
  {Costagliola}, {Labiano}, {Neri}, {Pereira-Santaella}, {Tacconi}, \& {van der
  Werf}}]{Garcia-Burillo+15}
---. 2015, \aap, 580, A35

\bibitem[{{Garc{\'{\i}}a-Rojas} \& {Esteban}(2007)}]{Garcia+07}
{Garc{\'{\i}}a-Rojas}, J., \& {Esteban}, C. 2007, \apj, 670, 457

\bibitem[{{Geach} {et~al.}(2014){Geach}, {Hickox}, {Diamond-Stanic}, {Krips},
  {Rudnick}, {Tremonti}, {Sell}, {Coil}, \& {Moustakas}}]{Geach+14}
{Geach}, J.~E., {Hickox}, R.~C., {Diamond-Stanic}, A.~M., {et~al.} 2014, \nat,
  516, 68

\bibitem[{{Genzel} {et~al.}(2014){Genzel}, {F{\"o}rster Schreiber}, {Rosario},
  {Lang}, {Lutz}, {Wisnioski}, {Wuyts}, {Wuyts}, {Bandara}, {Bender}, {Berta},
  {Kurk}, {Mendel}, {Tacconi}, {Wilman}, {Beifiori}, {Brammer}, {Burkert},
  {Buschkamp}, {Chan}, {Carollo}, {Davies}, {Eisenhauer}, {Fabricius},
  {Fossati}, {Kriek}, {Kulkarni}, {Lilly}, {Mancini}, {Momcheva}, {Naab},
  {Nelson}, {Renzini}, {Saglia}, {Sharples}, {Sternberg}, {Tacchella}, \& {van
  Dokkum}}]{Genzel+14}
{Genzel}, R., {F{\"o}rster Schreiber}, N.~M., {Rosario}, D., {et~al.} 2014,
  \apj, 796, 7

\bibitem[{{Granato} {et~al.}(2004){Granato}, {De Zotti}, {Silva}, {Bressan}, \&
  {Danese}}]{Granato+04}
{Granato}, G.~L., {De Zotti}, G., {Silva}, L., {Bressan}, A., \& {Danese}, L.
  2004, \apj, 600, 580

\bibitem[{{Grogin} {et~al.}(2011){Grogin}, {Kocevski}, {Faber}, {Ferguson},
  {Koekemoer}, {Riess}, {Acquaviva}, {Alexander}, {Almaini}, {Ashby}, {Barden},
  {Bell}, {Bournaud}, {Brown}, {Caputi}, {Casertano}, {Cassata}, {Castellano},
  {Challis}, {Chary}, {Cheung}, {Cirasuolo}, {Conselice}, {Roshan Cooray},
  {Croton}, {Daddi}, {Dahlen}, {Dav{\'e}}, {de Mello}, {Dekel}, {Dickinson},
  {Dolch}, {Donley}, {Dunlop}, {Dutton}, {Elbaz}, {Fazio}, {Filippenko},
  {Finkelstein}, {Fontana}, {Gardner}, {Garnavich}, {Gawiser}, {Giavalisco},
  {Grazian}, {Guo}, {Hathi}, {H{\"a}ussler}, {Hopkins}, {Huang}, {Huang},
  {Jha}, {Kartaltepe}, {Kirshner}, {Koo}, {Lai}, {Lee}, {Li}, {Lotz}, {Lucas},
  {Madau}, {McCarthy}, {McGrath}, {McIntosh}, {McLure}, {Mobasher},
  {Moustakas}, {Mozena}, {Nandra}, {Newman}, {Niemi}, {Noeske}, {Papovich},
  {Pentericci}, {Pope}, {Primack}, {Rajan}, {Ravindranath}, {Reddy}, {Renzini},
  {Rix}, {Robaina}, {Rodney}, {Rosario}, {Rosati}, {Salimbeni}, {Scarlata},
  {Siana}, {Simard}, {Smidt}, {Somerville}, {Spinrad}, {Straughn}, {Strolger},
  {Telford}, {Teplitz}, {Trump}, {van der Wel}, {Villforth}, {Wechsler},
  {Weiner}, {Wiklind}, {Wild}, {Wilson}, {Wuyts}, {Yan}, \& {Yun}}]{Grogin+11}
{Grogin}, N.~A., {Kocevski}, D.~D., {Faber}, S.~M., {et~al.} 2011, \apjs, 197,
  35

\bibitem[{{Guo} {et~al.}(2013){Guo}, {Ferguson}, {Giavalisco}, {Barro},
  {Willner}, {Ashby}, {Dahlen}, {Donley}, {Faber}, {Fontana}, {Galametz},
  {Grazian}, {Huang}, {Kocevski}, {Koekemoer}, {Koo}, {McGrath}, {Peth},
  {Salvato}, {Wuyts}, {Castellano}, {Cooray}, {Dickinson}, {Dunlop}, {Fazio},
  {Gardner}, {Gawiser}, {Grogin}, {Hathi}, {Hsu}, {Lee}, {Lucas}, {Mobasher},
  {Nandra}, {Newman}, \& {van der Wel}}]{Guo+13}
{Guo}, Y., {Ferguson}, H.~C., {Giavalisco}, M., {et~al.} 2013, \apjs, 207, 24

\bibitem[{{Hainline} {et~al.}(2011){Hainline}, {Shapley}, {Greene}, \&
  {Steidel}}]{Hainline+11}
{Hainline}, K.~N., {Shapley}, A.~E., {Greene}, J.~E., \& {Steidel}, C.~C. 2011,
  \apj, 733, 31

\bibitem[{{Hainline} {et~al.}(2012){Hainline}, {Shapley}, {Greene}, {Steidel},
  {Reddy}, \& {Erb}}]{Hainline+12}
{Hainline}, K.~N., {Shapley}, A.~E., {Greene}, J.~E., {et~al.} 2012, \apj, 760,
  74

\bibitem[{{Harrison} {et~al.}(2014){Harrison}, {Alexander}, {Mullaney}, \&
  {Swinbank}}]{Harrison+14}
{Harrison}, C.~M., {Alexander}, D.~M., {Mullaney}, J.~R., \& {Swinbank}, A.~M.
  2014, \mnras, 441, 3306

\bibitem[{{Harrison} {et~al.}(2016){Harrison}, {Alexander}, {Mullaney},
  {Stott}, {Swinbank}, {Arumugam}, {Bauer}, {Bower}, {Bunker}, \&
  {Sharples}}]{Harrison+16}
{Harrison}, C.~M., {Alexander}, D.~M., {Mullaney}, J.~R., {et~al.} 2016,
  \mnras, 456, 1195

\bibitem[{{Heckman} \& {Best}(2014)}]{Heckman+14}
{Heckman}, T.~M., \& {Best}, P.~N. 2014, \araa, 52, 589

\bibitem[{{Holt} {et~al.}(2008){Holt}, {Tadhunter}, \& {Morganti}}]{Holt+08}
{Holt}, J., {Tadhunter}, C.~N., \& {Morganti}, R. 2008, \mnras, 387, 639

\bibitem[{{Hopkins} {et~al.}(2006){Hopkins}, {Hernquist}, {Cox}, {Di Matteo},
  {Robertson}, \& {Springel}}]{Hopkins+06}
{Hopkins}, P.~F., {Hernquist}, L., {Cox}, T.~J., {et~al.} 2006, \apjs, 163, 1

\bibitem[{{Hopkins} {et~al.}(2008){Hopkins}, {Hernquist}, {Cox}, \& {Kere{\v
  s}}}]{Hopkins+08}
{Hopkins}, P.~F., {Hernquist}, L., {Cox}, T.~J., \& {Kere{\v s}}, D. 2008,
  \apjs, 175, 356

\bibitem[{{Hopkins} {et~al.}(2013){Hopkins}, {Kere{\v s}}, {Murray},
  {Hernquist}, {Narayanan}, \& {Hayward}}]{Hopkins+13}
{Hopkins}, P.~F., {Kere{\v s}}, D., {Murray}, N., {et~al.} 2013, \mnras, 433,
  78

\bibitem[{{Husemann} {et~al.}(2015){Husemann}, {Scharw{\"a}chter}, {Bennert},
  {Manieri}, {Woo}, \& {Kakkad}}]{Husemann+15}
{Husemann}, B., {Scharw{\"a}chter}, J., {Bennert}, V.~N., {et~al.} 2015, ArXiv
  e-prints, arXiv:1512.05595

\bibitem[{{Husemann} {et~al.}(2013){Husemann}, {Wisotzki}, {S{\'a}nchez}, \&
  {Jahnke}}]{Husemann+13}
{Husemann}, B., {Wisotzki}, L., {S{\'a}nchez}, S.~F., \& {Jahnke}, K. 2013,
  \aap, 549, A43

\bibitem[{{Ishibashi} \& {Fabian}(2012)}]{Ishibashi+12}
{Ishibashi}, W., \& {Fabian}, A.~C. 2012, \mnras, 427, 2998

\bibitem[{{Jenkins}(2009)}]{Jenkins09}
{Jenkins}, E.~B. 2009, \apj, 700, 1299

\bibitem[{{Juneau} {et~al.}(2013){Juneau}, {Dickinson}, {Bournaud},
  {Alexander}, {Daddi}, {Mullaney}, {Magnelli}, {Kartaltepe}, {Hwang},
  {Willner}, {Coil}, {Rosario}, {Trump}, {Weiner}, {Willmer}, {Cooper},
  {Elbaz}, {Faber}, {Frayer}, {Kocevski}, {Laird}, {Monkiewicz}, {Nandra},
  {Newman}, {Salim}, \& {Symeonidis}}]{Juneau+13}
{Juneau}, S., {Dickinson}, M., {Bournaud}, F., {et~al.} 2013, \apj, 764, 176

\bibitem[{{Karouzos} {et~al.}(2016){Karouzos}, {Woo}, \& {Bae}}]{Karouzos+16}
{Karouzos}, M., {Woo}, J.-H., \& {Bae}, H.-J. 2016, \apj, 819, 148

\bibitem[{{King} \& {Pounds}(2015)}]{KingPounds15}
{King}, A., \& {Pounds}, K. 2015, \araa, 53, 115

\bibitem[{{King} {et~al.}(2011){King}, {Zubovas}, \& {Power}}]{King+11}
{King}, A.~R., {Zubovas}, K., \& {Power}, C. 2011, \mnras, 415, L6

\bibitem[{{Koekemoer} {et~al.}(2011){Koekemoer}, {Faber}, {Ferguson}, {Grogin},
  {Kocevski}, {Koo}, {Lai}, {Lotz}, {Lucas}, {McGrath}, {Ogaz}, {Rajan},
  {Riess}, {Rodney}, {Strolger}, {Casertano}, {Castellano}, {Dahlen},
  {Dickinson}, {Dolch}, {Fontana}, {Giavalisco}, {Grazian}, {Guo}, {Hathi},
  {Huang}, {van der Wel}, {Yan}, {Acquaviva}, {Alexander}, {Almaini}, {Ashby},
  {Barden}, {Bell}, {Bournaud}, {Brown}, {Caputi}, {Cassata}, {Challis},
  {Chary}, {Cheung}, {Cirasuolo}, {Conselice}, {Roshan Cooray}, {Croton},
  {Daddi}, {Dav{\'e}}, {de Mello}, {de Ravel}, {Dekel}, {Donley}, {Dunlop},
  {Dutton}, {Elbaz}, {Fazio}, {Filippenko}, {Finkelstein}, {Frazer}, {Gardner},
  {Garnavich}, {Gawiser}, {Gruetzbauch}, {Hartley}, {H{\"a}ussler},
  {Herrington}, {Hopkins}, {Huang}, {Jha}, {Johnson}, {Kartaltepe},
  {Khostovan}, {Kirshner}, {Lani}, {Lee}, {Li}, {Madau}, {McCarthy},
  {McIntosh}, {McLure}, {McPartland}, {Mobasher}, {Moreira}, {Mortlock},
  {Moustakas}, {Mozena}, {Nandra}, {Newman}, {Nielsen}, {Niemi}, {Noeske},
  {Papovich}, {Pentericci}, {Pope}, {Primack}, {Ravindranath}, {Reddy},
  {Renzini}, {Rix}, {Robaina}, {Rosario}, {Rosati}, {Salimbeni}, {Scarlata},
  {Siana}, {Simard}, {Smidt}, {Snyder}, {Somerville}, {Spinrad}, {Straughn},
  {Telford}, {Teplitz}, {Trump}, {Vargas}, {Villforth}, {Wagner}, {Wandro},
  {Wechsler}, {Weiner}, {Wiklind}, {Wild}, {Wilson}, {Wuyts}, \&
  {Yun}}]{Koekemoer+11}
{Koekemoer}, A.~M., {Faber}, S.~M., {Ferguson}, H.~C., {et~al.} 2011, \apjs,
  197, 36

\bibitem[{{Kormendy} \& {Ho}(2013)}]{Kormendy+13}
{Kormendy}, J., \& {Ho}, L.~C. 2013, \araa, 51, 511

\bibitem[{{Kornei} {et~al.}(2012){Kornei}, {Shapley}, {Martin}, {Coil}, {Lotz},
  {Schiminovich}, {Bundy}, \& {Noeske}}]{Kornei+12}
{Kornei}, K.~A., {Shapley}, A.~E., {Martin}, C.~L., {et~al.} 2012, \apj, 758,
  135

\bibitem[{{Kriek} {et~al.}(2009){Kriek}, {van Dokkum}, {Labb{\'e}}, {Franx},
  {Illingworth}, {Marchesini}, \& {Quadri}}]{Kriek+09}
{Kriek}, M., {van Dokkum}, P.~G., {Labb{\'e}}, I., {et~al.} 2009, \apj, 700,
  221

\bibitem[{{Krug} {et~al.}(2010){Krug}, {Rupke}, \& {Veilleux}}]{Krug+10}
{Krug}, H.~B., {Rupke}, D.~S.~N., \& {Veilleux}, S. 2010, \apj, 708, 1145

\bibitem[{{Ku} {et~al.}(1980){Ku}, {Helfand}, \& {Lucy}}]{Ku+80}
{Ku}, W.~H.-M., {Helfand}, D.~J., \& {Lucy}, L.~B. 1980, \nat, 288, 323

\bibitem[{{Laird} {et~al.}(2009){Laird}, {Nandra}, {Georgakakis}, {Aird},
  {Barmby}, {Conselice}, {Coil}, {Davis}, {Faber}, {Fazio}, {Guhathakurta},
  {Koo}, {Sarajedini}, \& {Willmer}}]{Laird+09}
{Laird}, E.~S., {Nandra}, K., {Georgakakis}, A., {et~al.} 2009, \apjs, 180, 102

\bibitem[{{Law} {et~al.}(2012){Law}, {Steidel}, {Shapley}, {Nagy}, {Reddy}, \&
  {Erb}}]{Law+12}
{Law}, D.~R., {Steidel}, C.~C., {Shapley}, A.~E., {et~al.} 2012, \apj, 759, 29

\bibitem[{{Liu} {et~al.}(2014){Liu}, {Zakamska}, \& {Greene}}]{Liu+14}
{Liu}, G., {Zakamska}, N.~L., \& {Greene}, J.~E. 2014, \mnras, 442, 1303

\bibitem[{{Liu} {et~al.}(2013){Liu}, {Zakamska}, {Greene}, {Nesvadba}, \&
  {Liu}}]{Liu+13}
{Liu}, G., {Zakamska}, N.~L., {Greene}, J.~E., {Nesvadba}, N.~P.~H., \& {Liu},
  X. 2013, \mnras, 436, 2576

\bibitem[{{Martin} \& {Bouch{\'e}}(2009)}]{Martin+09}
{Martin}, C.~L., \& {Bouch{\'e}}, N. 2009, \apj, 703, 1394

\bibitem[{{Martin} {et~al.}(2012){Martin}, {Shapley}, {Coil}, {Kornei},
  {Bundy}, {Weiner}, {Noeske}, \& {Schiminovich}}]{Martin+12}
{Martin}, C.~L., {Shapley}, A.~E., {Coil}, A.~L., {et~al.} 2012, \apj, 760, 127

\bibitem[{{McElroy} {et~al.}(2015){McElroy}, {Croom}, {Pracy}, {Sharp}, {Ho},
  \& {Medling}}]{McElroy+15}
{McElroy}, R., {Croom}, S.~M., {Pracy}, M., {et~al.} 2015, \mnras, 446, 2186

\bibitem[{{Mineo} {et~al.}(2014){Mineo}, {Gilfanov}, {Lehmer}, {Morrison}, \&
  {Sunyaev}}]{Mineo+14}
{Mineo}, S., {Gilfanov}, M., {Lehmer}, B.~D., {Morrison}, G.~E., \& {Sunyaev},
  R. 2014, \mnras, 437, 1698

\bibitem[{{Morganti} {et~al.}(2005){Morganti}, {Tadhunter}, \&
  {Oosterloo}}]{Morganti+05}
{Morganti}, R., {Tadhunter}, C.~N., \& {Oosterloo}, T.~A. 2005, \aap, 444, L9

\bibitem[{{Morton}(2003)}]{Morton03}
{Morton}, D.~C. 2003, \apjs, 149, 205

\bibitem[{{Mullaney} {et~al.}(2013){Mullaney}, {Alexander}, {Fine}, {Goulding},
  {Harrison}, \& {Hickox}}]{Mullaney+13}
{Mullaney}, J.~R., {Alexander}, D.~M., {Fine}, S., {et~al.} 2013, \mnras, 433,
  622

\bibitem[{{M{\"u}ller-S{\'a}nchez} {et~al.}(2011){M{\"u}ller-S{\'a}nchez},
  {Prieto}, {Hicks}, {Vives-Arias}, {Davies}, {Malkan}, {Tacconi}, \&
  {Genzel}}]{Muller-Sanchez+11}
{M{\"u}ller-S{\'a}nchez}, F., {Prieto}, M.~A., {Hicks}, E.~K.~S., {et~al.}
  2011, \apj, 739, 69

\bibitem[{{Muratov} {et~al.}(2015){Muratov}, {Kere{\v s}},
  {Faucher-Gigu{\`e}re}, {Hopkins}, {Quataert}, \& {Murray}}]{Muratov+15}
{Muratov}, A.~L., {Kere{\v s}}, D., {Faucher-Gigu{\`e}re}, C.-A., {et~al.}
  2015, \mnras, 454, 2691

\bibitem[{{Nesvadba} {et~al.}(2008){Nesvadba}, {Lehnert}, {De Breuck},
  {Gilbert}, \& {van Breugel}}]{Nesvadba+08}
{Nesvadba}, N.~P.~H., {Lehnert}, M.~D., {De Breuck}, C., {Gilbert}, A.~M., \&
  {van Breugel}, W. 2008, \aap, 491, 407

\bibitem[{{Nesvadba} {et~al.}(2006){Nesvadba}, {Lehnert}, {Eisenhauer},
  {Gilbert}, {Tecza}, \& {Abuter}}]{Nesvadba+06}
{Nesvadba}, N.~P.~H., {Lehnert}, M.~D., {Eisenhauer}, F., {et~al.} 2006, \apj,
  650, 693

\bibitem[{{Newman} {et~al.}(2013){Newman}, {Cooper}, {Davis}, {Faber}, {Coil},
  {Guhathakurta}, {Koo}, {Phillips}, {Conroy}, {Dutton}, {Finkbeiner}, {Gerke},
  {Rosario}, {Weiner}, {Willmer}, {Yan}, {Harker}, {Kassin}, {Konidaris},
  {Lai}, {Madgwick}, {Noeske}, {Wirth}, {Connolly}, {Kaiser}, {Kirby},
  {Lemaux}, {Lin}, {Lotz}, {Luppino}, {Marinoni}, {Matthews}, {Metevier}, \&
  {Schiavon}}]{Newman+13}
{Newman}, J.~A., {Cooper}, M.~C., {Davis}, M., {et~al.} 2013, \apjs, 208, 5

\bibitem[{{Oke} {et~al.}(1995){Oke}, {Cohen}, {Carr}, {Cromer}, {Dingizian},
  {Harris}, {Labrecque}, {Lucinio}, {Schaal}, {Epps}, \& {Miller}}]{Oke+95}
{Oke}, J.~B., {Cohen}, J.~G., {Carr}, M., {et~al.} 1995, \pasp, 107, 375

\bibitem[{{Oppenheimer} {et~al.}(2010){Oppenheimer}, {Dav{\'e}}, {Kere{\v s}},
  {Fardal}, {Katz}, {Kollmeier}, \& {Weinberg}}]{Oppenheimer+10}
{Oppenheimer}, B.~D., {Dav{\'e}}, R., {Kere{\v s}}, D., {et~al.} 2010, \mnras,
  406, 2325

\bibitem[{{Peng} {et~al.}(2002){Peng}, {Ho}, {Impey}, \& {Rix}}]{Peng+02}
{Peng}, C.~Y., {Ho}, L.~C., {Impey}, C.~D., \& {Rix}, H.-W. 2002, \aj, 124, 266

\bibitem[{{Piconcelli} {et~al.}(2005){Piconcelli}, {Jimenez-Bail{\'o}n},
  {Guainazzi}, {Schartel}, {Rodr{\'{\i}}guez-Pascual}, \&
  {Santos-Lle{\'o}}}]{Piconcelli+05}
{Piconcelli}, E., {Jimenez-Bail{\'o}n}, E., {Guainazzi}, M., {et~al.} 2005,
  \aap, 432, 15

\bibitem[{{Pontzen} \& {Governato}(2012)}]{PontzenGovernato12}
{Pontzen}, A., \& {Governato}, F. 2012, \mnras, 421, 3464

\bibitem[{{Prochaska} {et~al.}(2011){Prochaska}, {Kasen}, \&
  {Rubin}}]{Prochaska+11}
{Prochaska}, J.~X., {Kasen}, D., \& {Rubin}, K. 2011, \apj, 734, 24

\bibitem[{{Prochaska} {et~al.}(2014){Prochaska}, {Lau}, \&
  {Hennawi}}]{Prochaska+14}
{Prochaska}, J.~X., {Lau}, M.~W., \& {Hennawi}, J.~F. 2014, \apj, 796, 140

\bibitem[{{Rodr{\'{\i}}guez} \& {Rubin}(2005)}]{Rodriguez+05}
{Rodr{\'{\i}}guez}, M., \& {Rubin}, R.~H. 2005, \apj, 626, 900

\bibitem[{{Roos} {et~al.}(2015){Roos}, {Juneau}, {Bournaud}, \&
  {Gabor}}]{Roos+15}
{Roos}, O., {Juneau}, S., {Bournaud}, F., \& {Gabor}, J.~M. 2015, \apj, 800, 19

\bibitem[{{Rowlands} {et~al.}(2015){Rowlands}, {Wild}, {Nesvadba}, {Sibthorpe},
  {Mortier}, {Lehnert}, \& {da Cunha}}]{Rowlands+15}
{Rowlands}, K., {Wild}, V., {Nesvadba}, N., {et~al.} 2015, \mnras, 448, 258

\bibitem[{{Rubin} {et~al.}(2014){Rubin}, {Prochaska}, {Koo}, {Phillips},
  {Martin}, \& {Winstrom}}]{Rubin+14}
{Rubin}, K.~H.~R., {Prochaska}, J.~X., {Koo}, D.~C., {et~al.} 2014, \apj, 794,
  156

\bibitem[{{Rupke} {et~al.}(2005{\natexlab{a}}){Rupke}, {Veilleux}, \&
  {Sanders}}]{Rupke+05}
{Rupke}, D.~S., {Veilleux}, S., \& {Sanders}, D.~B. 2005{\natexlab{a}}, \apj,
  632, 751

\bibitem[{{Rupke} {et~al.}(2005{\natexlab{b}}){Rupke}, {Veilleux}, \&
  {Sanders}}]{Rupke+05a}
---. 2005{\natexlab{b}}, \apjs, 160, 87

\bibitem[{{Rupke} {et~al.}(2005{\natexlab{c}}){Rupke}, {Veilleux}, \&
  {Sanders}}]{Rupke+05b}
---. 2005{\natexlab{c}}, \apjs, 160, 115

\bibitem[{{Rupke} \& {Veilleux}(2013)}]{Rupke+13}
{Rupke}, D.~S.~N., \& {Veilleux}, S. 2013, \apj, 768, 75

\bibitem[{{Sarzi} {et~al.}(2016){Sarzi}, {Kaviraj}, {Nedelchev}, {Tiffany},
  {Shabala}, {Deller}, \& {Middelberg}}]{Sarzi+16}
{Sarzi}, M., {Kaviraj}, S., {Nedelchev}, B., {et~al.} 2016, \mnras, 456, L25

\bibitem[{{Sato} {et~al.}(2009){Sato}, {Martin}, {Noeske}, {Koo}, \&
  {Lotz}}]{Sato+09}
{Sato}, T., {Martin}, C.~L., {Noeske}, K.~G., {Koo}, D.~C., \& {Lotz}, J.~M.
  2009, \apj, 696, 214

\bibitem[{{Scannapieco} \& {Oh}(2004)}]{Scannapieco+04}
{Scannapieco}, E., \& {Oh}, S.~P. 2004, \apj, 608, 62

\bibitem[{{Sell} {et~al.}(2014){Sell}, {Tremonti}, {Hickox}, {Diamond-Stanic},
  {Moustakas}, {Coil}, {Williams}, {Rudnick}, {Robaina}, {Geach}, {Heinz}, \&
  {Wilcots}}]{Sell+14}
{Sell}, P.~H., {Tremonti}, C.~A., {Hickox}, R.~C., {et~al.} 2014, \mnras, 441,
  3417

\bibitem[{{Shih} {et~al.}(2013){Shih}, {Stockton}, \& {Kewley}}]{Shih+13}
{Shih}, H.-Y., {Stockton}, A., \& {Kewley}, L. 2013, \apj, 772, 138

\bibitem[{{Silk} \& {Nusser}(2010)}]{Silk+10}
{Silk}, J., \& {Nusser}, A. 2010, \apj, 725, 556

\bibitem[{{Silk} \& {Rees}(1998)}]{Silk+98}
{Silk}, J., \& {Rees}, M.~J. 1998, \aap, 331, L1

\bibitem[{{Somerville} {et~al.}(2008){Somerville}, {Hopkins}, {Cox},
  {Robertson}, \& {Hernquist}}]{Somerville+08}
{Somerville}, R.~S., {Hopkins}, P.~F., {Cox}, T.~J., {Robertson}, B.~E., \&
  {Hernquist}, L. 2008, \mnras, 391, 481

\bibitem[{{Spoon} {et~al.}(2013){Spoon}, {Farrah}, {Lebouteiller},
  {Gonz{\'a}lez-Alfonso}, {Bernard-Salas}, {Urrutia}, {Rigopoulou},
  {Westmoquette}, {Smith}, {Afonso}, {Pearson}, {Cormier}, {Efstathiou},
  {Borys}, {Verma}, {Etxaluze}, \& {Clements}}]{Spoon+13}
{Spoon}, H.~W.~W., {Farrah}, D., {Lebouteiller}, V., {et~al.} 2013, \apj, 775,
  127

\bibitem[{{Springel} {et~al.}(2005){Springel}, {Di Matteo}, \&
  {Hernquist}}]{Springel+05}
{Springel}, V., {Di Matteo}, T., \& {Hernquist}, L. 2005, \mnras, 361, 776

\bibitem[{{Sturm} {et~al.}(2011){Sturm}, {Gonz{\'a}lez-Alfonso}, {Veilleux},
  {Fischer}, {Graci{\'a}-Carpio}, {Hailey-Dunsheath}, {Contursi}, {Poglitsch},
  {Sternberg}, {Davies}, {Genzel}, {Lutz}, {Tacconi}, {Verma}, {Maiolino}, \&
  {de Jong}}]{Sturm+11}
{Sturm}, E., {Gonz{\'a}lez-Alfonso}, E., {Veilleux}, S., {et~al.} 2011, \apjl,
  733, L16

\bibitem[{{Sun} {et~al.}(2014){Sun}, {Greene}, {Zakamska}, \&
  {Nesvadba}}]{Sun+14}
{Sun}, A.-L., {Greene}, J.~E., {Zakamska}, N.~L., \& {Nesvadba}, N.~P.~H. 2014,
  \apj, 790, 160

\bibitem[{{Tang} {et~al.}(2014){Tang}, {Giavalisco}, {Guo}, \&
  {Kurk}}]{Tang+14}
{Tang}, Y., {Giavalisco}, M., {Guo}, Y., \& {Kurk}, J. 2014, \apj, 793, 92

\bibitem[{{Thacker} {et~al.}(2006){Thacker}, {Scannapieco}, \&
  {Couchman}}]{Thacker+06}
{Thacker}, R.~J., {Scannapieco}, E., \& {Couchman}, H.~M.~P. 2006, \apj, 653,
  86

\bibitem[{{Tombesi} {et~al.}(2015){Tombesi}, {Mel{\'e}ndez}, {Veilleux},
  {Reeves}, {Gonz{\'a}lez-Alfonso}, \& {Reynolds}}]{Tombesi+15}
{Tombesi}, F., {Mel{\'e}ndez}, M., {Veilleux}, S., {et~al.} 2015, \nat, 519,
  436

\bibitem[{{Tremonti} {et~al.}(2007){Tremonti}, {Moustakas}, \&
  {Diamond-Stanic}}]{Tremonti+07}
{Tremonti}, C.~A., {Moustakas}, J., \& {Diamond-Stanic}, A.~M. 2007, \apjl,
  663, L77

\bibitem[{{Vasudevan} \& {Fabian}(2007)}]{Vasudevan+07}
{Vasudevan}, R.~V., \& {Fabian}, A.~C. 2007, \mnras, 381, 1235

\bibitem[{{Veilleux} {et~al.}(2005){Veilleux}, {Cecil}, \&
  {Bland-Hawthorn}}]{Veilleux+05}
{Veilleux}, S., {Cecil}, G., \& {Bland-Hawthorn}, J. 2005, \araa, 43, 769

\bibitem[{{Veilleux} {et~al.}(2013){Veilleux}, {Mel{\'e}ndez}, {Sturm},
  {Gracia-Carpio}, {Fischer}, {Gonz{\'a}lez-Alfonso}, {Contursi}, {Lutz},
  {Poglitsch}, {Davies}, {Genzel}, {Tacconi}, {de Jong}, {Sternberg}, {Netzer},
  {Hailey-Dunsheath}, {Verma}, {Rupke}, {Maiolino}, {Teng}, \&
  {Polisensky}}]{Veilleux+13}
{Veilleux}, S., {Mel{\'e}ndez}, M., {Sturm}, E., {et~al.} 2013, \apj, 776, 27

\bibitem[{{Villar-Mart{\'{\i}}n} {et~al.}(2016){Villar-Mart{\'{\i}}n},
  {Arribas}, {Emonts}, {Humphrey}, {Tadhunter}, {Bessiere}, {Cabrera Lavers},
  \& {Ramos Almeida}}]{Villar-Martin+16}
{Villar-Mart{\'{\i}}n}, M., {Arribas}, S., {Emonts}, B., {et~al.} 2016, \mnras,
  460, 130

\bibitem[{{Villar Mart{\'{\i}}n} {et~al.}(2014){Villar Mart{\'{\i}}n},
  {Emonts}, {Humphrey}, {Cabrera Lavers}, \& {Binette}}]{VillarMartin+14}
{Villar Mart{\'{\i}}n}, M., {Emonts}, B., {Humphrey}, A., {Cabrera Lavers}, A.,
  \& {Binette}, L. 2014, \mnras, 440, 3202

\bibitem[{{Weiner} {et~al.}(2009){Weiner}, {Coil}, {Prochaska}, {Newman},
  {Cooper}, {Bundy}, {Conselice}, {Dutton}, {Faber}, {Koo}, {Lotz}, {Rieke}, \&
  {Rubin}}]{Weiner+09}
{Weiner}, B.~J., {Coil}, A.~L., {Prochaska}, J.~X., {et~al.} 2009, \apj, 692,
  187

\bibitem[{{Whitaker} {et~al.}(2014){Whitaker}, {Franx}, {Leja}, {van Dokkum},
  {Henry}, {Skelton}, {Fumagalli}, {Momcheva}, {Brammer}, {Labb{\'e}},
  {Nelson}, \& {Rigby}}]{Whitaker+14}
{Whitaker}, K.~E., {Franx}, M., {Leja}, J., {et~al.} 2014, \apj, 795, 104

\bibitem[{{Wild} {et~al.}(2010){Wild}, {Heckman}, \& {Charlot}}]{Wild+10}
{Wild}, V., {Heckman}, T., \& {Charlot}, S. 2010, \mnras, 405, 933

\bibitem[{{Wuyts} {et~al.}(2011){Wuyts}, {F{\"o}rster Schreiber}, {Lutz},
  {Nordon}, {Berta}, {Altieri}, {Andreani}, {Aussel}, {Bongiovanni}, {Cepa},
  {Cimatti}, {Daddi}, {Elbaz}, {Genzel}, {Koekemoer}, {Magnelli}, {Maiolino},
  {McGrath}, {P{\'e}rez Garc{\'{\i}}a}, {Poglitsch}, {Popesso}, {Pozzi},
  {Sanchez-Portal}, {Sturm}, {Tacconi}, \& {Valtchanov}}]{Wuyts+11}
{Wuyts}, S., {F{\"o}rster Schreiber}, N.~M., {Lutz}, D., {et~al.} 2011, \apj,
  738, 106

\bibitem[{{Xue} {et~al.}(2011){Xue}, {Luo}, {Brandt}, {Bauer}, {Lehmer},
  {Broos}, {Schneider}, {Alexander}, {Brusa}, {Comastri}, {Fabian}, {Gilli},
  {Hasinger}, {Hornschemeier}, {Koekemoer}, {Liu}, {Mainieri}, {Paolillo},
  {Rafferty}, {Rosati}, {Shemmer}, {Silverman}, {Smail}, {Tozzi}, \&
  {Vignali}}]{Xue+11}
{Xue}, Y.~Q., {Luo}, B., {Brandt}, W.~N., {et~al.} 2011, \apjs, 195, 10

\bibitem[{{Yesuf} {et~al.}(2014){Yesuf}, {Faber}, {Trump}, {Koo}, {Fang},
  {Liu}, {Wild}, \& {Hayward}}]{Yesuf+14}
{Yesuf}, H.~M., {Faber}, S.~M., {Trump}, J.~R., {et~al.} 2014, \apj, 792, 84

\bibitem[{{Zahid} {et~al.}(2011){Zahid}, {Kewley}, \& {Bresolin}}]{Zahid+11}
{Zahid}, H.~J., {Kewley}, L.~J., \& {Bresolin}, F. 2011, \apj, 730, 137

\bibitem[{{Zakamska} \& {Greene}(2014)}]{Zakamska+14}
{Zakamska}, N.~L., \& {Greene}, J.~E. 2014, \mnras, 442, 784

\bibitem[{{Zakamska} {et~al.}(2016){Zakamska}, {Hamann}, {P{\^a}ris}, {Brandt},
  {Greene}, {Strauss}, {Villforth}, {Wylezalek}, {Alexandroff}, \&
  {Ross}}]{Zakamska+16}
{Zakamska}, N.~L., {Hamann}, F., {P{\^a}ris}, I., {et~al.} 2016, \mnras, 459,
  3144

\bibitem[{{Zhu} {et~al.}(2015){Zhu}, {Comparat}, {Kneib}, {Delubac},
  {Raichoor}, {Dawson}, {Newman}, {Y{\`e}che}, {Zhou}, \& {Schneider}}]{Zhu+15}
{Zhu}, G.~B., {Comparat}, J., {Kneib}, J.-P., {et~al.} 2015, \apj, 815, 48

\bibitem[{{Zubovas} \& {King}(2012)}]{Zubovas+12}
{Zubovas}, K., \& {King}, A. 2012, \apjl, 745, L34

\end{thebibliography}
\end{document}